\documentclass[longauth]{aa}
\usepackage{natbib}
\usepackage[varg]{txfonts}
\usepackage{graphicx}
\DeclareMathOperator{\sech}{sech}
\usepackage{booktabs}
\usepackage{multirow}
\usepackage{tabularx}
\usepackage{lscape}
\usepackage{longtable}
\usepackage{hyperref}
\usepackage{placeins}
\usepackage{longtable}

\newcolumntype{L}[1]{>{\raggedright\arraybackslash}p{#1}}
\newcolumntype{C}[1]{>{\centering\arraybackslash}p{#1}}
\newcolumntype{R}[1]{>{\raggedleft\arraybackslash}p{#1}}

\usepackage{color} 
\usepackage{amssymb,amsmath, graphicx}
\usepackage{tikz}
\newcommand{\sqdiamond}[1][fill=black]{\tikz [x=1.2ex,y=1.85ex,line width=.1ex,line join=round, yshift=-0.285ex] \draw  [#1]  (0,.5) -- (.5,1) -- (1,.5) -- (.5,0) -- (0,.5) -- cycle;}%
\newcommand{\MyDiamond}[1][fill=black]{\mathop{\raisebox{-0.275ex}{$\sqdiamond[#1]$}}}

\def\Msun{\ifmmode{\mathrm M_\odot}\else{$M_\odot$}\fi}

\defcitealias{GB+05}{GB05}
\defcitealias{Bolatto+13}{B13}
\defcitealias{Herrera-Endoqui+15}{HE15}

\defcitealias{Williams+21}{W21}

\defcitealias{Egusa+09}{E09}
\defcitealias{Sheth+02}{SH02}

\defcitealias{GB94}{GB94}
\defcitealias{Lindblad+Kristen1996}{LK96}
\defcitealias{Rautiainen2008}{R08}
\defcitealias{Treuthardt+08}{T08}
\defcitealias{LindbladLindbladAthanassoula1996}{LLA96}
\defcitealias{Tamburro+08}{T08}
\defcitealias{Wada+98}{W98}
\defcitealias{GB+98}{GB98}
\defcitealias{Sierra+15}{S15}
\defcitealias{Jogee+02}{J02}
\defcitealias{ScaranoLepine13}{SL13}
\defcitealias{Korchagin+05}{K05}
\defcitealias{ElmegreenElmegreenSeiden1989}{EES89}
\defcitealias{VandeVenFathi2010}{VF10}
\defcitealias{Chemin+03}{C03}
\defcitealias{JorsaterVanMoorsel1995}{JM95}
\defcitealias{Ryder+96}{R96}
\defcitealias{Buta2017}{B17}
\defcitealias{ColinaWada2000}{CW00}
\defcitealias{Hirota+09}{H09}
\defcitealias{Schmidt+2019}{S19}
\defcitealias{Buta+01}{B01}
\defcitealias{Teuben+86}{T86}
\defcitealias{Schinnerer+02}{S02}
\defcitealias{Salak+19}{S19}
\defcitealias{Comte83}{C83}
\defcitealias{Oey+03}{O03}
\defcitealias{Buta1986}{B86}
\defcitealias{Ondrechen+89_NGC1097}{O89-NGC1097}
\defcitealias{Ondrechen+89_NGC1365}{O89-NGC1365}
\defcitealias{SakhibovSmirnov89}{SS89}
\defcitealias{VilaCostasEdmunds92}{VCE92}
\defcitealias{Pinol-Ferrer+14}{PF14}
\defcitealias{Font+14}{F14}
\defcitealias{ButaZhang09}{BZ09}
\defcitealias{Elmegreen+92}{E92}
\defcitealias{MartinezGarciaPuerari2014}{MGP14}
\defcitealias{Sempere+95}{S95}
\defcitealias{CanzianAllen97}{CA97}
\defcitealias{GB+09}{GB09}
\defcitealias{Canzian1993}{C93}
\defcitealias{Elmegreen+96}{E96}
\defcitealias{Haan+09}{H09}
\defcitealias{Vera-Vilalmizar+01}{VM01}
\defcitealias{Devereux+92}{D92}
\defcitealias{Elmegreen+09}{E09}
\defcitealias{Hernandez+05}{H05}
\defcitealias{Rand+04}{RW04}
\defcitealias{Combes14}{C14}
\defcitealias{ZhangButa07}{ZB07}
\defcitealias{England1989}{E89}
\defcitealias{Casasola+11}{C11}

\begin{document}

\title{Dynamical resonances in PHANGS galaxies}

% AFFILIATIONS (add or correct if necessary):
% --------------------------------------------

\newcommand{\OSU}{\label{OSU} Department of Astronomy, The Ohio State University, 140 West 18th Avenue, Columbus, Ohio 43210, USA}

\newcommand{\Alberta}{\label{Alberta} Department of Physics, University of Alberta, Edmonton, AB T6G 2E1, Canada}

\newcommand{\ANU}{\label{ANU} Research School of Astronomy and Astrophysics, Australian National University, Canberra, ACT 2611, Australia}

\newcommand{\IPAC}{\label{IPAC} Caltech-IPAC, 1200 E. California Blvd. Pasadena, CA 91125, USA}

\newcommand{\Carnegie}{\label{Carnegi} Observatories of the Carnegie Institution for Science, 813 Santa Barbara Street, Pasadena, CA 91101, USA}

\newcommand{\CCAPP}{\label{CCAPP} Center for Cosmology and Astroparticle Physics, 191 West Woodruff Avenue, Columbus, OH 43210, USA}

\newcommand{\CfA}{\label{CfA} Harvard-Smithsonian Center for Astrophysics, 60 Garden Street, Cambridge, MA 02138, USA}

\newcommand{\CITEVA}{\label{CITEVA} Centro de Astronomía (CITEVA), Universidad de Antofagasta, Avenida Angamos 601, Antofagasta, Chile}

\newcommand{\CNRS}{\label{CNRS} CNRS, IRAP, 9 Av. du Colonel Roche, BP 44346, F-31028 Toulouse cedex 4, France}

\newcommand{\ESO}{\label{ESO} European Southern Observatory, Karl-Schwarzschild Stra{\ss}e 2, D-85748 Garching bei M\"{u}nchen, Germany}

\newcommand{\Heidelberg}{\label{Heidelberg} Astronomisches Rechen-Institut, Zentrum f\"{u}r Astronomie der Universit\"{a}t Heidelberg, M\"{o}nchhofstra\ss e 12-14, D-69120 Heidelberg, Germany}

\newcommand{\COOL}{\label{COOL} Cosmic Origins Of Life (COOL) Research DAO, coolresearch.io}

\newcommand{\ICRAR}{\label{ICRAR} International Centre for Radio Astronomy Research, University of Western Australia, 35 Stirling Highway, Crawley, WA 6009, Australia}

\newcommand{\IRAM}{\label{IRAM} Institut de Radioastronomie Millim\'{e}trique (IRAM), 300 Rue de la Piscine, F-38406 Saint Martin d'H\`{e}res, France}

\newcommand{\ITA}{\label{ITA} Universit\"{a}t Heidelberg, Zentrum f\"{u}r Astronomie, Institut f\"{u}r Theoretische Astrophysik, Albert-Ueberle-Str 2, D-69120 Heidelberg, Germany}

\newcommand{\IWR}{\label{IWR} Universit\"{a}t Heidelberg, Interdisziplin\"{a}res Zentrum f\"{u}r Wissenschaftliches Rechnen, Im Neuenheimer Feld 205, D-69120 Heidelberg, Germany}

\newcommand{\JHU}{\label{JHU} Department of Physics and Astronomy, The Johns Hopkins University, Baltimore, MD 21218, USA}

\newcommand{\Leiden}{\label{Leiden} Leiden Observatory, Leiden University, P.O. Box 9513, 2300 RA Leiden, The Netherlands}

\newcommand{\Maryland}{\label{Maryland} Department of Astronomy, University of Maryland, College Park, MD 20742, USA}

\newcommand{\MPE}{\label{MPE} Max-Planck-Institut f\"{u}r extraterrestrische Physik, Giessenbachstra{\ss}e 1, D-85748 Garching, Germany}

\newcommand{\MPIA}{\label{MPIA} Max-Planck-Institut f\"{u}r Astronomie, K\"{o}nigstuhl 17, D-69117, Heidelberg, Germany}

\newcommand{\Nagoya}{\label{Nagoya} Department of Physics, Nagoya University, Furo-cho, Chikusa-ku, Nagoya, Aichi 464-8602, Japan}

\newcommand{\NRAO}{\label{NRAO} National Radio Astronomy Observatory, 520 Edgemont Road, Charlottesville, VA 22903-2475, USA}

\newcommand{\OAN}{\label{OAN} Observatorio Astron\'{o}mico Nacional (IGN), C/Alfonso XII, 3, E-28014 Madrid, Spain}

\newcommand{\ObsParis}{\label{ObsParis} Sorbonne Universit\'{e}, Observatoire de Paris, Universit\'{e} PSL, CNRS, LERMA, F-75014, Paris, France}

\newcommand{\Princeton}{\label{Princeton} Department of Astrophysical Sciences, Princeton University, 4 Ivy Ln., Princeton, NJ 08544 USA}

\newcommand{\UToledo}{\label{UToledo} University of Toledo, 2801 W. Bancroft St., Mail Stop 111, Toledo, OH, 43606}

\newcommand{\Toulouse}{\label{Toulouse} Universit\'{e} de Toulouse, UPS-OMP, IRAP, F-31028 Toulouse cedex 4, France}

\newcommand{\UBonn}{\label{UBonn} Argelander-Institut f\"ur Astronomie, Universit\"at Bonn, Auf dem H\"ugel 71, 53121 Bonn, Germany}

\newcommand{\UChile}{\label{UChile} Departamento de Astronom\'{i}a, Universidad de Chile, Camino del Observatorio 1515, Las Condes, Santiago, Chile}

\newcommand{\UConn}{\label{UConn} Department of Physics, University of Connecticut, Storrs, CT, 06269, USA}

\newcommand{\UCSD}{\label{UCSD} Center for Astrophysics and Space Sciences, Department of Physics,  University of California, San Diego, 9500 Gilman Drive, La Jolla, CA 92093, USA}

\newcommand{\UCSDAA}{\label{UCSDAA} Department of Astronomy \& Astrophysics,  University of California, San Diego, 9500 Gilman Drive, La Jolla, CA 92093, USA}

\newcommand{\UGent}{\label{UGent} Sterrenkundig Observatorium, Universiteit Gent, Krijgslaan 281 S9, B-9000 Gent, Belgium}

\newcommand{\ULyon}{\label{ULyon} Univ Lyon, Univ Lyon 1, ENS de Lyon, CNRS, Centre de Recherche Astrophysique de Lyon UMR5574,\\ F-69230 Saint-Genis-Laval, France}

\newcommand{\UMass}{\label{UMass} University of Massachusetts—Amherst, 710 N. Pleasant Street, Amherst, MA 01003, USA}

\newcommand{\UWyoming}{\label{UWyoming} Department of Physics and Astronomy, University of Wyoming, Laramie, WY 82071, USA}

\newcommand{\LAM}{\label{LAM} Aix Marseille Univ, CNRS, CNES, LAM (Laboratoire d’Astrophysique de Marseille), Marseille, France}

\newcommand{\UHawaii}{\label{UHawaii} Institute for Astronomy, University of Hawaii, 2680 Woodlawn Drive, Honolulu, HI 96822, USA}

\newcommand{\UCM}{\label{UCM} Departamento de F\'{\i}sica de la Tierra y Astrof\'{\i}sica, Universidad Complutense de Madrid, E-28040, Spain}

\newcommand{\IPARC}{\label{IPARC} Instituto de F\'{\i}sica de Part\'{\i}culas y del Cosmos IPARCOS, Facultad de Ciencias F\'{\i}sicas, Universidad Complutense de Madrid, E-28040, Spain}

\newcommand{\STScI}{\label{STScI} Space Telescope Science Institute, 3700 San Martin Drive, Baltimore, MD 21218, USA}

\newcommand{\McMaster}{\label{McMaster} Department of Physics and Astronomy, McMaster University, 1280 Main Street West, Hamilton, ON L8S 4M1, Canada}

\newcommand{\INAF}{\label{INAF} INAF -- Osservatorio Astrofisico di Arcetri, Largo E. Fermi 5, I-50157, Firenze, Italy}

\newcommand{\Sydney}{\label{Sydney} Sydney Institute for Astronomy, School of Physics A28, The University of Sydney, NSW 2006, Australia}

\newcommand{\UA}{\label{UA} Centro de Astronomía (CITEVA), Universidad de Antofagasta, Avenida Angamos 601, Antofagasta, Chile}

\newcommand{\CITA}{\label{CITA} Canadian Institute for Theoretical Astrophysics (CITA), University of Toronto, 60 St George St, Toronto, ON M5S 3H8, Canada}

\newcommand{\ASIAA}{\label{ASIAA} Institute of Astronomy and Astrophysics, Academia Sinica, No. 1, Sec. 4, Roosevelt Road, Taipei 10617, Taiwan}

\newcommand{\TKU}{\label{TKU} Department of Physics, Tamkang University, No.151, Yingzhuan Rd., Tamsui Dist., New Taipei City 251301, Taiwan}

\newcommand{\PSMA}{\label{PSMA} Penn State Mont Alto, 1 Campus Drive, Mont Alto, PA  17237, USA}

\newcommand{\ILL}{\label{ILL} Institut Laue-Langevin, 71 avenue des Martyrs, F-38042 Grenoble, France}

\newcommand{\TUM}{\label{TUM} Technical University of Munich, School of Engineering and Design, Department of Aerospace and Geodesy, Chair of Remote Sensing Technology, Arcisstr. 21, 80333 Munich, Germany}

\newcommand{\Surrey}{\label{Surrey} Department of Physics, University of Surrey, Guildford GU2 7XH, UK}

\newcommand{\Oxford}{\label{Oxford} Sub-department of Astrophysics, Department of Physics, University of Oxford, Keble Road, Oxford OX1 3RH, UK}

\newcommand{\AIP}{\label{AIP} Leibniz-Institut for Astrophysik Potsdam (AIP), An der Sternwarte 16, 14482 Potsdam, Germany}

\newcommand{\Insubria}{\label{Insubria}{Universit{\`a} dell’Insubria, via Valleggio 11, 22100 Como, Italy}}

\newcommand{\StAndrews}{\label{StAndrews} School of Physics and Astronomy, University of St Andrews, North Haugh, St Andrews, KY16 9SS}

\newcommand{\IAC}{\label{IAC}{Instituto de Astrof\'isica de Canarias, C/ V\'ia L\'actea s/n, E-38205, La Laguna, Spain}}

\newcommand{\ULL}{\label{ULL}{Departamento de Astrof\'isica, Universidad de La Laguna, Av. del Astrof\'isico Francisco S\'anchez s/n, E-38206, La Laguna, Spain}}

\newcommand{\AIfA}{\label{AIfA}{
Argelander-Institut für Astronomie, Universität Bonn, Auf dem Hügel 71, 53121 Bonn, Germany}}

\newcommand{\Heidel}{\label{Heidel} Universit\"{a}t Heidelberg, Zentrum f\"{u}r Astronomie, Albert-Ueberle-Str. 2, 69120 Heidelberg, Germany}

\newcommand{\UMich}{\label{UMich} 
Department of Astronomy, University of Michigan, Ann Arbor, MI 48109, USA}

% =========================================================

% PAPER TEAM:
% ------------
\author{%
Marina Ruiz-García\inst{\ref{OAN}} %
\and Miguel~Querejeta\inst{\ref{OAN}}              % 0000-0002-0472-1011
\and Santiago García-Burillo\inst{\ref{OAN}} %
\and Eric~Emsellem\inst{\ref{ESO},\ref{ULyon}}
\and Sharon~E.~Meidt\inst{\ref{UGent}}        % 0000-0002-6118-4048
\and Mattia~C.~Sormani\inst{\ref{Insubria}}         %
\and Eva~Schinnerer\inst{\ref{MPIA}} 
\and Thomas~G.~Williams\inst{\ref{Oxford}}
% -----------------------------------
% PLEASE ADD YOUR NAME AND AFFILIATION HERE (same format as above)
% PHANGS team in alphabetical order:
% -----------------------------------
\and Zein~Bazzi\inst{\ref{UBonn}}
\and Dario~Colombo\inst{\ref{UBonn}}
\and Damian~R.~Gleis\inst{\ref{MPIA}}
\and Oleg~Y.~Gnedin\inst{\ref{UMich}}
\and Ralf~S.~Klessen\inst{\ref{Heidel}}
\and Adam~K.~Leroy\inst{\ref{OSU}}
\and Patricia~Sánchez-Blázquez\inst{\ref{UCM}}
%\and Kathryn~Grasha\inst{\ref{ANU}}
\and Sophia~K.~Stuber\inst{\ref{MPIA}} % 0000-0002-9333-387X
}

\institute{\OAN{} \and \ESO{} \and \ULyon{} \and \UGent{} \and \Insubria{} \and \MPIA{} 
\and \Oxford{} \and \UBonn{} \and \UMich{} \and \Heidel%\and \ANU{}
\and \OSU{} \and \UCM{} 
}
% \and \INAF{} \and \ITA{} \and \IWR{} \and \Princeton{} \and \Surrey{} \and \ObsParis{} \and \MPE{} \and \COOL{} \and \UBonn{} \and \UWyoming \and \ANU{} \and \ICRAR{} \and \CfA{} \and \TKU{} \and \AIP{} \and \IRAM{} \and \IAC{} \and \ULL{} \and \Alberta{} \and \and \StAndrews{} \and \STScI{} \and \UChile{} \and \OSU{}}

\date{Received ..... / Accepted .....}

\abstract {
Bars are remarkable stellar structures that can transport gas toward centers and drive the secular evolution of galaxies. In this context, it is important to locate dynamical resonances associated with bars. For this study, we used \textit{Spitzer} near-infrared images as a proxy for the stellar gravitational potential and the ALMA CO(J=2-1) gas distribution from the PHANGS survey to determine the position of the main dynamical resonances associated with the bars in the PHANGS sample of 74 nearby star-forming galaxies. We used the gravitational torque method to estimate the location of the bar corotation radius ($R_{\rm CR}$), where stars and gas rotate at the same angular velocity as the bar. Of the 46 barred galaxies in PHANGS, we have successfully determined the corotation (CR) for 38 of them. The mean ratio of the $R_{\rm CR}$ to the bar radius ($R_{\rm bar}$) is $\mathcal{R} = R_{\rm CR}/R_{\rm bar} = 1.12$, with a standard deviation of $0.39$. This is consistent with the average value expected from theory and suggests that bars are predominantly fast. We also compared our results with other bar CR measurements from the literature, which employ different methods, and find good agreement ($\rho = 0.64$). Finally, using rotation curves, we have estimated other relevant resonances such as the inner Lindblad resonance (ILR) and the outer Lindblad resonance (OLR), which are often associated with rings. This work provides a useful catalog of resonances for a large sample of nearby galaxies and emphasizes the clear connection between bar dynamics and morphology.}


\titlerunning{Dynamical resonances in PHANGS galaxies}
\authorrunning{M.~Ruiz-García et al.}

\maketitle 
\section{Introduction} 
\label{Sec:introduction}

A longstanding challenge in modern astrophysics is understanding and characterizing how galactic structures can drive galaxy evolution through the redistribution of gas. It is widely known that gas is drawn inward by asymmetries in the gravitational potential, such as those induced by bars \citep{Roberts+79, Simkin+80, vanAlbadaRoberts81, HasanNorman90, PfennigerNorman90, FriedliBenz93, Sakamoto+99, 1999MNRAS.304..475M, Athanassoula2000, 2006LNP...693..143J,  Haan+09}. Other non-axisymmetries in the gravitational potential can also result in gas inflow including the following: nuclear warps \citep{2000ApJ...533..850S}, nuclear spirals \citep{Combes+14}, secondary bars within large scale bars \citep{1989Natur.338...45S, Friedli&Martinet93, Maciejewski&Sparke00, Heller+01, Shlosman&Heller02, Englmaier&Shlosman04}, and $m=1$ perturbations \citep{Shu+90, Junqueira&Combes96, GB+00}. Other nongravitational mechanisms have also been advocated for to explain the radial redistribution of the gas, such as viscous torques and dynamical friction between large molecular clouds and stars \citep{ShakuraSunyaev73, Pringle1981, BalbusHawley98, Combes02, Combes04, GB+05}.

%%%%%%%%%%%%%%%%%%%%%%%%%%%%%%%%%%%%%%%%%%%%

Dynamical resonances, such as the corotation (CR) or the Lindblad resonances, represent important locations in spiral galaxies, where the angular frequency $\Omega$ of a particle rotating with the disk, such as a star, is related to the angular frequency of the rotating pattern (e.g., bar or spiral pattern) $\Omega_p$, as follows: 

\begin{equation}
    \Omega = \Omega_p + \frac{l}{m}\kappa,
    \label{eq:dynamical_resonances_equation}
\end{equation}

\noindent where $\kappa$ is the epicyclic frequency, and $m$ and $l$ are integer numbers\footnote{$m=2$ for a bar, and in that case, $l=1$ in the case of the outer Lindblad resonance (OLR), $l=-1$ in the case of the inner Lindblad resonance (ILR), and $l=0$ for the CR.} (\citealt{1987gady.book.....B}). 
At the CR, the stars and the rotating density pattern rotate at the same angular velocity, $\Omega_p$, known as the pattern speed, that is, $\Omega (\rm CR)= \Omega_p$. In this work we specifically focus on the bar.

The ratio between the corotation radius $R_{\rm CR}$ and the bar length $R_{\rm bar}$ is referred to as $\mathcal{R}$, and is typically used to differentiate between fast and slow bars. As the pattern speed $\Omega =v/R$ typically decreases with radius, for a fixed $R_{\rm bar}$, the distance-independent ratio $\mathcal{R}$ increases if the bar rotates slower (a lower $\Omega_{\rm bar}$, implying that CR is further out). {\cite{DebattistaSellwood2000} proposed a classification where fast bars are the ones that appear in galaxies with $1<\mathcal{R}<1.4$, while slow bars are the ones with $\mathcal{R}>1.4$. Later, \cite{ButaZhang09} and \cite{Aguerri+15} added ``ultra-fast bars'' ($\mathcal{R}<1$) to this classification.} This is the nomenclature we subsequently follow in the paper. We note that ultra-fast bars are, in most cases, compatible with fast bars within error bars in the literature (see references in Table \ref{table:Appendix-Table_Literature}).

The existence of fast and slow bars has cosmological implications for the braking of bars over time (\citealt{DebattistaSellwood2000}), as the dynamical friction with dark matter can produce a ``slowdown'' of the bars, increasing the $R_{\rm CR}$ (\citealt{TremaineWeinberg1984, Chiba2023}). {Even though one could think the amount of dark matter in the inner parts of galaxies can be inferred from $\mathcal{R}$, this determination is far from straightforward (see \citealt{Athanassoula2013}).} According to the simulation work of \cite{Athanassoula1992}, the range $\mathcal{R} \sim 1.2\pm 0.2$ is favored. This is also supported by other simulations (e.g., \citealt{DebattistaSellwood2000}) and by observational results that show {that most bars have} $1<\mathcal{R}<1.4$ ({e.g., \citealt{Corsini2011, Guo+19, Cuomo+21} or \citealt{Garma+2022}; all of them studied} a sample focused on late-type barred galaxies, using the Tremaine-Weinberg method, \citealt{Tremaine&Weinberg1984}, and relied on rotation curves to infer $R_{\rm CR}$ from $\Omega_p$). It is important to note, however, that recent studies such as \cite{Font+17} do not agree on the interpretation that $1<\mathcal{R}<1.4$ implies that bars necessarily rotate faster than bars with a larger $\mathcal{R}$, as explained in Sect.~\ref{Sec:discussion}.  

The $R_{\rm CR}$ (and other resonances) have been estimated through many different methods. The most common ones include (1) the Tremaine-Weinberg method (hereafter TW method or TW), which {is a kinematic method that considers measurements of the velocity field along a set of long slits and assumes that the tracer used satisfies the equation of continuity (\citealt{Tremaine&Weinberg1984})}; (2) the comparison with hydrodynamical simulations {tailored to reproduce the observed features in the galaxy, where the pattern speed can be obtained directly from the simulation} (e.g., {\citealt{Weiner+01, Perez+2004, Rautiainen+05, Salo+07, Zanmar-Sanchez+2008, Lin+13, 2015MNRAS.454.1818S, Fragkoudi+2017, Feng+23}}); (3) calculating offsets between tracers of gas and star formation along spiral arms{, as such offsets vary radially as a result of the local difference between $\Omega$ and $\Omega_p$} (\citealt{Egusa+09, Sierra+15}); (4) studying the streaming motion caused by a density wave{, as the residual velocity field changes shape from inside to outside CR} (e.g., \citealt{Canzian1993}); or (5) studying gravitational torques (\citealt{GB+05}) {as we explain next and in more detail in Sect.~\ref{Sec:gravitational_torques}}. The main objective of this paper is to apply this last method and compare it with the results obtained with other methods in the literature.

In this work, we focus on gravitational torques as a mechanism to inform us about the expected redistribution of the angular momentum of the gas in a galaxy (\citealt{GB+05}). In order to quantify the effect of these gravitational torques on molecular gas, we need to combine stellar mass maps and molecular gas observations. Torques are mathematically defined as the cross product $\boldsymbol{\tau} = \mathbf{r} \times \mathbf{F}$, where $\mathbf{r}$ is the relative position of the particle with respect to the center, and $\mathbf{F}$ is the gravitational force exerted on the particles. Torques can also be defined as the rate of gained or lost angular momentum, $\boldsymbol{\tau} = \text{d}\mathbf{L}/\text{dt}$, where $\mathbf{L}$ is the angular momentum, and $t$ is time. It is important to note that, for the torque to be nonzero, the force must have a non-radial component. In the case of gravitational torques, they are exerted (on the gas) due to the tangential forces that the non-axisymmetric potential creates. The efficiency with which the angular momentum of the gas is drained by the gravitational torques depends, in the first instance, on the strengths of the non-axisymmetric perturbations of the potential ($m > 0$, as in Eq.~\ref{eq:dynamical_resonances_equation}), but also on the existence of phase shifts between gas and stellar distributions (\citealt{Quillen+95, GB+05, Haan+09, Casasola+11, Meidt+13, Combes+14, Querejeta+16}){, because if there is no phase shift, the net effect of the torques on the gas will be zero.} The accurate estimation of these phase shifts requires images at high spatial resolution ($\sim1''$, corresponding to $\sim100 \ \rm pc$ in the PHANGS sample of galaxies), tracing the distribution of the stars and the gas, as otherwise any offsets would get diluted.
We estimate the $R_{\rm CR}$ as the position in the disk showing a well-defined change of sign (from negative to positive) of the azimuthally averaged torques weighted by the gas surface density (i.e., CO intensity), represented by $\tau(R)$. This is because, inside CR, we expect gas to lose angular momentum ($\boldsymbol{\tau} = \text{d}\mathbf{L}/\text{dt}<0$), while the opposite is expected outside CR ($\boldsymbol{\tau} = \text{d}\mathbf{L}/\text{dt}>0$), resulting in a zero crossing at CR, where phase shifts vanish .%(see Fig.~\ref{fig:NGC1097-case_study} for a visual example).

Other significant resonances are, for instance, the inner Lindblad resonances (ILR) and the outer Lindblad resonances (OLR). The ILR has been traditionally assumed to be related to central rings in the galaxy (traditionally also called nuclear rings, \citealt{Buta1986-rings}); meanwhile, an OLR can be related to an outer ring (\citealt{ButaCombes96}). As explained in \cite{ButaCombes96}, at these resonances gas accumulates due to angular momentum transfer. For instance, on average we find negative torques, $\tau(R)<0$, from the CR toward the ILR, implying radial gas inflow, while from the center toward the ILR we usually find net positive torques, $\tau(R)>0$, which implies radial gas outflow, leading to gas accumulation at the Lindblad resonance. An analogous process happens at the OLR. Torques are on average positive, $\tau(R)>0$, from the CR toward the OLR, implying radial gas outflow, while we expect net negative torques, $\tau(R)<0$, beyond the OLR, which imply radial gas inflow. Again, this torque pattern leads to an accumulation of gas at the OLR. More recently, in \cite{Sormani+23}, they propose that central rings do not necessarily coincide with ILR, but with the inner edge of a gap in the gas disk, formed around it. Either way, there is an intimate connection between dynamical resonances, stellar morphology and gas distribution. In fact, rings are visible mainly, though not exclusively, in the young stars, formed out of the gas reservoirs built up in the rings themselves (most particularly in the central rings associated with an ILR) (see \citealt{Comeron+14}).

In the present work, we study a sample of 74 galaxies, of which 46 are barred galaxies, taken from the PHANGS-ALMA\footnote{Physics at High Angular resolution in Nearby GalaxieS; \url{http://www.phangs.org}} survey (see \citealt{2021ApJS..255...19L}). In this paper, we focus our efforts on those 46 barred galaxies. %(see Table \ref{table:complete_sample})
We use \mbox{CO(2--1)} maps and Near-Infrared (NIR) stellar mass maps from the PHANGS--ALMA and S$^4$G survey\footnote{\textit{Spitzer} Survey of Stellar Structure in Galaxies, \url{https://irsa.ipac.caltech.edu/data/SPITZER/S4G/overview.html}} respectively. 

We describe the data sample used in this paper in Sect.~\ref{Sec:data}. A description of our method (\citealt{GB+05}) can be found in Sect.~\ref{Sec:gravitational_torques}. We present the results obtained using the gravitational torque method in Sect.~\ref{Sec:results} and compare the torque-based results with different methods used in the literature. Finally, we discuss our findings in Sect.~\ref{Sec:discussion}. A summary of the results and conclusions is given in Sect.~\ref{Sec:summary}.

\section{Data} 
\label{Sec:data}

\subsection{Sample}
\label{Sec:sample}

The PHANGS--ALMA survey includes massive star-forming galaxies observed at a resolution of $\lesssim 1''$ at various wavelengths. The sample was selected based on: small distances ($D \approx 17$\,Mpc), low inclinations ($i<75^{\circ}$), relatively high masses ($\log_{10}{\rm M_{\star}[M_{\odot}]}\gtrsim 9.75$), and active star formation (SFR$/M_{\star}>10^{-11} \ \rm yr^{-1}$). Readers can refer to \cite{2021ApJS..257...43L} for a more detailed explanation of the selection strategy. Because of this selection strategy, the main PHANGS--ALMA sample of 74 galaxies is not entirely a survey of spiral galaxies (Sa-Sd), but also contains, for example, four lenticular galaxies (S0), namely NGC\,1546, NGC\,2775, NGC\,4694A, and NGC\,5128. 

From the parent sample of 74 galaxies, 46 are barred (\citealt{Querejeta+21}, Table A.2). Out of these barred galaxies, $39\%$ are strongly barred (SB) and $61\%$ are weakly barred (SAB\footnote{Barred galaxies of type SA$\underline{\rm B}$ have been classified as SAB.}). This bar classification is based on infrared data from \cite{Buta+15}. NGC\,4941 is classified as non-barred (SA) in \cite{Buta+15}, however, we classify it as SAB and consider it in this study, following \cite{Menendez-Delmestre+07, Querejeta+21} and \cite{Stuber+23} (the latter classification is based on the clear bar lanes visible in the CO data).

The corresponding galaxy parameters such as distances, centers, and orientations are adopted from PHANGS sample table version v1.6 (see \citealt{2021ApJS..257...43L}). The inclination ($i$) and position angle (PA) are obtained from the CO kinematic analysis in \cite{Lang+20}, and the distances are listed in \cite{2021MNRAS.501.3621A}. 

\subsection{PHANGS--ALMA data (gas maps)} 
\label{Sec:PHANGS--ALMA}

For the CO maps, we use the zeroth-order moment maps (integrated intensity) based on the public "strict" masks presented in \citet{2021ApJS..257...43L}, as published in PHANGS--ALMA version~4.0. These maps include single dish measurements to correct for missing short spacings, a correction which is expected to have an impact on weighted torque profiles, as a larger amount of diffuse molecular gas is recovered when including these kinds of measurements (\citealt{GB+09, Haan+09, vanderLaan+11}). 

These CO maps were calculated first considering only positions in the cubes where emission exceeds a signal-to-noise ratio of four over two consecutive velocity channels. Then, these positions were extended by including any surrounding areas where the S/N is greater than two over two consecutive velocity channels in terms of signal-to-noise ratio. This results in a set of strict masks, which are also referred as ``high confidence'' maps because of the great likelihood of primarily true emission. 

The CO(2-1) transition has been used in PHANGS to trace the cold molecular gas H$_2$. %Since H$_2$ has no permanent dipole moment, there are no dipolar rotational transitions. 
Hence, it is essential to establish a ``CO-to-H$_2$'' conversion factor $\alpha_{\rm CO}$ as follows:

\begin{equation}
    \Sigma_{\rm mol} = \alpha_ {\rm CO}I_{\rm CO}\cos{(i)}.
\end{equation}

\noindent Since $I_{\rm CO}$ refers to the CO(1-0) transition and has units of [K km s$^{-1}$], multiplication by $\alpha_{\rm CO}= 4.3\ \rm M_{\odot} (\rm K \ km \ s^{-1} pc^2)^{-1}$ results in units of $[\rm M_{\odot}/\rm pc^2]$. We note that this value is the Milky Way-like $\alpha_{\rm CO}$. Furthermore, $\Sigma_{\rm mol}$ is the total mass surface density of molecular gas, which leads us to consider $\alpha_{\rm CO}$ to be analogous to a mass-to-light ratio, as stated in \cite{Bolatto+13}.

Whenever the CO molecule is used as a tracer of H$_2$, it is necessary to take into account which transition is considered. In the present work, observations were carried out through the study of the CO(2-1) instead of CO(1-0). We assumed a constant line ratio $R_{21}=(2-1)/(1-0)\sim 0.65$ for PHANGS, see \cite{2021ApJS..257...43L}. 

\subsection{NIR images (stellar mass maps)}
\label{Sec:stellar_maps}

Our stellar mass maps rely on NIR imaging, which traces old stars and is not affected by dust extinction as much as the optical. We use a constant mass-to-light ratio of $M/L=0.6$ at $3.6 \ \mu\rm m$, based on \cite{Meidt+14}, who argues that this constant $M/L$ can be applied to $3.6 \ \mu \rm m$ maps corrected for dust emission with an uncertainty of $\sim$0.1 dex. \cite{Querejeta+2024} explicitly compares these stellar mass maps corrected with Independent Component Analysis (ICA, see \citealt{Meidt+12}) with fully independent estimates derived via stellar population fitting to PHANGS-MUSE spectra (\citealt{Pessa+21, Pessa+22}). They find very good agreement in terms of stellar mass surface density, $\Sigma_{\star}$, arm/interarm contrasts. Furthermore, any significant $M/L$ variations are expected to be mostly radial, which should not have a significant effect on the torque profiles calculated here.

Most stellar mass maps are obtained from the S$^4$G survey (\citealt{Sheth+2010, Querejeta+15}). The S$^4$G survey has a resolution of $\sim 1.7''$ and is limited in size, volume, and magnitude ($D_{25}>1'$, d$<$40 Mpc, $|b|>30^{\circ}$ and $m_{\rm Bcorr}<15.5$). Images from this survey used in the present work were obtained using the 3.6$\mu$m band of IRAC\footnote{Infrared Array Camera.}. However, some galaxies (marked with an asterisk in Table \ref{table:complete_sample}) were not observed by this specific survey, but they werw obtained from {other archival \textit{Spitzer} observations, as described in \cite{Querejeta+21}. These maps were downloaded from IRSA\footnote{Infrared Science Archive} and include galaxies from the SINGS\footnote{\textit{Spitzer} Infrared Nearby Galaxies Survey} survey (\citealt{Kennicutt2003}), as well as other individual observations.}

The ICA technique, first described in \cite{Meidt+12}, is a method of blind source separation which maximizes the statistical independence of the sources. Essentially, this method can distinguish the old stellar population from any additional emission at $3.6 \ \mu \rm m$ (mostly ``diffuse dust''). In some cases, the separation can be improved by running ICA again and generating a more corrected mass map known as ICA2. Hereafter, IRAC will denote the mass map before the ICA correction (i.e., IRAC is the original NIR image). Further information can be found in \cite{Querejeta+15}. 

There are some galaxies in our sample for which it is not plausible to apply the ICA method under optimal conditions. Some of the galaxies have $[3.6]-[4.5]$ global colors, contained in the range of expected colors $-0.2 < [3.6]-[4.5] < 0$ for an old stellar population. This means, these galaxies are compatible with an old stellar population, so ICA is not applied to this kind of galaxies (as it can lead to artifacts). Therefore, we adopt the map recommended in \cite{Querejeta+15} when available and the original (IRAC) map otherwise.

\section{Gravitational torques}
\label{Sec:gravitational_torques}

\begin{figure*}
    \centering
    \includegraphics[width=1\textwidth]{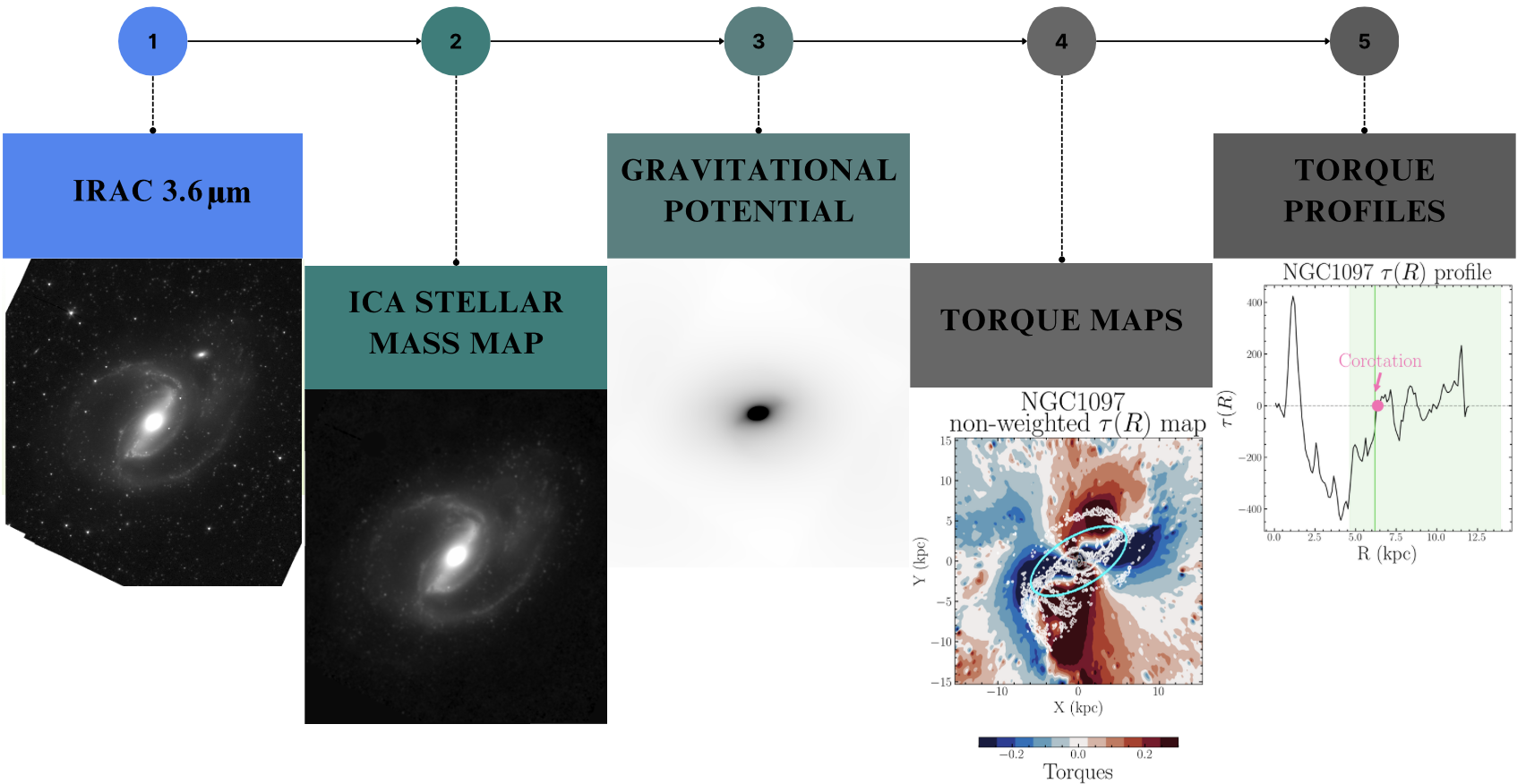}
    \caption{Workflow showing the different steps involved in this study. From left to right, first, we start from the NIR image from \textit{Spitzer}. Next, we apply the ICA technique to this image, in order to derive the gravitational potential in the plane of the galaxy (third step). In the fourth step, we compute forces to create a non-weighted torque map. In that image we represent, for clarity, the bar (as a cyan-contoured ellipse) and white contours corresponding to $[5\sigma, 15\sigma, 45\sigma, ..., 0.9\sigma_{\rm max}]$ in the CO map, with $\sigma$ the mean value of the gas map and $\sigma_{\rm max}$ its maximum value. Finally, in the fifth step, by weighting the torques by the gas and azimuthally averaging the result, we obtain a torque profile, where we estimate the CR location (pink dot) inside the $0.75R_{\rm bar}-2.25R_{\rm bar}$ region (green shaded region).}
    \label{fig:explanatory_workflow}
\end{figure*}

This section summarizes the method from \cite{GB+05} used to study gravitational torques exerted by the stellar potential on the gas distribution and estimate the location of CR. For clarity, the most important steps of this process are presented in a flow chart, in Fig.~\ref{fig:explanatory_workflow}.

Prior to obtaining the gravitational potential, our images are deprojected according to the PA and $i$ from Table \ref{table:complete_sample}, using the code {\tt pydisc}\footnote{\url{https://github.com/emsellem/pydisc}} implemented in Python. Readers can refer to Sect.~\ref{sec:Method_limitations} to see caveats related to the deprojection process. 

\subsection{Gravitational potential}
\label{sec:Gravitational_potential}

Following \citet[][Section 2.8]{1987gady.book.....B}, it is possible to obtain the gravitational potential $\Phi (\mathbf{r})$ as a convolution of the three-dimensional mass density $\rho(\mathbf{r'})$ and the function $1/r$:

\begin{equation}
    \Phi (\mathbf{r}) = -G \int \frac{\rho (\mathbf{r'})d^3\mathbf{r'}}{|\mathbf{r}-\mathbf{r'}|}.
\end{equation}

However, in order to take into account the true three-dimensional mass density, it is necessary to consider the true thickness of the galactic disk. {This implicitly assumes that the disk is baryon-dominated, as expected for the inner regions of high surface brightness spiral galaxies (e.g., \citealt{Kranz+03}), where we have CO observations. However, this can still introduce a bias, which we discuss in Sect.~\ref{sec:Method_limitations}.} To take this thickness into account, we adopt a vertical profile of constant scale-height (see \citealt{ 1989ApJ...337..163W, 1992AJ....103...41B}):

\begin{equation}
    \Phi (x,y,z=0) = -G \int\Sigma (x', y') g(x-x', y-y')dx'dy',
\end{equation}

\noindent where $g(x,y,z=0)$ is the convolution function:

\begin{equation}
    g(x,y,z=0) = \int_{-\infty}^{\infty} \frac{\rho_z(z)dz}{\sqrt{x^2+y^2+z^2}},
\end{equation}

\noindent and, according to \cite{1989ApJ...337..163W} and \cite{1992AJ....103...41B}, we assume our galaxies have a vertical distribution similar to that of an isothermal disk: 

\begin{equation}
    \rho_z (z) = \rho_0 \sech^2{(z/h)},
\end{equation} 

\noindent with $h$ proportional to the disk scale-length $H_{\rm disk}$ as follows:

\begin{equation}
    h\sim 1/12 H_{\rm disk},
\end{equation}

\noindent where $H_{\rm disk}$ is obtained from photometric decompositions of NIR images (\citealt{2015ApJS..219....4S, Querejeta+21}). The convolution for the potential is implemented using a Fourier transform approach (see \citealt{GB+05} or \citealt{Querejeta+16} for more details).

\subsection{Torques}

Once the gravitational potential is obtained, the forces can be calculated as the gradient of the gravitational potential in Cartesian coordinates $F_{x,y}(x,y)= -\nabla_{x,y}\Phi(x,y)$. Then, the torques per unit gas mass, in units of $\rm km^{2}s^{-2}$, associated with the gravitational potential are obtained as follows: 

\begin{equation}
    {t(x,y)} = xF_y - y F_x  \ .
    \label{eq:tau_xy}
\end{equation}

The instantaneous rate of change (gain or loss) of angular momentum experienced by the gas over time, this is, $\tau = \text{d}\mathbf{L} / \text{d}t$, is obtained by multiplying the torque by the total amount of molecular gas in each position. 

\subsection{Gas flows}
\label{sec:gas_flows}

The next step consists of using the torque field to derive the angular momentum variations as a function of radius. We assume that the measured CO gas column density, $N(x, y)$, is representative of the locations where gas spends more time as it orbits around the galaxy, this is, we assume steady state. In this statistical approach, we implicitly average over all possible orbits of gaseous particles and take into account the time spent by the gas clouds along the orbit paths. Thus, we compute the azimuthally averaged torques weighted by the present-day gas distribution as in Eq.~\ref{eq:azimuthal_average}. This is done by an integration of pixels in an annulus defined as $[R+\delta R]$, where $\delta R = 1.5 \ \rm arcsec$.

\begin{equation}
    \tau (R) = \frac{\int_{\theta} [N(x,y)\cdot (xF_y - y F_x)]}{\int_{\theta} N(x,y)}.
    \label{eq:azimuthal_average}
\end{equation}

We adopt the convention that negative torques imply gas inflow and positive torques correspond to gas outflow. Thus, the sense of circulation of the gas affects the sign of $\tau(R)$. Assuming a dextro-rotatory $[x,y,z]$ reference frame, we need to reverse the sign of $\tau(R)$ if the galaxy rotates clockwise, while the sign remains unchanged if the galaxy rotates counter-clockwise. Since all spirals in PHANGS are trailing (see \citealt{Lang+20}), we assign the corresponding signs based on the orientation of spiral arms (as listed in Table \ref{table:complete_sample}).

The net effect of torques is expected to result in a %secular
gas inflow inside CR and outflow immediately outside of it. However, this picture can become more complex if there are additional density wave modes in the disk that compete with the bar potential, which can affect the torque profile around the CR (e.g., \citealt{GB+09, Meidt+09}) and also favor gas inflow inside the ILR region, like the presence of trailing spirals (e.g., \citealt{Haan+09,Combes+14, Audibert+21}).

We identify the CR of the bar at the location where $\tau(R)$ changes its sign from negative ($R<R_{\rm CR}$) to positive ($R>R_{\rm CR}$). {As bars are not isolated structures, we expect to find fluctuations in the torque profiles, due to the influence of other components such as spiral arms. This often results in multiple zero crossings, and we identify the most significant crossing as the one with the largest amplitude, which we associate with $R_{\rm CR}$. The amplitude is automatically calculated inside a range that spans from the %centered in the negative-to-positive crossing up to both 
previous to the following crossing (regardless of whether these are negative-to-positive crossings or vice versa). So the amplitude is calculated as $\rm abs({\tau_{\rm max}-\tau_{\rm min}})$, where $\tau_{\rm max}$ is the maximum $\tau$ value reached within this range and $\tau_{\rm min}$ the minimum value. This process is repeated for each galaxy.} The CR is expected to lie in the radial range defined by $R\sim[R_{\rm bar}, 2R_{\rm bar}]$, where $R_{\rm bar}$ is obtained from \cite{Querejeta+21}%\action{ref for the $R\sim[R_{\rm bar}, 2R_{\rm bar}]$ range?}
. This is because we do not anticipate to find the CR position inside the bar because,
at least according to classical theory,
stable periodic orbits become chaotic beyond the CR, implying that a bar cannot go beyond this location (\citealt{ContopoulosPapayannopoulos80, Contopoulos80, Elmegreen1996}). {The upper limit on this radial range is motivated by the previous findings in the literature, that obtain a mean value for the range of $R\simeq (1.5\pm0.5) \ R_{\rm bar}$. Moreover, the choice of extending this range much further out than $2R_{\rm bar}$ would only be justified in the case of slow bars suspected of having been misclassified by the gravitational torque method. These are, nevertheless, virtually absent in our sample (see Sect.~\ref{sec:caveats_misclassifying_bars}).} However, we extend this interval further, from $0.75R_{\rm bar}$ to $2.25R_{\rm bar}$, considering that $\Delta = 0.25R_{\rm bar}$ is a reasonable margin given typical uncertainties (discussed in Sect.~\ref{sec:bootstrap}).

We assessed the impact of extending the radial range to $0.3R_{\rm bar}-2.25R_{\rm bar}$, but this change mainly affects galaxies with a non-robust quality flag $QF$ (i.e., $QF=3$, see Sect.~\ref{sec:quality_flag_explanation}). In some particular cases, this change can affect galaxies with robust $QF$, but after a visual inspection, we believe the selected crossings in these particular cases are artifacts. Therefore, we establish $0.75R_{\rm bar}-2.25R_{\rm bar}$ as the range in which we search for the CR of the primary bar.

\subsection{{Identifying other dynamical resonances}}
\label{sec:identifying_other_dynamical_resonances}

We use rotation curves based on CO kinematics from \citet{Lang+20} as a proxy for the circular velocity. These rotation curves approximate the circular velocity $v_{\rm c}$ {in 150 pc wide fixed radial bins}. Bayesian Markov Chain Monte Carlo (MCMC) analysis was used to fit the galaxy center, $i$, and PA. After that, the rotational velocity within a sequence of radial annuli was modeled using harmonic decomposition and least-squares fitting. 

{The rotation curves from \cite{Lang+20} are defined as the azimuthally averaged tangential velocity of the gas, this is, $v_c=\langle v_{\theta}\rangle(R)$. It is important to note that the epicyclic theory is only valid when the perturbations are small. Thus, the calculated dynamical resonances may not be entirely correct, when studying bars (especially strong ones).}

We have tried a number of alternatives (see Appendix \ref{sec:appendix-rotation_curves}) 
before choosing our nominal rotation curves and find significant differences. Thus, we recommend to use the derived Lindblad resonances with care, especially for the cases where the derived rotation curves are subject to large uncertainties. 

We can infer other dynamical resonances (such as the Lindblad resonances) using the rotation curve derived from the PHANGS-ALMA observations in \cite{Lang+20}. We note that our choice of rotation curves introduces some intrinsic uncertainty in the location inferred for the ILR and OLR. From our derived $R_{\rm CR}$, we estimate the pattern speed $\Omega_p$ as the intersection of the position of the $R_{\rm CR}$ and the angular velocity curve, since $\Omega = v / R$, and estimate other important resonances such as the Lindblad resonances. 

Apart from CR, using the epicyclic theory, we infer the ILR and OLR. We calculate $\Omega\pm \kappa/2$ curves, where $\kappa$ is the epicyclic frequency calculated as follows: 

\begin{equation}
    \kappa^2 = R \frac{\rm{d}\Omega^2}{\rm{d}R} + 4 \Omega^2.
\end{equation}

Since for a bar ($m=2$), $\Omega = \Omega_p \pm \kappa/2$ (Eq.~\ref{eq:dynamical_resonances_equation}), 
the intersection of $\Omega\pm \kappa/2$ with the pattern speed $\Omega_p$ provides the location of the ILR and OLR, respectively. Thus, the inferred position of ILR and OLR depend on the derivative of the rotation curve, which makes results highly sensitive to local wiggles. Results for all galaxies are listed in Table \ref{table:Appendix-Table_ILR_OLR} and visually shown in Appendix \ref{sec:Appendix-All_galaxies-with_CR} and Figs.~\ref{fig:NGC1097-case_study}-\ref{fig:NGC4579-case_study} in Sect.~\ref{Sec:results}. We note that there are cases where we retrieve two ILR. In those cases we have an inner and outer ILR, designated as $i{\rm ILR}$, $o\rm ILR$, which leads to an ILR region.

\subsection{Uncertainties and quality flags}
\label{sec:bootstrap}

A change of $i$, PA or center position may either artificially reinforce or lower the strength of the bar mode present in the disk and/or also introduce ``fake'' $m=1$ terms in the gravitational potential. We estimate the uncertainty and robustness (or $QF$) of the $R_{\rm CR}$ via bootstraping.

\begin{figure}[t!]
    \centering
    \includegraphics[width=0.43\textwidth]{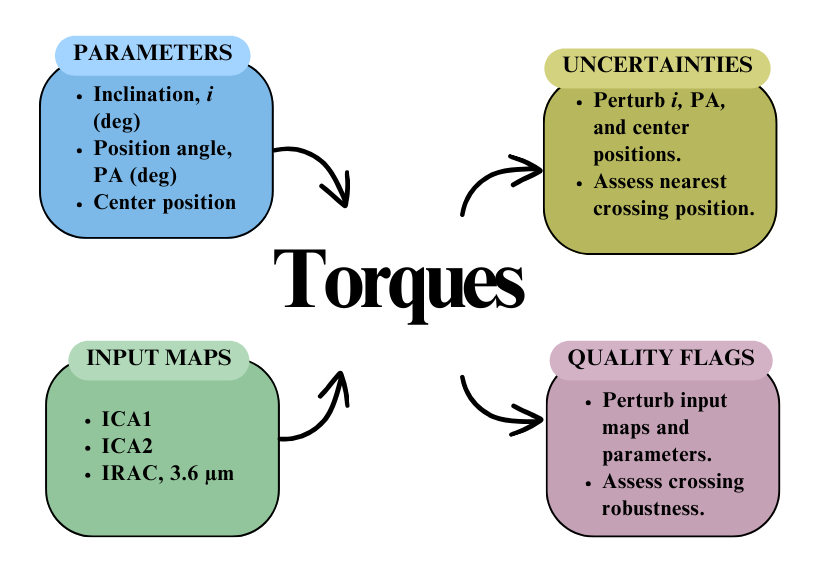}
    \caption{Schematic diagram that shows the process involved in obtaining CR uncertainties and quantifying $QF$.}
    \label{fig:quality_flags_explanations}
\end{figure}

We derive the nominal CR following the $i$ and PA from Table~\ref{table:complete_sample} (which also shows their uncertainties, which we use for bootstrapping) in addition to the center position and its uncertainty ($\pm 0.5 \ \rm px$). For each iteration, a value of the input parameters is chosen following a Gaussian distribution according to its error bars. Therefore, we obtain a new {torque (weighted by the gas)} profile in each iteration, for different input parameters, and repeat this process $N=100 \ \rm times$. Figure \ref{fig:quality_flags_explanations} schematically shows the process.

\subsubsection{Uncertainties}

To estimate the uncertainties in $R_{\rm CR}$, we perturb each input parameter ($i$, PA, center) following a Gaussian distribution within its error bars $N=100 \ \rm times$. For each iteration, we reassess the position of a given CR as the nearest crossing from negative to positive. We take the 16th-84th percentile range on the resulting CR as the error bar ($\sigma_{\rm CR}^i$, $\sigma_{\rm CR}^{PA}$ or $\sigma_{\rm CR}^{\rm center}$). Then we combine all errors by adding them in quadrature as in Eq.~\ref{eq:sigmas_unc}. The resulting uncertainties are listed in Table \ref{table:sample+CR_appendix}. 

\begin{equation}
    \sigma = \sqrt{\left(\sigma_{\rm CR}^i\right)^2 + \left(\sigma_{\rm CR}^{PA}\right)^2 + \left(\sigma_{\rm CR}^{\rm center}\right)^2}.
    \label{eq:sigmas_unc}
\end{equation}

\subsubsection{Quality flags}
\label{sec:quality_flag_explanation}

It is possible that when varying the input $i$, PA or center position, a different crossing is chosen, in the sense that a different crossing has a larger amplitude from negative to positive torques after perturbing the input parameters. {The amplitude of the crossings is automatically calculated inside a range (from the previous to the following crossing) as $\rm abs(\tau_{\rm max} - \tau_{\rm min}$). Readers can refer to Sect.~\ref{sec:gas_flows} for more details.}

Therefore, to assess the crossing robustness, we again perturb each input parameter ($i$, PA, and center) within their error bars (following a Gaussian distribution) and look for the preferred crossing 
in each iteration{. We note that, when perturbing the parameters, the amplitude of the crossings may change from one iteration to another. Therefore, the chosen crossing in a certain iteration may be different than the nominal one, giving an idea of the robustness of the nominal crossing. As before, the chosen crossing in each iteration is defined as the location in the disk inside the adopted radial range $[0.75R_{\rm bar}, 2.25R_{\rm bar}]$ where the negative to positive sign change shows the largest amplitude.} This process is repeated $N=100 \ \rm times$ for each parameter, taking the 16th-84th percentile range on the resulting CR as the error bar ($err_{i}$, $err_{PA}$ or $err_{\rm center}$).

Next, we vary the stellar mass maps and calculate the CR position for each map (ICA1, ICA2 and IRAC, see Sect.~\ref{Sec:stellar_maps}). Once we have the three values for each galaxy, we compare the CR obtained with the nominal map with the other ones and select the maximum difference as the $err_{map}$. Once we have {errors} associated with $i$, PA, center and stellar mass map variations, we conservatively assume: $err = \max\{err_i,\, err_{PA}, \, err_{\rm center}, \,err_{mass\ map}\}$. {We do this to showcase the worst-case scenario in terms of the location of the CR, as this procedure will show if the crossing is more sensitive to perturbations in the input parameters.} Then we assign $QF$ following the criteria:

\begin{equation}
    \text{If} \
    \left\{
    \begin{aligned}
    err < 30 \% \ R_{\rm bar} \hspace{13mm} \Longrightarrow QF = 1  \\
    30 \% \ R_{\rm bar} < err < 80 \% \ R_{\rm bar} \Longrightarrow QF = 2 \\
    err > 80 \% \ R_{\rm bar} \hspace{13mm} \Longrightarrow QF = 3 \\
    \end{aligned}
    \right. .
    \label{eq:QF_classification}
\end{equation}

Therefore, galaxies with an assigned $QF=1$ are the ones that present the most reliable solutions, this is, the negative-to-positive crossing with the largest amplitude remains stable to perturbations of the input parameters within their error bars and to perturbations of the stellar mass maps. The ones with $QF=2$ represent intermediate cases, with a slight change in the position of the largest crossing. Finally, the ones with $QF=3$ represent the least reliable cases which are excluded from our plots and subsequent calculations. The $QF$ are further validated by visual inspection of the results, as they may eventually depend on other factors such as S/N of CO maps (Sect.~\ref{Sec:PHANGS--ALMA}) or insufficient coverage of the bar region (Sect.~\ref{Sec:PHANGS--ALMA}). In {21} galaxies {($46\%$)} (see Table \ref{table:sample+CR_appendix}) we degrade the $QF$ to $QF=3$ due to various reasons, such as unreliable or insufficient CO coverage, unreliable radial torque profiles, or the extent of the error bars (see Appendix \ref{sec:Appendix-All_galaxies-with_CR} for individual galaxies profiles and explanation of the downgrading). While in one galaxy ($2\%$) we upgrade the $QF$ as we believe both the gas response and stellar potential behave as expected (see Appendix \ref{sec:Appendix-All_galaxies-with_CR}).

\section{Results} 
\label{Sec:results}

We start by presenting four case studies (Sect.~\ref{sec:case_study-NGC1097} to \ref{sec:case_study-NGC4579}), before shifting to results on CR based on the whole sample (Sect.~\ref{Sec:sample_statistics}). {We choose these four case studies based on their different bar strengths (SAB vs SB), the presence of a nuclear bar (NGC\,4321), or the presence of a central ring (NGC\,1097), to showcase the connection with the ILR. The number of case studies is chosen to demonstrate how these and other factors cause variations in torque profiles, as well as to have a more comprehensive comparison with the literature.} In Sect.~\ref{Sec:sample_statistics} we also consider other dynamical resonances and compare our results with the literature.

\subsection{NGC\,1097}
\label{sec:case_study-NGC1097}

\begin{figure*}
    \begin{center}
    \includegraphics[trim=0 0 0 0, clip,width=1\textwidth]{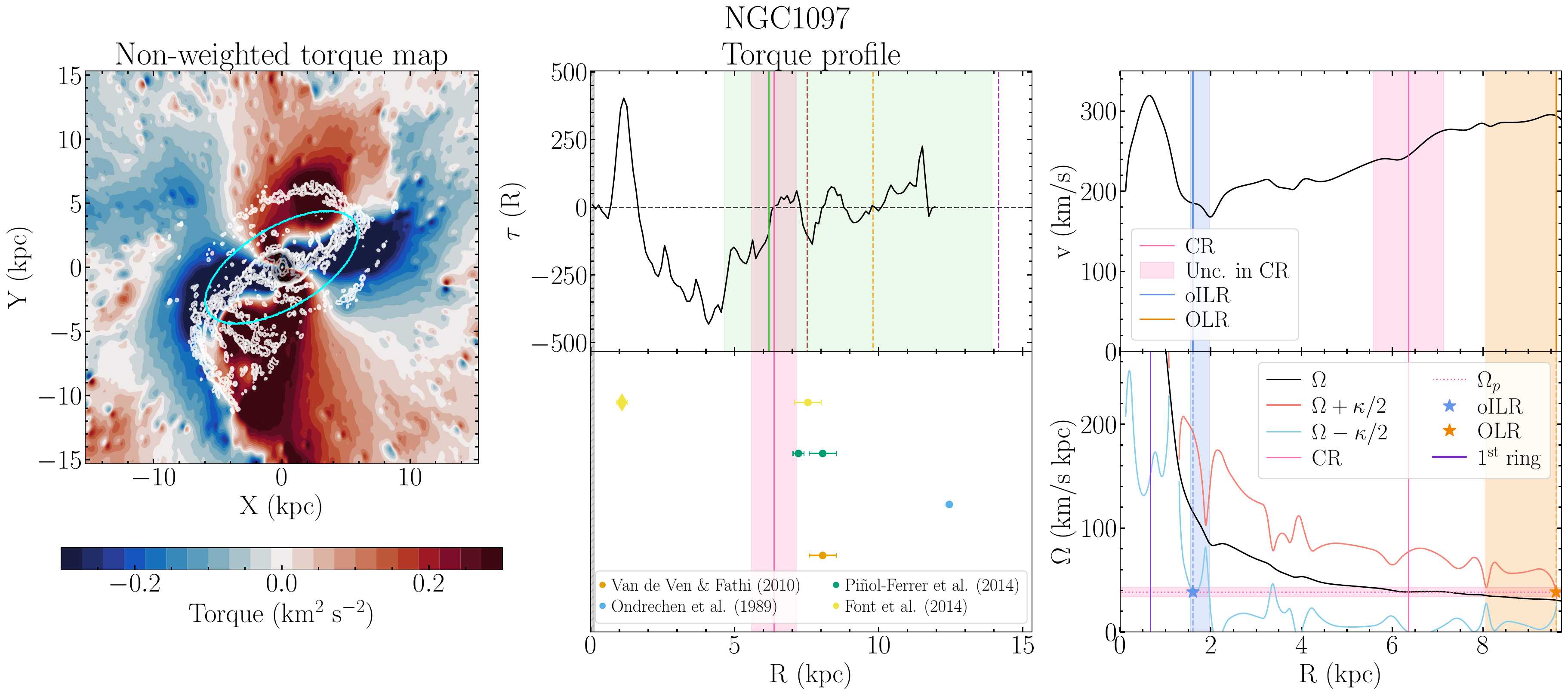}
    \end{center}

    \caption{\textbf{NGC\,1097 (SB)} non-weighted deprojected torque map (left panel), torque profile (upper central panel), comparison with values from the literature (lower central panel), rotation curve (upper right panel), and angular rotation curve (lower right panel). The cyan-contoured ellipse in the left panel indicates the bar extent. We show contours corresponding to $[5\sigma, 15\sigma, 45\sigma, \dots, 0.9\sigma_{\rm max}]$ in the CO map, where $\sigma$ is the mean value of the gas map $\sigma = 0.59\ \rm K \ km \ s^{-1}$, and $\sigma_{\rm max} = 934.83 \ \rm K \ km \ s^{-1}$. In the central and right panels, CR is represented as a vertical pink line, together with its uncertainties (pink-shaded area). This is a statistical uncertainty due to bootstrapping for $i$, PA and center position (see Sect.~\ref{sec:bootstrap}). The solid green line represents the bar length (from \citealt{Querejeta+21}), while the shaded green region represents the region where we search for the CR. Brown dashed line marks the radius at which the coverage of CO starts to be nonuniform ($R_{\rm 100 \%\ CO}$ in Table \ref{table:complete_sample}), orange dashed line is the radius at which the coverage of CO is uniform about $50\%$ ($R_{\rm 50 \%\ CO}$ in Table \ref{table:complete_sample}), and purple dashed line represents the end of CO coverage ($R_{\rm End \ CO}$ in Table \ref{table:complete_sample}). The shaded gray region represents the inner region inside which we cannot say anything on $\tau(R)$ due to the limited spatial resolution of our observations. Finally, in the lower central panel, each dot represents a different measure of the CR (of the main bar) from the literature, and the $\MyDiamond$ symbol represents a measure of a nuclear bar CR. For both right panels, the solid black line represents the rotation curve (upper panel) and the angular rotation curve $\Omega$ (lower panel). Solid light pink and light blue lines represent $\Omega + \kappa /2$ and $\Omega - \kappa /2$, respectively. The OLR and its uncertainties are represented in orange and the oILR (and its uncertainties) in dark blue. Purple vertical line represents the central ring detected by \cite{Querejeta+21}.} 
    \label{fig:NGC1097-case_study}
\end{figure*}

Figure \ref{fig:NGC1097-case_study} shows the torque map and torque profile for NGC\,1097. In the central part of this galaxy, we observe the expected butterfly pattern (a change of signs in the bars' consecutive quadrants). %We find 
Another butterfly pattern on larger scales beyond the bar has a different orientation. This could be explained by a scenario where the stellar spiral structure has a different pattern speed than the bar (e.g., \citealt{GB+09, Meidt+09}). In this scenario, our profile probes the area inside the CR resonance of the spiral structure. This secondary spiral mode might be characterized by a lower pattern speed and results in negative torques inside the spiral CR. These negative torques can lower the positive torques associated with the bar immediately outside the bar CR. Furthermore, in this map, we can see departures from a smooth map (e.g., dipoles out of the butterfly pattern). These can arise from the imperfect subtraction of dusty regions in the stellar mass map, or from image artifacts that are not properly masked but they should have a limited impact on the torque profiles after azimuthally averaging. 

Now, if the gas response is canonical inside (outside) the bar CR, the gas feels negative (positive) torques and as a consequence, the $\tau(R)$ profile will show negative (positive) values at $R<R_{\rm CR}$ ($R>R_{\rm CR}$), as shown in the upper {central} panel of Fig.~\ref{fig:NGC1097-case_study}, where we show the azimuthal average torque profile of NGC1097. In the particular case of NGC\,1097, we find negative torques along the leading sides of the bar, as expected. In this panel, the change from negative to positive torques at $R_{\rm CR}=6.4 \ \rm kpc$, is identified as the CR. There are two more crossings from negative to positive within the range where we look for CR, but their overall amplitude is smaller, which is why the first crossing is {automatically} chosen {(as explained in Sect.~\ref{sec:gas_flows})}. A visual inspection of the profile confirms that weighted torques are clearly negative inside this crossing, as expected inside the bar CR. 
This figure also contains a comparison with other estimates of the $R_{\rm CR}$ obtained from the literature using a range of different methods (lower-{middle} panel), obtaining a reasonable agreement between those values (and their uncertainties) and the one we calculated. In Fig.~\ref{fig:NGC1097-case_study}, well inside the CR, we clearly see a second crossing from negative to positive $\tau(R)$ at $R\sim0.7 \ \rm kpc$, which could be indicative of an ILR, and is potentially related to the presence of a ring. Indeed, by visual inspection, we find a ring both in deprojected CO and deprojected $3.6\ \mu \rm m$ at $R\sim 10.4"\sim 0.7\ \rm kpc$ and $R\sim 13.6"\sim 0.9\ \rm kpc$ respectively, which is consistent with the results from \cite{Comeron+14}, where $R_{\rm ring} = 9.3-14.1 "$ (deprojected).

We note that \cite{Font+14} identified a secondary inner bar. With our analysis, we tentatively find the signature of the nuclear bar on the sign of the $\tau(R)$ profile which switches to negative values for $R<0.7\ \rm kpc$ (see Fig. \ref{fig:NGC1097-case_study}). This is the expected behavior for the torque profile if the gas lies inside the CR of the nuclear bar at these inner radii.

Finally, we measure a pattern speed value of $\Omega_{\rm p} = 38.5\pm 4.7 \ \rm km \ s^{-1}\ kpc^{-1}$. This leads to an oILR at $R = 1.0\ \rm kpc$ and an OLR at $R = 9.6\ \rm kpc$. When inferring the ILRs, the iILR shows a positive slope in the $\Omega-\kappa/2$ curve, while the oILR shows a negative slope. This way, we are able to distinguish them. {In this particular case, one could argue the iILR should be at $R = 1.0\ \rm kpc$ and the oILR at $R = 2.0\ \rm kpc$. However, we believe this is not correct, as this bump in $\Omega-\kappa/2$ seems to be the result of an artifact in the rotation curve. That is why}, when interpreting these results, it is important to keep in mind that the input rotation curves are not circular velocities and that this may have a significant impact on the derived locations of resonances, as seen in some examples in Appendix \ref{sec:appendix-rotation_curves}. 

\subsection{NGC\,3627}
\label{sec:Case_study-NGC3627}

\begin{figure*}[t]
\begin{center}
    \includegraphics[trim=0 0 0 0, clip,width=1\textwidth]{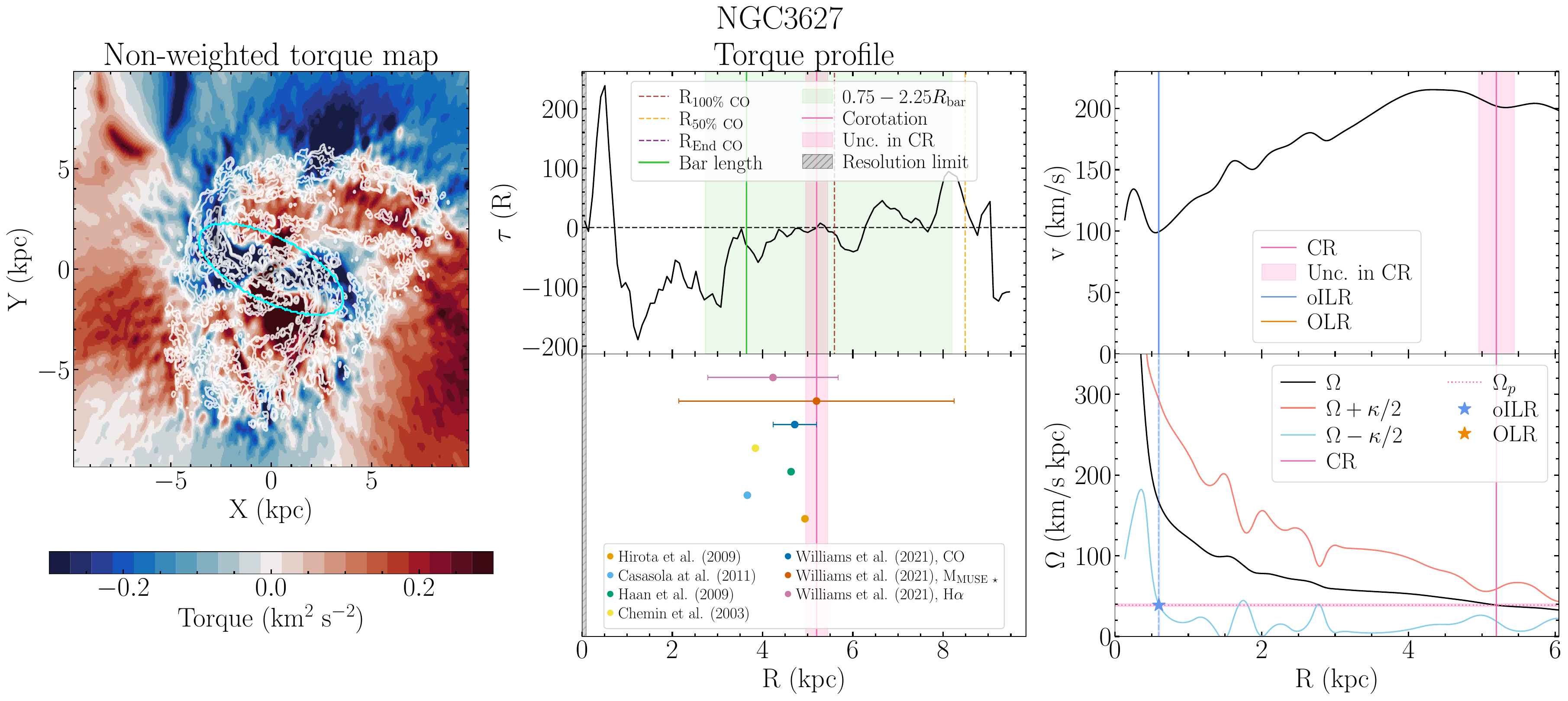}
    \end{center}
    \caption{\textbf{NGC\,3627 (SB)} non-weighted deprojected torque map (left panel), torque profile (upper central panel), comparison with values from the literature (lower central panel), rotation curve (upper right panel), and angular rotation curve (lower right panel). We show contours corresponding to $[5\sigma, 15\sigma, 45\sigma, \dots, 0.9\sigma_{\rm max}]$ in the CO map, where $\sigma$ is the mean value of the gas map $\sigma = 1.21\ \rm K \ km \ s^{-1}$ and $\sigma_{\rm max} = 1394.02\ \rm K \ km \ s^{-1}$. Symbols as in Fig.~\ref{fig:NGC1097-case_study}. {The CR is automatically selected as $R_{\rm CR} = 5.2 \ \rm kpc$, due to the change of torques signs at this position and the amplitude of the crossing. $R= 3.5 \ \rm kpc$ is not selected as the CR because the torque profile does not change sign at this position. $R= 6.3 \ \rm kpc$ is not selected as the CR because, even though the torque profile changes sign, it oscillates around zero (a behavior usually associated with the presence of a spiral) and presents a lower amplitude (compared to $R_{\rm CR} = 5.2 \ \rm kpc$).}}%\action{RECORDAR cómo hemos sacado la CR (el corte con mayor amplitud). Hacer mucho hincapié. DEcir que se espera que haya fluctuaciones (barras no aisladas, espirales.) Decir que el primer pico NO corta}} 
    \label{fig:NGC3627-case_study}
\end{figure*}

Figure \ref{fig:NGC3627-case_study} shows the torque map, torque profile and rotation curves for NGC\,3627. In this figure there are two different butterfly patterns, similar to those in NGC\,1097. We distinguish the expected butterfly pattern inside the bar region and a second one on larger scales, which has a different orientation (as explained in Sect.~\ref{sec:case_study-NGC1097}). %, explained by the presence of a spiral mode or an effect of deprojection. 
As expected, there are negative torques along the leading sides of the bar. Taking a look at this torque profile (upper central panel of Fig.~\ref{fig:NGC3627-case_study}), we consider the CR to be at $R_{\rm CR}=5.2 \ \rm kpc$. 

It can be argued that {either the peak at $R= 3.5 \ \rm kpc$ or the crossing at $R= 6.3 \ \rm kpc$ are more plausible crossings, however they are not chosen due to various reasons. On the one hand, the peak at $R= 3.5 \ \rm kpc$ cannot be selected as $R_{\rm CR}$ because the torque profile does not change sign at this position. On the other hand,} the crossing at $R= 6.3 \ \rm kpc$ {could be} more plausible, because torques become positive after $\sim 6.3 \ \rm kpc$ and, at smaller radii, torques are only mildly positive, with oscillations around zero. However, this behavior could result from the presence of the spiral, which decreases intrinsically positive torques due to the bar. {Taking this into account, the selected crossing is $R_{\rm CR} = 5.2 \ \rm kpc$ because its amplitude is larger than the amplitude of $R= 6.3 \ \rm kpc$.} We note that there is also a peak at $R\sim R_{\rm bar}$ that approaches zero without crossing. This could be another plausible candidate for the CR. Finally, we measure a pattern speed value of $\Omega_{\rm p} = 38.8\pm 1.8 \ \rm km \ s^{-1} \ kpc ^{-1}$, leading to an oILR at $R = 0.6 \ \rm kpc$ and an OLR at $R = 6.8 \ \rm kpc$. 

\subsection{NGC\,4321}

\begin{figure*}[t]
\begin{center}
    \includegraphics[trim=0 0 0 0, clip,width=1\textwidth]{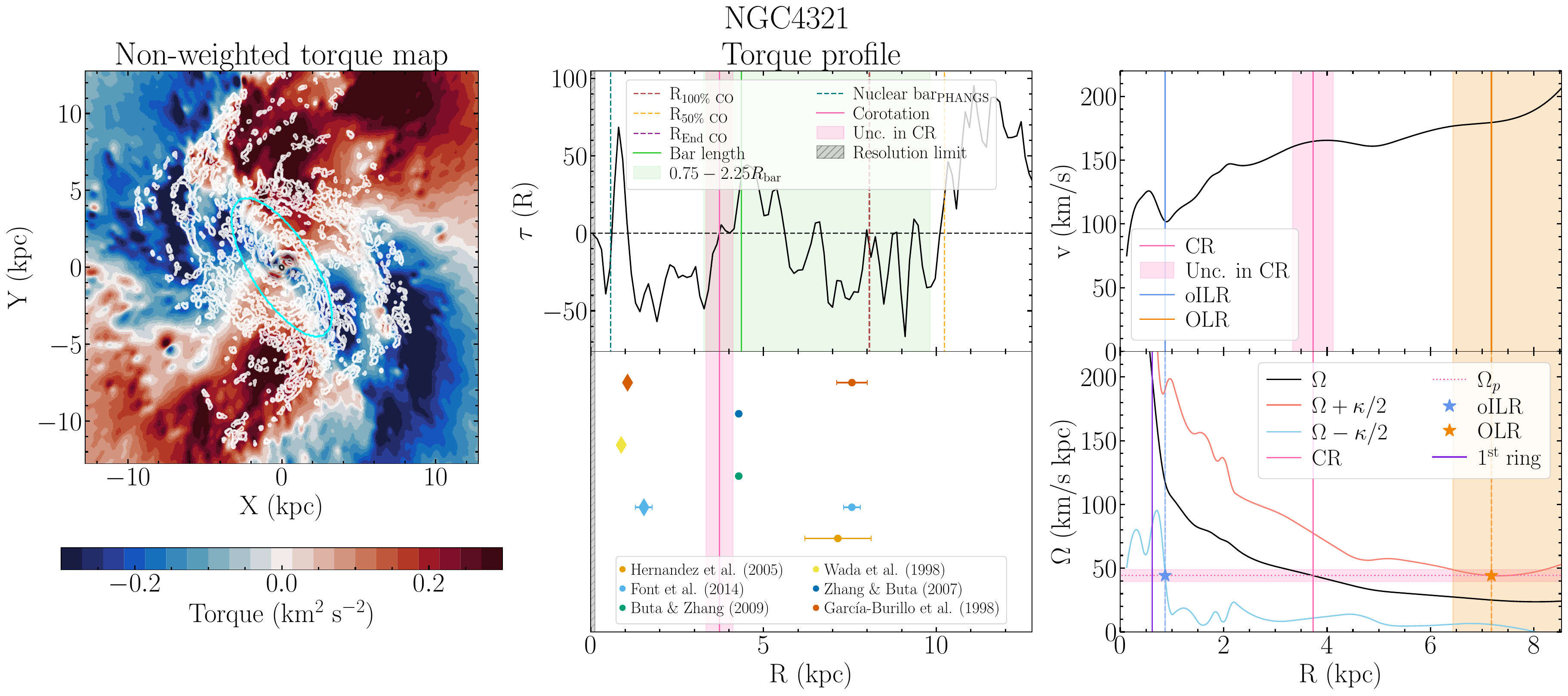}
    \end{center}
    \caption{\textbf{NGC\,4321 (SAB)} non-weighted deprojected torque map (left panel), torque profile (upper central panel), comparison with values from the literature (lower central panel), rotation curve (upper right panel), and angular rotation curve (lower right panel). We show contours corresponding to $[5\sigma, 15\sigma, 45\sigma, \dots, 0.9\sigma_{\rm max}]$ in the CO map, where $\sigma$ is the mean value of the gas map $\sigma = 1.08 \ \rm K \ km \ s^{-1}$ and $\sigma_{\rm max}=885.64 \ \rm K \ km \ s^{-1}$. The teal dashed line represents the nuclear bar radius registered by \cite{Querejeta+21}. Symbols as in Fig.~\ref{fig:NGC1097-case_study}.}
    \label{fig:NGC4321-case_study}
\end{figure*}

Figure \ref{fig:NGC4321-case_study} shows the torque map and torque profile for NGC\,4321. A butterfly pattern is clearly visible in the torque map, that alignes quite well with the bar. In addition, inside the bar, we find negative torques along the leading side of the bar edges. Based on the {method from \cite{GB+05}}, the selected crossing ($R_{\rm CR}=3.7 \ \rm kpc$) clearly represents the CR of the bar. However, several literature studies place the CR further out in the disk, and that would be consistent with a coupling of the $R_{\rm CR_{\rm bar}}$ with the $R_{\rm ILR_{\rm spiral}}$ or the inner 4:1 resonance of the spiral (\citealt{Font+14-base}). Readers can refer, for example, to \cite{Gnedin1995}, that found a peak in the torque profile at $R\sim11 \ \rm kpc$, in agreement with our profile, and which would almost certainly correspond to the spiral.

Visually inspecting the central region of the torque profile, we find a clear negative-to-positive crossing at $R\sim 0.6\ \rm kpc$, %which  could be indicative of an ILR, 
leading to gas accumulation between $R\sim 0.6\ \rm kpc$ and $R\sim 1.1\ \rm kpc$ (positive-to-negative crossing), where we expect to find a ring. Indeed, upon visual inspection, we find a ring both in CO and $3.6 \ \mu\rm m$ at $R \sim 7.4" \sim 0.5 \ \rm kpc$ (CO deprojected) and $R \sim 8.2" \sim 0.6 \ \rm kpc$ ($3.6 \ \mu\rm m$ deprojected), which is coherent with the results from \cite{Comeron+14}, where $R_{\rm ring} = 8.2-9.0"$ (deprojected) and also with the measurements from \cite{Querejeta+21}, for which $R_{\rm inner\ ring} = 8.4"$ (deprojected). Finally, we measure a pattern speed of $\Omega_{\rm p} = 44.2\pm4.5 \ \rm km \ s^{-1} \ kpc^{-1}$. We have found an oILR at $R = 0.9 \ \rm kpc$ and an OLR at $R = 7.2 \ \rm kpc$. 

\subsection{NGC\,4579}
\label{sec:case_study-NGC4579}

\begin{figure*}[t]
\begin{center}
    \includegraphics[trim=0 0 0 0, clip,width=1\textwidth]{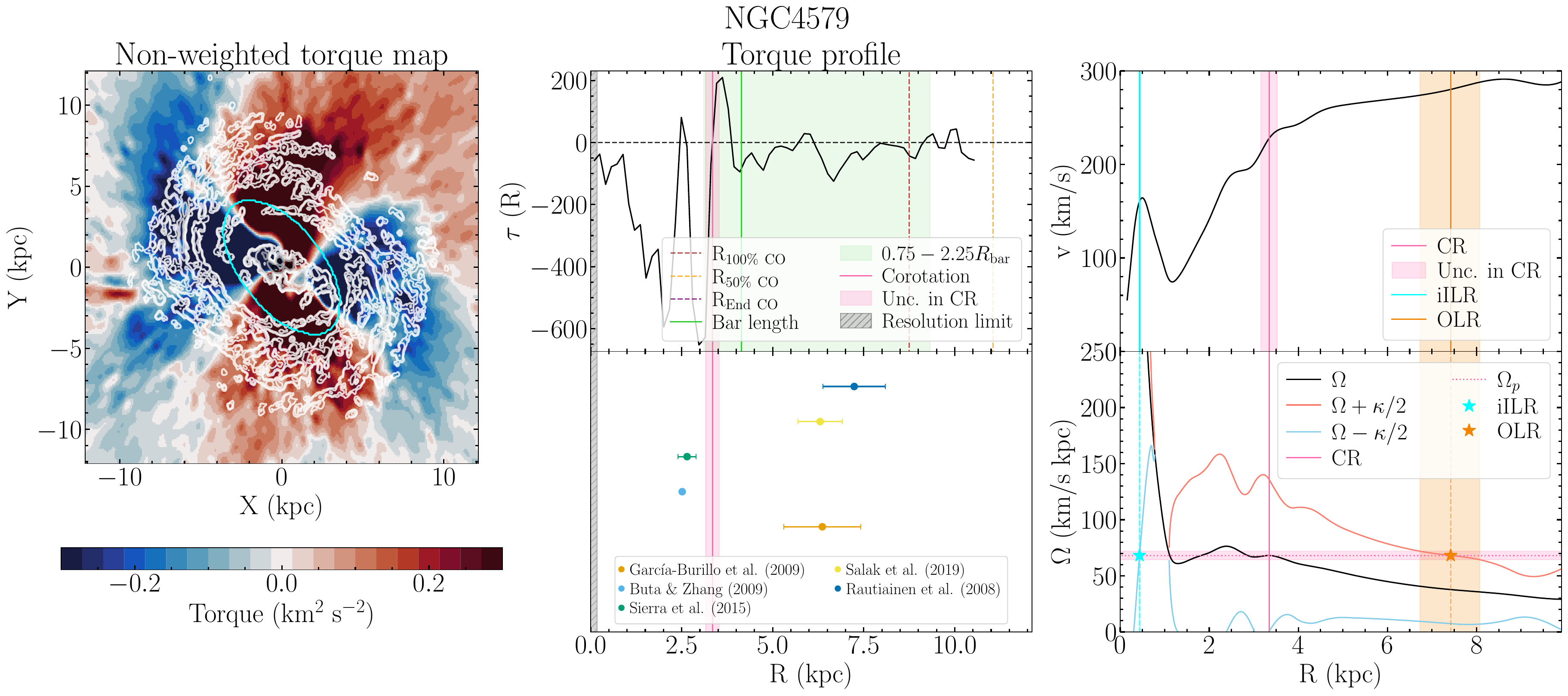}
    \end{center}
    \caption{\textbf{NGC\,4579 (SB)} non-weighted deprojected torque map (left panel), torque profile (upper central panel), comparison with values from the literature (lower central panel), rotation curve (upper right panel), and angular rotation curve (lower right panel). We show contours corresponding to $[5\sigma, 15\sigma, 45\sigma, \dots, 0.9\sigma_{\rm max}]$ in the CO map, where $\sigma$ is the mean value of the gas map $\sigma = 0.56 \ \rm K \ km \ s^{-1}$ and $\sigma_{\rm max} = 258.19 \ \rm K \ km \ s^{-1}$. The iILR and its uncertainties are represented in cyan. There is no oILR due to artifacts on the rotation curve. Symbols as in Fig.~\ref{fig:NGC1097-case_study}.} 
    \label{fig:NGC4579-case_study}
\end{figure*}

Figure \ref{fig:NGC4579-case_study} shows the torque map and profile for NGC\,4579, where a butterfly pattern, perfectly aligned with the orientation of the bar, is clearly seen. This states that the large-scale bar dominates the torque field. Inside the bar, we find gas preferentially along the leading edges of the bar, this is, negative torques along this side of the bar. 

For NGC\,4579, the torques robustly point to a CR at $R_{\rm CR} = 3.3 \ \rm kpc $ because torques are mostly negative at $R<3.3 \ \rm kpc $, even though they oscillate around zero afterwards. As commented before, this could result from the presence of the spiral structure. Most studies from the literature place the $R_{\rm CR}$ further out, these results could be explained due to the coupling of the $R_{\rm CR_{\rm bar}}$ with the $R_{\rm ILR_{\rm spiral}}$. We note that there is another negative-to-positive crossing at $R\sim 2.5 \ \rm kpc$. This crossing is not selected as the CR, as it is outside the region where we choose to find the CR ($0.75R_{\rm bar}-2.25R_{\rm bar}$, see Sect.~\ref{sec:gas_flows}), even though it is in better agreement with other measurements from the literature (\citealt{ButaZhang09, Sierra+15}).

We measure a pattern speed $\Omega_{\rm p} = 68.2\pm3.5 \ \rm km \ s^{-1} \ kpc^{-1}$, and the positions of the iILR and OLR, at $R = 0.4 \ \rm kpc$ and $R = 7.4 \ \rm kpc$, respectively. We note that in this particular case, there is no oILR, due to an artifact in the rotation curve, possibly due to limited CO coverage. 

\subsection{Statistics} 
\label{Sec:sample_statistics}

The analysis explained in the previous subsections is carried out for our sample of 46 barred galaxies. Figures analogous to Figs. \ref{fig:NGC1097-case_study}, \ref{fig:NGC3627-case_study}, \ref{fig:NGC4321-case_study} and \ref{fig:NGC4579-case_study} for the remaining galaxies can be found in Appendix \ref{sec:Appendix-All_galaxies-with_CR}. 
We derived the $R_{\rm CR}$ for 38 out of 46 galaxies ($83\%$ out of the barred sample, Table \ref{table:sample+CR_appendix}). For the remaining eight galaxies we did not find a negative-to-positive crossing in the torque profiles within the $[0.75R_{\rm bar} - 2.25R_{\rm bar}]$ range (possibly due to limited CO coverage). 

As explained in Sect. \ref{sec:quality_flag_explanation}, we have established automated $QF$ that indicate the reliability of the identified CR. We visually inspected the results and manually downgraded some $QF$, for example due to limited CO coverage around the identified CR, high $i$, or insuficient FoV coverage (Table \ref{table:sample+CR_appendix}). Out of 46 galaxies, 22 were modified ($\sim 48\%$): 21 galaxies ($\sim 46\%$) were downgraded and one galaxy ($\sim 2\%$) was upgraded. Once this double-check is done, we find that {$20$} out of $46$ barred galaxies are reliable, this is, a {$43\%$} of the barred sample have a $QF$ of either $QF=1$ ({8 galaxies, {$\sim17\%$}}) or $QF=2$ (12 galaxies, {$\sim26\%$}). In Sect.~\ref{Sec:CR_comparison_w_literature} we compare our estimations of the $R_{\rm CR}$ (with $QF=1,2$) with previous estimates, while in Sect.~\ref{Sec:discussion_sample_statistics} we examine the ratio $\mathcal{R} = R_{\rm CR}/R_{\rm bar}$. 

Finally, knowing the rotation curves for these galaxies, we can obtain the angular rotation curve for each galaxy (see Sect.~\ref{sec:identifying_other_dynamical_resonances}). Then, as we have already calculated the CR position (see Table \ref{table:sample+CR_appendix}), we can obtain the pattern speed value $\Omega_p$, and the corresponding Lindblad resonances (iILR, oILR, OLR). Table \ref{table:Appendix-Table_ILR_OLR} lists the different pattern speeds and Lindblad resonances found for this sample. 

From our sample, we have a total of 19 ILRs (i.e., a $41\%$ of the sample) deduced from the inferred CR (see Table \ref{table:Appendix-Table_ILR_OLR}). As discussed in Sect.~\ref{sec:identifying_other_dynamical_resonances}, the choice of the rotation curve affects the ILR/OLR calculation, for example some rotation curves lead to a nonexistence of an ILR. In \cite{Querejeta+21}, a total of 12 galaxies have a central ring. Therefore, there are five galaxies in common between our sample of ILRs and the subsample of \cite{Querejeta+21} that present a central ring. If we cross-check these two sets, we can check if the central ring lies inside the ILR, as predicted by \cite{Sormani+23}\footnote{Note that the rotation curves we are using are defined as $v_c=<v_{\Phi}>$, while the approach followed by \cite{Sormani+23} defines rotation curves as $v_c=\sqrt{R\cdot \rm{d}\Phi/\text{d}R}$.}. Examining these five galaxies (NGC\,1097, NGC\,1300, NGC\,2903, NGC\,4321 and NGC\,5248) we notice that in all these galaxies ($100\%$), the central ring lies inside the registered ILR (see Table \ref{table:rings}). In order to double-check these results, we visually obtain the size of the ring in both CO and $3.6\ \mu \rm m$ images, and also take into account results from \cite{Comeron+14}, confirming that the rings form inside\footnote{For NGC\,1097, the ring forms at $R_{\rm ring}\sim 0.47R_{\rm ILR}$; for NGC\,1300 inside the $R_{\rm ring}\sim 0.29R_{\rm ILR}$; for NGC\,2903 inside the $R_{\rm ring}\sim 0.43R_{\rm ILR}$; for NGC\,4321 inside the $R_{\rm ring}\sim 0.69R_{\rm ILR}$ and for NGC\,5248 inside the $R_{\rm ring}\sim 0.57R_{\rm ILR}$.} the inferred positions of the ILR, in agreement with the expectation from \cite{Sormani+23}. 

\begin{table*}[t!]
\begin{center}
    \caption[h!]{{Central ring positions.} \label{table:rings}}
    \begin{tabular}{ccccccc}
    \hline\hline
    Galaxy & $R_{\rm CO, \ depro}$ & $R_{\rm 3.6\mu m, \ depro}$ & $R_{\rm PHANGS}$ & $R_{\rm ARRAKIS}$ & iILR & oILR\\
    % & (arcsec) & (arcsec) & (arcsec) & (arcsec) & (arcsec) & (arcsec) \\
    (1) & (2) & (3) & (4) & (5) & (6) & (7) \\ \hline 
    \multirow{2}{*}{NGC\, 1097} & $10.4 \ \rm arcsec$ & $13.6 \ \rm arcsec$ & $10.1 \ \rm arcsec$ & $9.3-14.1 \ \rm arcsec$ & $\dots$ & $24.4  \ \rm arcsec$\\ 

    & $0.7 \ \rm kpc$ & $0.9 \ \rm kpc$ & $0.7 \ \rm kpc$ & $0.6-0.9 \ \rm kpc$ & $\dots$ & $1.6 \ \rm kpc$\\ \hline 

    \multirow{2}{*}{NGC\, 1300} & $\rm 4.4 \ arcsec$ & $\dots$ & $\rm 3.8 \ arcsec$ & $\rm 4.5-5.1 \ arcsec$ & $\dots$ & $\rm 14.8 \ arcsec$\\ 

    & $\rm 0.4 \ kpc$ & $\dots$ & $\rm 0.3 \ kpc$ & $\rm 0.4-0.5 \ kpc$ & $\dots$ & $1.4 \ \rm kpc$ \\ \hline

    \multirow{2}{*}{NGC\, 2903} & $\dots$ & $\dots$ & $5.3 \ \rm arcsec$ & $4.8-5.7 \ \rm arcsec$ & $\dots$ & $12.2 \ \rm arcsec$\\ 

    & $\dots$ & $\dots$ & $0.3 \ \rm kpc$ & $0.2-0.3 \ \rm kpc$ & $\dots$ & $0.6 \ \rm kpc$\\ \hline
    
    \multirow{2}{*}{NGC\, 4321} & $7.4 \ \rm arcsec$ & $8.2 \ \rm arcsec$ & $8.4 \ \rm arcsec$ & $8.2-9.0 \ \rm arcsec$ & $\dots$ & $\rm 11.8 \ arcsec$ \\  

    & $0.5 \ \rm kpc$ & $0.6 \ \rm kpc$ & $0.6\ \rm kpc$ & $0.6-0.7 \ \rm kpc$ & $\dots$ & $0.9\ \rm kpc$\\ \hline
    
    \multirow{2}{*}{NGC\, 5248} & $5.9 \ \rm arcsec$ & $8.2 \ \rm arcsec$ & $6.6 \ \rm arcsec$ & $\dots$ & $\dots$ & $12.0 \ \rm arcsec$\\  

     & $0.4 \ \rm kpc$ & $0.6 \ \rm kpc$ & $0.5 \ \rm kpc$ &  $\dots$ & $\dots$ & $0.9 \ \rm kpc$\\  
    \hline
    \end{tabular}
\end{center}  
\tablefoot{Column (1) contains the galaxy identifier, Cols.~(2) and (3) contain the measurement, in arcsec, of the deprojected ring in CO and $3.6\ \mu\rm m$ respectively. Columns (4) and (5) contain the central ring positions registered in \cite{Querejeta+21} and \cite{Comeron+14} respectively. Columns (6) and (7) contain the positions of the iILR and oILR respectively. ``$\dots$'' in ~(2) and (3) means it is not possible to visually determine the ring, 
while ``$\dots$'' in Col.~(5) means either there is no ring registered in the ARRAKIS survey (\citealt{Comeron+14}) or the galaxy is not included, and ``$\dots$'' in Cols.~(6) or (7) imply there is no iILR/oILR calculated.}
\end{table*}

\section{Discussion} 
\label{Sec:discussion}

%\action{Una vez miremos las galaxias en la que nos podríamos haber equivocado, lo ponemos aquí con el razonamiento teórico}

\subsection{$R_{\rm CR}$ comparison} 
\label{Sec:CR_comparison_w_literature}

\begin{figure}[t!]
\begin{center}
    \includegraphics[trim=0 0 0 0, clip,width=.5\textwidth]{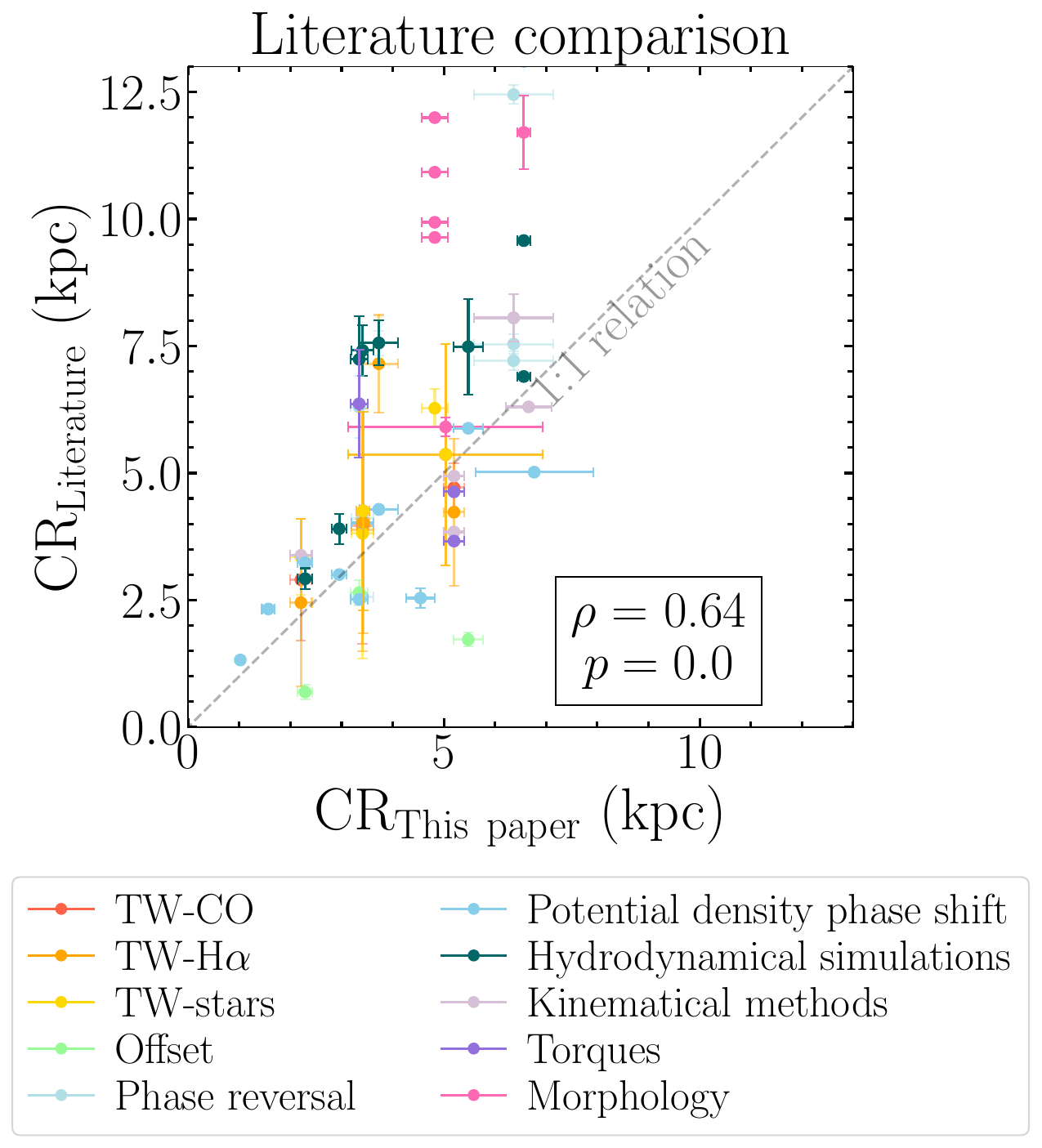}
    \end{center}

    \caption{CR measurements from the literature compared to CR measurements obtained through the gravitational torque method (\citealt{GB+05}). Readers can refer to Table~\ref{table:method-quotients} for statistics on each method. Vertically aligned data points correspond to the same galaxy presenting different measurements from the literature. References on each method can be found in Table~\ref{table:Appendix-Table_Literature}.}
    \label{fig:all_methods_at_once}
\end{figure}

Here we compare our estimations of the $R_{\rm CR}$ (with $QF=1,2$) with previous estimates from the literature. Figure \ref{fig:all_methods_at_once} %and \ref{fig:Appendix-Literature_comparison} 
summarizes these comparisons. We quantify the differences using a Spearman correlation coefficient (see Table \ref{table:method-quotients}). 

\begin{table*}[t!]
\begin{center}
    \caption[h!]{{Comparison of the methods of the literature.} \label{table:method-quotients}}
    \begin{tabular}{cccccc}
    \hline\hline
    Method & $\cfrac{|\rm CR - \rm CR_{\rm Method}|}{\rm CR}$ & $\cfrac{\rm CR}{\rm CR_{\rm Method}}$ &$\rho$ & $p$-value & $N$\\
    (1) & (2) & (3) & (4) & (5) & (6)\\ \hline 
            All methods & $0.46\pm0.44$ & $0.90\pm0.50$ & 0.64 & <0.001 & 63\\ 
            
            Tremaine-Weinberg (general) & $0.23\pm0.21$ & $0.87\pm0.16$ & 0.79 & <0.001 & 16 \\ 
            
            Tremaine-Weinberg (gas=CO+H$\alpha$) & $0.23\pm0.23$ & $0.89\pm0.17$ & 0.77 & 0.005 & 11\\ 
            
            Tremaine-Weinberg (CO) & $0.18\pm0.09$ & $0.89\pm0.12$ & 0.89 & 0.037 & 5\\ 
            
            Tremaine-Weinberg (H$\alpha$) & $0.27\pm0.29$ & $0.89\pm0.21$ & 0.77 & 0.072 & 6\\ 
            
            Tremaine-Weinberg (MUSE-M$_{\star}$) & $0.25\pm0.16$ & $0.81\pm0.10$ & 0.89 & 0.037 & 5\\
            
            Offset method & $0.53\pm0.23$ & $2.58\pm0.94$ & 0.50 & 0.667 & 3\\
            
            Phase-reversal method & $0.52\pm0.38$ & $0.77\pm0.28$ & 0.69 & 0.085 & 7\\ 
            
            Potential-density phase-shift & $0.31\pm0.25$ & $0.97\pm0.34$ & 0.83 & 0.007 & 12\\
            
            Hydrodynamical simulations & $0.86\pm0.63$ & $0.60\pm0.20$ & 0.67 & 0.034 & 10\\ 
            
            Kinematics & $0.24\pm0.16$ & $0.95\pm0.23$ & 0.79 & 0.059 & 6\\ 
            
            Morphology & $0.96\pm0.41$ & $0.54\pm0.15$ & -0.07 & 0.899 & 6\\ 
            
            Torques & $0.44 \pm 0.34$ & $1.02 \pm 0.37$ & -0.87 & 0.333 & $3^{\star}$ \\ 
            \hline
    \end{tabular}
\end{center}  
\tablefoot{Column (1) states the method used to calculate the CR. Column (2) reports the quotient ${|\rm CR_{\rm This \ \rm paper} - \rm CR_{\rm Method}|}/{\rm CR_{\rm This \ \rm paper}}$ while Col.~(3) contains the relative quotient ${\rm CR_{\rm This \ \rm paper}}/{\rm CR_{\rm Method}}$. Column (4) contains the Spearman correlation coefficient ($\rho$), Col.~(5) contains its associated $p$-value, and Col.~(6) contains the number of galaxies in each method (only taking into account galaxies with $QF=1,2$). $N^{\star}$ means the number of galaxies is higher (as seen in Fig.~\ref{fig:all_methods_at_once}) but we chose not to include some of them for the reasons explained in Sect.~\ref{Sec:CR_comparison_w_literature} for each particular case. References on each method can be found in Table~\ref{table:Appendix-Table_Literature}.}
\end{table*}

First we study the relation between our results and the CR measurements (for galaxies with $QF = 1,2$) from the literature, without differentiating the various methods, to see in general terms if our results do or do not agree with the literature. We obtain a Spearman coefficient $\rho=0.64\ (p\rm-value<0.001)$. This pair of values indicates a moderate positive relationship between our CR measurements and the ones from the literature, and a high significance of such correlation. The measurements from the literature are, on average, reasonably coincident with the CR measurements of this paper, as the mean quotient of all datapoints $\rm CR_{\rm This \ paper}/\rm CR_{\rm Literature}\sim 0.9$ and the median is $\sim 0.8$ (see Table \ref{table:method-quotients}). {However the scatter among methods (not just with our results) suggests typical uncertainties, in practice, are very large.}

We also study the same statistics splitting the literature measurements based on the methods used to obtain the CR measurement, namely: Tremaine-Weinberg (\citealt{Tremaine&Weinberg1984}) hereafter TW, offset method (\citealt{Egusa+09}), phase-reversal method (\citealt{Font+11}), potential-density phase-shift method (\citealt{ZhangButa07, ButaZhang09}), hydrodynamical simulations (such as \citealt{GB+98, Rautiainen2008}), kinematics (such as \citealt{Chemin+03, VandeVenFathi2010}), morphological methods (such as \citealt{Buta1986, Elmegreen+96}), and gravitational torques (\citealt{GB+05}). A comparison of our results to the literature, classified by methods, can be found in Table \ref{table:Appendix-Table_Literature} while a collection of correlations between all methods is provided in Table \ref{table:Appendix-all_vs_all} and Fig.~\ref{fig:Appendix-all_vs_all}. 

\subsubsection{Tremaine-Weinberg method comparison}

In general, our measurements agree quite well with the results derived from the TW application. Hereby, we refer to the TW method applied to a particular tracer as TW-tracer, namely: CO (\citealt{Williams+21}), H$\alpha$ (\citealt{Hernandez+05, Williams+21}), MUSE M$_{\star}$ (\citealt{Williams+21}). The Spearman coefficient ($\rho$) yielded a value of $0.89$, $0.77$, and $0.89$ for TW-CO, TW-H$\alpha$, and TW-MUSE M$_{\star}$ respectively, indicating a strong positive relationship between the variables. Meanwhile the $p$-value for each correlation is $0.037$, $0.072$, and $0.037$, indicating quite a high significance. We note that the number of datapoints available for comparison is low\footnote{{Five datapoints for TW-MUSE M$_{\star}$, and 11 datapoints for gaseous tracers (five measurements for TW-CO, and six measurements for TW-H$\alpha$).}}.

The TW methods overestimate by little the CR measurements compared to ours, as $\rm CR_{\rm This \ paper} / CR_{\rm TW-CO}=\rm CR_{\rm This \ paper} / CR_{\rm TW-H\alpha}\sim 0.9$ and $\rm CR_{\rm This \ paper} / CR_{\rm TW-MUSE \ M_{\star}}\sim 0.8$. This could be explained based on the third of the three main assumptions of the TW method: the satisfaction of the continuity equation, as this assumption may not be met with the tracers we are working with, namely H$\alpha$ (\citealt{Hernandez+05, Williams+21}) and CO ($J=2-1$) (\citealt{Williams+21}). As discussed in \cite{Rand+04}, the formal validity of this ``continuity equation'' may be compromised due to the clumpy nature of CO and H$\alpha$ emissions, as they can introduce fake signals. At high resolution, this effect becomes particularly noticeable when the clumpiness associated with the tracer morphology (CO and H$\alpha$) is arranged around specific areas (see \citealt{Kreckel+18, Schinnerer+19, Meidt+21}). This clumpiness may lead to measurements composed of both pattern speed and average velocity field, this is, an overestimated pattern speed value, as shown in \cite{Williams+21} and \cite{Borodina+23}. We note that \cite{Williams+21} already reports that using CO and H$\alpha$ as tracers may not be reliable when applying TW. However, \cite{Hernandez+05} states that the stellar population (e.g., using MUSE stellar mass surface density from \citealt{Williams+21} as a tracer) may satisfy the continuity equation, as long as the star formation efficiency is low. We note that in this analysis, %neither NGC\,1087 nor 
NGC\,3627 has not been taken into account\footnote{In the case of %NGC\,1087 (\citealt{Williams+21}), the CR measurement could be contaminated by the spiral arm. While in the case of 
NGC\,3627, the $R_{\rm CR}$ value from \cite{Williams+21} is far inside the bar, which could be related to rotation curve issues.}.

\subsubsection{Offset method comparison}

{The offset method consists of using different tracers of gas and star formation along spiral arms in order to calculate $\Omega_p$. These offsets vary radially as a result of the local difference between $\Omega$ and $\Omega_p$.} 

When comparing with the measurements obtained from the application of the offset method (\citealt{Sierra+15}), we see that their measurements exhibit limited agreement with our results. While $\rho = 0.5$ indicates a positive relationship, the $p\rm -value$ of $0.667$ suggests that the trend is not statistically significant (due to the small number of datapoints: only {three}). This method tends to underestimate the CR compared to our results, as $\rm CR_{\rm This \ paper}/CR_{\rm Offset}\sim 2.6$, as seen in Table \ref{table:method-quotients}. The difference between our results and the offset estimations could be due to various reasons. On the one hand, as explained in Sect.~\ref{sec:Method_limitations}, the deprojection step is crucial, and dependent on the $i$ values, which are not identical to ours. \cite{Sierra+15} obtain a range of $i$ for each galaxy, which differ a lot from the $i$ from \cite{Lang+20}. On the other hand, for three out of four galaxies, \cite{Sierra+15} give two values of CR, as there are two nearby crossings ($<5"$) and they cannot differentiate between them; in two cases (NGC\,4548, NGC\,4654) we have adopted the mean value as the CR estimation. 

\subsubsection{Potential-density phase-shift method comparison}

{The potential and density spirals are azimuthally shifted form each other. This method assumes the CR to be at the location where the phase shift changes %For an S-shaped spiral or bar, the sign convention is to ``assume the phase shift is positive when the potential lags the density pattern in the direction of galactic rotation'' 
(\citealt{ZhangButa07}).}

The CR estimations {derived} from the potential density phase-shift method (\citealt{ZhangButa07, ButaZhang09}) on average coincide with our CR calculations, as $\rm CR_{\rm This\ \rm paper}/\rm CR_{\rm PD \ phase-shift}\sim 0.9$. We find a strong and statistically significant correlation ($\rho=0.83$, $p\rm-value<0.001$). The potential density phase-shift method (\citealt{ZhangButa07}) relies on the morphological evidence in H-band images that have been deprojected and decomposed assuming a spherical bulge. The main idea is that the sense of angular momentum exchange between a density wave mode and the fundamental state of a galaxy is indicated by the sign of the phase-shift, which has to change at the CR of the bar, where the phase-shift changes from positive to negative. Even though the assumption of a spherical bulge could lead to what \cite{ButaZhang09} call a ``decomposition pinch'' (isophotes showing a pinched shape in a deprojected image), they prove the assumption of a spherical bulge does not lead to the presence of a false positive-to-negative crossing. Moreover, the main assumption of this approach is that the wave modes are quasi-steady. If this assumption is not satisfied, the phase-shift plot will be more noisy, and thus the $R_{\rm CR}$ measurement will not be reliable. 

\subsubsection{Hydrodynamical simulations comparison}

Now, turning our attention to the comparison with hydrodynamical (HD) simulation measurements, we observe that their results tend to be overestimated compared to our results ($\rm CR_{\rm This \ \rm paper}/\rm CR_{\rm HD}\sim0.6$). We find a strong and statistically significant correlation ($\rho=0.67$, $p$-value$=0.034$). It is important to take into account that in most of these references they first calculate the pattern speed from their model, and only then they can calculate the CR using the modeled rotation curve. This indirect measurement of the CR is very sensitive to changes in the used rotation curve (as seen in Sect.~\ref{Sec:Resonances_comparison}). Furthermore, for the galaxy NGC\,4321, they might be obtaining the CR of the spiral instead of the CR of the bar (\citealt{GB+98}).

\subsubsection{Kinematics methods comparison}

The CR estimations from methods involving kinematics on average coincide with our results, as $\rm CR_{\rm This \ paper}/\rm CR_{\rm Kinematics}\sim 0.9$. These methods measure $R_{\rm CR}$ by first deriving $\Omega_p$, and from this value (knowing or deriving the rotation curve) calculating $R_{\rm CR}$ (\citealt{Chemin+03, Hirota+09, VandeVenFathi2010, Pinol-Ferrer+14}). However, it is important to note that \cite{Chemin+03} and \cite{VandeVenFathi2010} combine photometric information with kinematic information, this is, they assume a certain structure of the galaxy corresponds to a dynamical resonance, and from there they obtain $\Omega_p$ with the rotation curve. We obtain a Spearman coefficient of $\rho=0.79$ and an associated $p\rm -value=0.059$ which imply a positive irrelevant correlation between both datasets. The differences between our results and the measurements from the literature involving kinematical methods may rely on the differences on the chosen rotation curves because, as explained in Sect.~\ref{Sec:Resonances_comparison}, the calculated resonances (including the CR) are very sensitive to the differences between rotation curves. 

\subsubsection{Phase-reversal method comparison}

Following another kinematic approach, the phase-reversal method (\citealt{Font+11}) overestimates the CR position compared to our results ($\rm CR_{\rm This \ \rm paper}/\rm CR_{\rm Phase-reversal}\sim0.8$). We find a moderate correlation between our measurements and the phase-reversal method, even though its statistical significance is marginal ($\rho=0.69$, $p-\rm value=0.085$). The phase-reversal approach involves determining the galactocentric radius at which a $180^{\circ}$ phase shift in the gas response is observed, this is, a phase-reversal. This way, plotting (in histogram format) the phase-reversals of the noncircular velocities as a function of galactocentric radius, the CR can be identified as the strongest ``peak'' (located near the end of the bar) in this radial distribution histogram, assuming that streaming velocities change sign at the CR. This method requires sufficient resolution and good azimuthal coverage (\citealt{Font+14}). 

\subsubsection{Gravitational torque method comparison}

Now, comparing our CR measurements with the literature, where gravitational torques have also been used, we observe that, on average, the results are coincident as $\rm CR_{\rm This \ paper}/\rm CR_{\rm Torques}\sim 1.0$. However, calculating the Spearman correlation coefficient, we obtain $\rho = -0.87$ and its associated $p\rm -value=0.333$, which are results that imply a strong negative and unreliable correlation, due to the low number of available datapoints (only three reliable datapoints). We note that in this particular case, we do not take into account galaxy NGC\,5248%(marked as a cross in Fig.~\ref{fig:Appendix-Literature_comparison2})
, as in \cite{Haan+09} they refer to this CR as related to the ``outer spiral/bar''. Therefore, the number of datapoints we use in this comparison is three. \cite{GB+09} used CO(2-1), CO(1-0) and HI maps to study gravitational torques. Comparing their CR measurements for NGC\,4579 with ours we find different torque profiles, this could be the reason of the difference between both measurements. On the other hand, \cite{Haan+09} also used CO(2-1), CO(1-0) and HI maps. They used CO until $R\sim 0.8 \ \rm kpc$ and HI from $R\sim0.8 \ \rm kpc$ until $R\sim 6 \ \rm kpc$. Comparing their NGC\,3627 results with ours, we see a good agreement. We also realize that following our procedure applied to their torque profiles, we would obtain the CR near the bump mentioned in Sect.~\ref{sec:Case_study-NGC3627}. Finally, in \cite{Casasola+11} they used CO(2-1) and CO(1-0) maps for NGC\,3627. Again, we find a good agreement between torque profiles, but their selected CR is related to the bump we find in our torque profiles (which is not our CR).

\subsubsection{Morphological method comparison}

Examining an alternative approach, we classified as ``morphological'' methods those that identify stellar structures such as rings, associate them with ILRs, and then calculate the CR position based on such assumption. This approach tends to overestimate the CR calculations according to our results ($\rm CR_{\rm This\ paper}/CR_{\rm Morphology}\sim 0.5$). We find a weak and insignificant negative correlation ($\rho = -0.07$, $p\rm -value = 0.899$). Differences can be attributed to the choice of rotation curve, as this method critically depends on that measurement, as they assume that the central ring lies at the ILR, they calculate the CR based on that pattern speed (see Sect.~\ref{Sec:Resonances_comparison}).

\subsection{Fast and slow bars} 
\label{Sec:discussion_sample_statistics}

Following \cite{DebattistaSellwood2000} we classify bars according to $\mathcal{R}$ as follows:

\begin{equation}
    \left. 
    \begin{aligned}
    \hspace{26mm} \mathcal{R}<1, \ \hspace{3mm} \text{ Ultra-fast bars}\\
    1<\mathcal{R}<1.4, \ \hspace{5mm} \text{Fast bars} \hspace{4.4mm} \\
    \mathcal{R}>1.4, \ \hspace{5mm} \text{Slow bars} \hspace{3mm}\\
    \end{aligned}
    \right. .
    \label{eq:fancyR-eq}
\end{equation} 

\begin{figure}[t!]
\begin{center}
    \includegraphics[trim=0 0 0 0, clip,width=0.5\textwidth]{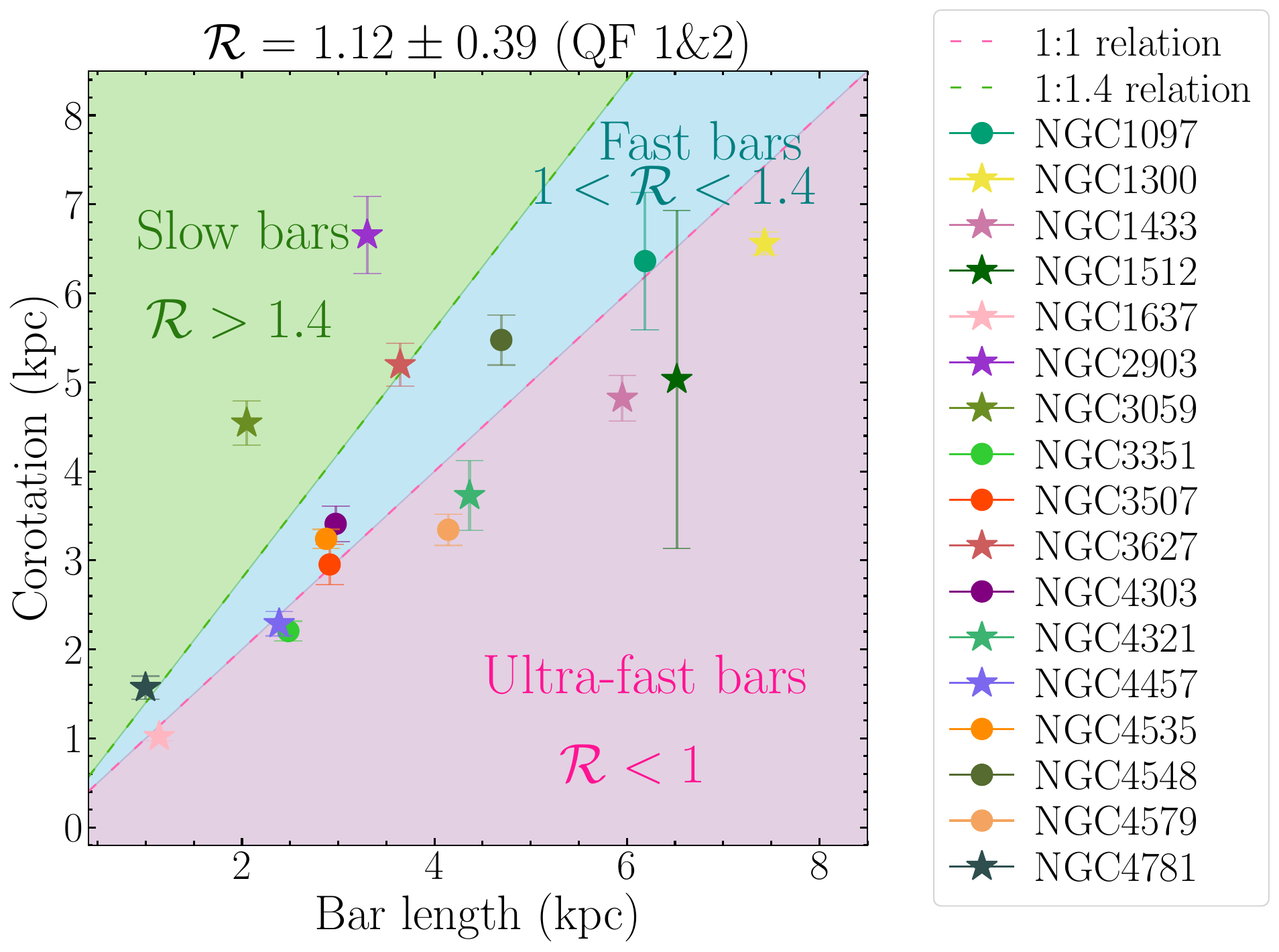}
    \end{center}
    \caption{Slow, fast, and ultra-fast bars containing exclusively galaxies with $QF = 1$ and $QF = 2$. Galaxies with $QF = 2$ are represented with stars, while galaxies with $QF = 1$ are represented with dots.}
    \label{fig:QF1&2}
\end{figure}

We calculate the ratio $\mathcal{R} = R_{\rm CR} / R_{\rm bar}$, using the $R_{\rm bar}$ values compiled by \cite{Querejeta+21}. Our average value, if we exclude those galaxies with $QF=3$, is $\mathcal{R}\sim 1.12 \pm 0.39$, which is in agreement with the value from \cite{Athanassoula1992}: $\mathcal{R}=1.2\pm 0.2$. Figure \ref{fig:QF1&2} shows that most bars are indeed fast, or compatible with the fast region within uncertainties. In this reduced sample {of 20 galaxies} (only taking into account galaxies with $QF=1,\ 2$) we find six galaxies (30\%) presenting fast bars, four galaxies (20\%) presenting slow bars and ten galaxies (50\%) presenting ultra-fast bars, but in six cases, these galaxies are close to the transition line between these two regimes of $\mathcal{R}\sim 1$. This is in line with previous findings that bars in the local universe are predominantly fast (e.g., \citealt{Corsini2011, Aguerri+15, Guo+19, Cuomo+21} using the TW method), which would argue against the braking of bars by dark matter haloes. Interestingly, the vast majority of measurements quite closely follow the one-to-one line in Fig.~\ref{fig:QF1&2}. This suggests that, to first order, CR lies at the end of the bar for most galaxies and we can estimate the bar pattern speed as $\Omega_p = \Omega(R_{\rm bar})$. There are only a handful of galaxies with clearly higher $\mathcal{R}$, which drive the average close to the standard $\mathcal{R}=1.2$ value. A graphic analog to Fig. \ref{fig:QF1&2} for $QF=1,\, 2,\, 3$ can be found in Fig.~\ref{fig:FancyR-all_galaxies-appendix}. 

\cite{ContopoulosPapayannopoulos80} and \cite{Contopoulos80, Contopoulos81} argue against the existence of ultra-fast bars ($\mathcal{R}<1$) because stable periodic orbits beyond CR become chaotic, implying that a self-consistent bar cannot go beyond its own CR. However, the existence of this kind of bars might be possible in the framework of the manifold theory (\citealt{Romeor-Gomez+06, Romero-Gomez+07, Athanassoula+09-I, Athanassoula+09-II, Athanassoula+10}). While we find several ultra-fast bars in our sample, these lie relatively close to the transition region, and given the large error bars, most of these do not necessarily pose a challenge to the classical theory.

The existence of fast and slow bars has cosmological consequences, as for the braking of bars over time (\citealt{DebattistaSellwood2000}. The dynamical friction with the dark matter (DM) halo (\citealt{TremaineWeinberg1984, Chiba2023}) can reduce the $\Omega_{\rm p}$ of the bar without increasing its length (\citealt{DebattistaSellwood2000}), causing the CR to move outward (higher $\mathcal{R}$). However, we find a mean $\mathcal{R} = 1.12\pm 0.39$, this is, a majority of fast bars. The fact that galaxies with fast bars are more common than the ones with slow bars (see also {\citealt{Corsini2011, Aguerri+15, Guo+19, Cuomo+21, Garma+2022}}), could suggest that bars do not suffer much dynamical friction produced by DM haloes, as explained by \cite{Fragkoudi+21}, and therefore DM haloes are not likely to be dominant in the inner regions of galaxies. Alternatively, our results might imply that bars do slow down via dynamical friction but at the same time grow in mass and length over time, therefore evolving with $\mathcal{R}\sim 1$ (\citealt{Athanassoula2003}).

It is important to note that recent studies such as \cite{Font+17} argue that bars with $\mathcal{R}<1.4$ might be rotating slower than previously thought because the bar grows faster than the CR moves outward when evolving. So the bars traditionally called ``fast'' are not necessarily the fastest bars in absolute terms (bars with a larger $\mathcal{R}$ can be faster if their bar length is smaller). Therefore, \cite{Font+17} propose to use the bar pattern speed relative to the angular velocity of the disk in order to describe fast and slow rotating bars instead of using $\mathcal{R}$. However, in this paper we follow the traditional nomenclature mentioned above (Eq.~\ref{eq:fancyR-eq}).

\subsection{Derivation of Lindblad resonances} 
\label{Sec:Resonances_comparison}

Central rings have traditionally been associated with the ILR. The simplest line of thought is that rings are formed at the Lindblad resonances (central rings at ILRs and outer rings at OLRs) because, under the constant influence of gravity torques from the bar potential, gas accumulates in such locations (\citealt{Schwarz1981, Combes88, Combes96, ButaCombes96}). However, there are alternative models. For example, \cite{Sormani+23} propose that central rings are instead an accumulation of gas at the inner edge of a gap in the gas disk formed around the ILR. This gap develops around the ILR because a bar potential produces the excitation of waves, which remove angular momentum from the gas disk, transporting it inward, and thus forming the ring at the inner edge of the gap. In this framework, the size of the central ring is related to the local sound speed. Moreover, this approach is compatible with rings being inside the ILR, as seen in simulations (\citealt{Englmaier97, Patsis00, Kim+12, Sormani+15, Li+15}). 

As seen in Sect.~\ref{Sec:sample_statistics}, for the five galaxies\footnote{NGC\,1097, NGC\,1300, NGC\,2903, NGC\,4321 and NGC\,5248} that present both an iILR/oILR and a ring (as compiled in \citealt{Querejeta+21}), the central ring lies inside the iILR/oILR we obtain (see Table \ref{table:rings}). This also happens when visually obtaining the size of the ring in both CO and $3.6\ \mu \rm m$ images, and when comparing to the results from \cite{Comeron+14}. This agreement confirms that the rings in our reduced sample lie inside the iILR/oILR, as expected from \cite{Sormani+23}. 

As already mentioned in Sect.~\ref{sec:identifying_other_dynamical_resonances}, we expect the identification of resonances to be sensitive to our choice to use the observed rotation velocity curve as a tracer of the circular velocity. This approach follows \cite{Lang+20}, and is based on a spline fitting to these CO-based rotation curves \citep{Sun+2020-splines}. However, circularity may not always be maintained, as there are local wiggles (e.g., in the case of bar streaming motions or in the centers of galaxies, as seen in \citealt{Hayashi&Navarro2006}) which can originate from streaming motions, and which can therefore bias the derived Lindblad resonances. In addition, as this is a first-order approximation, we use the observed velocity profile as a proxy for the galaxy's real circular motion. The observed velocity profile will overestimate the true circular velocity where the non-axisymmetry of the potential predominates, leading to resonance locations that may not be correct. The resonance identification will thus also be sensitive to the chosen rotation curve and its behavior (see Sect.~\ref{sec:identifying_other_dynamical_resonances}). 

\subsection{Methodological choices and caveats}
\label{sec:Method_limitations}

%\action{\textbf{Añadir aquí (en algún punto de los caveats) que en XX casos hemos impuesto un ``tope'' extra (a mano) a partir de donde NO buscar la CR?}}

%%%%%%%%%%%%%%%%%%%%%%%%%%%%%%%%%%%%%%%%%%
%%%% Maps + lack of HI observations
\subsubsection{Choice of gas tracer}

We favor strict mask maps over the ``broad'' masks (see Sect.~\ref{Sec:PHANGS--ALMA}), which have high completeness at the expense of higher noise, because noise spikes often bias the azimuthal averages when calculating torque profiles. Also we note that in this analysis, we neglect HI because gas is mostly molecular in the inner parts of galaxies, where bars exist. But also, for practical reasons, because we do not yet have HI observations at sufficiently high resolution for this sample of galaxies. We expect this choice to have a limited impact on our results.

%%%%%%%%%%%%%%%%%%%%%%%%%%%%%%%%%%%%%%%%%%
%%%% alphaCO + M/L caveats
\subsubsection{$\alpha_{\rm CO}$ and $M/L$}

The $\alpha_{\rm CO}$ conversion factor is known to significally depend on metallicity (\citealt{Bolatto+13}), implying that $\alpha_{\rm CO}$ changes between galaxies. There is not strong evidence that $\alpha_{\rm CO}$ varies azimuthally%between arm and inter-arm
. While arm-to-interarm metallicity variations have been found in some galaxies to be typically on the order of $\sim$0.05\,dex in $12+\log({\rm O/H})$ (e.g., \citealt{Ho+17, Ho+18}), in other cases such arm-to-interarm variations were found to be even more limited or absent (e.g., \citealt{Kreckel+19}). However, by definition (see Eq.~\ref{eq:azimuthal_average}), our results are not going to be affected by radial variations of $\alpha_{\rm CO}$ or $R_{21}$, and would only be affected by azimuthal variations, which are expected to be very limited, or at least much more limited than any radial dependencies (\citealt{denBrok+21, Leroy+22}).

Another caveat is that, in this analysis, we assume a constant stellar $M/L$. The $M/L$ is expected to display some radial variations, with much more limited azimuthal trends (e.g., \citealt{Leroy+19}). Thus, radial variations could affect the relative normalization of the torque profiles, and have an impact on inferred gas inflow rates. Since we do not calculate gas inflow rates in this paper, we expect variations of $M/L$ to have a limited impact on our results.

%%%%%%%%%%%%%%%%%%%%%%%%%%%%%%%%%%%%%%%%%%
%%%% Deprojection issues
\subsubsection{Deprojection issues}

When deprojecting the 2D stellar and gas maps, we assume that any central structures are as flattened as the disk. Therefore, we do not account for spherical bulges, which would result in an unrealistic elongated structure (similar to a bar) after deprojection. This deprojection effect is more critical when working with objects whose $i$ is large. For some galaxies, this deprojection effect may imply an overestimation of the radial forces by a factor of up to two (see \citealt{Buta&Block01}). However, this is likely not a large concern, as bulges in PHANGS galaxies are expected to be rather pseudo-bulges, and therefore mostly flat (\citealt{Querejeta+21}). 

%%%%%%%%%%%%%%%%%%%%%%%%%%%%%%%%%%%%%%%%%%
%%%% Non-interaction + static state
\subsubsection{Other assumptions}

When using the gravitational torque method, we do not take into account the inclusion of gas self-gravity. This effect could produce perturbations that could induce a departure from axisymmetry. However, it is more important at high redshifts, where galaxies contain higher gas fractions. 

{As seen in Sect.~\ref{sec:Gravitational_potential}, a constant scale-height is adopted for each galaxy, to take into account the thickness of the disk when calculating the gravitational potential. Even though the scale-height can affect the derived forces, its effect is negligible, as seen in \cite{Querejeta+16} for M\,51, and it mostly affects the scale of the profiles, not the location of the CR. The existence of a Boxy/Peanut (B/P) bulge {can be} taken into account through the scale-height function. However, we assume there are no B/P bulges, which has an impact on potential forces and bar strengths (e.g., \citealt{Fragkoudi+2015, Tahmasebzadeh+2021}), as this is beyond the scope of the present paper.}

{We are implicitly assuming that the disks from our sample are baryon-dominated, as we derived the potential from NIR images without accounting for DM. This assumption does not present an issue because DM is expected to be considerably more axisymmetric than the baryonic component studied here. So, if DM were to be dominant, its effect would only change the amplitude of the torque profiles, not the locations at which they change signs. Furthermore, we only study regions where we have a full azimuthal CO coverage, typically excluding the outskirts of galaxies where DM dominates. Yet, our profiles in some cases go out to radii where the contribution from DM might not be negligible. However, as mentioned before, the inclusion of an axisymmetric DM halo when computing the torque profiles will only affect the amplitude of the torques, but not their sign.}

{Also, as seen in \cite{Verwilghen+2024}, it appears that gas bars are mostly occurring in galaxies with $\rm M_{\star} \lesssim 10^{10} \ \rm M_{\odot}$, while more defined bars (and structures such as bar lanes or rings) are formed in more massive galaxies. %with $\rm M_{\star} > 10^{10} \ \rm M_{\odot}$. 
So there could be an observational bias, as most of our work is focused on more massive galaxies ($78.26\%$ of the sample, 36 galaxies).}

Another limitation comes from the assumption that the gas response to the stellar potential is roughly stationary with respect to the potential reference frame during a few rotation periods. In order to have a representative gas response for the dynamical resonances, we implicitly average over all orbits of gas at each position and account for the time spent by the gas clouds along all the possible orbit paths rather than following particles along individual orbits. In other words, we assume that the potential does not change much over time. This is a resonable assumption for isolated galaxies (\citealt{GB+09}), but could be compromised for interacting systems, where dynamical timescales are shorter.  

\subsubsection{{Determining $R_{\rm CR}$ in slow bars: Caveats on  the use of the torque method}}
\label{sec:caveats_misclassifying_bars}

{Slow bar candidates identified by the gravitational torque method represent a small percentage of the galaxies analyzed in this paper: only four out of 20 galaxies with $QF= 1,\ 2$ in our sample appear to host slow bars. This result is consistent with previous work using alternative observational methods to determine the $\mathcal{R}$ parameter in different galaxy samples (\citealt{Aguerri+15, Cuomo19, Cuomo20}). On the other hand, evidence found in different numerical simulations suggests that a significant fraction of barred galaxies should show slow bars (\citealt{Algorri17, Roshan21}). Fast bars are expected to live in baryon-dominated disks, whereas slow bars could experience a significant drag from the dynamical friction due to a dense DM halo (\citealt{Buttitta22}). The small fraction of slow bars identified in observations has been interpreted as being due to the low concentration of DM halos (\citealt{DebattistaSellwood2000}) or to the inefficient angular momentum exchange with this component, which is unable to reduce the pattern speed of bars in real galaxies as a result of secular evolution (\citealt{Athanassoula02, Athanassoula2013}). Alternatively, the disagreement between observations and simulations could be explained if observational methods were somehow systematically biased against the detection of slow bars. Also we note that $QF=3$ denotes the CR is not reliable, including the cases in which a bar may have been misclassified into a fast or ultra-fast bar, instead of a slow one.} 

{Next we explore %this scenario below, regarding 
the validity of the gravitational torque method used in this work. \cite{Combes93} analyzed the expected gas response to a slow stellar bar using self-consistent numerical simulations. In particular, \cite{Combes93} studied bar formation and pattern speeds in galaxies with different bulge-to-disk mass ratios. Slow bars form in galaxies with low bulge-to-disk mass ratios, typical of late Hubble-type disks, which have low values of $\Omega-\kappa/2$ everywhere and correspondingly low bar pattern speeds. In our sample, we have few late Hubble-type galaxies. CR is pushed well beyond the ends of the slow stellar bars, which can only grow up to their ILR. The characteristic gas response to a slow stellar bar consists of a small-scale gas bar that extends to the ILR. Outside the gas bar, a two-armed spiral pattern develops between the ILR and the CR of the bar. For large values of $\mathcal{R}$, the slow bar is therefore substantially shorter than its $R_{\rm CR}$, and the offset dust lane gas response inside the bar, characteristic of fast bars, disappears (see also \citealt{Collier18}). The gravitational torques exerted on the gas by a slow bar are expected to be negative inside the ILR. On the other hand, the torques between the ILR and the CR of a slow bar are expected to be positive: the two-armed gas spiral lies in the quadrants where the bar-driven torques are expected to be positive. The spiral torques are also expected to be positive in the gas outside the ILR, as the gas in these regions appears systematically upstream relative to the stellar arms. This is in stark contrast to the radial trend pattern expected for the torques in the fast bar scenario, where $\tau(R)<0$ systematically within $R_{\rm CR}$ and down to the ILR locus.}

{The validity of the torque diagnostic tool used in this paper is based on the fact that the CR of the bar defines the transition from negative ($\tau(R)<$0, $R\leq R_{\rm CR}$) to positive ($\tau(R)<$0, $R\geq R_{\rm CR}$) torques in the disk. Overall, while this is true for ultrafast, fast and moderately fast bars where $\mathcal{R}\leq1.4$, our method may misidentify the ILR of a slow bar as the CR locus. We have investigated the existence of potentially misidentified CR loci among the galaxies in our sample on a case-by-case basis.} %Suspect cases have been sought in those galaxy disks that have a CO gas bar within the stellar bar, 
{We suspect that this may be the case in galaxies that have a CO bar within the stellar bar, with no offset CO gas/dust lanes and no conspicuous nuclear gas ring morphology. These cases correspond to galaxies with a smooth $\Omega-\kappa/2$ curve without a pronounced maximum in the nuclear region. The result of this analysis shows that CR may have been incorrectly assigned by the torque method in virtually none of the galaxies of our sample satisfying that $QF=1, \ 2$. In sum, none of the galaxies suspected to have been misclassified as fast bars passed the $QF$ test ($QF < 3$).}

%%%%%%%%%%%%%%%%%%%%%%%%%%%%%%%%%%%%%%%%%%
%%%% Rotation curves + epicyclic approximation
\subsubsection{Rotation curves}

{As explained in Sect.~\ref{sec:identifying_other_dynamical_resonances}, the rotation curves from \cite{Lang+20} are defined as $v_c=\langle v_{\theta}\rangle(R)$. In simulations, the rotation curves are usually defined
as $v_c=\sqrt{R\cdot \rm{d}\Phi(R)/\rm{d}R}$} where $\Phi(R)$ is the axisymmetrized gravitational potential (used for example in \citealt{Sormani+23}). For an ideal case in an axisymmetric gravitational potential, the results from both approaches are essentially the same. However, when working with real observational data, inferring $v_c$ from the gravitational potential relies on strong assumptions and is more indirect than estimating the rotation curve directly from kinematic CO data, despite possible biases introduced, for example, by streaming motions. Given the high-quality of the CO kinematic information available in PHANGS, we prefer to use \cite{Lang+20} rotation curves, but we note that other definitions of $v_c$ could lead to differences in the measurements. Also we note that, strictly speaking, the epicyclic theory is only valid when the perturbations (departures from axisymmetry) are small, therefore it might fail when calculating dynamical resonances associated to a strong bar.

\section{Summary and conclusions} 
\label{Sec:summary}

We studied a sample of 46 barred galaxies (from the parent sample PHANGS-ALMA, of 74 galaxies) with the gravitational torque method (\citealt{GB+05}). We estimated the position of the dynamical resonance of CR by studying the gravitational torques imprinted by the stellar bar on the gas. We examined gravitational torques $\tau(R)$ along the galactocentric radius and assumed that the $R_{\rm CR}$ must be inside the $0.75R_{\rm bar}-2.25R_{\rm bar}$ range, and that it must be seen as a negative-to-positive crossing in $\tau(R)$.

We present here an unprecedented catalog of resonances obtained through the gravitational torque method, containing the bar CR of 38 galaxies out of the 46 barred galaxies analyzed. In this catalog, we include the uncertainties in the CR resonance of the bar, as well as the $QF$ assigned to each galaxy, as a marker of the reliability of the results. We also present in this catalog other interesting dynamical resonances (ILR, OLR), but these must be taken carefully, as they are dependent of the chosen rotation curves. Along with this resonances catalog, we present an inspection and comparison of the bar $R_{\rm CR}$ present in the literature. After this extensive literature research (ten different methods, 33 different papers), we generally find a good agreement between our CR measurements and the ones calculated in the literature ($\rho = 0.64$, $p\rm-value<0.001$). However, it is important to note that the uncertainties in the CR literature measurements are very large. We also find that our results are most compatible with those from the literature that are calculated using the potential-density phase-shift method.

Once the value of the $R_{\rm CR}$ is known, and also the approximate $R_{\rm bar}$, we calculated the ratio between those two quantities, $\mathcal{R}=R_{\rm CR}/R_{\rm bar}$, obtaining an average value of $\mathcal{R}=1.12\pm 0.39$ (fast bars) when studying the more reliable portion of the sample (i.e., $QF=1, \ 2$). While doing the statistical study, we also classify our sample into galaxies with slow bars, fast bars, and ultra-fast bars. Most of the galaxies from our sample present fast and ultra-fast bars in the region of transition from fast to ultra-fast bars (i.e., $\mathcal{R}\sim 1$).

\section*{Data availability}

Supplementary material is available at: {\url{https://zenodo.org/records/13898598}}

%In conclusion, ....

% Do not delete the next line
\small  % Do not delete
%
%%% Comment the following line if you do not have acknowledgments.
\begin{acknowledgements}   % Do not delete if you declare acknowledgments
{We thank the anonymous referee for carefully reviewing this paper.} This work was carried out as part of the PHANGS collaboration. 
%\mod{We would like to thank the anonymous referee for very detailed and valuable comments that helped us improve the manuscript.}

MRG, MQ and SGB acknowledge support from the Spanish grant PID2022-138560NB-I00, funded by MCIN/AEI/10.13039/501100011033/FEDER, EU. MCS acknowledges financial support from the European Research Council under the ERC Starting Grant ``GalFlow'' (grant 101116226). ES acknowledges funding from the European Research Council (ERC) under the European Union’s Horizon 2020 research and innovation programme (grant agreement No. 694343). TGW acknowledges funding from the European Research Council (ERC) under the European Union’s Horizon 2020 research and innovation programme (grant agreement No. 694343). ZB gratefully acknowledges the Collaborative Research Center 1601 (SFB 1601 sub-project B3) funded by the Deutsche Forschungsgemeinschaft (DFG, German Research Foundation) – 500700252. RSK acknowledges financial support from the German Research Foundation (DFG) via the collaborative research center (SFB 881, Project-ID 138713538) “The Milky Way System” (subprojects A1, B1, B2, and B8). He also thanks for funding from the Heidelberg Cluster of Excellence ``STRUCTURES'' in the framework of Germany’s Excellence Strategy (grant EXC-2181/1, Project-ID 390900948) and for funding from the European Research Council via the ERC Synergy Grant ``ECOGAL'' (grant 855130). The work of AKL is partially supported by the National Science Foundation under Grants No. 1615105, 1615109, and 1653300. SKS acknowledges financial support from the German Research Foundation (DFG) via Sino-German research grant SCHI 536/11-1. PSB acknowledges support from the Spanish grant   PID2022-138855NB-C31, funded by MCIN/AEI/10.13039/501100011033/FEDER, EU.

This paper makes use of the following ALMA data:  ADS/JAO.ALMA\#2012.1.00650.S,  % (N628/M74) 
ADS/JAO.ALMA\#2013.1.00803.S,  % (N5128/CenA)
ADS/JAO.ALMA\#2013.1.01161.S,  % (N1365 + N5236/M83)
ADS/JAO.ALMA\#2015.1.00121.S,  % (N5236/M83) % 
ADS/JAO.ALMA\#2015.1.00782.S,  % (N1313 + N7793) 
ADS/JAO.ALMA\#2015.1.00925.S,  % (pilot low mass) 
ADS/JAO.ALMA\#2015.1.00956.S,  % (pilot high mass) 
ADS/JAO.ALMA\#2016.1.00386.S, % (N5236/M83) 
ADS/JAO.ALMA\#2017.1.00886.L,  % (large program) %ADS/JAO.ALMA\#2018.1.01321.S, \linebreak % (N253, N300, Circinus)
ADS/JAO.ALMA\#2018.1.01651.S.  % (main sample follow-up) % ADS/JAO.ALMA\#2018.A.00062.S. \linebreak % (ACA-only nearby) 
ALMA is a partnership of ESO (representing its member states), NSF (USA) and NINS (Japan), together with NRC (Canada), MOST and ASIAA (Taiwan), and KASI (Republic of Korea), in cooperation with the Republic of Chile. The Joint ALMA Observatory is operated by ESO, AUI/NRAO and NAOJ. The National Radio Astronomy Observatory is a facility of the National Science Foundation operated under cooperative agreement by Associated Universities, Inc.

\end{acknowledgements}

% =============================================================
% Bibliography
% =============================================================
\normalsize
\newpage

\bibliography{AAbibilio}{}
\bibliographystyle{aa}{}

\newpage

% =============================================================
% Appendix
% =============================================================
\begin{appendix}

\onecolumn

%%%%%%%%%%%%%%%%%%%%%%%%%%%%%%%%%%%%%%%%%%%%%%%
%%%% Galaxy sample + parameters (Appendix) %%%%
%%%%%%%%%%%%%%%%%%%%%%%%%%%%%%%%%%%%%%%%%%%%%%%

\section{Galaxy sample}
\label{sec:Appendix-all_sample}

{
\begin{longtable}{ccccccccccc}
\caption{PHANGS sample of nearby galaxies. \label{table:complete_sample}}\\ 
\noalign{\smallskip}
\hline
\hline
Object & RA & DEC & $PA$ & $i$ & $d$ & Rotation & Bar & $R_{\rm \ 100\% \ CO}$ & $R_{\rm \ 50\% \ CO}$ & $R_{\rm End \ CO}$\\ 
& (deg) & (deg) & (deg) & (deg) & (Mpc) & & & (kpc) & (kpc) & (kpc) \\ 
(1) & (2) & (3) & (4) &(5)&(6) & (7) & (8) & (9) & (10) & (11) \\
\hline
\noalign{\smallskip}
\endfirsthead
\caption{continued}\\
\noalign{\smallskip}
\hline
\hline

Object & RA & DEC & $PA$ & $i$ & $d$ & Rotation & Bar & $R_{\rm \ 100\% \ CO}$ & $R_{\rm \ 50\% \ CO}$ & $R_{\rm End \ CO}$\\ 
& (deg) & (deg) & (deg) & (deg) & (Mpc) & & & (kpc) & (kpc) & (kpc)\\ 
(1) & (2) & (3) & (4) &(5)&(6) & (7) & (8) & (9) & (10) & (11)  \\
\hline
\endhead
\hline
\endfoot
\hline
\endlastfoot
IC\,1954 & 52.88 & $-51.90$ & 63.4{$\pm 0.2$} & 57.1{$\pm 0.7$}& 12.8{$\pm 0.068$} & $+1$ & 1 & 4.49 & 5.58 & 6.96\\ 
    
    IC\,5273 & 344.86 & $-37.70$ & 234.1{$\pm 2.0$}& 52.0{$\pm 2.1$}& 14.18{$\pm 0.061$} & $+1$ & 1 & 4.12 & 5.41 & 7.43\\

    IC\,5332 & 353.61 & $-36.10$ & 74.4{$\pm 10.0$} & 26.9{$\pm 5.0$} & 9.01{$\pm 0.019$} & $+1$ & 0 & $\dots$ & $\dots$ & $\dots$\\
    
    NGC\,0628 & 24.17 & $15.78$ & 20.7{$\pm 1.0$} & 8.9{$\pm 12.2$} & 9.84{$\pm 0.027$} & $-1$ & 0 & $\dots$ & $\dots$ & $\dots$\\
    
    NGC\,0685 & 26.93 & $-52.76$ & 100.9{$\pm 2.8$} & 23.0{$\pm 43.4$} & 19.94{$\pm 0.061$} & $+1$ & 1 & 6.15 & 7.41 & 9.49\\
    
    NGC\,1087 & 41.60 & $-0.49$ & 359.1{$\pm 1.2$} & 42.9{$\pm 3.9$} & 15.85{$\pm 0.057$} & $+1$ & 1 & 6.55 & 7.85 & 9.43 \\
    
    NGC\,1097 & 41.58 & $-30.27$ & 122.4{$\pm 3.6$} & 48.6{$\pm 6.0$} & 13.58{$\pm 0.061$} & $-1$ & 1 & 7.52 & 9.81 & 14.16\\
    
    NGC\,1300 & 49.92 & $-19.41$ & 278.0{$\pm 1.0$} & 31.8{$\pm 6.0$} & 18.99{$\pm 0.061$} & $-1$ & 1 & 9.85 & 12.07 & 15.46\\
    
    NGC\,1317* & 50.68 & $-37.10$ & 221.5{$\pm 2.9$} & 23.2{$\pm 7.5$} & 19.11{$\pm 0.019$} & $+1$ & 1 & 3.61 & 4.18 & 5.03\\
    
    NGC\,1365 & 53.40 & $-36.14$ & 201.1{$\pm 7.5$} & 55.4{$\pm 6.0$} & 19.57{$\pm 0.017$} & $-1$ & 1 & 6.17 & 9.57 & 19.61 \\
    
    NGC\,1385 & 54.37 & $-24.50$ & 181.3{$\pm 4.8$} & 44.0{$\pm 7.6$} & 17.22{$\pm 0.061$} & $+1$ & 0 & $\dots$ & $\dots$ & $\dots$\\
    
    NGC\,1433 & 55.51 & $-47.22$ & 199.7{$\pm 0.3$} & 28.6{$\pm 6.0$} & 18.63{$\pm 0.041$} & $-1$ & 1 & 7.74 & 10.83 & 15.07 \\
    
    NGC\,1511 & 59.90 & $-67.63$ & 297.0{$\pm 2.1$} & 72.7{$\pm 1.2$} & 15.28{$\pm 0.06$} & 0 & 0 & $\dots$ & $\dots$ & $\dots$ \\
    
    NGC\,1512 & 60.97 & $-43.35$ & 261.9{$\pm 4.2$} & 42.5{$\pm 6.0$} & 18.83{$\pm 0.041$} & $-1$ & 1 & 7.30 & 9.86 & 14.26\\
    
    NGC\,1546 & 63.65 & $-56.06$ & 147.8{$\pm 0.4$} & 70.3{$\pm 0.6$} & 17.69{$\pm 0.047$} & 0 & 0 & $\dots$ & $\dots$ & $\dots$\\
    
    NGC\,1559 & 64.40 & $-62.78$ & 244.5{$\pm 3.0$} & 65.4{$\pm 8.4$}  & 19.44{$\pm 0.01$}  & $-1$ & 1 & 8.63 & 12.37 & 21.29 \\
    
    NGC\,1566 & 65.00 & $-54.94$ & 214.{$\pm 4.1$} 7 & 29.5{$\pm 10.6$}  & 17.69{$\pm 0.047$}  & $-1$ & 1 & 7.95 & 10.46 & 14.14 \\
    
    NGC\,1637 & 70.37 & $-2.86$ & 20.61{$\pm 10.0$}  & 31.1{$\pm 5.0$}  & 11.7{$\pm 0.036$}  & $+1$ & 1 &  3.67 & 5.07 & 6.59 \\

    NGC\,1672 & 71.43 & $-59.25$ & 134.3{$\pm 0.4$}  & 42.6{$\pm12.9$}  & 19.4{$\pm 0.061$}  &$+1$ & 1 &  5.38 & 9.95 & 17.18 \\
    
    NGC\,1792 & 76.31 & $-37.98$ & 318.9 $\pm$ 0.9 & 65.1 $\pm$ 1.1 & 16.2 $\pm$ 0.061 & $-1$ & 0 & $\dots$ & $\dots$ & $\dots$\\
    
    NGC\,1809 & 75.52 & $-69.57$ & 138.2 $\pm$ 8.9 & 57.6 $\pm$ 23.6 & 19.95 $\pm$ 0.108 & 0 & 0 & $\dots$ & $\dots$ & $\dots$\\
    
    NGC\,2090* & 86.76 & $-34.25$ & 192.46 $\pm$ 0.6 & 64.5 $\pm$ 0.2 & 11.75 $\pm$ 0.03 & $+1$ & 0 & $\dots$ & $\dots$ & $\dots$\\
    
    NGC\,2283* & 101.47 & $-18.21$ & -4.1$\pm$1.0 & 43.7$\pm$3.6 & 13.68$\pm$0.061 & $-1$ & 1 & 4.96 &  6.06 & 8.45 \\
    
    NGC\,2566* & 124.69 & $-25.49$ & 312.0$\pm$2.0 & 48.5$\pm$6.0 & 23.44$\pm$0.061 &$-1$ & 1 & 6.89 &  11.27 & 17.77\\
    
    NGC\,2775 & 137.58 & 7.04 & 156.5$\pm$0.1 & 41.2$\pm$0.6 & 23.15$\pm$0.061 & 0 & 0 & $\dots$ & $\dots$ & $\dots$\\
    
    NGC\,2835* & 139.47 & $-22.35$ & 1.0 $\pm$1.0 & 41.3$\pm$5.3 & 12.22$\pm$0.032 &$-1$ & 1 & 5.45 & 6.45 & 7.79 \\
    
    NGC\,2903 & 143.04 & $21.50$ & 203.7$\pm$2.0 & 66.8$\pm$3.1 & 10.0$\pm$0.079 & $+1$ & 1 & 7.02 & 10.09 & 16.25\\
    
    NGC\,2997* & 146.41 & $-31.19$ & 108.1$\pm$0.7 & 33.0$\pm$9.0 & 14.06$\pm$0.079 &$-1$ & 0 & $\dots$ & $\dots$ & $\dots$\\
    
    NGC\,3059* & 147.53 & $-73.92$ & -14.8$\pm$2.9 & 29.4$\pm$11.0 & 20.23$\pm$0.079 &$-1$ & 1 & 7.81 & 9.64 & 11.91\\

    NGC\,3137* & 152.28 & $-29.06$ & -0.3$\pm$0.5 & 70.3 $\pm$ 1.2& 16.37$\pm$0.058 &$+1$ & 0 & $\dots$ & $\dots$ & $\dots$\\
    
    NGC\,3239 & 156.27 & $17.16$ & 72.9$\pm$10.0 & 60.3$\pm$5.0 & 10.86$\pm$0.04 &0 & 0 & $\dots$ & $\dots$ & $\dots$\\
    
    NGC\,3351 & 160.99 & $11.70$ & 139.2 $\pm$2.0 & 45.1$\pm$6.0 & 9.96$\pm$0.014 & $+1$ & 1 & 4.09 & 5.29 & 6.93\\
    
    NGC\,3507 & 165.86 & $18.13$ & 55.8 $\pm$1.3 & 21.7$\pm$11.3 & 23.55$\pm$0.068 & $-1$ & 1 & 7.55 & 8.95 & 10.94\\
    
    NGC\,3511 & 165.85 & $-23.09$ & 256.8$\pm$0.8 & 75.1$\pm$2.2 & 13.94$\pm$0.061 &$+1$ & 1 & 7.77 & 11.07 & 14.77\\  
    
    NGC\,3521 & 166.86 & $-0.03$ & 343.0$\pm$0.6 & 68.8$\pm$0.3 & 13.24$\pm$0.06 &$+1$ & 0 & $\dots$ & $\dots$ & $\dots$\\
    
    NGC\,3596 & 168.77 & $14.79$ & 78.4 $\pm$1.0 & 25.1$\pm$11.0 & 11.3 $\pm$0.038&$-1$ & 0 & $\dots$ & $\dots$ & $\dots$\\
    
    NGC\,3621* & 169.57 & $-32.81$ & 343.8$\pm$0.3 & 65.8$\pm$1.8 & 7.06$\pm$0.017 & $-1$ & 0 & $\dots$ & $\dots$ & $\dots$\\
    
    NGC\,3626 & 170.01 & $18.36$ & 165.2$\pm$2.0 & 46.6$\pm$6.0 & 20.05$\pm$0.048 & $+1$ & 1 & 4.48 & 5.43 & 6.38\\
    
    NGC\,3627 & 170.06 & $12.99$ & 173.1 $\pm$3.6& 57.3$\pm$1.0 & 11.32$\pm$0.018 & $+1$ & 1 & 5.59 & 8.49 & 12.16\\
    
    NGC\,4207 & 183.88 & $9.58$ & 121.9 $\pm$ 2.0 & 64.5$\pm$ 6.0 & 15.78$\pm$0.06 & 0 & 0 & $\dots$ & $\dots$ & $\dots$ \\
    
    NGC\,4254 & 184.71 & $14.42$ & 68.1 $\pm$ 0.5 & 34.4 $\pm$ 1.0 & 13.1 $\pm$ 0.062 & $-1$ & 0 & $\dots$ & $\dots$ & $\dots$ \\
    
    NGC\,4293 & 185.30 & $18.38$ & 48.3$\pm$2.0 & 65.0$\pm$6.0 & 15.76 $\pm$ 0.061 & $+1$ & 1 & 4.45 & 6.71 & 12.77\\
    
    NGC\,4298 & 185.39 & $14.61$ & 313.9 $\pm$ 0.7 & 59.2 $\pm$ 0.8 & 14.92 $\pm$ 0.038 & $+1$ & 0 & $\dots$ & $\dots$ & $\dots$\\
    
    NGC\,4303 & 185.78 & $4.47$ & 312.4 $\pm$ 2.5 & 23.5 $\pm$ 9.2  & 16.99 $\pm$ 0.071& $-1$ & 1 & 7.15 & 7.97 & 9.73\\
    
    NGC\,4321 & 185.73 & $15.82$ & 156.2 $\pm$ 1.7 & 38.5 $\pm$ 2.4 & 15.21 $\pm$ 0.014 & $+1$ & 1 & 8.06 & 10.24 & 14.88 \\
    
    NGC\,4424 & 186.79 & $9.42$ & 88.3 $\pm$ 2.0 & 58.2 $\pm$ 6.0 & 16.2 $\pm$ 0.018 & $+1$ & 0 & $\dots$ & $\dots$ & $\dots$\\
    
    NGC\,4457 & 187.24 & $3.57$ & 78.7 $\pm$ 2.0 & 17.4 $\pm$ 6.0  & 15.1 $\pm$ 0.054 &$-1$ & 1 & 3.39 & 3.70 & 4.34\\
    
    NGC\,4496A & 187.91 & $3.94$ & 51.1 $\pm$ 4.1 & 53.8 $\pm$ 3.5 & 14.86 $\pm$ 0.03 & $-1$ & 1 & 4.32 & 6.26 & 11.30 \\
    
    NGC\,4535 & 188.58 & $8.19$ & 179.7$\pm$1.6 & 44.7$\pm$10.8 & 15.77$\pm$0.01 & $+1$ & 1 & 6.36 & 8.52 & 11.32\\
    
    NGC\,4536 & 188.61 &$2.19$ & 305.6 $\pm$2.3& 66.0 $\pm$2.9 & 16.25 $\pm$ 0.029&$-1$ & 1 & 11.02 & 13.17& 18.23 \\
    
    NGC\,4540 & 188.71 & $15.55$ & 12.8 $\pm$ 4.3 & 28.7$\pm$28.7 & 15.76$\pm$0.061 &$+1$ & 1 & 2.99 & 3.69& 4.42 \\
    
    NGC\,4548 & 188.86 & $14.49$ & 138.0$\pm$2.0 & 38.3 $\pm$ 6.0& 16.22$\pm$0.01&$-1$ & 1 & 5.74 & 7.44 & 9.19 \\
    
    NGC\,4569 & 189.21 & $13.16$ & 18.0$\pm$2.0 & 70.0$\pm$6.0 & 15.76$\pm$0.061 &$+1$ & 1 & 7.79 & 10.73 & 14.46 \\
    %\endfirstfoot
    
    NGC\,4571 & 189.23 & $14.22$ & 217.5$\pm$0.6 & 32.7$\pm$2.1 & 14.9$\pm$0.03 &$-1$ & 0 & $\dots$ & $\dots$ & $\dots$ \\
    
    NGC\,4579 & 189.43 & $11.82$ & 91.3 $\pm$1.6& 40.22$\pm$5.6 & 21.0 $\pm$0.04 &$+1$ & 1 & 8.75 & 11.05 & 15.31\\
    
\noalign{\smallskip}

    NGC\,4654 & 190.98 & $13.13$ & 132.2 $\pm$ 1.0 & 55.6$\pm$5.9  & 21.98$\pm$ 0.022 &$+1$ & 1 & 11.30 & 13.50 & 17.57\\
    
    NGC\,4689 & 191.94 & $13.76$ & 164.1$\pm$0.3 & 38.7$\pm$2.65 & 15.0$\pm$0.061 &$-1$ & 0 & $\dots$ & $\dots$ & $\dots$ \\
    
    NGC\,4694 & 192.06 & $10.98$ & 143.3 $\pm$2.0 & 60.7$\pm$6.0 & 15.76$\pm$0.061 &0 & 0 & $\dots$ & $\dots$ & $\dots$ \\
    
    NGC\,4731 & 192.75 & $-6.39$ & 255.4$\pm$2.0 & 64.0$\pm$6.0 & 13.28$\pm$0.064 &$+1$ & 1 & 3.31 & 5.09 & 14.68\\
    
    NGC\,4781 & 193.59 & $-10.54$ & 290.0$\pm$1.3 & 59.0$\pm$3.8 & 11.31$\pm$0.043 & $+1$ & 1 & 5.12 & 5.83 & 7.93\\
    
    NGC\,4826 & 194.18 & $21.68$ & 293.6$\pm$1.2 & 59.1$\pm$0.9 & 4.41$\pm$0.018 &$-1$ & 0 & $\dots$ & $\dots$ & $\dots$ \\
    
    NGC\,4941 & 196.05 & $-5.55$ & 202.2 $\pm$0.6 & 53.4$\pm$1.1 & 15.0$\pm$0.125 &$+1$ & 1 & 5.15 & 5.85 & 7.14\\
    
    NGC\,4951 & 196.28 & $-6.49$ & 91.2$\pm$0.5 & 70.2$\pm$2.2 & 15.0$\pm$0.107 &$+1$ & 0 & $\dots$ & $\dots$ & $\dots$\\
    
    NGC\,5042 & 198.88 & $-23.98$ & 190.6$\pm$0.8 & 49.4 $\pm$ 8.6 & 16.78$\pm$0.061 &$+1$ & 0 & $\dots$ & $\dots$ & $\dots$ \\
    
    NGC\,5068 & 199.73 & $-21.04$ & 342.4$\pm$3.2 & 35.7$\pm$10.9 & 5.2$\pm$0.018 &$+1$ & 1 & 1.00 & 3.47 & 5.75 \\
    
    NGC\,5128* & 201.36 & $-43.02$ & 32.17$\pm$10.0 & 45.33$\pm$5.0 & 3.69$\pm$0.015 &$-1$ & 0 & $\dots$ & $\dots$ & $\dots$\\
    
    NGC\,5134 & 201.33 & $-21.13$ &311.6$\pm$2.0 & 22.7$\pm$6.0 & 19.92$\pm$0.055 &$+1$ & 1 & 3.95 & 5.52 & 7.34\\
    
    NGC\,5248 & 204.38 & $8.88$ & 109.2 $\pm$3.5 & 47.4$\pm$16.3 & 14.87 $\pm$0.037 &$-1$ & 1 & 6.38 & 8.52 & 12.48\\
    
    NGC\,5530* & 214.61 & $-43.39$ & 305.4$\pm$1.0 & 61.9$\pm$2.6 & 12.27$\pm$0.061 &$+1$ & 0 & $\dots$ & $\dots$ & $\dots$\\
    
    NGC\,5643* & 218.17 & $-44.17$ & 318.7 $\pm$2.0 & 29.9$\pm$6.0 & 12.68$\pm$0.018 &$+1$ & 1 & 5.86 & 7.20 & 9.34\\
    
    NGC\,6300* & 259.25 & $-62.82$ & 105.4$\pm$2.3 & 49.6$\pm$5.8 & 11.58$\pm$0.061 & $-1$ & 1 & 5.35 & 6.51 & 8.87 \\
    
    NGC\,6744* & 287.44 & $-63.86$ & 14.0$\pm$0.2 & 52.7$\pm$2.2 & 9.39$\pm$0.019 &$+1$ & 1 & 0.01 & 6.35 & 12.79\\
    
    NGC\,7456 & 345.54 & $-39.57$ & 16.0 $\pm$ 2.9 & 67.3$\pm$4.3 & 15.7 $\pm$ 0.06 &$-1$ & 0 & $\dots$ & $\dots$ & $\dots$\\
    
    NGC\,7496* & 347.45 & $-43.43$ & 193.7$\pm$4.2 & 35.9$\pm$6.0 & 18.72$\pm$0.061 & $-1$ & 1 & 5.26 & 7.09 & 9.99\\
\end{longtable}
}
\tablefoot{{Column (1) contains the identifiers of each object. Galaxies marked with $*$ are non-S$^4$G galaxies. Columns (2) and (3) contain the coordinates in the equatorial coordinate system: right ascension (RA) and declination (DEC), respectively. Columns from (4) to (6) contain relevant parameters, namely, the position angle (PA), inclination ($i$) and distance ($d$). Column (7) stands for the direction of rotation of the galaxy: -1 stands for clockwise rotation, +1 stands for counter clockwise rotation and 0 represents discarded galaxies. {Column (8) represents the existence of a bar (1) or its nonexistence (0)}. Columns (9), (10) and (11) contain the radius at which the coverage of CO starts to be nonuniform ($R_{\rm 100\% \ CO}$); the radius at which
the coverage of CO is uniform about 50\% ($R_{\rm 50\% \ CO}$); and the end of CO coverage ($R_{\rm End\ CO}$) respectively. If Cols.~(9)-(11) contain $\dots$ it means those values have not been determined, as the galaxies are non-barred.\\}}% End longtab

%%%%%%%%%%%%%%%%%%%%%%%%%%%%%%%%%%%%%%%%%%%%%%%
%% CR radius for galaxies (Appendix) %%
%%%%%%%%%%%%%%%%%%%%%%%%%%%%%%%%%%%%%%%%%%%%%%%

\section{$R_{\rm CR}$}
\label{sec:Appendix-Corotation_radius-TABLE}

\begin{longtable}{cccccc | cccccc}
\caption{Sample of galaxies and their $R_{\rm CR}$. \label{table:sample+CR_appendix}}\\ 
\noalign{\smallskip}
\hline
\hline
\noalign{\smallskip}
Object & $R_{\rm CR}$ & $R_{\rm bar}$ & {$\mathcal{R}$} & $QF$ & Nominal map & Object & $R_{\rm CR}$ & $R_{\rm bar}$ & {$\mathcal{R}$} & $QF$ & Nominal map \\ 
 & (kpc) & (kpc) & & & & & (kpc) & (kpc) & & &\\
 (1) & (2) & (3) & (4) & (5) & (6) & (1) & (2) & (3) & (4) & (5) & (6)\\ 
\hline
\endfirsthead

\caption{continued}\\
\noalign{\smallskip}
\hline
\hline
Object & $R_{\rm CR}$ & $R_{\rm bar}$ & {$\mathcal{R}$} & $QF$ & Nominal map & Object & $R_{\rm CR}$ & $R_{\rm bar}$ & {$\mathcal{R}$} & $QF$ & Nominal map \\ 
 & (kpc) & (kpc) & & & & & (kpc) & (kpc) & & &\\
 (1) & (2) & (3) & (4) & (5) & (6) & (1) & (2) & (3) & (4) & (5) & (6)\\
\hline
\endhead
\hline
\endfoot
\hline
\endlastfoot
IC\,1954 & $\dots$ & 0.40 & $\dots$ & 3 & ICA2 & 
    NGC\,3627 & $5.2\pm0.2$ & 3.64 & 1.43 & $2$ & ICA2\\

    IC\,5273 & $2.6\pm0.2$ & 1.36 & 1.90 & 3 & ICA2 & 
    NGC\,4293 & $\dots$ & 5.97 & $\dots$ & 3 & ICA2\\

    NGC\,0685 & $2.1\pm0.8$ & 1.92 & 1.11 & $3^{\dagger}$ & ICA1 &
    NGC\,4303 & $3.4\pm0.2$ & 2.97 & 1.15 & 1 & ICA2\\

    NGC\,1087 & $1.6\pm0.1$ & 1.29 & 1.24 & $3^{\dagger}$ & ICA2 & 
    NGC\,4321 & $3.7\pm0.4$ & 4.37 & 0.85 & $2^{\ddagger}$ & ICA2\\

    NGC\,1097 & $6.4\pm0.8$ & 6.19 & 1.03 & 1 & ICA2 &
    NGC\,4457 & $2.3\pm0.1$ & 2.39 & 0.96 & $2^{\dagger}$ & IRAC\\
    
    NGC\,1300 & $6.6\pm0.1$ & 7.43 & 0.88 & 2$^{\dagger}$ & ICA2 & 
    NGC\,4496A & $3.8\pm0.3$ & 1.69 & 2.22 & $3^{\dagger}$ & ICA2\\

    NGC\,1317 & $\dots$ & 3.89 & $\dots$ & 3 & IRAC & 
    NGC\,4535 & $3.2\pm0.1$ & 2.87 & 1.13 & 1 & ICA2\\
    
    NGC\,1365 & 1$4.5\pm7.4$ & 8.58 & 1.69 & $3^{\dagger}$ & ICA2 &
    NGC\,4536 & $5.0\pm0.4$ & 2.64 & 1.89 & 3 & ICA2\\
    
    NGC\,1433 & $4.8\pm0.3$ & 5.95 & 0.81 & 2 & IRAC & 
    NGC\,4540 & $\dots$ & 1.38 & $\dots$ & 3 & ICA1\\
    
    NGC\,1512 & $5.0\pm0.1$ & 6.52 & 0.77 & $2^{\dagger}$ & IRAC & 
    NGC\,4548 & $5.5\pm0.3$ & 4.69 & 1.17 & 1 & ICA2\\
    
    NGC\,1559 & $\dots$ & 1.05 & $\dots$ & 3 & ICA1 & 
    NGC\,4569 & $11.8\pm2.9$ & 7.50 & 1.57 & $3^{\dagger}$ & ICA2\\
    
    NGC\,1566 & $6.2\pm0.6$ & 3.07 & 2.01 & $3^{\dagger}$ & ICA2 & 
    NGC\,4579 & $3.3\pm0.2$ & 4.14 & 0.81 & 1 & ICA2\\
    
    NGC\,1637 & $1.0\pm0.1$ & 1.14 & 0.89 & $2^{\dagger}$ & ICA2 & 
    NGC\,4654 & $2.7\pm0.2$ & 3.07 & 0.86 & 3 & ICA2\\

    NGC\,1672 & $11.9\pm1.9$ & 7.00 & 1.70 & $3^{\dagger}$ & ICA1 & 
    NGC\,4731 & $6.8\pm1.0$ & 3.70 & 1.82 & $3^{\dagger}$ & ICA2\\
    
    NGC\,2283* & $1.3\pm0.0$ & 0.80 & 1.59 & $3^{\dagger}$ & IRAC & 
    NGC\,4781 & $1.6\pm0.1$ & 1.00 & 1.57 & 2 & ICA2\\
    
    NGC\,2566 & $6.4\pm0.6 $ & 6.82 & 0.94 & $3^{\dagger}$ & IRAC & 
    NGC\,4941 & $\dots$ & 6.76 & $\dots$ & 3 & ICA2\\
    
    NGC\,2835 & $2.4\pm0.0$ & 1.24 & 1.96 & $3^{\dagger}$ & IRAC & 
    NGC\,5068 & $1.9\pm0.1$ & 0.88 & 2.19 & $3^{\dagger}$ & ICA2\\
    
    NGC\,2903 & $6.7\pm0.5$ & 3.30 & 2.02 & 2 & ICA2 & 
    NGC\,5134 & $4.3\pm0.0$ & 4.38 & 0.99 & $3^{\dagger}$ & IRAC\\ 
    
    NGC\,3059 & $4.5\pm0.3$ & 2.05 & 2.22 & 2 & IRAC & 
    NGC\,5248 & $6.8\pm0.7$ & 6.85 & 0.99 & $2^{\dagger}$ & ICA1\\
    
    NGC\,3351 & $2.2\pm0.1$ & 2.48 & 0.89 & 1 & ICA2 & 
    NGC\,5643 & $2.9\pm0.1$ & 3.32 & 0.86 & $2^{\dagger}$ & IRAC\\ 
    \noalign{\smallskip}
    NGC\,3507 & $3.0\pm0.1$ & 2.91 & 1.02 & 1 & ICA1 & 
    NGC\,6300 & $2.7\pm0.2$ & 2.02 & 1.31 & $3^{\dagger}$ & IRAC\\ 
    
    NGC\,3511 & $\dots$ & 1.27 & $\dots$& 3 & ICA2 & 
    NGC\,6744 & $6.4\pm0.1$ & 4.10 & 1.56 & $3^{\dagger}$ & IRAC\\
    
    NGC\,3626 & $\dots$ & 1.92 & $\dots$ & 3 & IRAC & 
    NGC\,7496* & $3.4\pm0.1$ & 3.41 & 1.00 & 1 & IRAC\\
\end{longtable}
\tablefoot{{This table only includes barred galaxies from the sample. Column (1) contains the identifiers of each object. Galaxies marked with $*$ are non-S$^4$G galaxies. Column (2) contains this work's $R_{\rm CR}$ estimations and its uncertainties. Column (3) contains $R_{\rm bar}$, obtained from \cite{Querejeta+21}. In addition, Col.~(4) contains $\mathcal{R} = R_{\rm CR}/R_{\rm bar}$. Column (5) shows the $QF$, which indicates the reliability of the $R_{\rm CR}$ determination according to this method, that is, 1 means high reliability while 3 means low reliability. Finally, Col.~(6) indicates the stellar mass map chosen for the calculation of the gravitational torques. $QF=3^{\dagger}$ in Col.~(5) means the $QF$ has been degraded by hand for one of the reasons listed on Sect.~\ref{sec:quality_flag_explanation}, while galaxies with ${\ddagger}$ symbol are galaxies that have been upgraded. \\}}

%%%%%%%%%%%%%%%%%%%%%%%%%%%%%%%%%%%%%%%%
%%%% Other relevant resonances (Appendix):
\section{Other relevant resonances}
\label{sec:Appendix-Other_resonances}

{
\begin{longtable}{ccccc|ccccc}
\caption{Other relevant resonances. \label{table:Appendix-Table_ILR_OLR}}\\ 
\noalign{\smallskip}
\hline
\hline
\noalign{\smallskip}
Object & $\Omega_{\rm p}$ & i${\rm ILR}$ & o${\rm ILR}$ & ${\rm OLR}$ & Object & $\Omega_{\rm p}$ & i${\rm ILR}$ & o${\rm ILR}$ & ${\rm OLR}$ \\
& $(\rm km \ s^{-1} kpc^{-1})$ & (kpc) & (kpc) & (kpc) & & $(\rm km \ s^{-1} kpc^{-1})$ & (kpc) & (kpc) & (kpc)\\
(1) & (2) & (3) & (4) & (5) & (1) & (2) & (3) & (4) & (5) \\ 
\hline
\endfirsthead

\caption{continued}\\

\hline
Object & $\Omega_{\rm p}$ & i${\rm ILR}$ & o${\rm ILR}$ & ${\rm OLR}$ & Object & $\Omega_{\rm p}$ & i${\rm ILR}$ & o${\rm ILR}$ & ${\rm OLR}$ \\
& $(\rm km \ s^{-1} kpc^{-1})$ & (kpc) & (kpc) & (kpc) & & $(\rm km \ s^{-1} kpc^{-1})$ & (kpc) & (kpc) & (kpc)\\
(1) & (2) & (3) & (4) & (5) & (1) & (2) & (3) & (4) & (5) \\ 
\hline
\endhead
\hline
\endfoot
\hline
\endlastfoot
IC\,1954 & $\dots$ & $\dots$ & $\dots$ & $\dots$ & 
        NGC\,3627 & $38.8\pm1.8$ & $\dots$ & $0.6\pm0.0$ & $6.8\pm 0.0$ \\
        
        %%%%%%%%%%%%%%%%%%
        %%%% New line %%%%
        IC\,5273 & $43.5\pm3.7$ & $\dots$ & $\dots$ & $\dots$ &
        NGC\,4293 & $\dots$ & $\dots$ & $\dots$ & $\dots$ \\

        %%%%%%%%%%%%%%%%%%
        %%%% New line %%%%
        NGC\,0685 & $38.5\pm199.7$ & $\dots$ & $\dots$ & 5.3$\pm \dots$ & 
        NGC\,4303 & $61.9\pm4.5$ & $\dots$ & $0.7\pm0.0$ & 5.9$_{-1.2}^{+0.1}$ \\ 
        
        %%%%%%%%%%%%%%%%%%   
        %%%% New line %%%%
        NGC\,1087 & $89.8\pm6.4$ & $\dots$ & $\dots$ & $2.5\pm0.1$ & 
        NGC\,4321 & $44.2\pm 4.5$ & $\dots$ & $0.9\pm0.0$ & 7.2$_{-0.7}^{+\dots}$ \\  
        
        %%%%%%%%%%%%%%%%%%   
        %%%% New line %%%%
        NGC\,1097 & $38.5\pm4.7$ & $\dots$ & 1.0$_{-0.1}^{+0.4}$ & 9.6$_{-1.5}^{+0.0}$ &
        
        NGC\,4457 & $88.1\pm5.9$ & $0.4\pm0.0$ & $1.0\pm0.0$ & $\dots$ \\ 
        
        %%%%%%%%%%%%%%%%%%   
        %%%% New line %%%%
        NGC\,1300 & $27.2\pm7.7$ & $\dots$ & $1.4^{+\dots}_{-1.3}$ & $\dots$ & 
        
        NGC\,4496A & $\dots$ & $\dots$ & $\dots$ & $\dots$ \\ 

        %%%%%%%%%%%%%%%%%%   
        %%%% New line %%%%
        NGC\,1317 & $\dots$ & $\dots$ & $\dots$ & $\dots$ &
        
        NGC\,4535 & $38.3\pm2.1$ & $\dots$ & $0.9\pm0.0$ & $\dots$ \\

        %%%%%%%%%%%%%%%%%%   
        %%%% New line %%%%
        NGC\,1365 & $\dots$ & $\dots$ & $\dots$ & $\dots$ & 
        
        NGC\,4536 & $32.9\pm2.7$ & $\dots$ & $0.8\pm0.0$ & $8.8_{-0.6}^{+0.4}$ \\

        %%%%%%%%%%%%%%%%%%   
        %%%% New line %%%%
        NGC\,1433 & $\dots$ & $\dots$ & $\dots$ & $\dots$ & 
        
        NGC\,4540 & $\dots$ & $\dots$ & $\dots$ & $\dots$ \\ 

        %%%%%%%%%%%%%%%%%%   
        %%%% New line %%%%
        NGC\,1512 & $\dots$ & $\dots$ & $\dots$ & $\dots$ & 
        
        NGC\,4548 & $30.8\pm1.7$ & $0.3\pm0.0$ & $0.6\pm0.0$ & $6.3_{-0.0}^{+\dots}$ \\ 

        %%%%%%%%%%%%%%%%%%   
        %%%% New line %%%%
        NGC\,1559 & $\dots$ & $\dots$ & $\dots$ & $\dots$ &
        
        NGC\,4569 & $\dots$ & $\dots$ & $\dots$ & $\dots$ \\

        %%%%%%%%%%%%%%%%%%   
        %%%% New line %%%%
        NGC\,1566 & $36.1\pm4.1$ & $\dots$ & $1.6\pm0.0$ & $7.3\pm0.0$ & 
        
        NGC\,4579 & $68.2\pm3.5$ & $0.4\pm0.0$ &$\dots$ & $7.4\pm 0.6$ \\ 

        %%%%%%%%%%%%%%%%%%   
        %%%% New line %%%%
        NGC\,1637 & $\dots$ & $\dots$ & $\dots$ & $\dots$ & 
        
        NGC\,4654 & $40.1\pm3.8$ & $\dots$ & $\dots$ & $6.4_{-0.6}^{+1.2}$ \\ 

        %%%%%%%%%%%%%%%%%%   
        %%%% New line %%%%
        NGC\,1672 & $\dots$ & $\dots$ & $\dots$ & $\dots$ &
        
        NGC\,4731 & $\dots$ & $\dots$ & $\dots$ & $\dots$\\

        %%%%%%%%%%%%%%%%%%   
        %%%% New line %%%%
        NGC\,2283 & $79.0\pm1.9$ & $\dots$ & $\dots$ & $2.1\pm0.0$ &
        
        NGC\,4781 & $64.1\pm5.9$ & $\dots$ & $\dots$ & $2.9\pm0.1$ \\ 

        %%%%%%%%%%%%%%%%%%   
        %%%% New line %%%%
        NGC\,2566 & $\dots$ & $\dots$ & $\dots$ & $\dots$ & 
        NGC\,4941 & $\dots$ & $\dots$ & $\dots$ & $\dots$\\

        %%%%%%%%%%%%%%%%%%   
        %%%% New line %%%%
        NGC\,2835 & $44.6 \pm 0.5$ & $0.3\pm0.0$ & $0.4\pm0.0$ & $\dots$ &
        
        NGC\,5068 & $26.0\pm1.2$ & $\dots$ & $\dots$ & 0.7$_{-0.4}^{+0.0}$ \\ 
        
        %%%%%%%%%%%%%%%%%%   
        %%%% New line %%%%
        NGC\,2903 & $33.3\pm2.4$ & $\dots$ & $0.6\pm0.0$ & $\dots$ & 
        
        NGC\,5134 & $34.2\pm0.3$ & $\dots$ & $\dots$ & $\dots$ \\

        %%%%%%%%%%%%%%%%%%   
        %%%% New line %%%%
        NGC\,3059 & $27.6\pm1.8$ & $\dots$ & $0.4\pm0.0$ & $\dots$ & 
        
        NGC\,5248 & $28.1\pm4.7$ & $\dots$ & $0.9\pm0.0$ & $\dots$ \\

        %%%%%%%%%%%%%%%%%%   
        %%%% New line %%%%
        NGC\,3351 & $\dots$ & $\dots$ & $\dots$ & $\dots$ & 
        
        NGC\,5643 & $57.5\pm2.6$ & $\dots$ & $0.3\pm0.0$ & $5.6\pm 0.1$ \\

        %%%%%%%%%%%%%%%%%%   
        %%%% New line %%%%
        NGC\,3507 & $48.7\pm4.4$ & $0.6\pm0.0$ & $0.9\pm0.0$ & 6.5$^{+0.4}_{-1.6}$ & 
        
        NGC\,6300 & $63.0\pm4.9$ & $\dots$ & $0.9\pm0.1$ & $\dots$ \\
        
        NGC\,3511 & $\dots$ & $\dots$ & $\dots$ & $\dots$ & 
        NGC\,6744 & $28.3\pm1.4$ & $\dots$ & $\dots$ & $\dots$ \\

        NGC\,3626 & $\dots$ & $\dots$ & $\dots$ & $\dots$ & 
        NGC\,7496 & $28.7\pm1.1$ & $0.4\pm0.0$ & $0.7\pm0.0$ & $\dots$ \\ 

        \noalign{\smallskip}
\end{longtable}
}
\tablefoot{{Column (2) contains the pattern speed velocity; Col.~(3) contains the position of the inner ILR (or alternatively, of the unique ILR); Col.~(4) contains the position of the outer ILR, and Col.~(5) contains the position of the OLR. The symbol $\dots$ means the resonance, pattern speed or uncertainty could not be determined.\\}}

%%%%%%%%%%%%%%%%%%%%%%%%%%%%%%%%%%%%%%%%
% List of papers + methods + stats (Appendix):
\newpage
\section{Literature}
\label{Sec:literature_appendix}

{
\begin{longtable}{ccccc}
\caption{Literature comparison \label{table:Appendix-Table_Literature}}\\ 
\noalign{\smallskip}
\hline
\hline
\noalign{\smallskip}
Method & $\cfrac{|\rm CR - \rm CR_{\rm Method}|}{\rm CR}$ & $\cfrac{\rm CR}{\rm CR_{\rm Method}}$ & Abbreviation & Reference \\
(1) & (2) & (3) & (4) & (5)\\ 
\hline
\endfirsthead

\caption{continued}\\

\hline
Method & $\cfrac{|\rm CR - \rm CR_{\rm Method}|}{\rm CR}$ & $\cfrac{\rm CR}{\rm CR_{\rm Method}}$ & Abbreviation & Reference \\
    (1) & (2) & (3) & (4) & (5)\\ 
    
\hline
\endhead
\hline
\endfoot
\hline
\endlastfoot

\noalign{\smallskip}
    Tremaine-Weinberg (CO, H$\alpha$) & $0.16\pm0.08$ & $0.89\pm0.12$ & \citetalias{Williams+21} & \cite{Williams+21} \\ 
    
    Tremaine-Weinberg (stars) & $0.25\pm0.16$ & $0.81\pm0.10$ & \citetalias{Williams+21} & \cite{Williams+21} \\ 
            
    Tremaine-Weinberg (H$\alpha$) & 0.92  & 0.52 & \citetalias{Hernandez+05} & \cite{Hernandez+05}\\

    Offset method & $0.53\pm0.23$ & $2.58\pm0.94$ & \citetalias{Sierra+15} & \cite{Sierra+15}\\

    Phase-reversal & $0.47\pm0.39$ & $0.71\pm0.16$ & \citetalias{Font+14} & \cite{Font+14}\\
            
    Phase-reversal & 0.13 & 0.88 & \citetalias{Pinol-Ferrer+14} & \cite{Pinol-Ferrer+14}\\

    Phase-reversal & 0.88 & 0.53 & \citetalias{Salak+19} & \cite{Salak+19} \\

    Phase-reversal & 0.96 & 0.51 & \citetalias{Ondrechen+89_NGC1097} & \cite{Ondrechen+89_NGC1097} \\

    Phase-reversal & \multirow{2}{*}{-} & \multirow{2}{*}{-} & \multirow{2}{*}{\citetalias{Ondrechen+89_NGC1365}} & \multirow{2}{*}{\cite{Ondrechen+89_NGC1365}} \\
    (unclear if bar or spiral)* & & & & \\
    
    Potential-density phase-shift & $0.32\pm0.26$ & $0.98\pm0.35$ & \citetalias{ButaZhang09} & \cite{ButaZhang09}\\ 

    Potential-density phase-shift & 0.15 & 0.87 & \citetalias{ZhangButa07} & \cite{ZhangButa07}\\ 

    Hydrodynamical simulations & $\dots(NB)$ & $\dots \ (NB)$ & \citetalias{Wada+98} & \cite{Wada+98}\\
    
    Hydrodynamical simulations & $0.66\pm0.42$ & $0.64\pm0.15$ & \citetalias{Rautiainen2008} & \cite{Rautiainen2008}\\ 
    
    Hydrodynamical simulations & $1.08\pm0.62$ & $0.53\pm0.16$ & \citetalias{Lindblad+Kristen1996} & \cite{Lindblad+Kristen1996}\\
    
    % - significa que hay datos de la literatura pero que el CR que tenemos nosotras es de QF = 3

    Hydrodynamical simulations & - & - &\citetalias{LindbladLindbladAthanassoula1996} & \cite{LindbladLindbladAthanassoula1996}\\
    
    Hydrodynamical simulations & 1.03 & 0.49 &  \citetalias{GB+98} & \cite{GB+98}\\
    
    Hydrodynamical simulations & 2.04 & 0.33 & \citetalias{Treuthardt+08} & \cite{Treuthardt+08}\\
    
    Hydrodynamical simulations & $\dots \ (NB)$ & $\dots \ (NB)$ & \citetalias{ColinaWada2000} & \cite{ColinaWada2000}\\

    Hydrodynamical simulations & 0.05 & 0.95 & \citetalias{England1989} & \cite{England1989}\\

    Kinematics & 0.26 & 1.35 & \citetalias{Chemin+03} & \cite{Chemin+03} \\

    Kinematics & 0.27 & 0.79 & \citetalias{Pinol-Ferrer+14} & \cite{Pinol-Ferrer+14}\\

    {Kinematics} & 0.27 & 0.79 & \citetalias{VandeVenFathi2010} & \cite{VandeVenFathi2010} \\
    
    Kinematics & $0.05\pm0.00$ & $1.05\pm0.00$ & \citetalias{Hirota+09} & \cite{Hirota+09} \\

    Kinematics  & - & - & \citetalias{JorsaterVanMoorsel1995} & \cite{JorsaterVanMoorsel1995}\\

    Kinematics & 0.53 & 0.65 & \citetalias{Devereux+92} & \cite{Devereux+92} \\
    
    Torques & 0.11 & 1.12 & \citetalias{Haan+09} & \cite{Haan+09}\\
    
    Torques & 0.9 & 0.53 & \citetalias{GB+09} & \cite{GB+09}\\

    Torques & 0.30 & 1.42 & \citetalias{Casasola+11} & \cite{Casasola+11}\\
    
    Morphology & 0.78 & 0.56 & \citetalias{Elmegreen+96} & \cite{Elmegreen+96}\\

    Morphology & 0.17 & 0.85 & \citetalias{Schmidt+2019} & \cite{Schmidt+2019} \\

    Morphology & 1.06 & 0.49 & \citetalias{Buta1986} & \cite{Buta1986} \\

    Morphology & - & - & \citetalias{Combes14} & \cite{Combes14} \\

    Morphology & 1.00 & 0.50 & \citetalias{Ryder+96} & \cite{Ryder+96} \\

    Morphology & 1.26 & 0.44 & \citetalias{Buta+01} & \cite{Buta+01} \\

    Morphology & 1.49 & 0.40 & \citetalias{Buta2017} & \cite{Buta2017} \\

    Morphology & - & - & \citetalias{Teuben+86} & \cite{Teuben+86} \\

            \noalign{\smallskip}
\end{longtable}
}
\tablefoot{{Column (1) contains the method used to calculate the pattern speed; Col.~(2) contains the quotient $|\rm CR - CR_{\rm Method}|/\rm CR$ together with its standard deviation, while Col.~(3) contains the quotient $\rm CR/\rm  CR_{\rm Method}$ and its standard deviation. If these quotients are marked as $\dots$, it means the value could not be obtained because we have no CR registered, while if there is no standard deviation registered, it means there is only one data point. If, instead, a $-$ is shown, it means we have available data for comparison, but choose not to represent it because $QF = 3$ for that galaxy. Column (4) contains the abbreviation of the reference and Col.~(5) contains the reference to the article.\\}}

\twocolumn

\section{Rotation curves}
\label{sec:appendix-rotation_curves}

As explained in Sect.~\ref{sec:identifying_other_dynamical_resonances}, the choice of the rotation curve is critical when calculating other dynamical resonances such as the Lindblad resonances. 
As seen below, in Figs.~\ref{fig:Appendix-NGC3627_rotation_curves-comparison}-\ref{fig:Appendix-NGC1300_rotation_curves-comparison}, the resonance is subject to large changes in its location. Figure \ref{fig:Appendix-NGC3627_rotation_curves-comparison} represents what we consider a reliable rotation curve. While Figs.~\ref{fig:Appendix-NGC1566_rotation_curves-comparison} and \ref{fig:Appendix-NGC1300_rotation_curves-comparison} represent an intermediate case and a non-reliable case respectively. 

In said figures, four different rotation curve approaches are represented, namely: our nominal choice, from \cite{Lang+20} (first row), an $\arctan$ approach (second row), and two approaches based on MUSE-$M_{\star}$ and MUSE-H$\alpha$ (third and fourth rows respectively). 

The usage of the $\arctan$ approximation relies on the assumption that $V_{\rm rot} = V_0 \cdot (2/\pi) \cdot \arctan{(R/r_t)}$, but this approximation is too smooth and shows large departures from the real rotation curve.

On the one hand, the approaches based on MUSE-$M_{\star}$ and MUSE-H$\alpha$ were obtained following a similar process than the one explained in \cite{Lang+20}, but based on MUSE data (see \citealt{Emsellem+22}). From the MUSE cubes, stellar kinematics were derived using stellar emission lines. On the other hand, the information on ionized gas kinematics was extracted using the H$\alpha$ line. These velocity fields were fitted using a tilted ring approach analogous to \cite{Lang+20}.

\begin{figure}[h!]
    \begin{center}
        \includegraphics[scale=0.24]{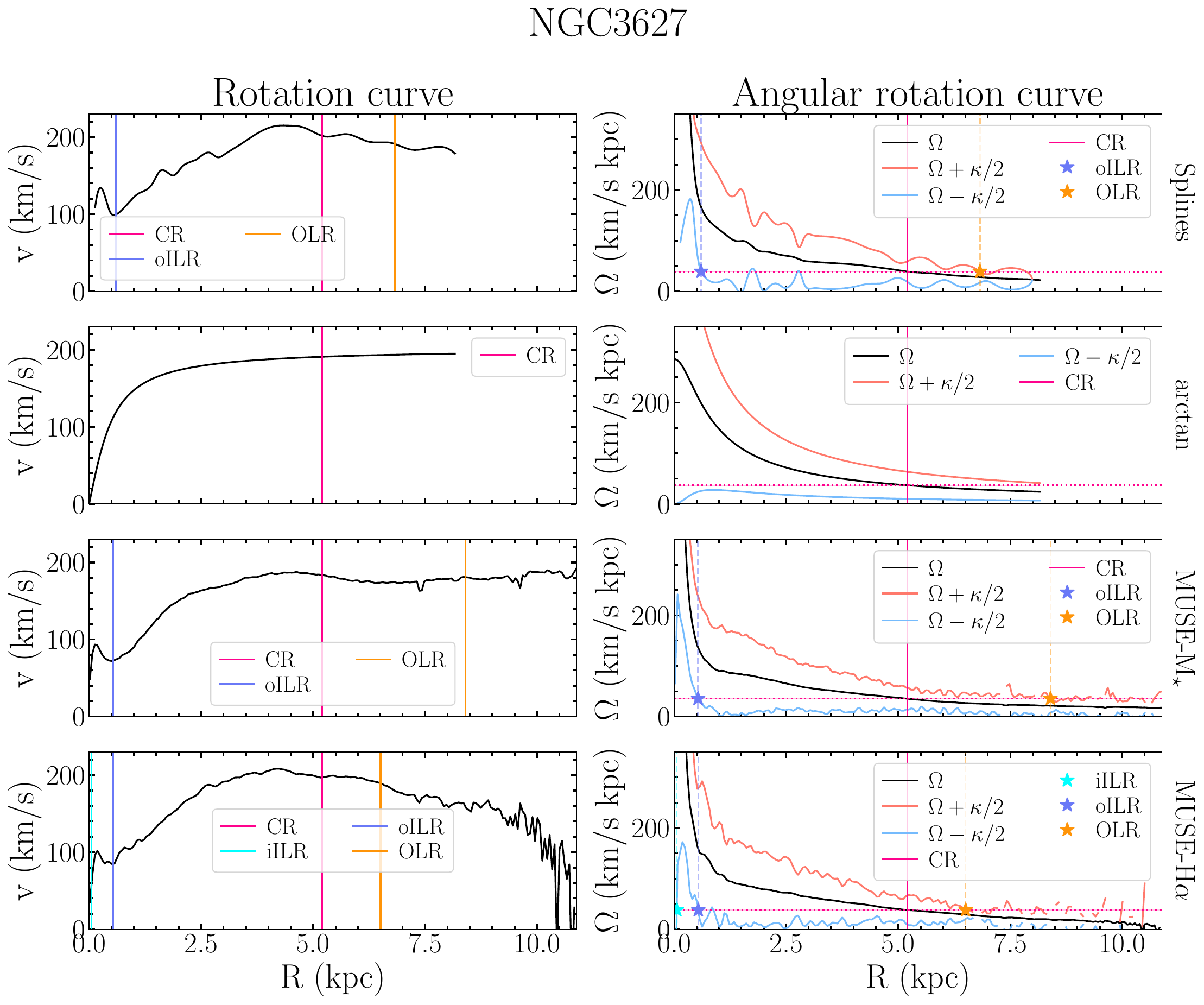}
    \end{center}
    \caption{Different approaches for the rotation curve and subsequent dynamical resonances calculation for \textbf{NGC\,3627}. This is an example of a reliable rotation curve. First row shows the rotation curve obtained with splines model, second row represents the rotation curve obtained using the $\arctan$ approach, while third and fourth rows represent MUSE-$M_{\star}$ and MUSE-H$\alpha$ rotation curves. Solid black line represents the rotation curve (left panels) and the angular rotation curve $\Omega$ (right panels). Solid light pink and blue lines represent both $\Omega+\kappa/2$ and $\Omega-\kappa/2$ respectively. The CR is represented as a vertical pink line, while the OLR is represented in orange and the oILR in blue.}
    \label{fig:Appendix-NGC3627_rotation_curves-comparison}
\end{figure}

\begin{figure}[h!]
    \begin{center}
        \includegraphics[scale=0.24]{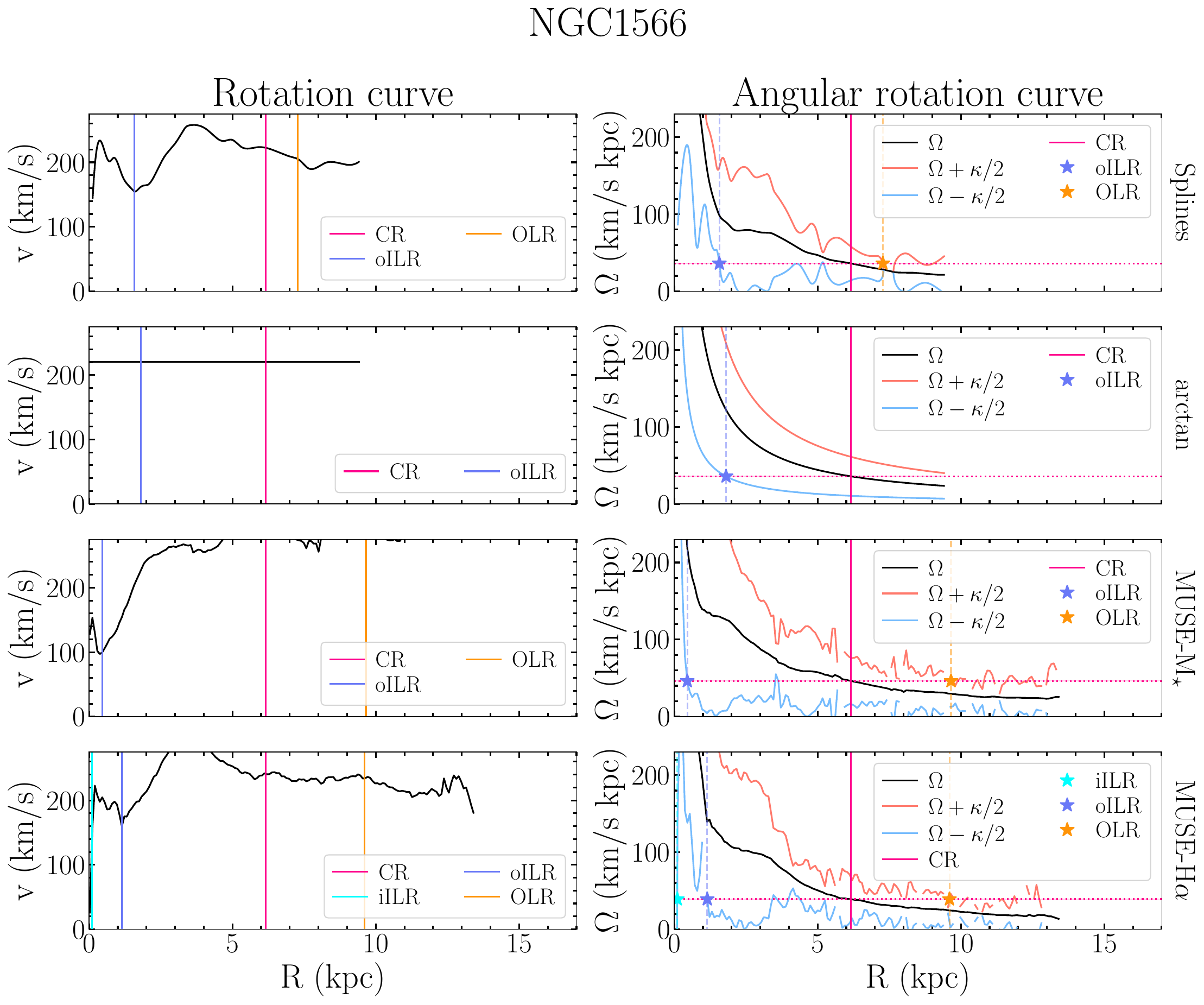}
    \end{center}
    \caption{Different approaches for the rotation curve and subsequent dynamical resonances calculation for \textbf{NGC\,1566}. This is an example of a slightly reliable rotation curve. Legend as in Fig.~\ref{fig:Appendix-NGC3627_rotation_curves-comparison}.} 
    \label{fig:Appendix-NGC1566_rotation_curves-comparison}
\end{figure}

\begin{figure}[h!]
    \begin{center}
        \includegraphics[scale=0.24]{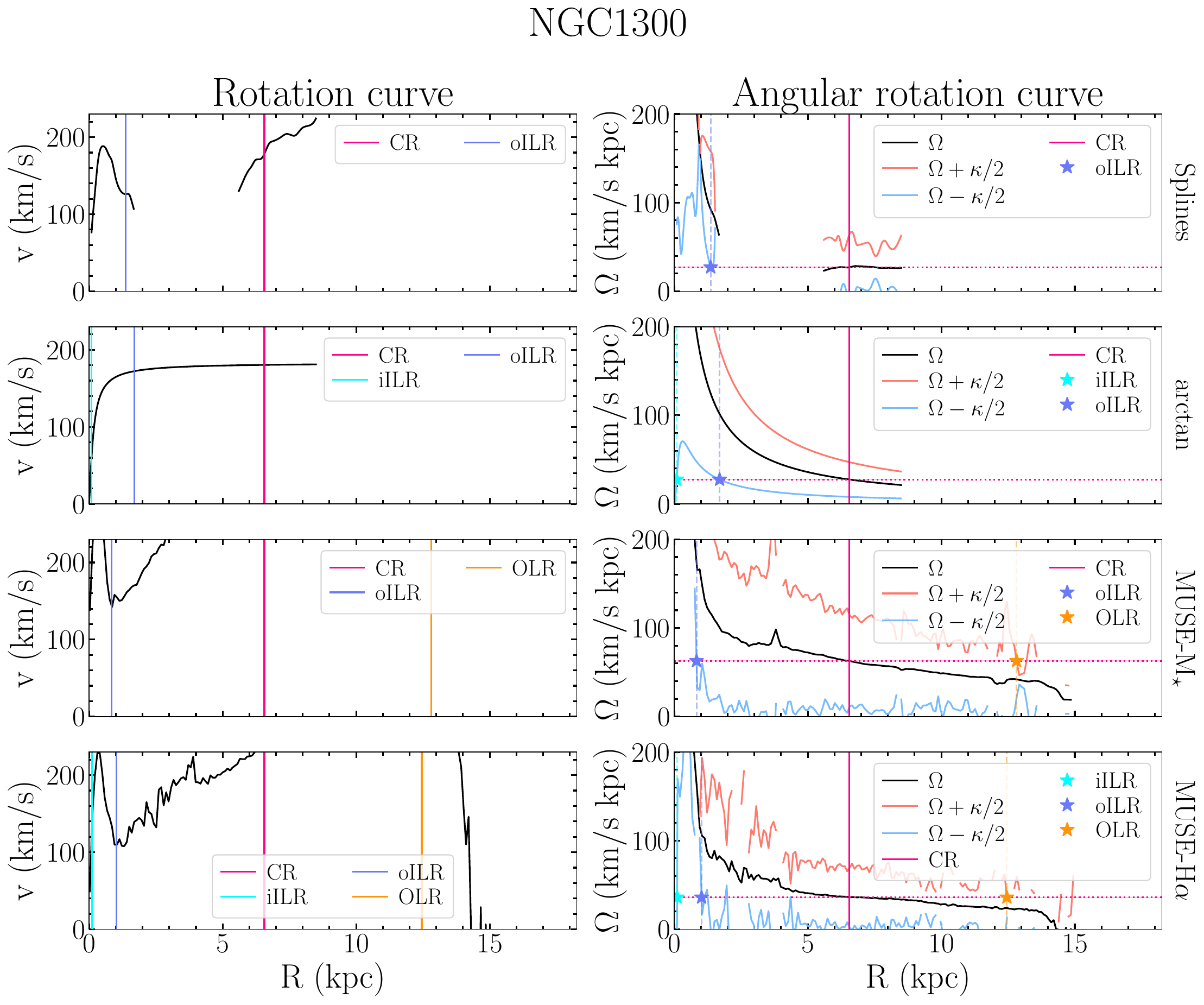}
    \end{center}
    \caption{Different approaches for the rotation curve and subsequent dynamical resonances calculation for \textbf{NGC\,1300}. This is an example of a non-reliable rotation curve. Legend as in Fig.~\ref{fig:Appendix-NGC3627_rotation_curves-comparison}.}
    \label{fig:Appendix-NGC1300_rotation_curves-comparison}
\end{figure}

\FloatBarrier

\section{$\mathcal{R}$}
\label{sec:fancyR-appendix}

\begin{figure*}[h]
\begin{center}
    \includegraphics[trim=0 0 0 0, clip,width=.8\textwidth]{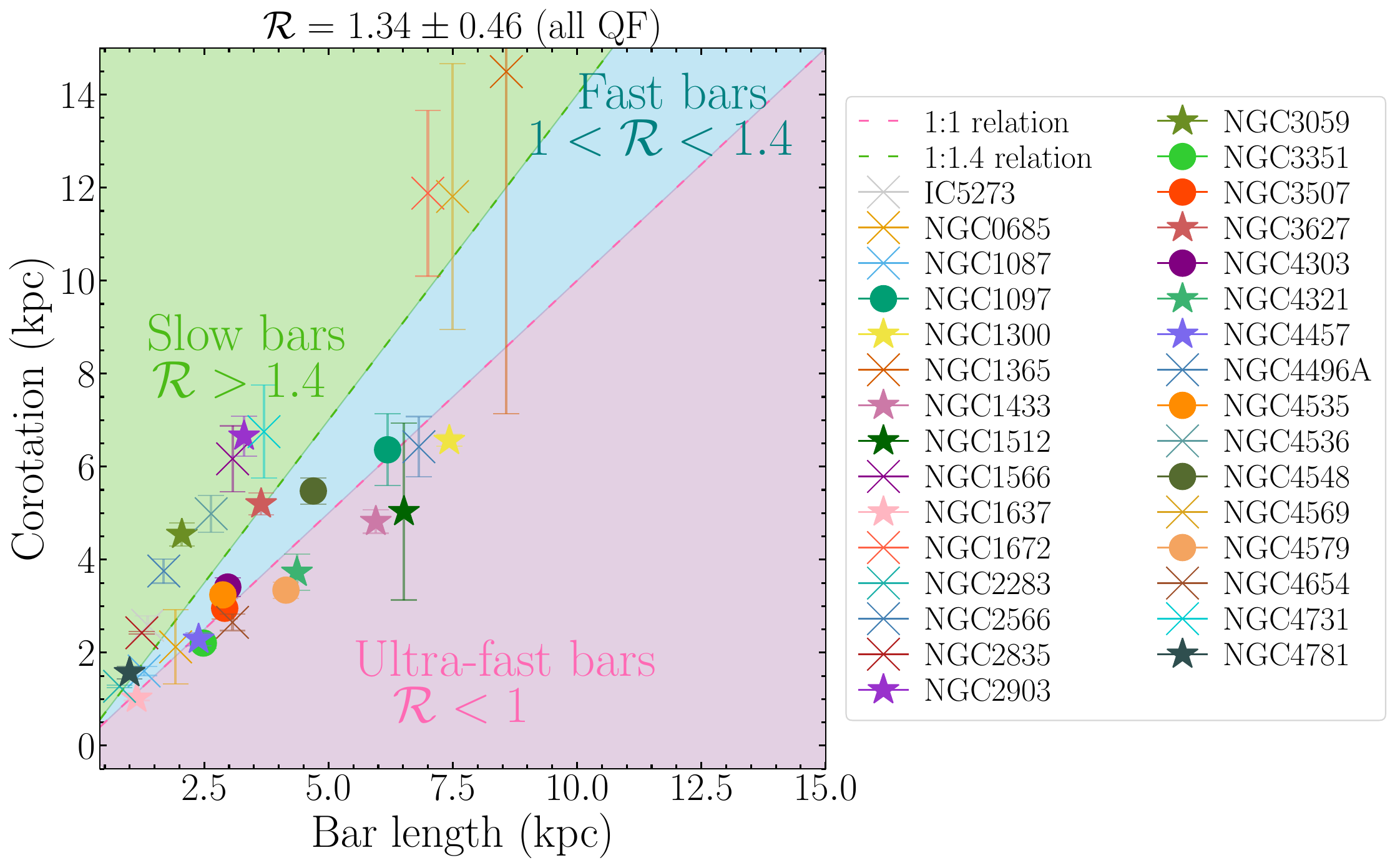}
    \end{center}
    \caption{Slow, fast and ultra-fast bars containing galaxies with all kinds of $QF$. Galaxies with $QF = 3$ are represented with a cross, while $QF = 2$ are represented with stars and galaxies with $QF = 1$ are represented with dots.}
    \label{fig:FancyR-all_galaxies-appendix}
\end{figure*}

In Fig.~\ref{fig:FancyR-all_galaxies-appendix} we have a graphic similar to Fig.~\ref{fig:QF1&2}, but including all galaxies from the sample, regardless of their $QF$. 

\FloatBarrier

%%%%%%%%%%%%%%%%%%%%%%%%%%%%%%%%%%%%%%%%%%%%
%%% Uncertainties section (Appendix):
\section{Uncertainties}
\label{sec:AppendixE_uncertainties}

\begin{figure}[h]
\begin{center}
    \includegraphics[trim=0 0 0 0, clip,width=0.45\textwidth]{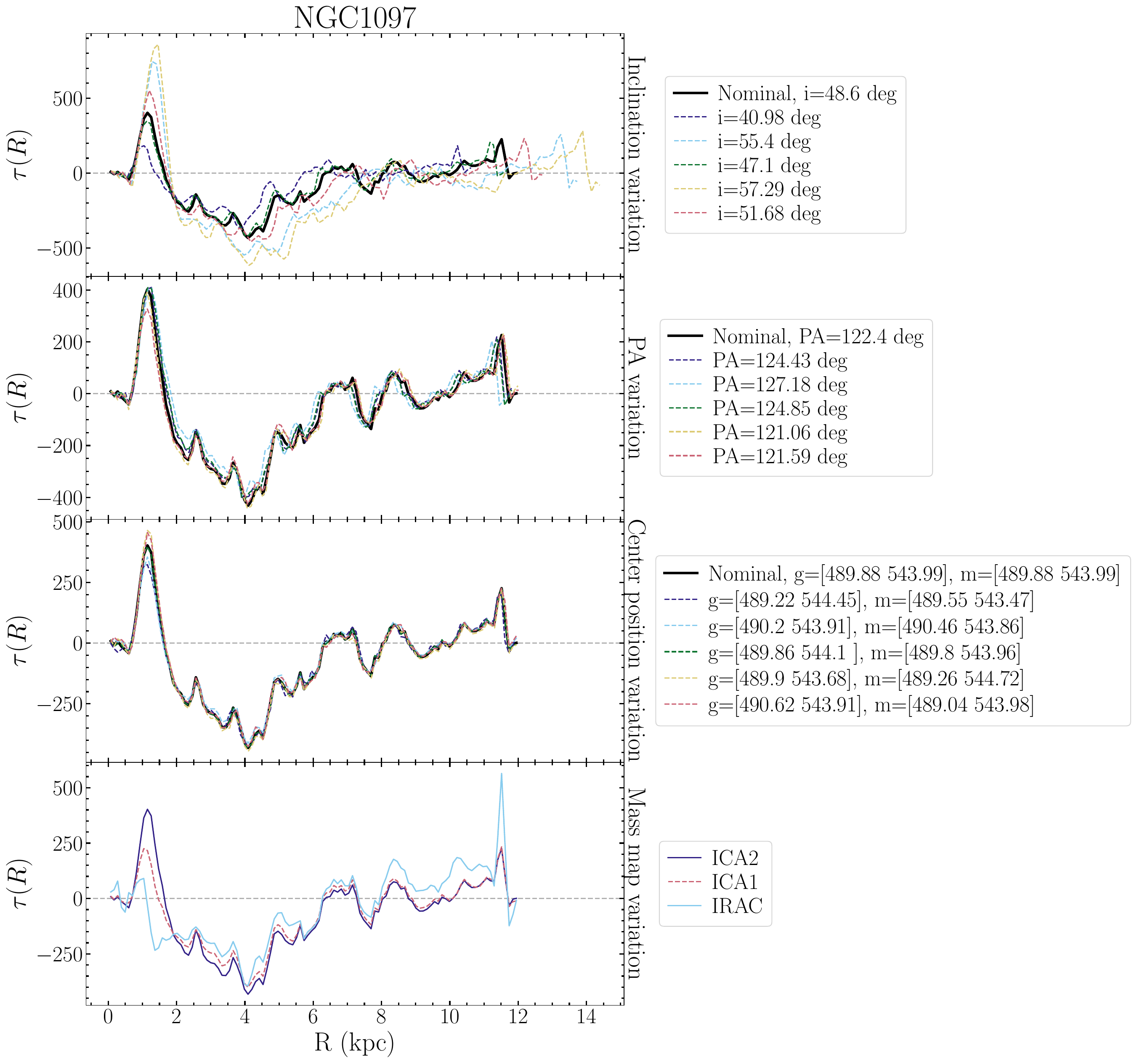}
    \end{center}
    \caption{\textbf{NGC\,1097} torque profile variation with different input parameters. Upper left panel represents the profiles obtained while varying the $i$ (dashed lines) and the nominal profile (solid black line). Idem for upper right panel (PA variation) and lower left panel (center position variation). In the lower right panel the stellar mass map variation can be found: each line represents the nominal torque profile for each mass map.} % Code --> "pruebas" folder
    \label{fig:porfiles_unc}
\end{figure}

Figure \ref{fig:porfiles_unc} shows the differences in torque profiles due to the variation of different parameters ($i$, PA, center position and stellar mass map). In order to compute these differences, we have followed the bootstrap process explained in Sect.~\ref{sec:bootstrap}. 

\FloatBarrier

%%%%%%%%%%%%%%%%%%%%%%%%%%%%%%%%%%%%%%%%%%%%%
%% Literature comparison section (Appendix):
\section{Comparison with other methods from the literature}
\label{sec:literature_comparison-Appendix}
\begin{figure}[h!]
\begin{center}
    \includegraphics[trim=0 0 0 0, clip,width=.2\textwidth]{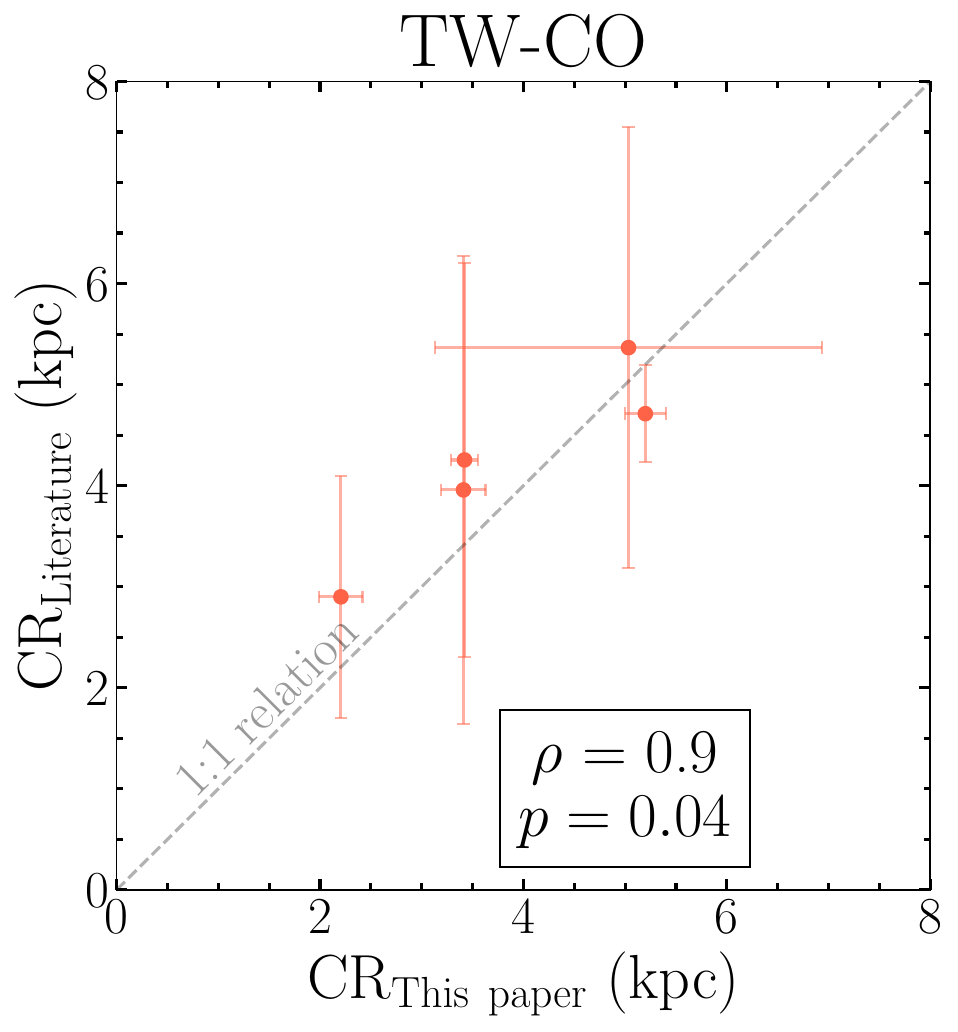}
    \includegraphics[trim=0 0 0 0, clip,width=.2\textwidth]{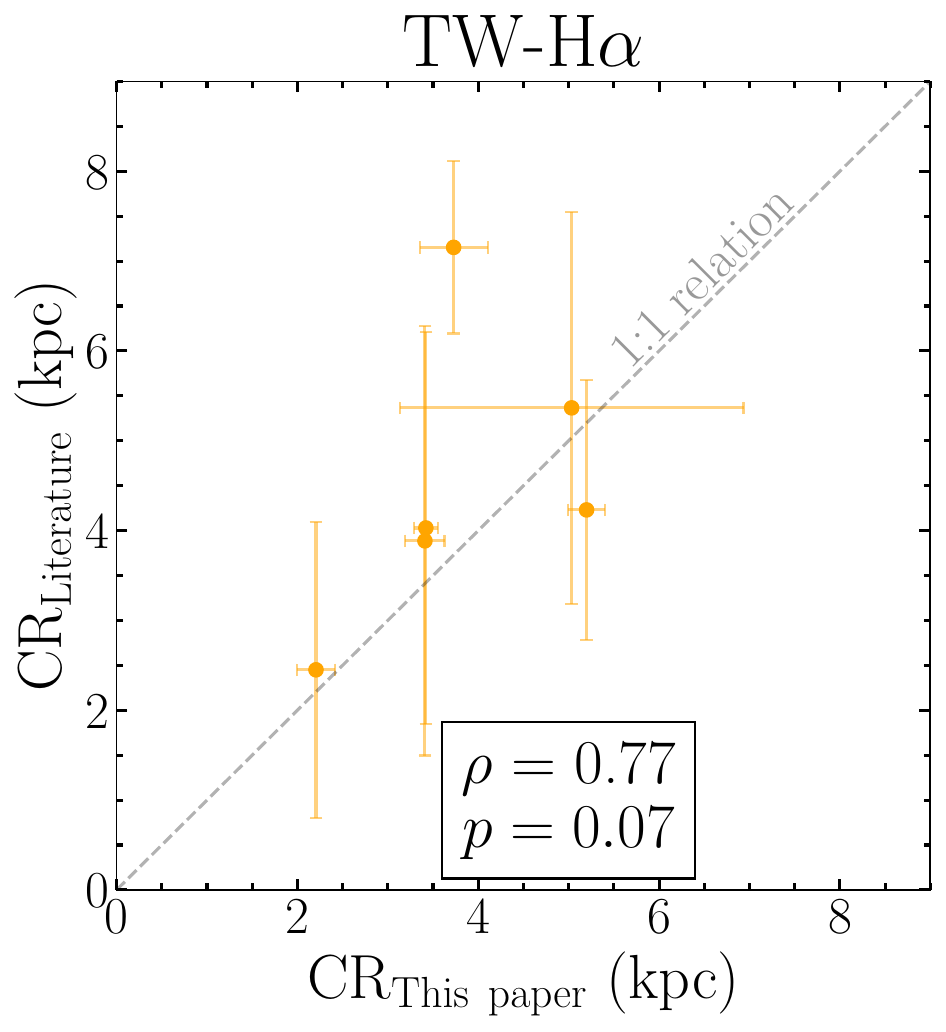}
    \includegraphics[trim=0 0 0 0, clip,width=.2\textwidth]{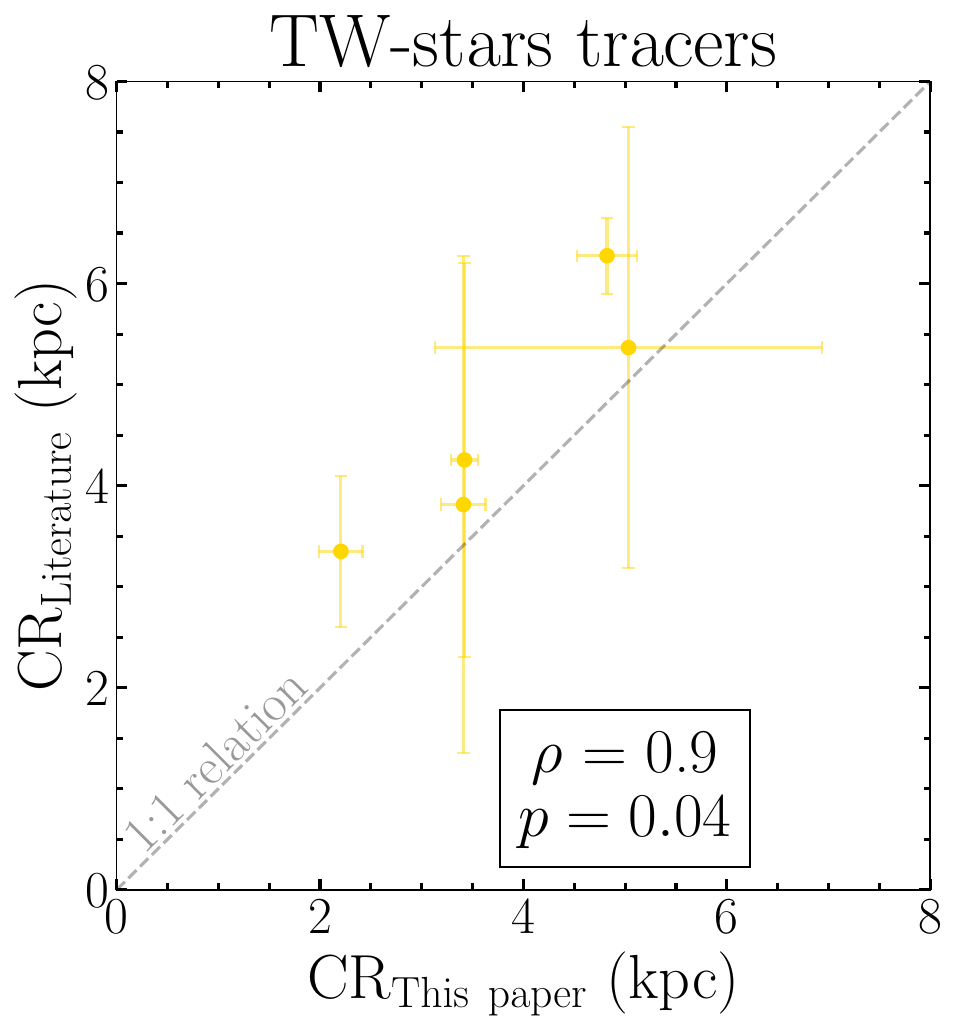}
    \includegraphics[trim=0 0 0 0, clip,width=.2\textwidth]{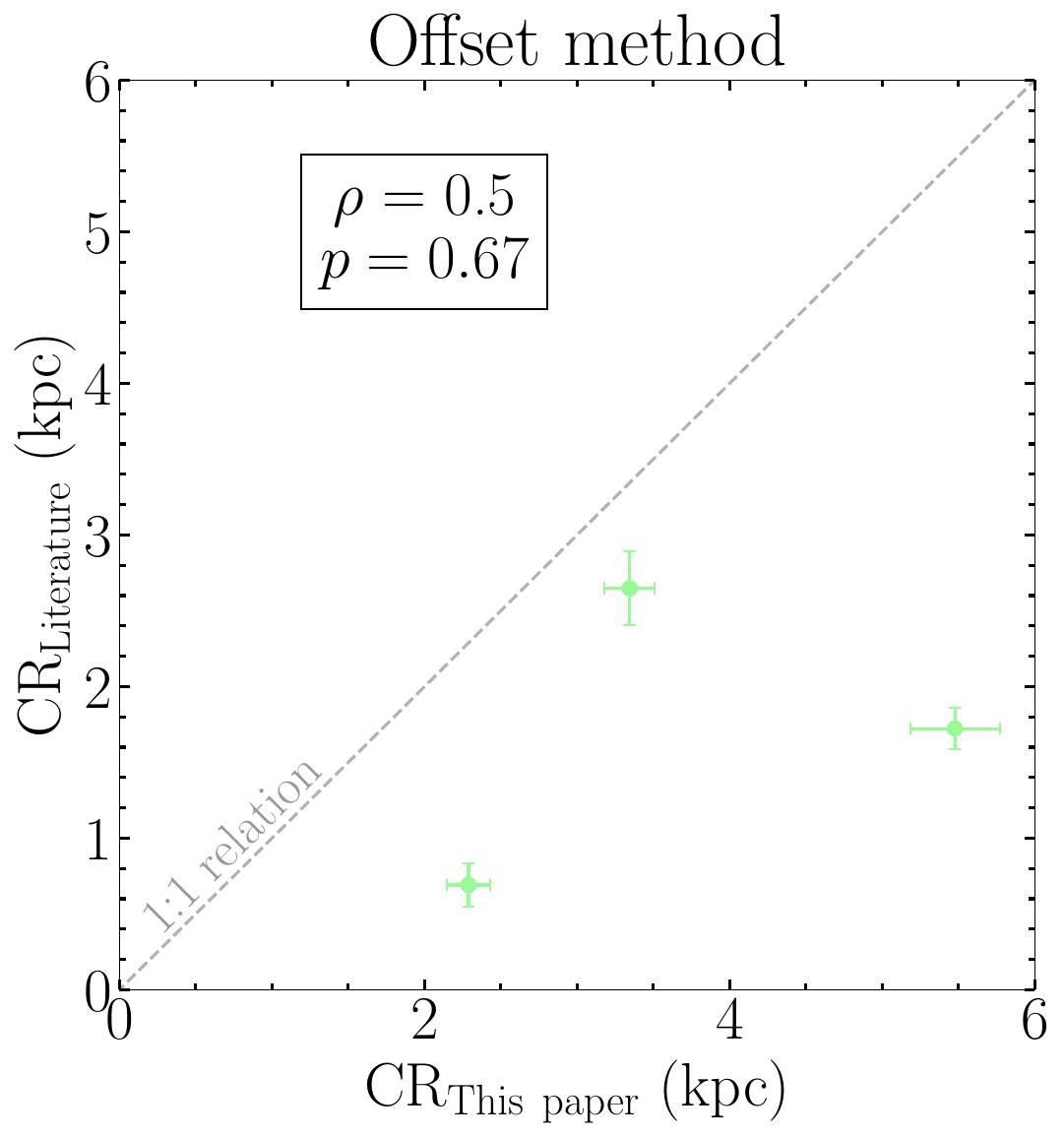}
    \end{center}
    \caption{\textbf{Literature comparison}. From left to right (and top to bottom): TW, offset method, phase reversal method, potential-density phase-shift method, hydrodynamical simulations, kinematics (including Canzian test), torques and morphology.}
    \label{fig:Appendix-Literature_comparison2}
\end{figure}

An individual comparison of the results obtained with the different methods from the literature to our results can be found in Fig.~\ref{fig:Appendix-Literature_comparison2}.

\begin{figure}[h!]
\begin{center}
    \includegraphics[trim=0 0 0 0, clip,width=.2\textwidth]{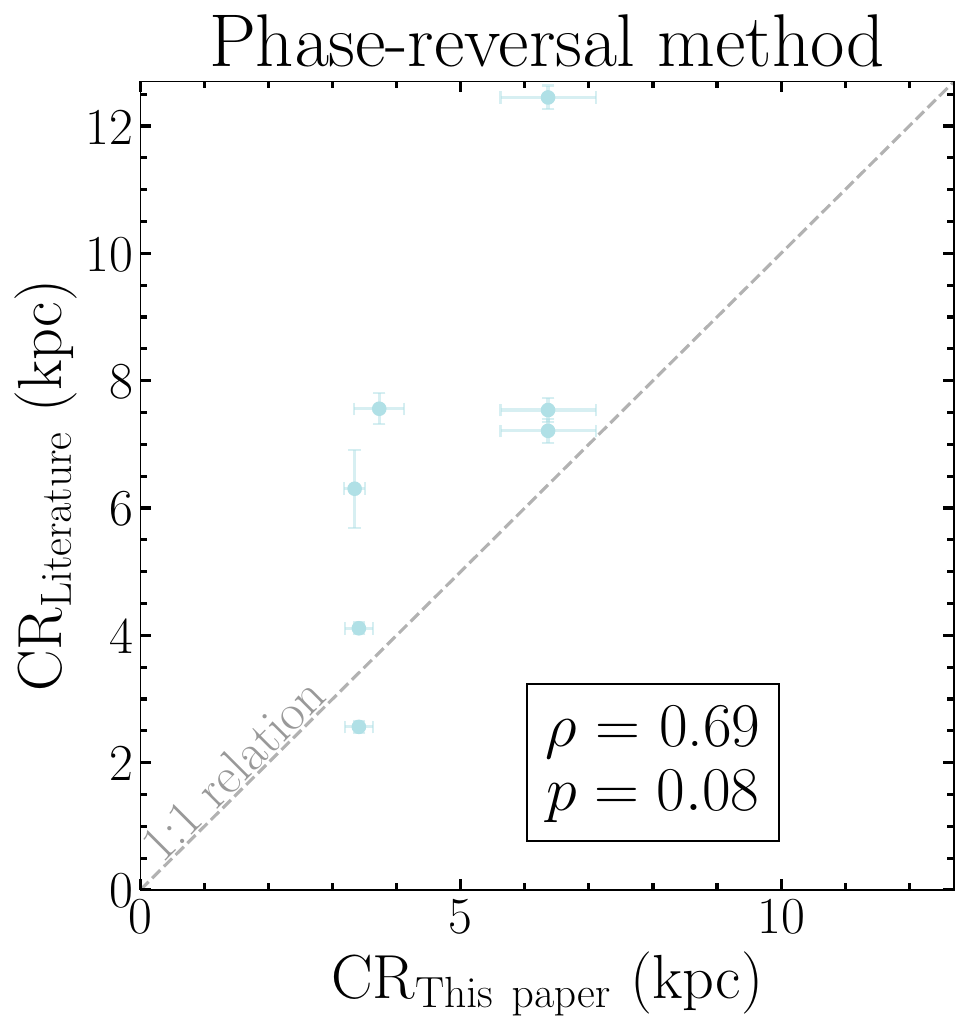}
    \includegraphics[trim=0 0 0 0, clip,width=.2\textwidth]{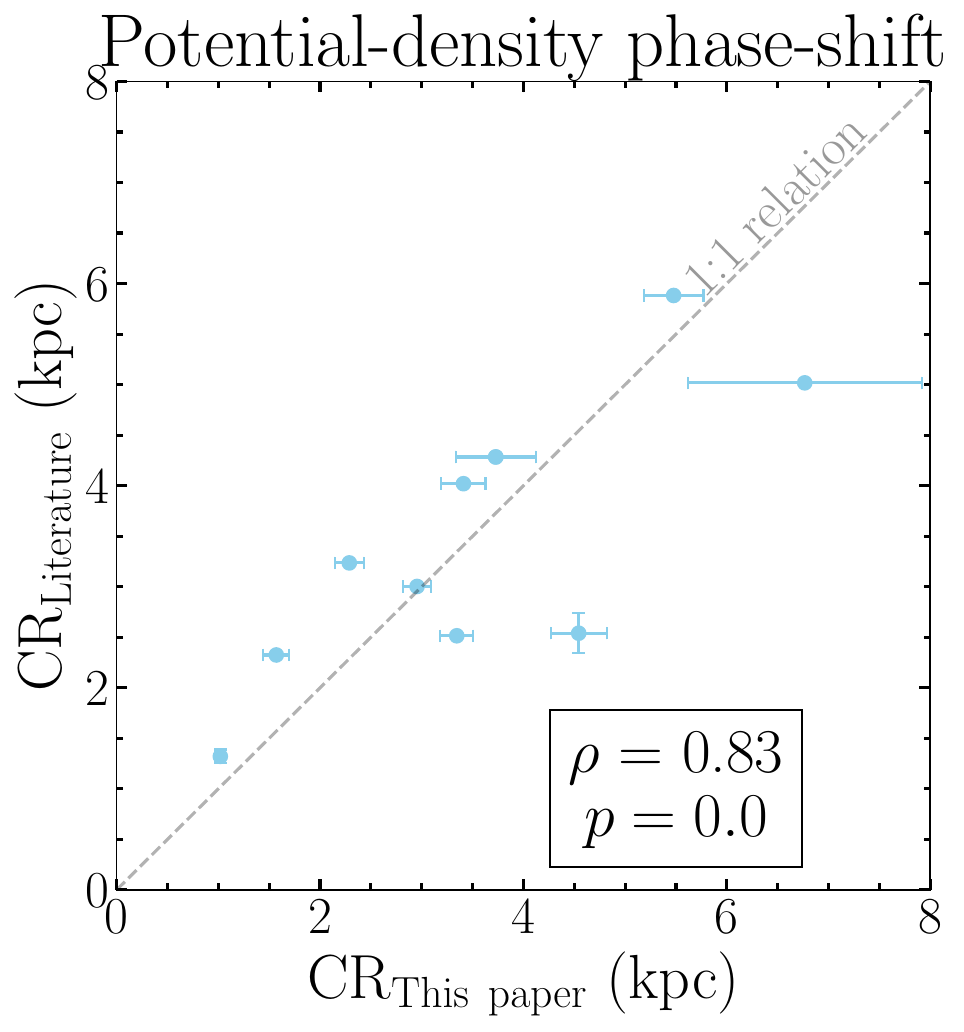}
    \includegraphics[trim=0 0 0 0, clip,width=.2\textwidth]{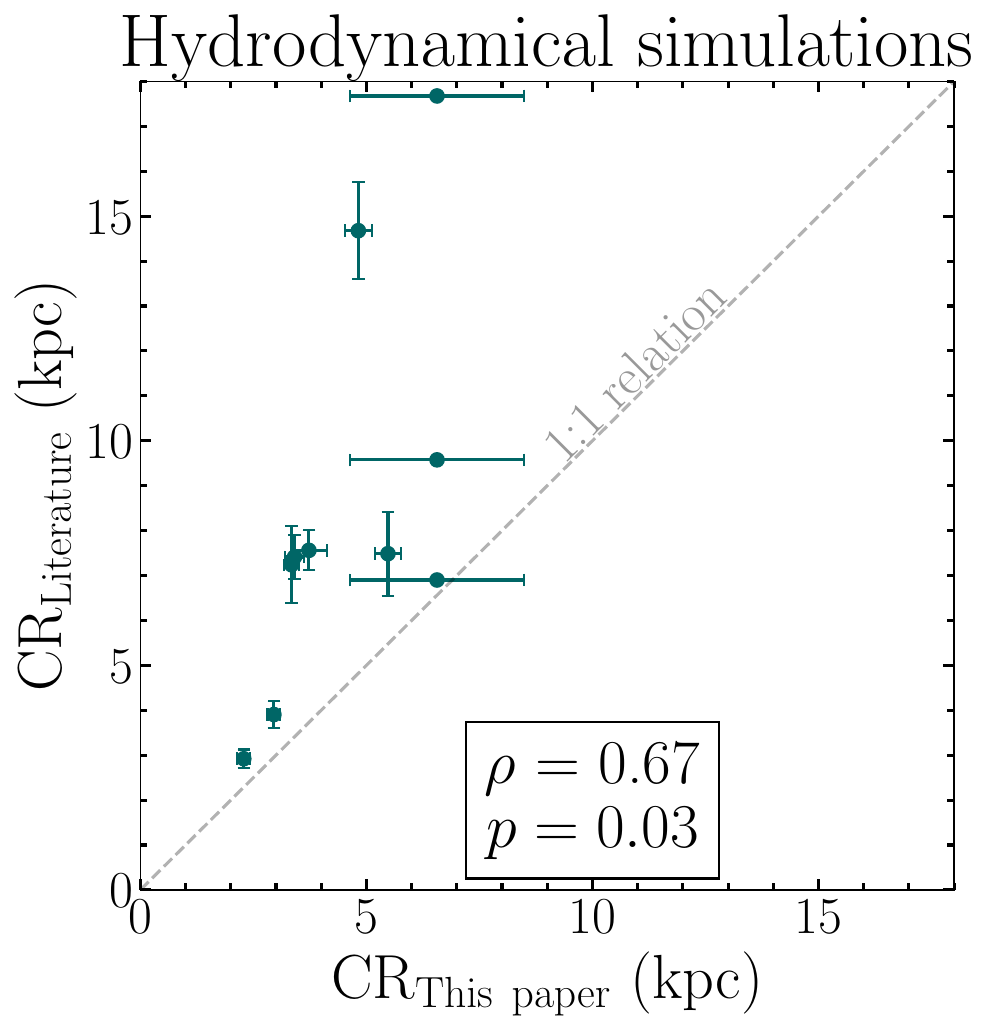}
    \includegraphics[trim=0 0 0 0, clip,width=.2\textwidth]{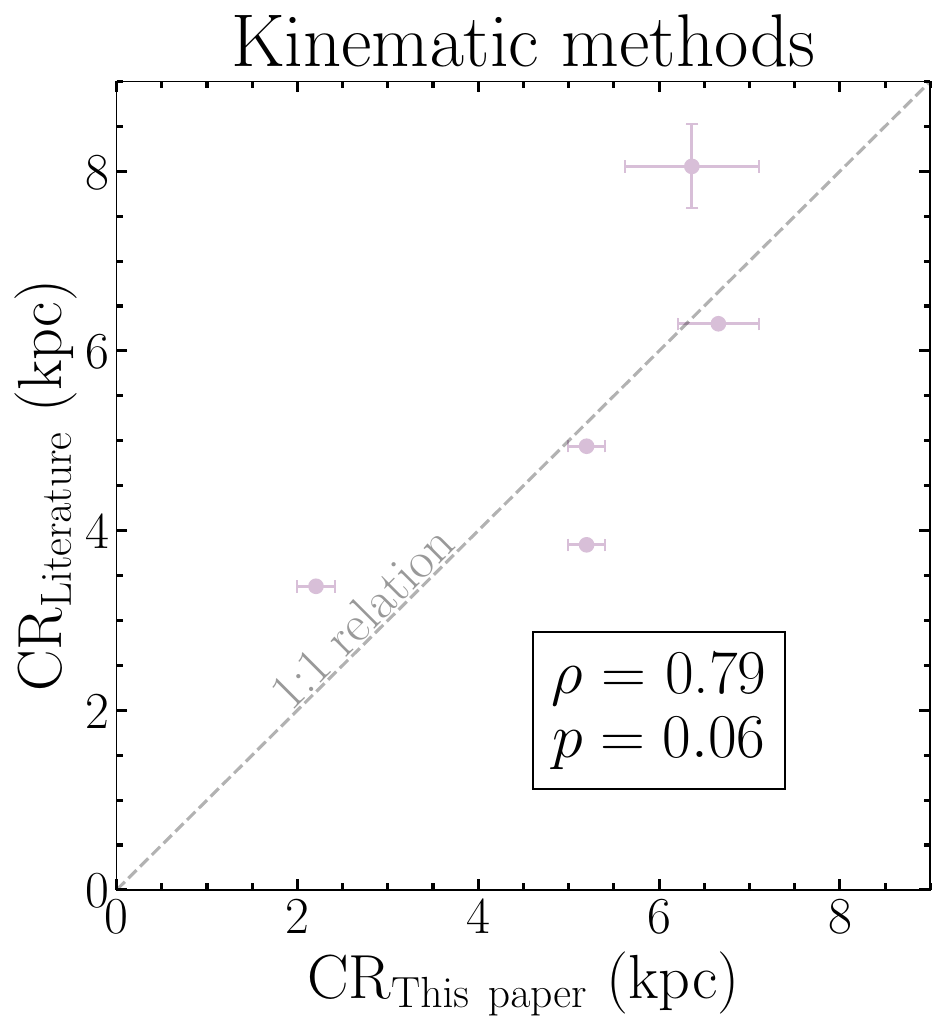}
    \includegraphics[trim=0 0 0 0, clip,width=.2\textwidth]{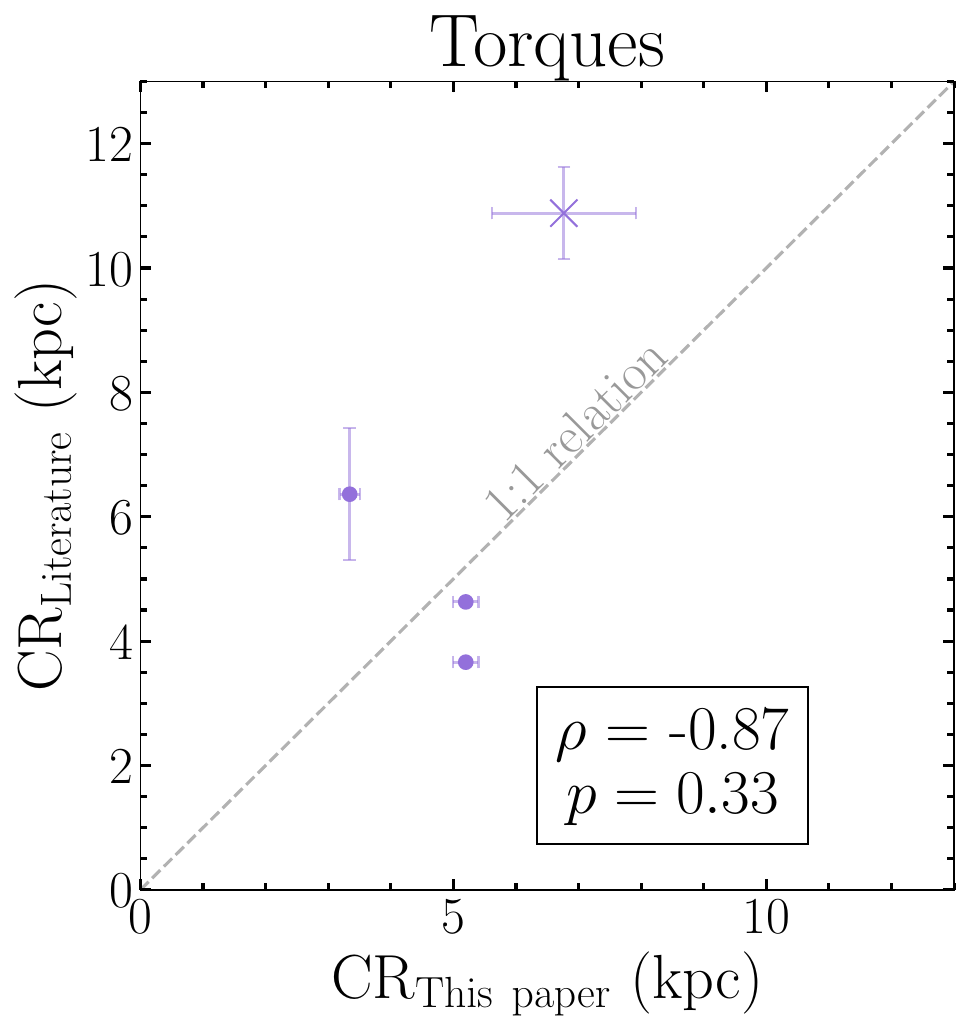}
    \includegraphics[trim=0 0 0 0, clip,width=.2\textwidth]{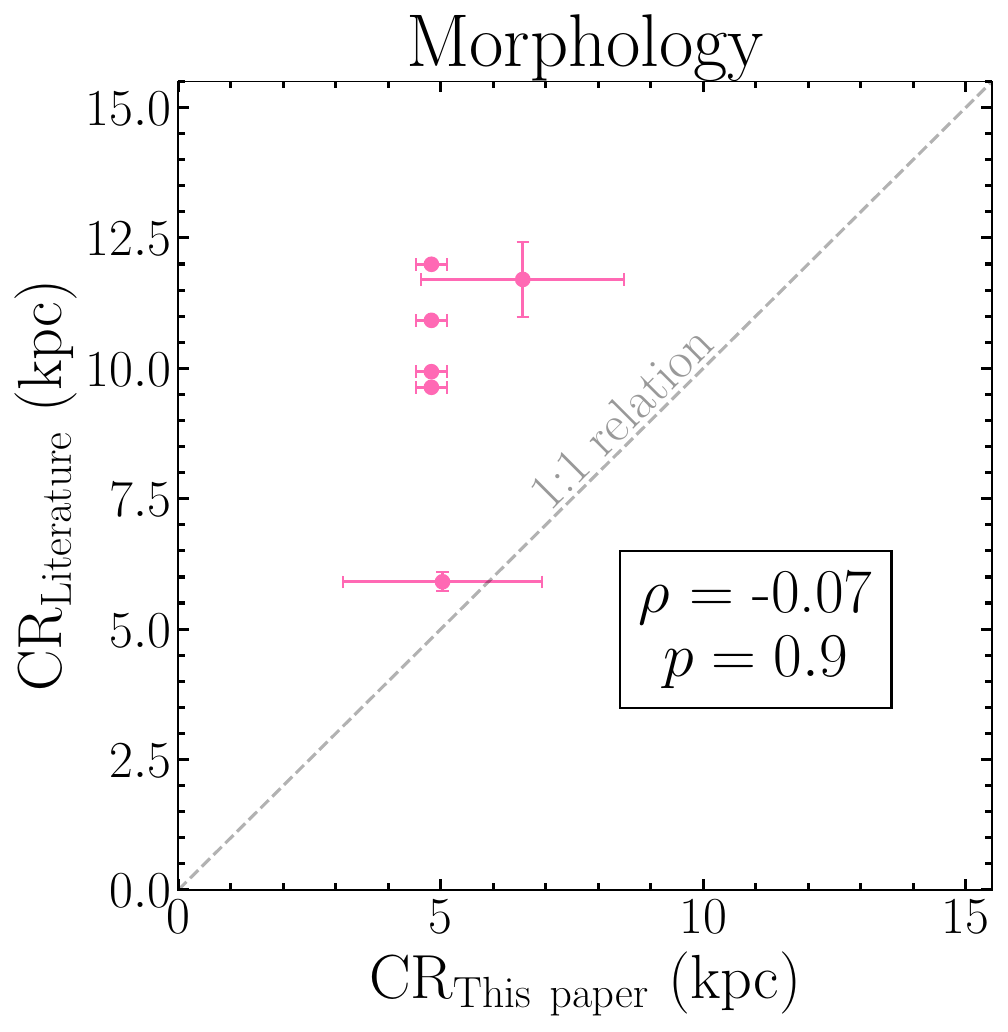}
    \end{center}
    \caption{Fig.~\ref{fig:Appendix-Literature_comparison2} continued.}
    \label{fig:Appendix-Literature_comparison}
\end{figure}

\FloatBarrier

\begin{figure*}[h!]
\begin{center}
    \includegraphics[trim=0 0 0 0, clip,width=1\textwidth]{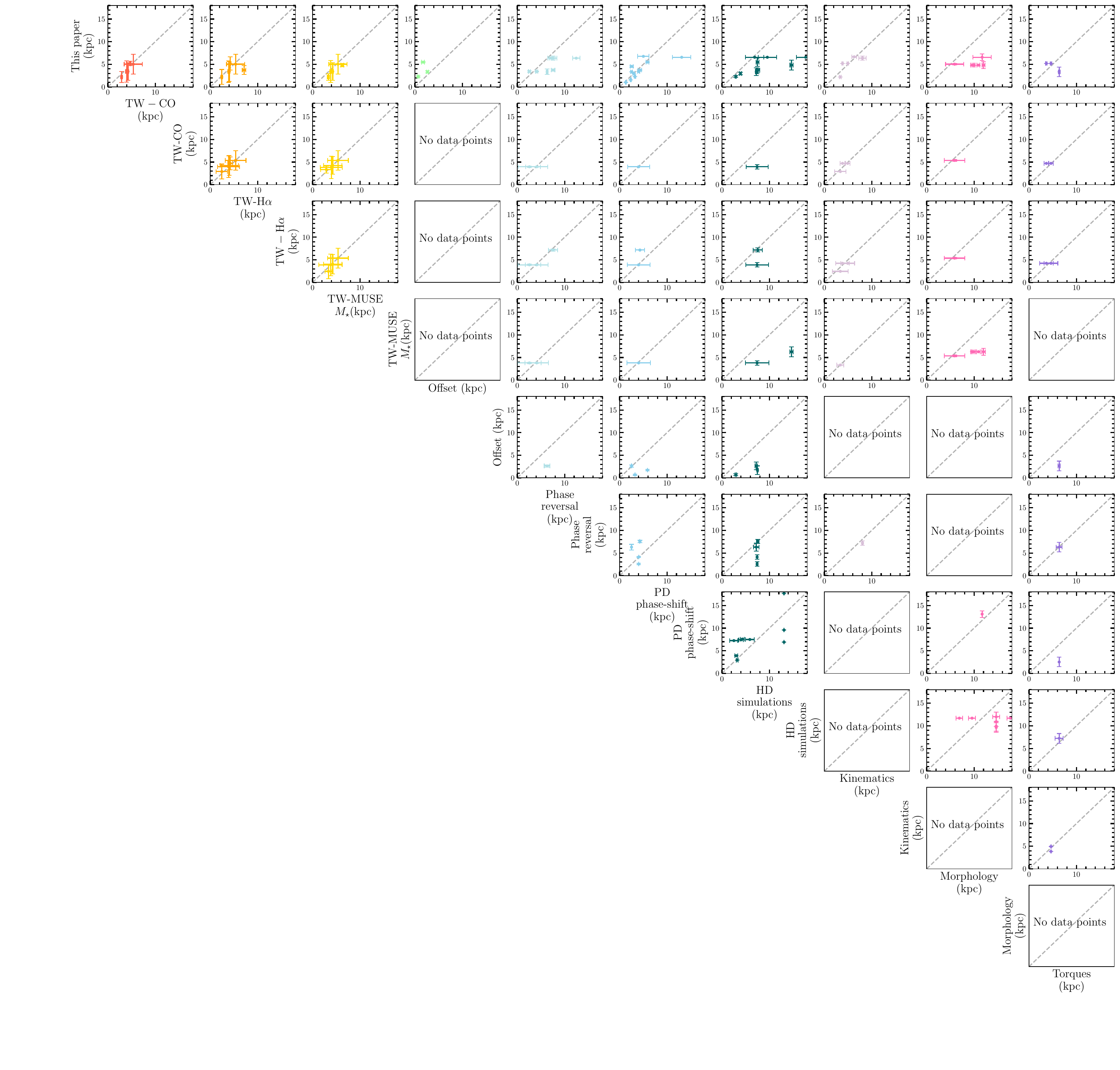}
    \end{center}
    \caption{Statistics on the comparison of all the gathered CR locations.}
    \label{fig:Appendix-all_vs_all}
\end{figure*}

A comparison of all results (including our results and the ones from the literature) to all results is represented in Fig.~\ref{fig:Appendix-all_vs_all}. 

Finally, the statistics of the comparison in Fig.~\ref{fig:Appendix-all_vs_all} can be found in table format in Table \ref{table:Appendix-all_vs_all}. 

\onecolumn
\begin{landscape}
\begin{center}
\begin{longtable}{|c|c|c|c|c|c|c|c|c|c|c|c|}
\caption{All methods compared to all methods \label{table:Appendix-all_vs_all}}\\ 
\hline%\hline 
         & {This} & {TW-} & {TW-} & {TW-} & \multirow{2}{*}{Offset} & {Phase} & {PD} &  \multirow{2}{*}{Simulations} & \multirow{2}{*}{Kinematics} & \multirow{2}{*}{Morphology} & \multirow{2}{*}{Torques} \\ 
        & work & CO & H$\alpha$ & M$_{\star}$ & & reversal & phase-shift & & & & \\
        \hline
        
        This & 1 & 0.89 & 0.77 & 0.89 & 0.50 & 0.69 & 0.83 & 0.67 & 0.79 & -0.07 & -0.87 \\ 
        work & 0 & 0.037 & 0.070 & 0.037 & 0.667 & 0.080 & <0.001 & 0.034 & 0.059 & 0.899 & 0.333 \\ \hline
        
        TW- & 0.89 & 1 & 1 & 1 & - & - & - & - & 0.87 & - & - \\ 
        CO & 0.037 & 0 & 0 & 0 & - & - & - & - & 0.333 & - & - \\ \hline
        
        TW- & 0.77 & 1 & 1 & 1 & - & 0.87 & 1 & - & 0.87 & - & - \\
        H$\alpha$ & 0.070 & 0 & 0 & 0 & - & 0.333 & 0 & - & 0.333 & - & - \\ \hline
        
        TW- & 0.89 & 1 & 1 & 1 & - & - & - & - & - & 0.71 & - \\ 
        M$_{\star}$ & 0.037 & 0 & 0 & 0 & - & - & - & - & - & 0.182 & - \\ \hline
        
        \multirow{2}{*}{Offset} & 0.50 & - & - & - & 1 & - & -0.50 & 0.50 & - & - & - \\ 
        ~  & 0.667 & - & - & - & 0 & - & 0.667 & 0.667 & - & - & - \\ \hline
        
        Phase & 0.69 & - & 0.87 & - & - & 1 & 0.65 & 0.32 & - & - & - \\ 
        reversal & 0.080 & - & 0.333 & - & - & 0 & 0.236 & 0.684 & - & - & - \\ \hline
        
        PD & 0.83 & - & 1 & - & -0.50 & 0.65 & 1 & 0.63 & - & - & - \\ 
        phase-shift & <0.001 & - & 0 & - & 0.667 & 0.236 & 0 & 0.051 & - & - & - \\ \hline
        
        \multirow{2}{*}{Simulations} & 0.67 & - & - & - & 0.50 & 0.32 & 0.63 & 1 & - & -0.14 & - \\ 
        ~ & 0.034 & - & - & - & 0.667 & 0.684 & 0.051 & 0 & - & 0.759 & - \\ \hline
        
        \multirow{2}{*}{Kinematics} & 0.79 & 0.87 & 0.87 & - & - & - & - & - & 1 & - & - \\ 
        ~ & 0.059 & 0.333 & 0.333 & - & - & - & - & - & 0 & - & - \\ \hline
        
        \multirow{2}{*}{Morphology} & -0.07 & - & - & 0.71 & - & - & - & -0.14 & - & 1 & - \\ 
        ~ & 0.899 & - & - & 0.182 & - & - & - & 0.759 & - & 0 & - \\ \hline
        
        \multirow{2}{*}{Torques} & -0.87 & - & - & - & - & - & - & - & - & - & 1\\ 
        ~ & 0.333 & - & - & - & - & - & - & - & - & - & 0 \\ \hline
\end{longtable}
\tablefoot{{Statistics on the comparison of all the CR locations. The upper row of each method represents the Spearman rank coefficient $\rho$, while the lower row of each method contains its associated $p\rm - value$. For clarity, we include under the symbol $-$ the cases where we have: only one point in common, only two points in common (therefore always correlated), or no data at all.\\}}
\end{center}
\end{landscape}

\twocolumn

\section{$R_{\rm CR}$ and literature}
\label{sec:Appendix-All_galaxies-with_CR}

\begin{figure*}[h!]
\begin{center}
    \includegraphics[trim=0 0 0 0, clip,width=1\textwidth]{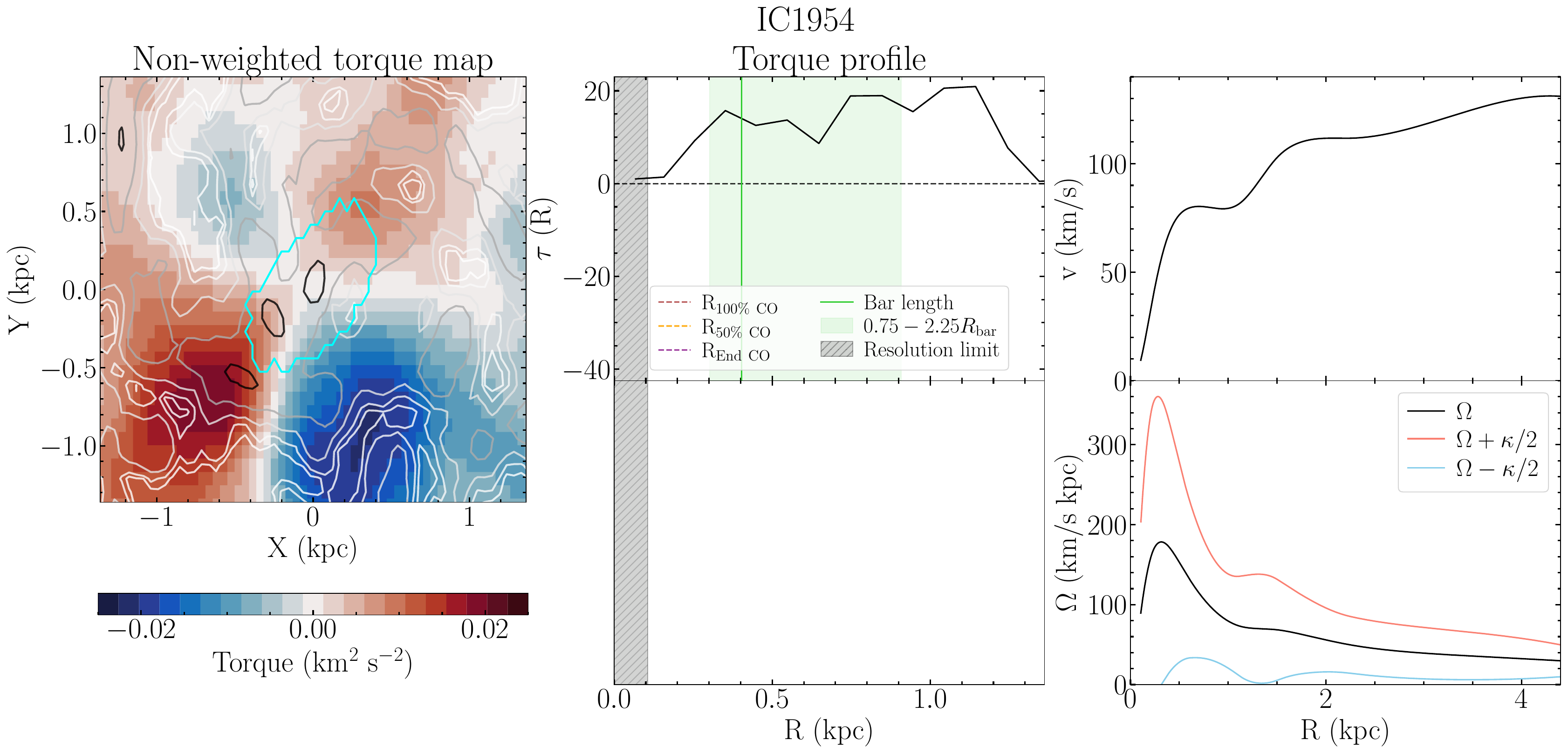}
    \end{center}

    \caption{\textbf{IC\,1954} (SB, $QF=3$) non-weighted deprojected torque map (left panel), torque profile (upper central panel), comparison with values from the literature (lower central panel), rotation curve (upper right panel), and angular rotation curve (lower right panel). The cyan-contoured ellipse indicates the bar extent. We show contours corresponding to $[5\sigma, 15\sigma, 45\sigma, ..., 0.9\sigma_{\rm max}]$ in the CO map, where $\sigma$ is the mean value of the gas map and $\sigma_{\rm max}$ its maximum value. CR (if determined) is represented as a vertical pink line, together with its uncertainties (pink shaded area). This is a statistical uncertainty due to bootstrapping for $i$, PA and center position (see Sect.~\ref{sec:bootstrap}). Solid green line represents the bar length (from \citealt{Querejeta+21}), while shaded green region represents the region where we search for the CR. Brown dashed line represents the radius at which the coverage of CO starts to be non-uniform ($R_{\rm 100\% \ CO}$ in Table~\ref{table:complete_sample}), orange dashed line is the radius at which the coverage of CO is uniform about 50\% ($R_{\rm 50\% \ CO}$ in Table~\ref{table:complete_sample}), and purple dashed line represents the end of CO coverage ($R_{\rm End \ CO}$ in Table~\ref{table:complete_sample}). Teal dashed line represents the nuclear bar registered by \cite{Querejeta+21} (if detected). The shaded gray region shown in this graphic represents the inner region inside which we cannot say anything on $\tau(R)$ due to the limited spatial resolution of our observations. In addition, on the lower central panel, each dot represents a different measure of the CR (of the main bar) from literature. Finally, on the upper right panel we show the rotation curve from \cite{Lang+20}, while on the lower right panel we represent in black solid line the angular rotation curve ($\Omega$) and the curves derived as $\Omega \pm \kappa/2$ (light pink and blue curves), together with the derived pattern speed ($\Omega_p$) and its uncertainties, shown as a dotted pink line and a pink-filled region. In this panel we also represent the Lindblad Resonances if present, as two shades of blue (iILR and oILR) and orange (OLR) dotted lines.} 
    \label{fig:Appendix-All_galaxies-IC1954}
\end{figure*}

\begin{figure*}[h!]
\begin{center}
    \includegraphics[trim=0 0 0 0, clip,width=1\textwidth]{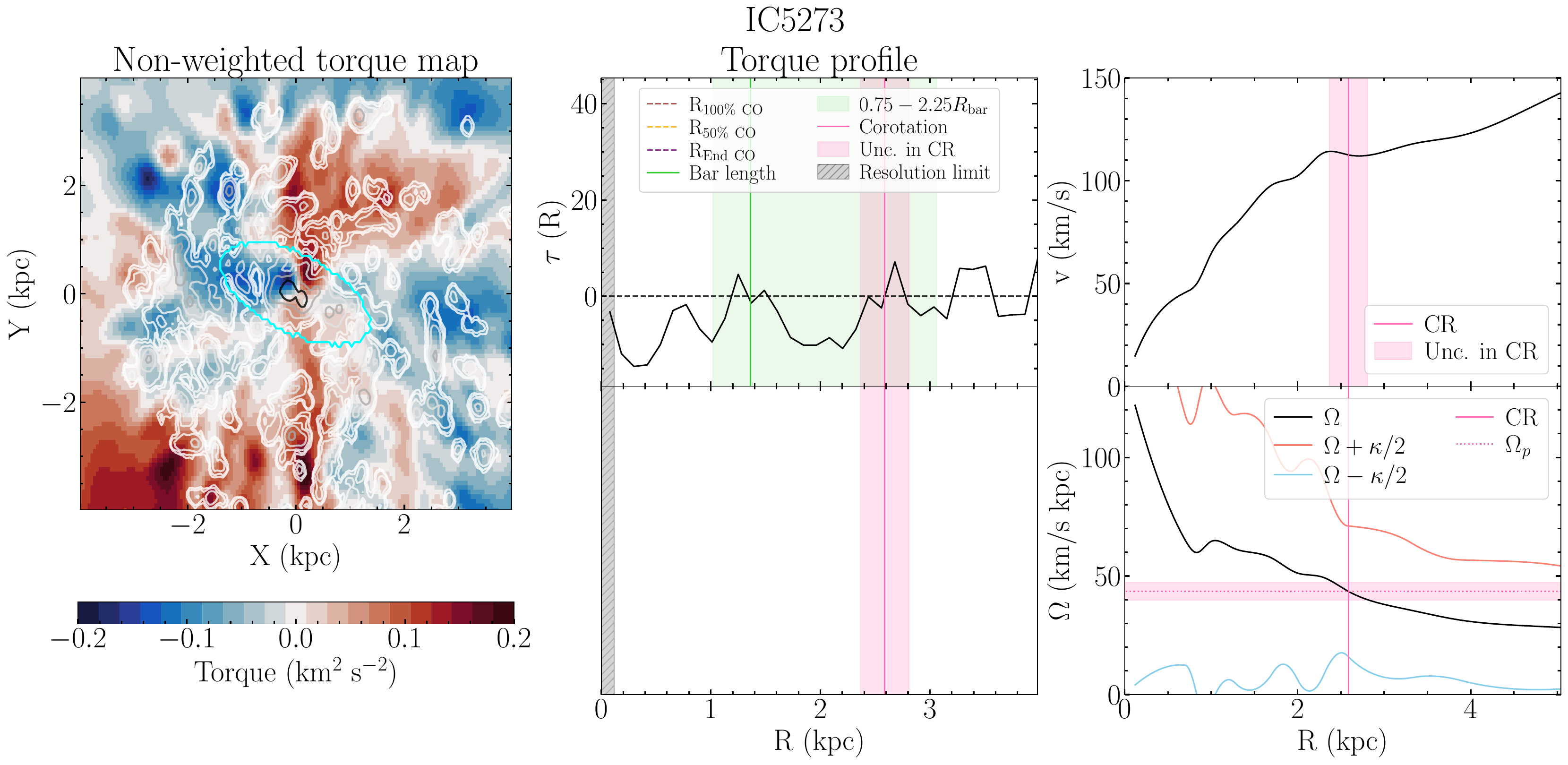}
    \end{center}
    \caption{\textbf{IC\,5273} (SB, $QF=3$). Legend as in Figure \ref{fig:Appendix-All_galaxies-IC1954}.}%This galaxy has been marked as $QF=3$ because we found little gas inside the bar. }
    \label{fig:Appendix-All_galaxies-IC5273}
\end{figure*}

\begin{figure*}[h!]
    \begin{center}
        \includegraphics[trim=0 0 0 0, clip,width=1\textwidth]{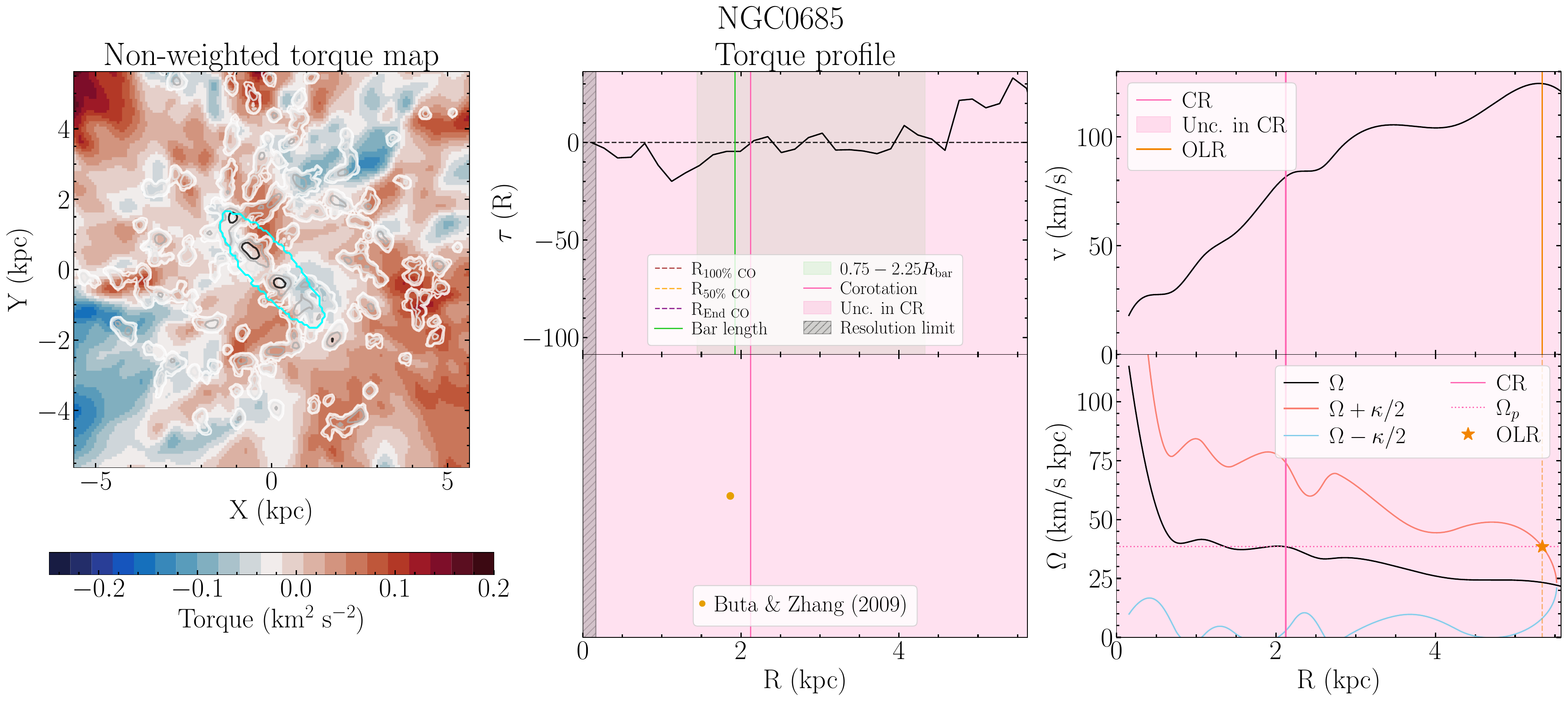}
    \end{center}
    \caption{\textbf{NGC\,0685} (SB, $QF=3^{\dagger}$). This galaxy has been marked as $QF=3$ instead of $QF=2$ because there is few information of the gas inside the bar. Legend as in Figure \ref{fig:Appendix-All_galaxies-IC1954}.} %Huge error bar \textbf{+} huge uncertainty on inclination (which is the parameter that most affects the deprojection and the uncertainty calculation)}
    \label{fig:Appendix-All_galaxies-NGC0685}
\end{figure*}

\begin{figure*}[t]
\begin{center}
    \includegraphics[trim=0 0 0 0, clip,width=1\textwidth]{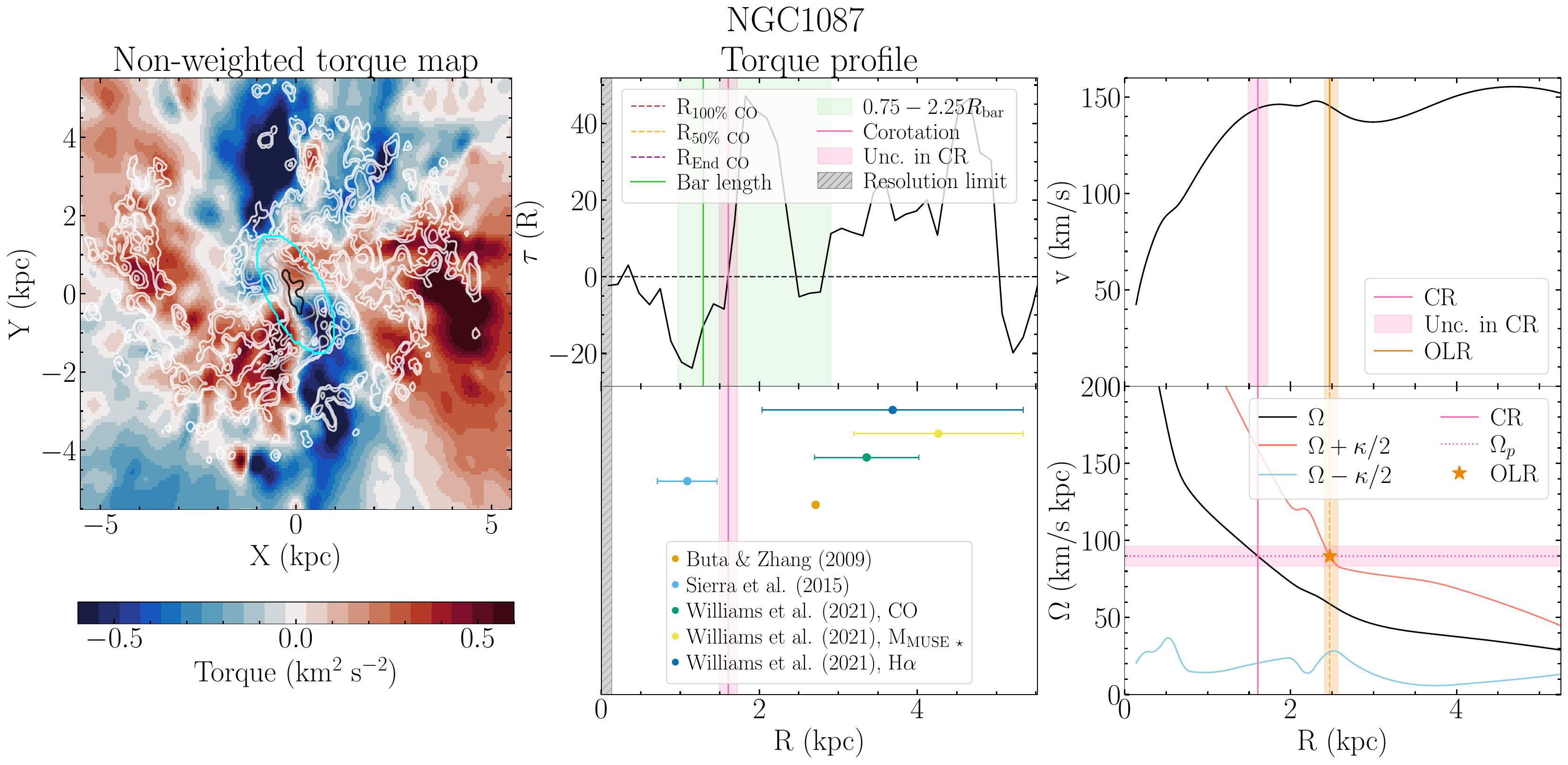}
    \end{center}
    \caption{\textbf{NGC\,1087} (SB, $QF=3^{\dagger}$). This galaxy has been marked as $QF=3$ instead of $QF=1$ because this is a floculent galaxy. Also, the gas response inside the bar is not the one expected. Legend as in Figure \ref{fig:Appendix-All_galaxies-IC1954}.}
    \label{fig:Appendix-All_galaxies-NGC1087}
\end{figure*}

\begin{figure*}[t]
\begin{center}
    \includegraphics[trim=0 0 0 0, clip,width=1\textwidth]{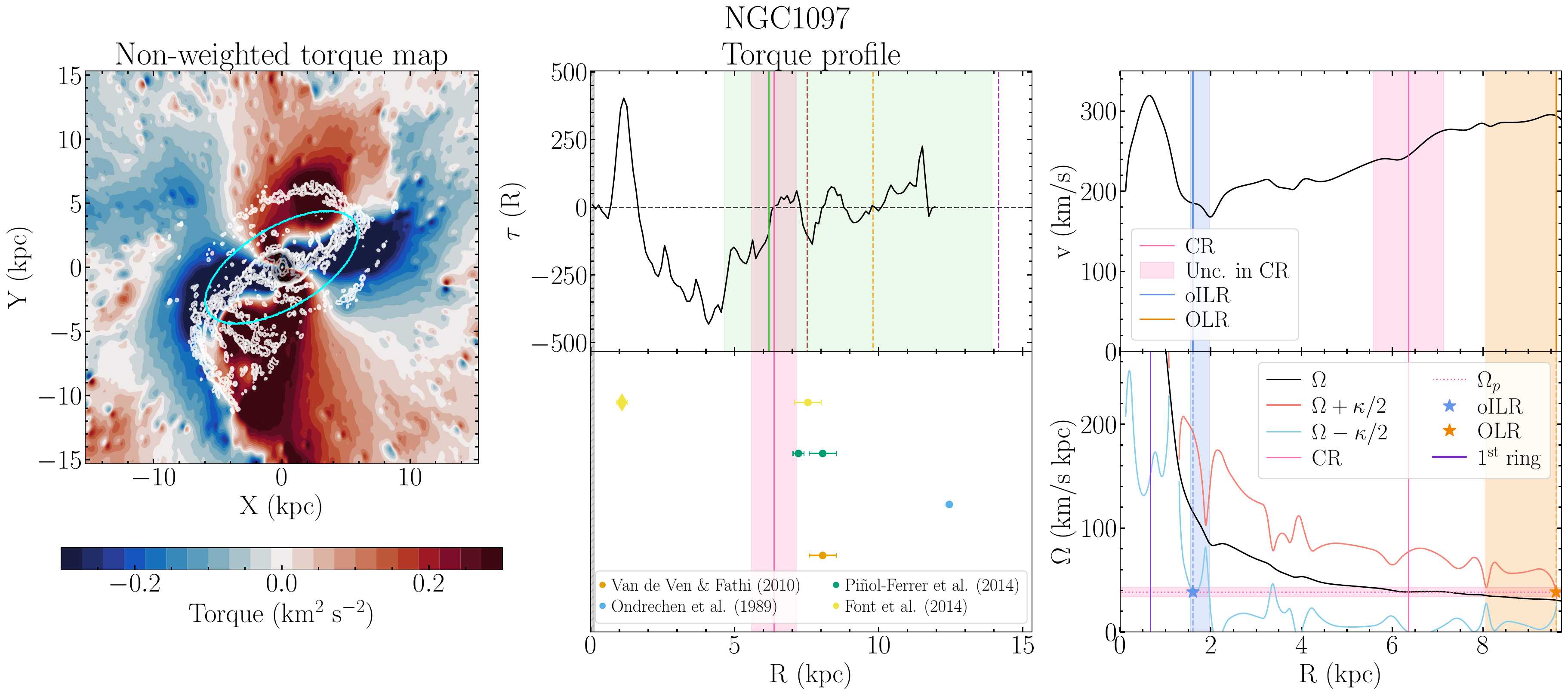}
    \end{center}
    \caption{\textbf{NGC\,1097} (SB, $QF=1$). Legend as in Figure \ref{fig:Appendix-All_galaxies-IC1954}.}
    \label{fig:Appendix-All_galaxies-NGC1097}
\end{figure*}

\begin{figure*}[t]
\begin{center}
    \includegraphics[trim=0 0 0 0, clip,width=1\textwidth]{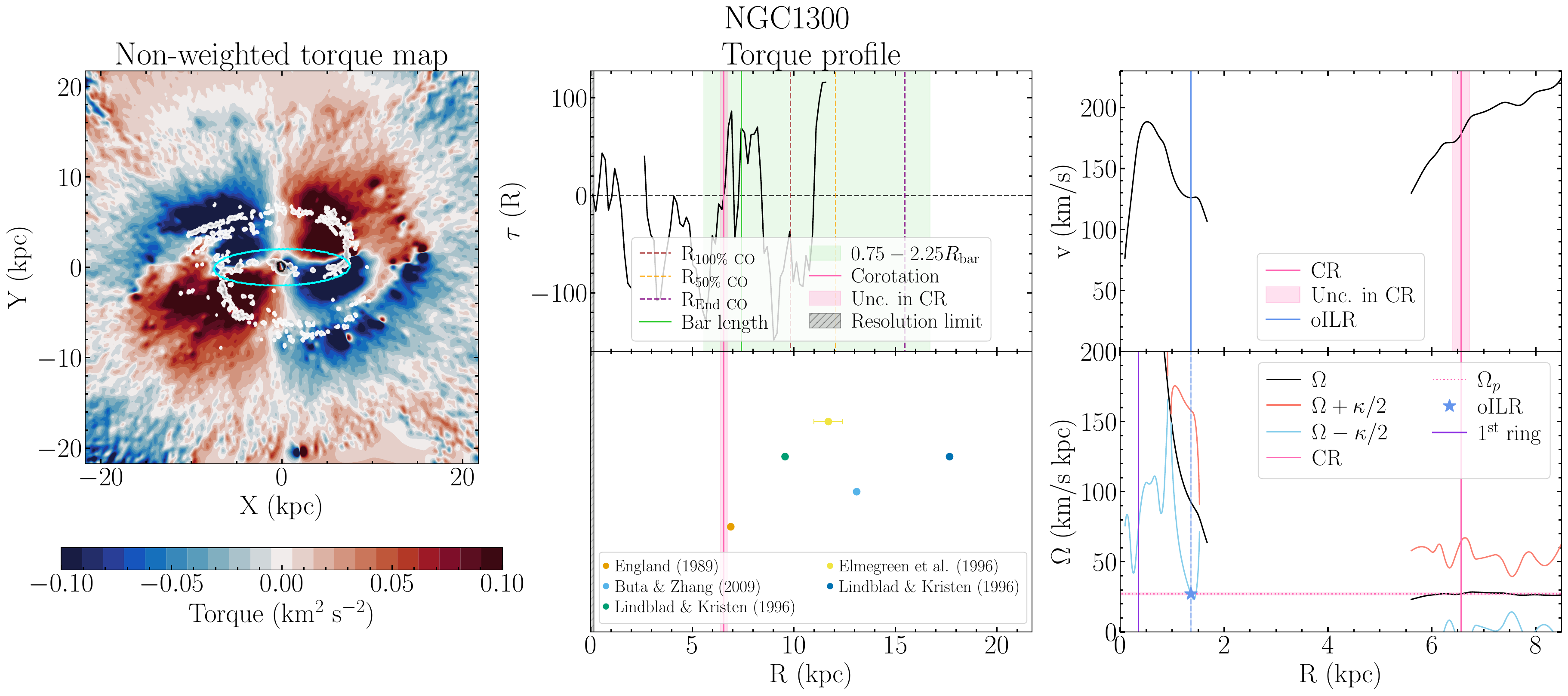}
    \end{center}
    \caption{\textbf{NGC\,1300} (SB, $QF=2^{\dagger}$). This galaxy has been marked as $QF =2$ instead of $QF=1$ because we do not to believe the automatically detected CR. Therefore we force the code to choose a different CR (the one shown here), which is more likely to be the CR of the bar. Legend as in Figure \ref{fig:Appendix-All_galaxies-IC1954}.}
    \label{fig:Appendix-All_galaxies-NGC1300}
\end{figure*}

\begin{figure*}[t]
\begin{center}
    \includegraphics[trim=0 0 0 0, clip,width=1\textwidth]{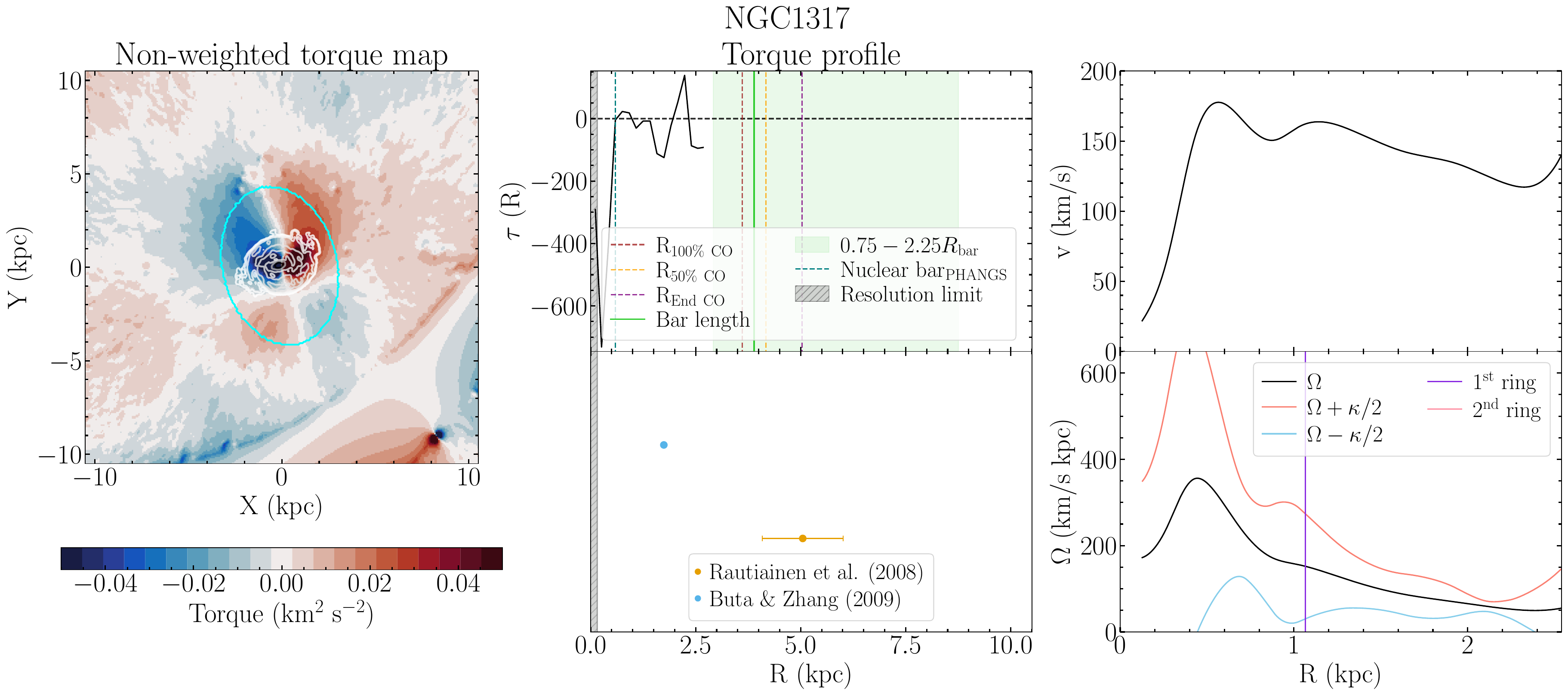}
    \end{center}
    \caption{\textbf{NGC\,1317} (SB, $QF=3$). Legend as in Figure \ref{fig:Appendix-All_galaxies-IC1954}.}
    \label{fig:Appendix-All_galaxies-NGC1317}
\end{figure*}

\begin{figure*}[t]
\begin{center}
    \includegraphics[trim=0 0 0 0, clip,width=1\textwidth]{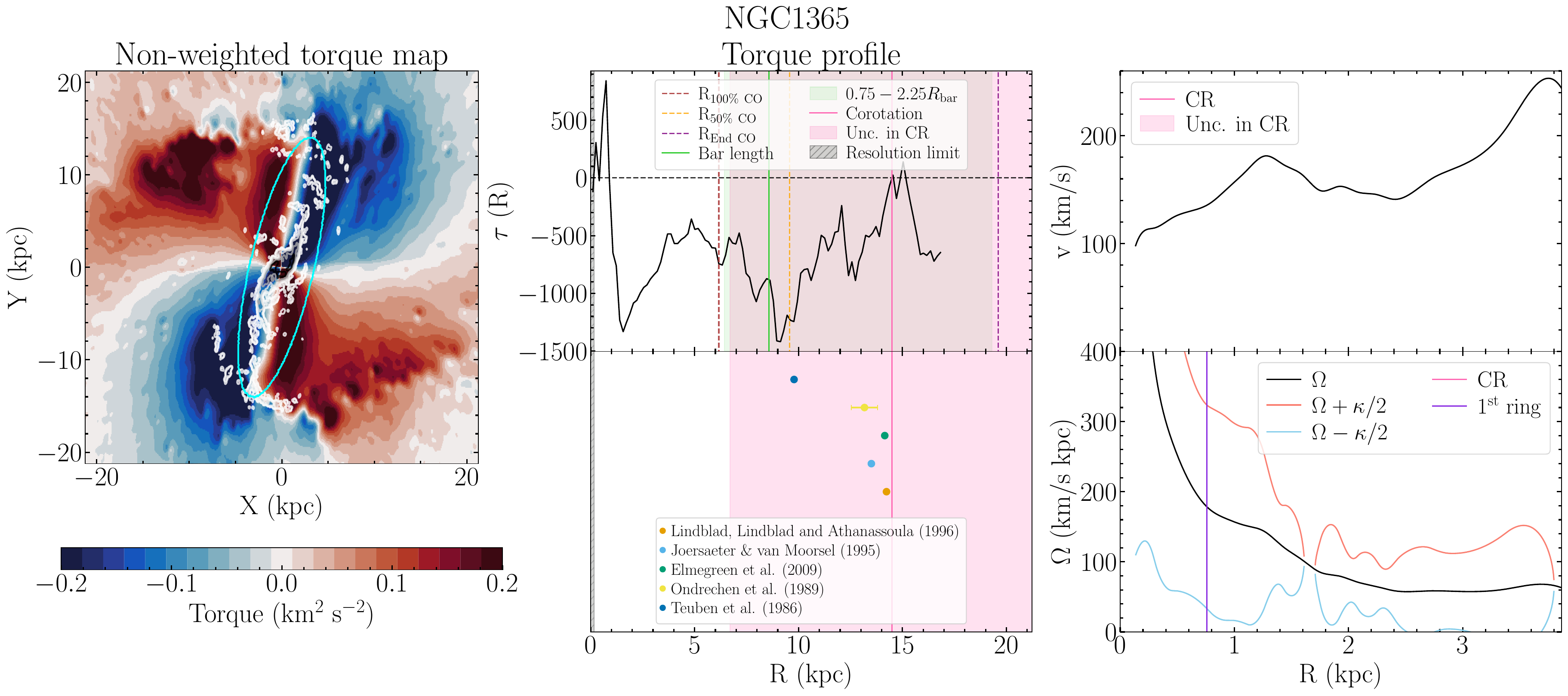}
    \end{center}
    \caption{\textbf{NGC\,1365} (SB, $QF=3^{\dagger}$). This galaxy has been marked as $QF=3$ instead of $QF=1$ because even though the stellar potential behaves as expected, we find very little gas in the bar. Legend as in Figure \ref{fig:Appendix-All_galaxies-IC1954}.}
    \label{fig:Appendix-All_galaxies-NGC1365}
\end{figure*}

\begin{figure*}[t]
\begin{center}
    \includegraphics[trim=0 0 0 0, clip,width=1\textwidth]{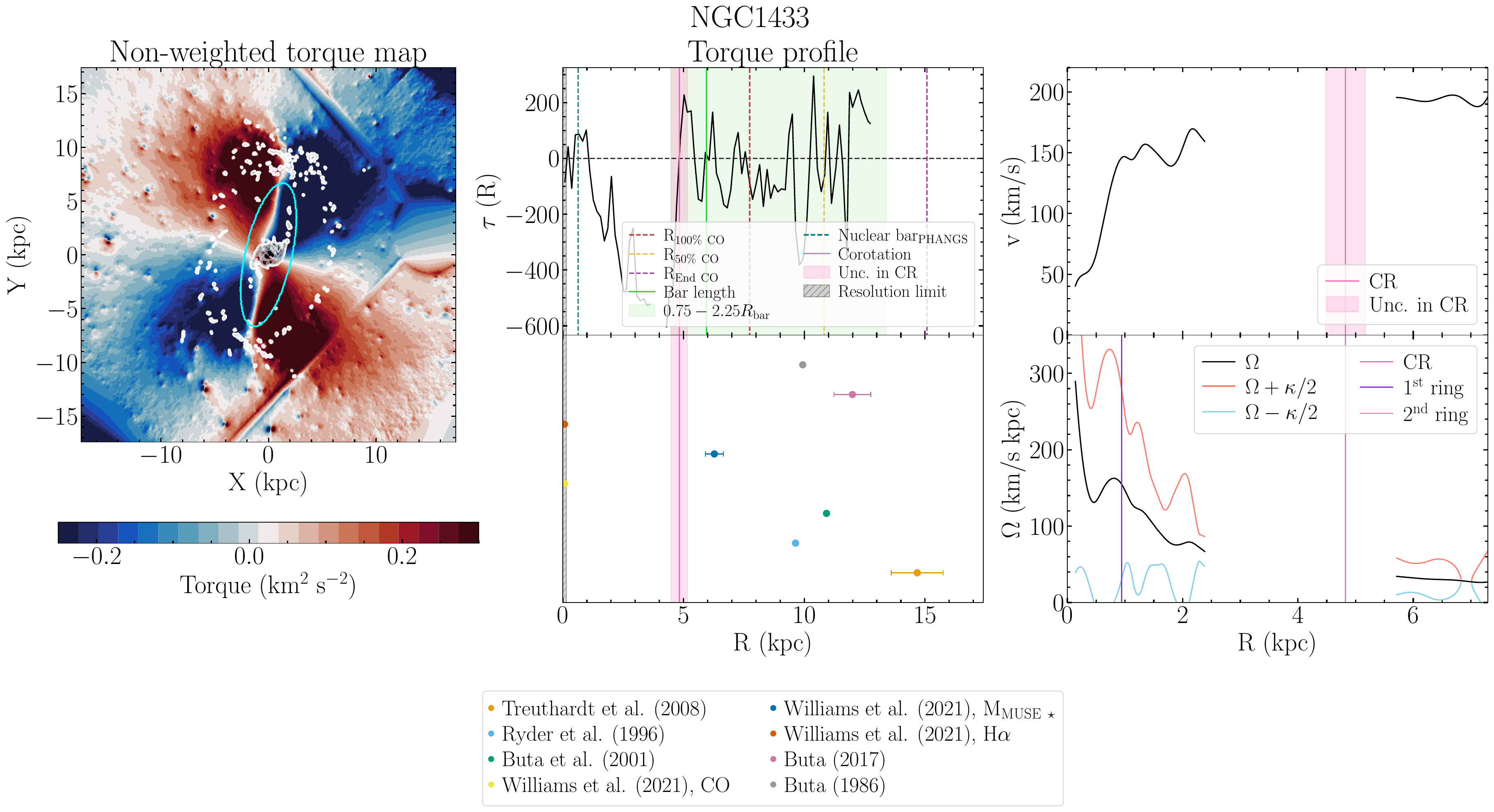}
    \end{center}
    \caption{\textbf{NGC\,1433} (SB, $QF=2$). %This galaxy has been marked as $QF=2$ instead of $QF=3$ because the stellar potential behaves as expected. 
    Legend as in Figure \ref{fig:Appendix-All_galaxies-IC1954}.}
    \label{fig:Appendix-All_galaxies-NGC1433}
\end{figure*}

\begin{figure*}[t]
\begin{center}
    \includegraphics[trim=0 0 0 0, clip,width=1\textwidth]{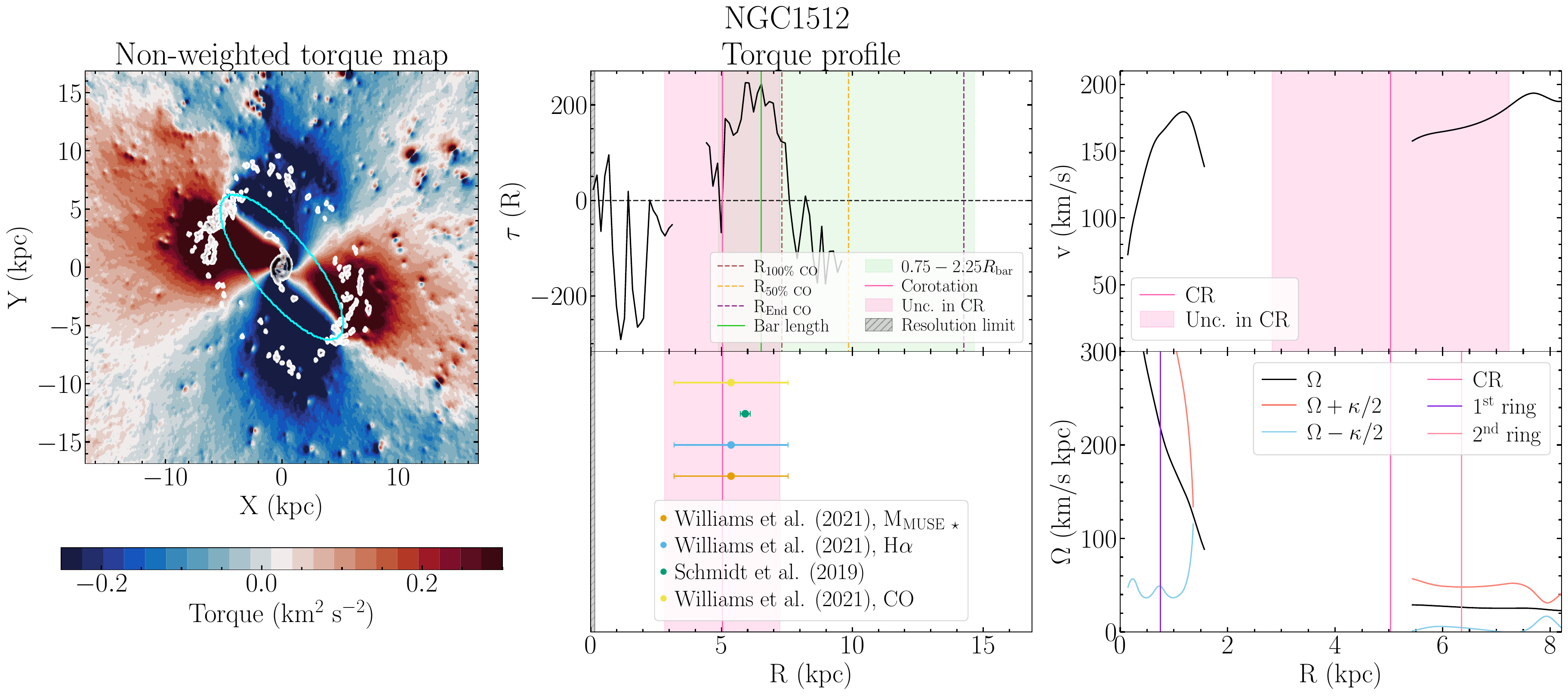}
    \end{center}

    \caption{\textbf{NGC\,1512} (SB, $QF=2^{\dagger}$). This galaxy has been marked as $QF=2$ instead of $QF=1$ because we do not believe the automatically detected CR. Therefore we force the code to choose a different CR (the one shown here), which is more likely to be the CR of the bar. Legend as in Figure \ref{fig:Appendix-All_galaxies-IC1954}.}
    \label{fig:Appendix-All_galaxies-NGC1512}
\end{figure*}

\begin{figure*}[t]
\begin{center}
    \includegraphics[trim=0 0 0 0, clip,width=1\textwidth]{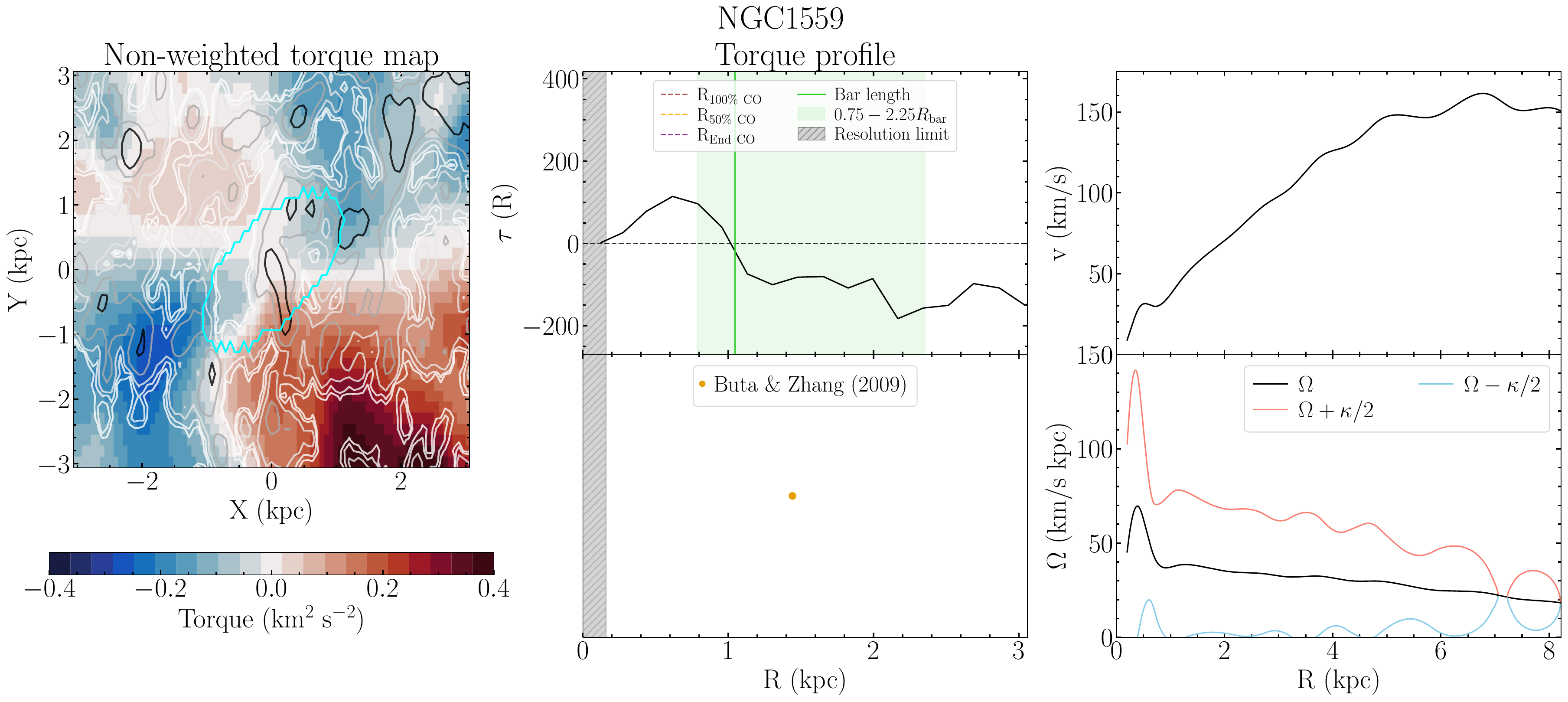}
    \end{center}
    \caption{\textbf{NGC\,1559} (SB, $QF=3$). Legend as in Figure \ref{fig:Appendix-All_galaxies-IC1954}.}
    \label{fig:Appendix-All_galaxies-NGC1559}
\end{figure*}

\begin{figure*}[t]
\begin{center}
    \includegraphics[trim=0 0 0 0, clip,width=1\textwidth]{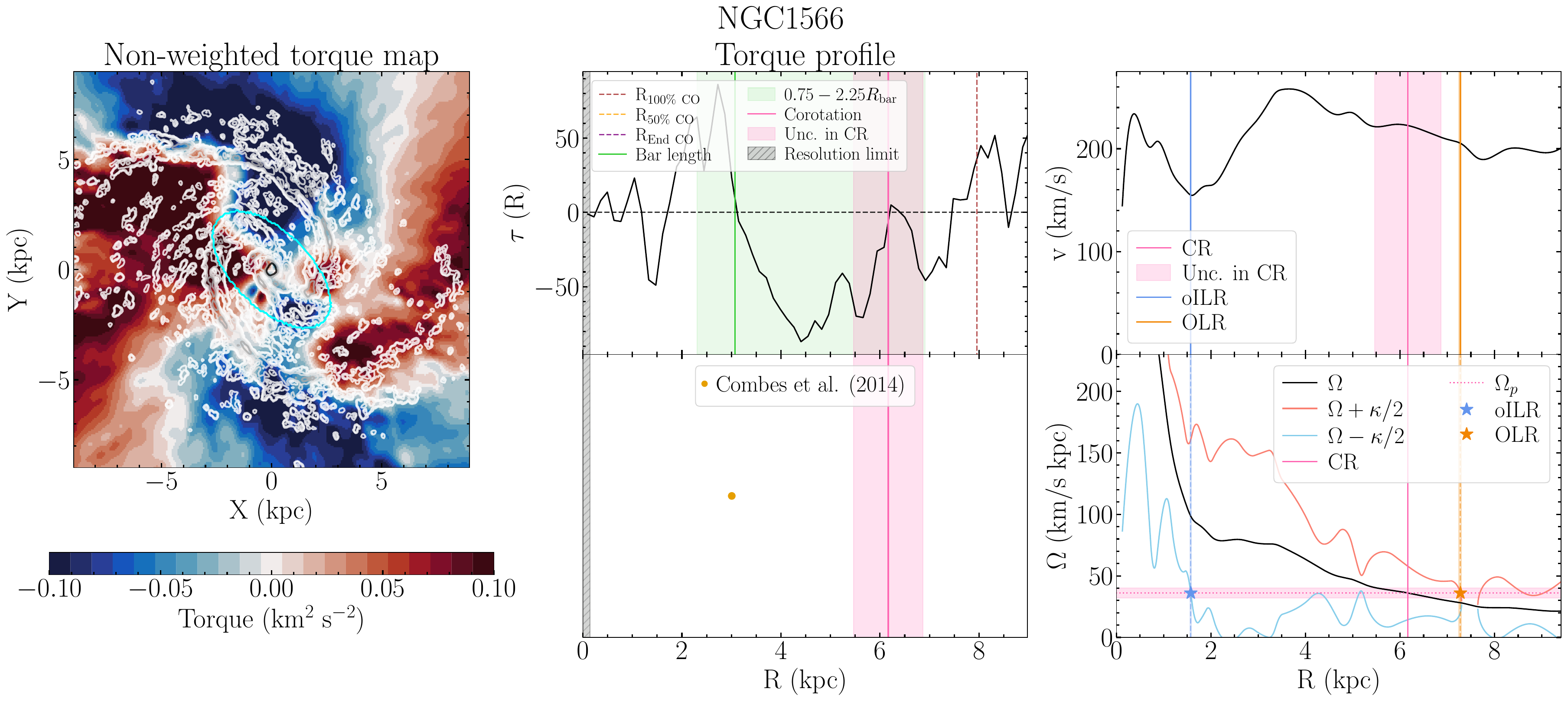}
    \end{center}
    \caption{\textbf{NGC\,1566} (SAB, $QF=3^{\dagger}$). This galaxy is marked as $QF=3$ instead of $QF=1$ because upon visual inspection, we realize the automatically selected CR might be the spiral CR, not the bar CR. Legend as in Figure \ref{fig:Appendix-All_galaxies-IC1954}.}
    \label{fig:Appendix-All_galaxies-NGC1566}
\end{figure*}

\begin{figure*}[t]
\begin{center}
    \includegraphics[trim=0 0 0 0, clip,width=1\textwidth]{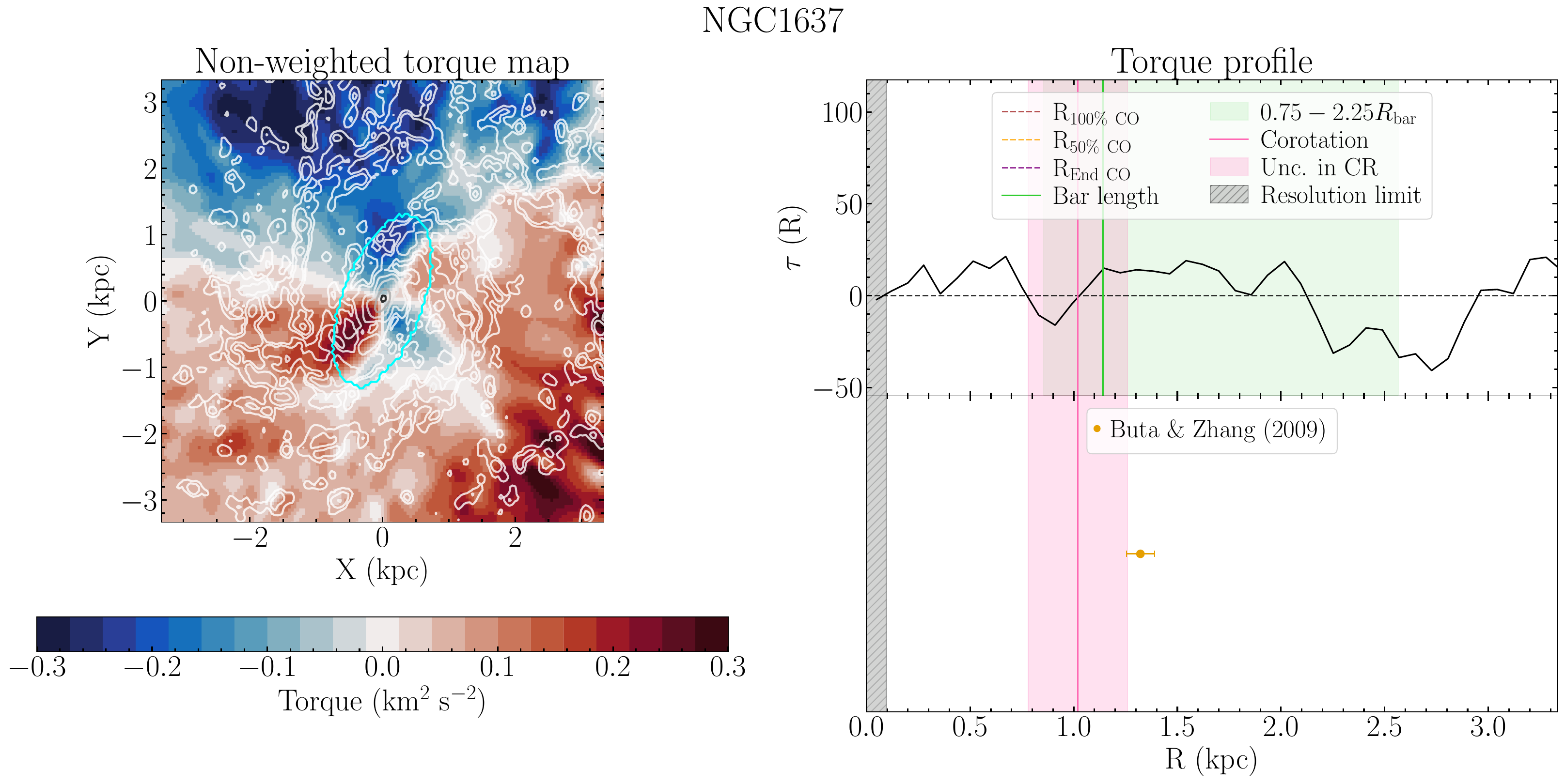}
    \end{center}
    \caption{\textbf{NGC\,1637} (SAB, $QF=2^{\dagger}$). This galaxy is marked as $QF=2$ instead of $QF=1$ because there is few information of the gas response inside the bar. We do not have the rotation curve for this galaxy. Legend as in Figure \ref{fig:Appendix-All_galaxies-IC1954}.}
    \label{fig:Appendix-All_galaxies-NGC1637}
\end{figure*}

\begin{figure*}[t]
\begin{center}
    \includegraphics[trim=0 0 0 0, clip,width=1\textwidth]{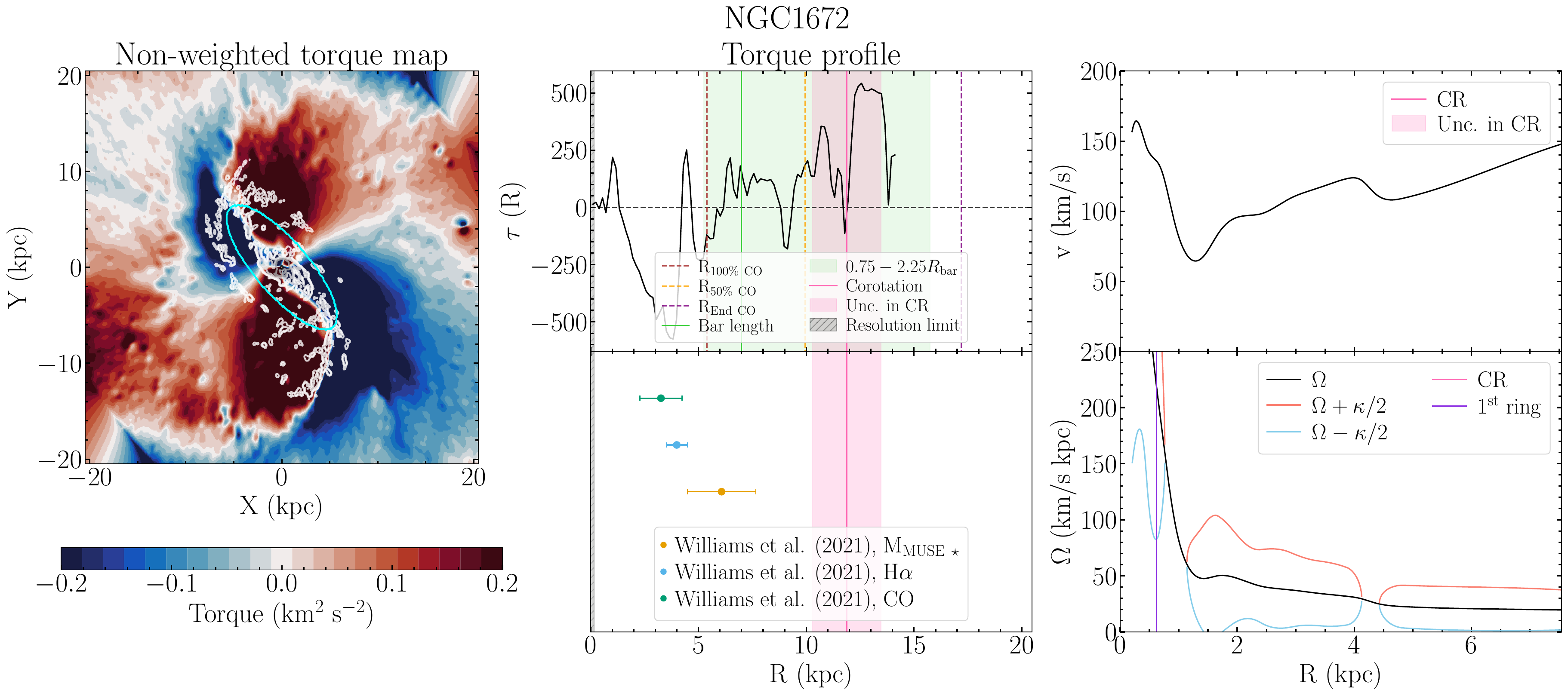}
    \end{center}
    \caption{\textbf{NGC\,1672} (SAB, $QF=3^{\dagger}$). This galaxy has been marked as $QF=3$ instead of $QF=2$ because the CR is located at a region where the CO coverage is insufficient. Legend as in Figure \ref{fig:Appendix-All_galaxies-IC1954}.}
    \label{fig:Appendix-All_galaxies-NGC1672}
\end{figure*}

\begin{figure*}[t]
\begin{center}
    \includegraphics[trim=0 0 0 0, clip,width=1\textwidth]{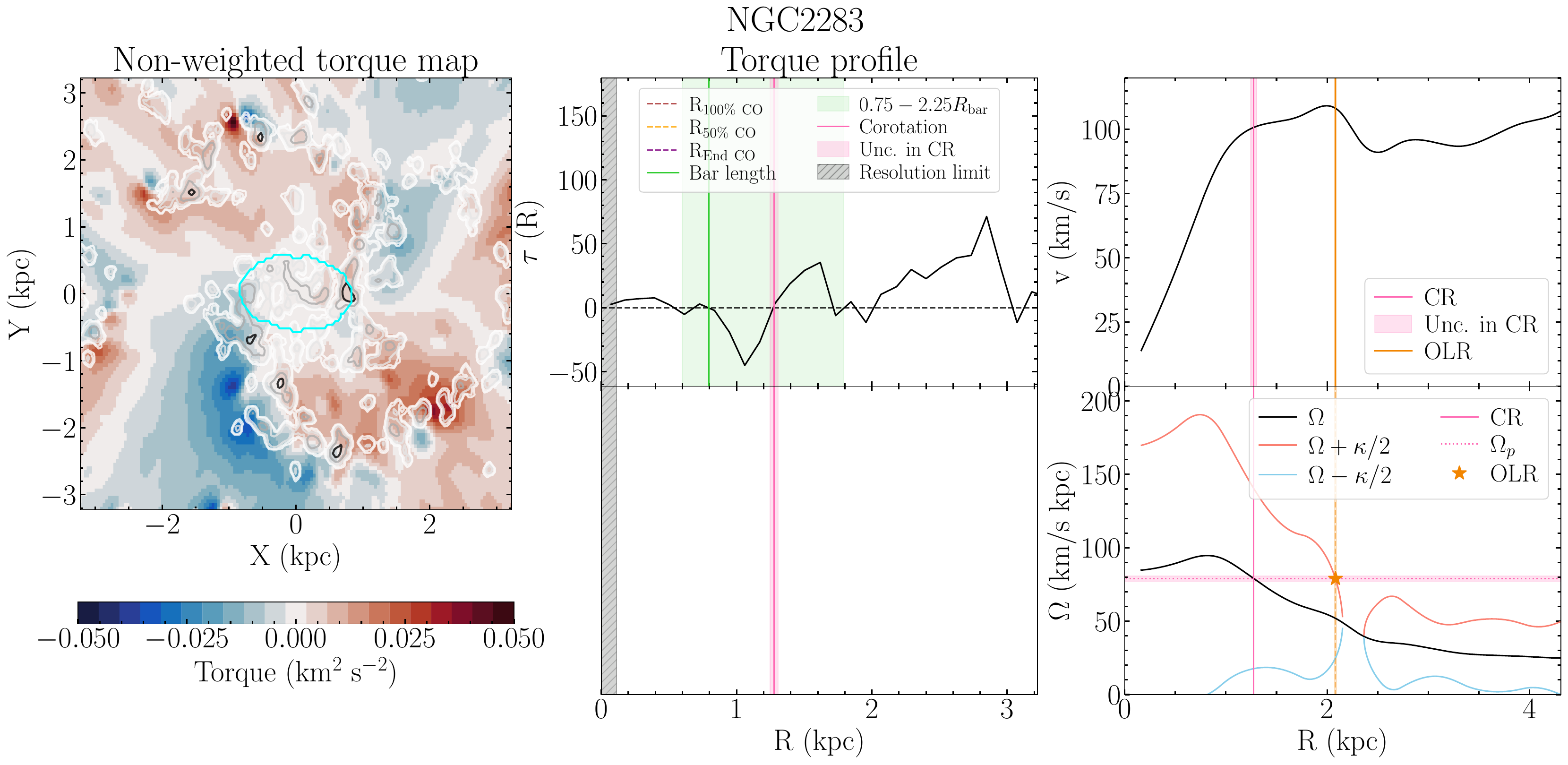}
    \end{center}
    \caption{\textbf{NGC\,2283} (SB, $QF = 3^{\dagger}$). This galaxy has been marked as $QF=3$ instead of $QF=1$ because it presents a discussable bar. Legend as in Figure \ref{fig:Appendix-All_galaxies-IC1954}.}
    \label{fig:Appendix-All_galaxies-NGC2283}
\end{figure*}

\begin{figure*}[t]
\begin{center}
    \includegraphics[trim=0 0 0 0, clip,width=1\textwidth]{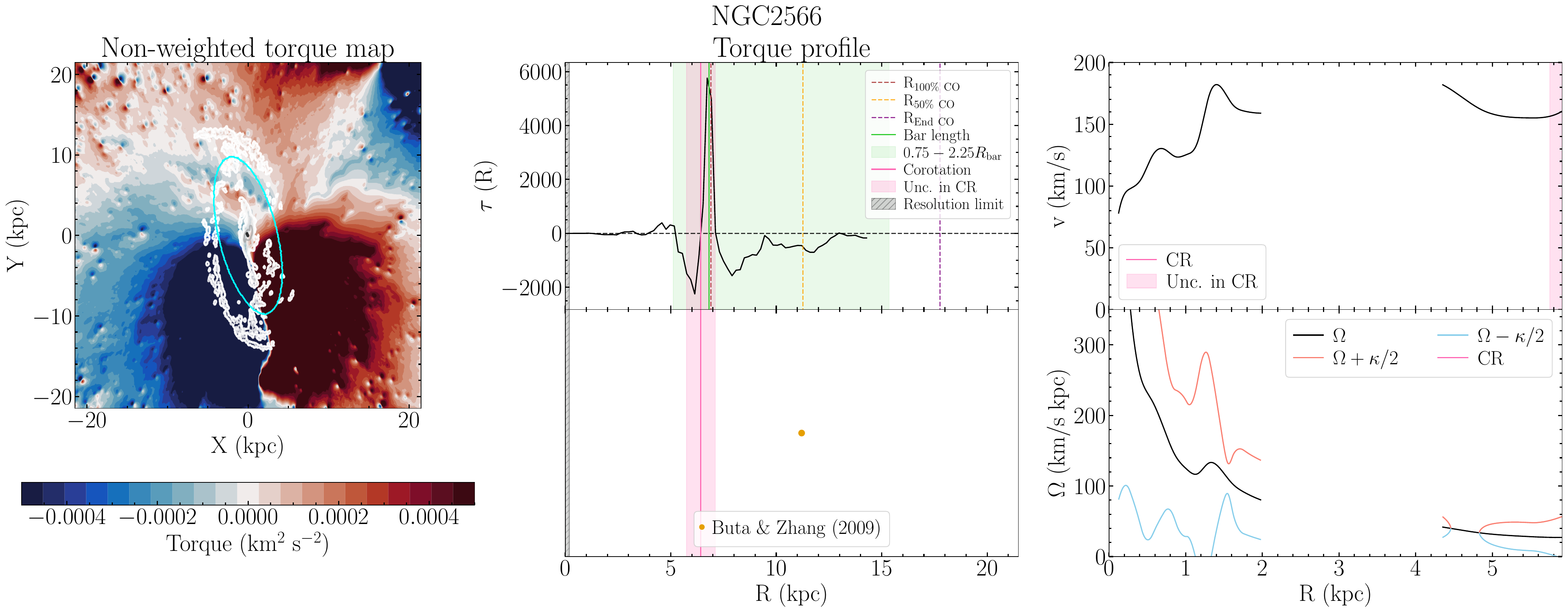}
    \end{center}
    \caption{\textbf{NGC\,2566} (SB, $QF=3^{\dagger}$). This galaxy has been marked as $QF=3$ instead of $QF=1$ because it is almost in the galactic plane, therefore there are many point sources (stars) all over the field, which can create false torques. Legend as in Figure \ref{fig:Appendix-All_galaxies-IC1954}.}
    \label{fig:Appendix-All_galaxies-NGC2566}
\end{figure*}

\begin{figure*}[t]
\begin{center}
    \includegraphics[trim=0 0 0 0, clip,width=1\textwidth]{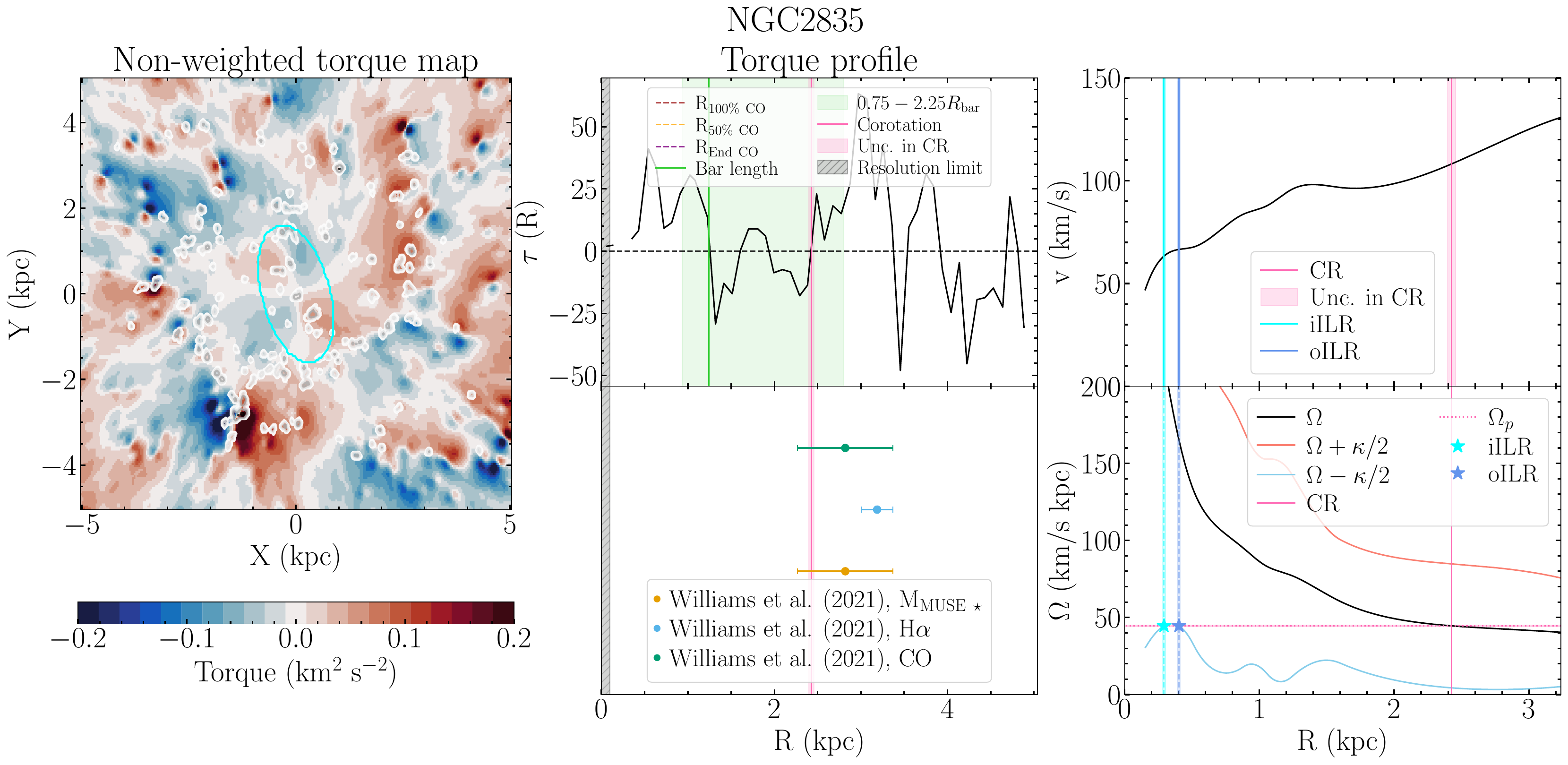}
    \end{center}
    \caption{\textbf{NGC\,2835} (SB, $QF=3^{\dagger}$). This galaxy has been marked as $QF=3$ instead of $QF=1$ because there is few information of the gas response inside the bar, moreover the gas seems clumpy. Legend as in Figure \ref{fig:Appendix-All_galaxies-IC1954}.}
    \label{fig:Appendix-All_galaxies-NGC2835}
\end{figure*}

\begin{figure*}[t]
\begin{center}
    \includegraphics[trim=0 0 0 0, clip,width=1\textwidth]{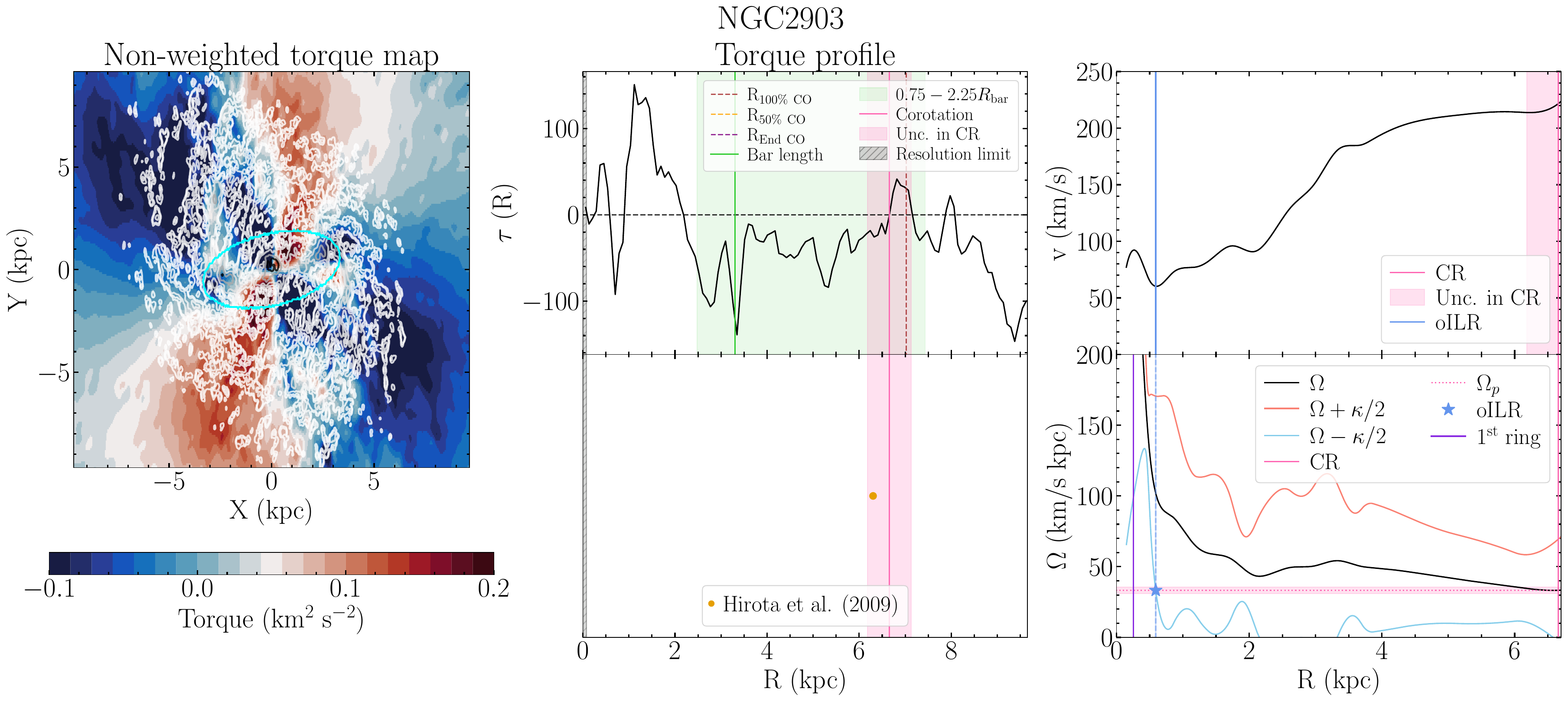}
    \end{center}
    \caption{\textbf{NGC\,2903} (SB, $QF=2$). %$QF=2^{\ddagger}$). This galaxy has been marked as $QF=2$ instead of $QF=3$ because we believe both the gas and stellar potential behave as expected. 
    Legend as in Figure \ref{fig:Appendix-All_galaxies-IC1954}.}
    \label{fig:Appendix-All_galaxies-NGC2903}
\end{figure*}

\begin{figure*}[t]
\begin{center}
    \includegraphics[trim=0 0 0 0, clip,width=1\textwidth]{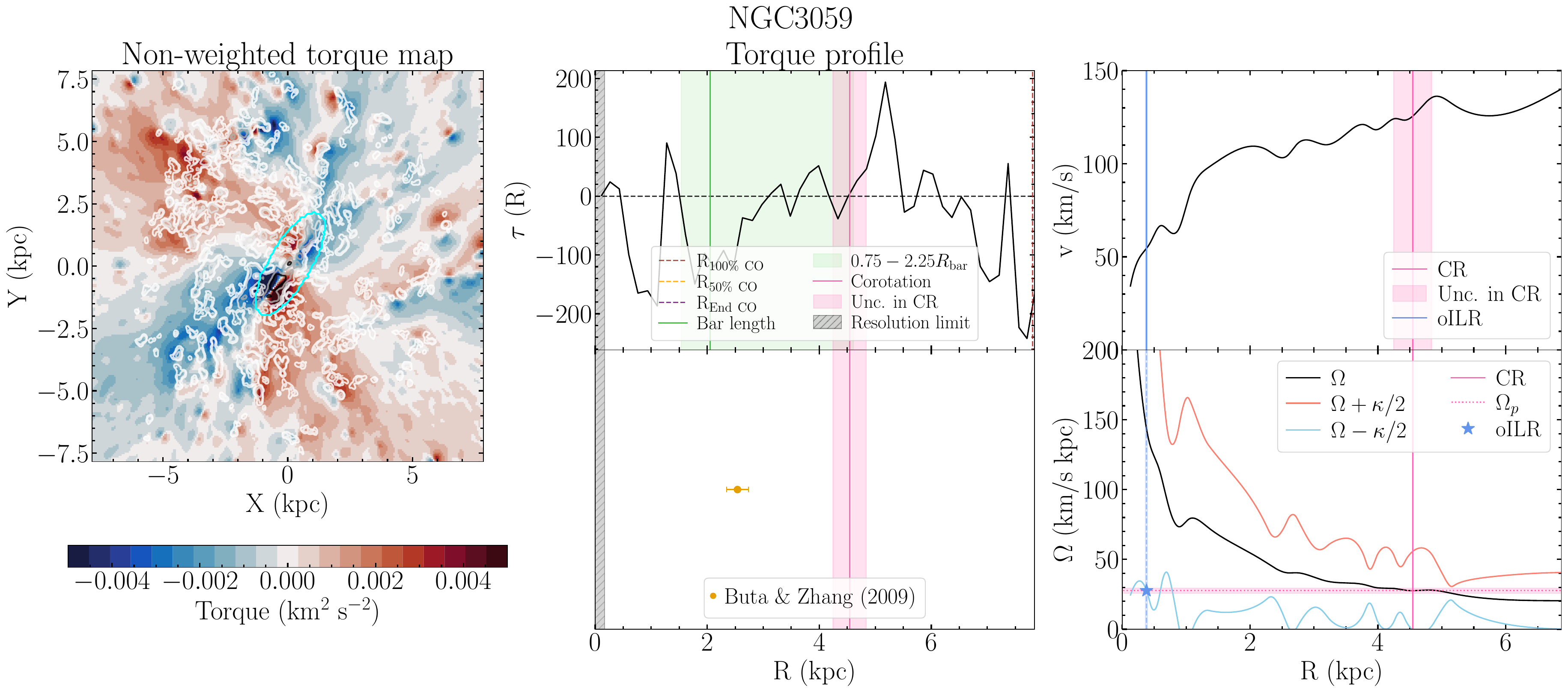}
    \end{center}
    \caption{\textbf{NGC\,3059} (SB, $QF=2$). % $QF=2^{\ddagger}$). This galaxy has been marked as $QF=2$ instead of $QF=3$ because we believe both the gas and stellar potential behave as expected. 
    Legend as in Figure \ref{fig:Appendix-All_galaxies-IC1954}.}
    \label{fig:Appendix-All_galaxies-NGC3059}
\end{figure*}

\begin{figure*}[t]
\begin{center}
    \includegraphics[trim=0 0 0 0, clip,width=1\textwidth]{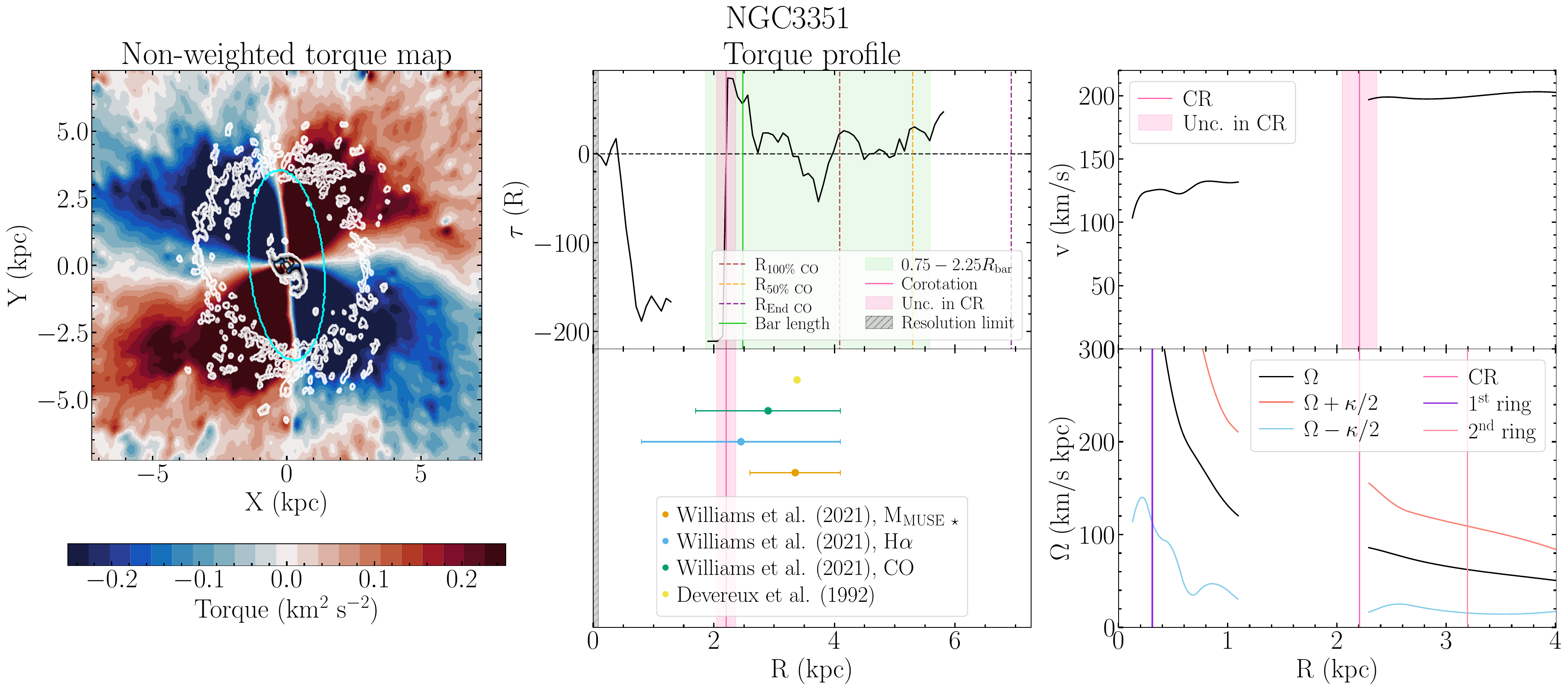}
    \end{center}
    \caption{\textbf{NGC\,3351} (SB, $QF=1$). Legend as in Figure \ref{fig:Appendix-All_galaxies-IC1954}.}
    \label{fig:Appendix-All_galaxies-NGC3351}
\end{figure*}

\begin{figure*}[t]
\begin{center}
    \includegraphics[trim=0 0 0 0, clip,width=1\textwidth]{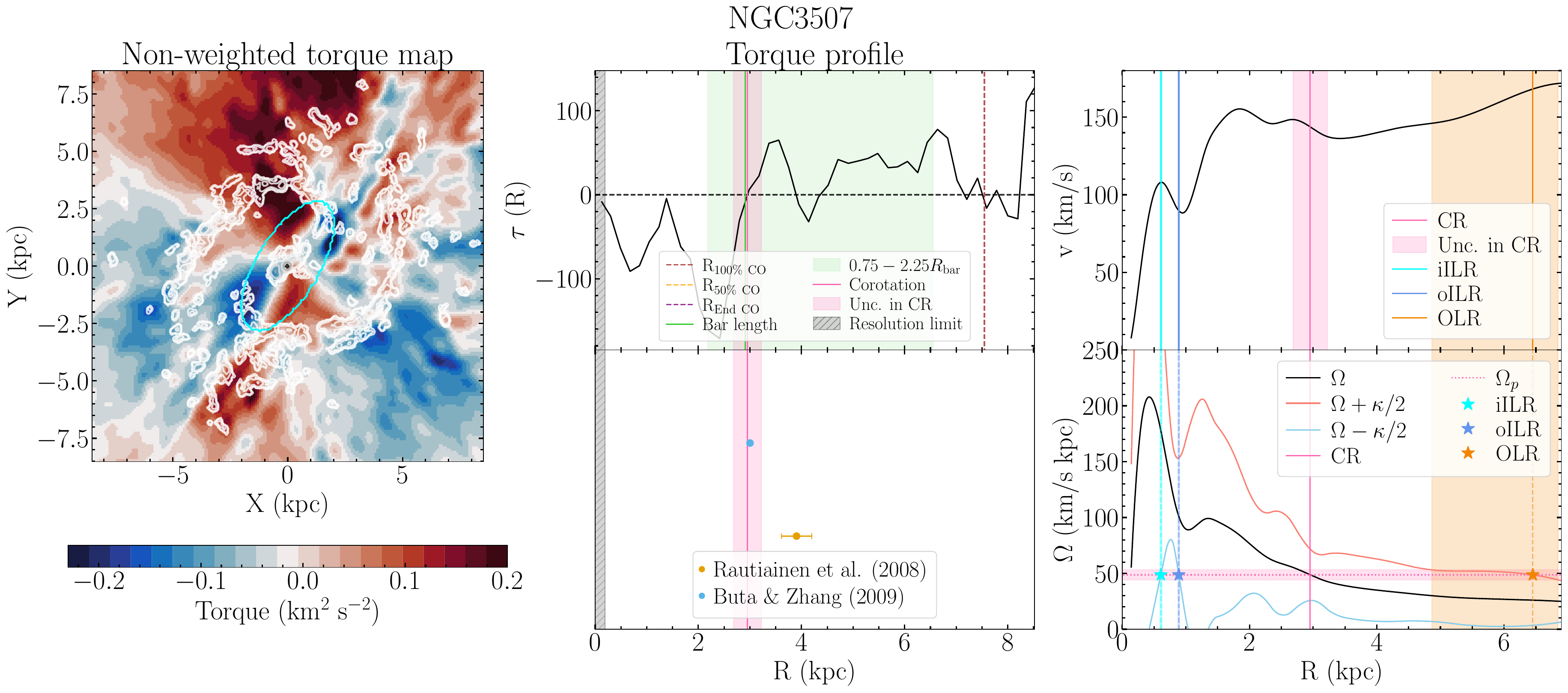}
    \end{center}
    \caption{\textbf{NGC\,3507} (SAB, $QF=1$). Legend as in Figure \ref{fig:Appendix-All_galaxies-IC1954}.}
    \label{fig:Appendix-All_galaxies-NGC3507}
\end{figure*}

\begin{figure*}[t]
\begin{center}
    \includegraphics[trim=0 0 0 0, clip,width=1\textwidth]{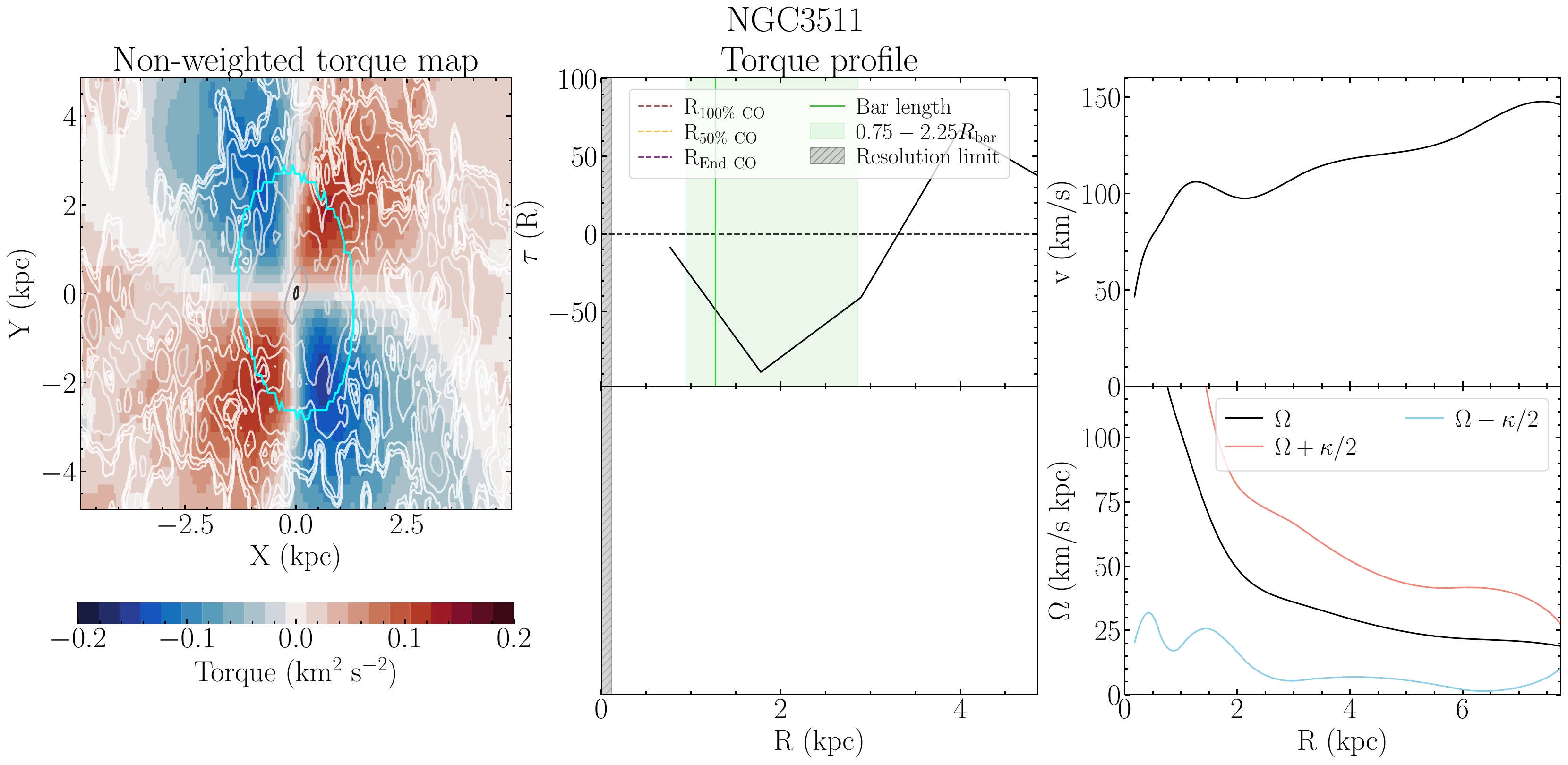}
    \end{center}
    \caption{\textbf{NGC\,3511} (SAB, $QF=3$). Legend as in Figure \ref{fig:Appendix-All_galaxies-IC1954}.}
    \label{fig:Appendix-All_galaxies-NGC3511}
\end{figure*}

\begin{figure*}[t]
\begin{center}
    \includegraphics[trim=0 0 0 0, clip,width=1\textwidth]{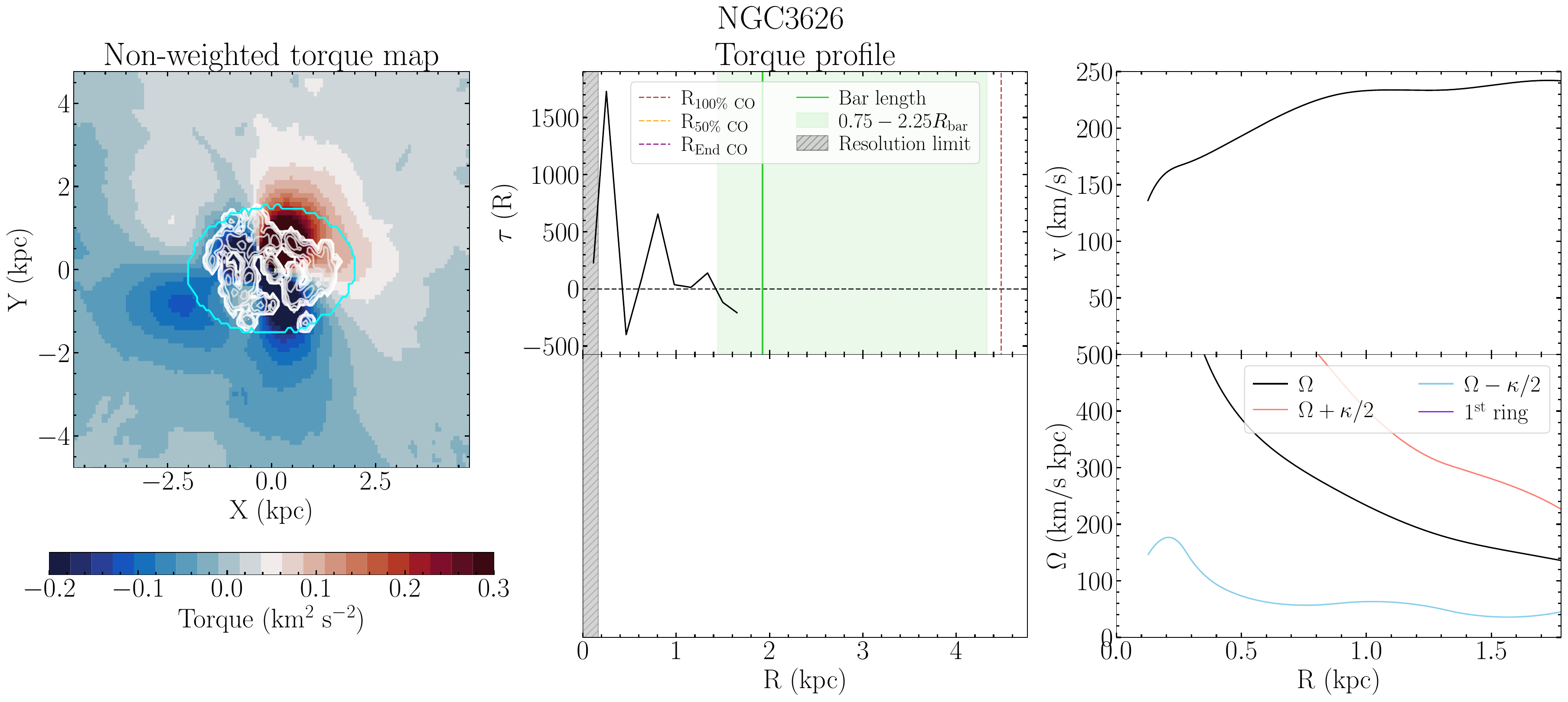}
    \end{center}
    \caption{\textbf{NGC\,3626} (SAB, $QF=3$). Legend as in Figure \ref{fig:Appendix-All_galaxies-IC1954}.}
    \label{fig:Appendix-All_galaxies-NGC3626}
\end{figure*}

\begin{figure*}[t]
\begin{center}
    \includegraphics[trim=0 0 0 0, clip,width=1\textwidth]{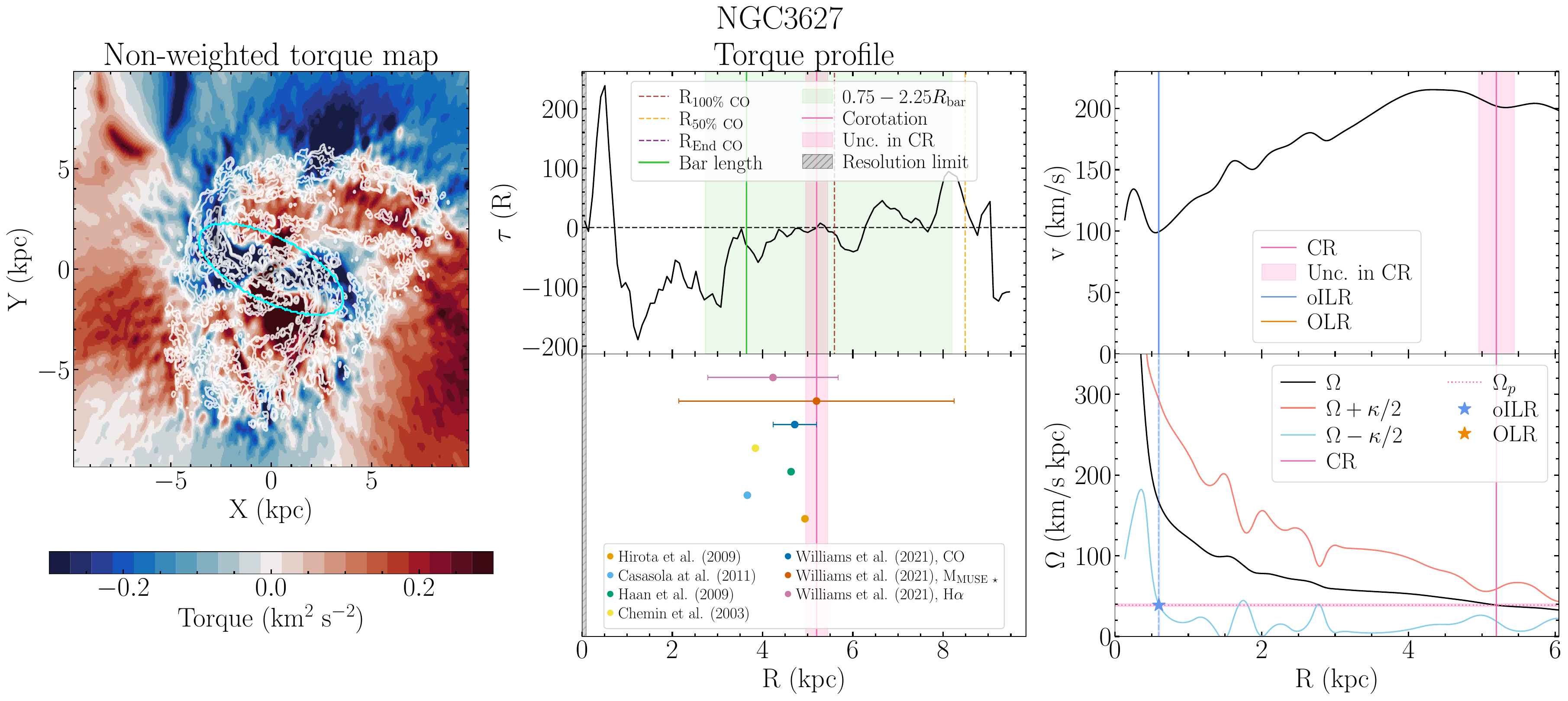}
    \end{center}
    \caption{\textbf{NGC\,3627} (SB, $QF=2$). Legend as in Figure \ref{fig:Appendix-All_galaxies-IC1954}.}
    \label{fig:Appendix-All_galaxies-NGC3627}
\end{figure*}

\begin{figure*}[t]
\begin{center}
    \includegraphics[trim=0 0 0 0, clip,width=1\textwidth]{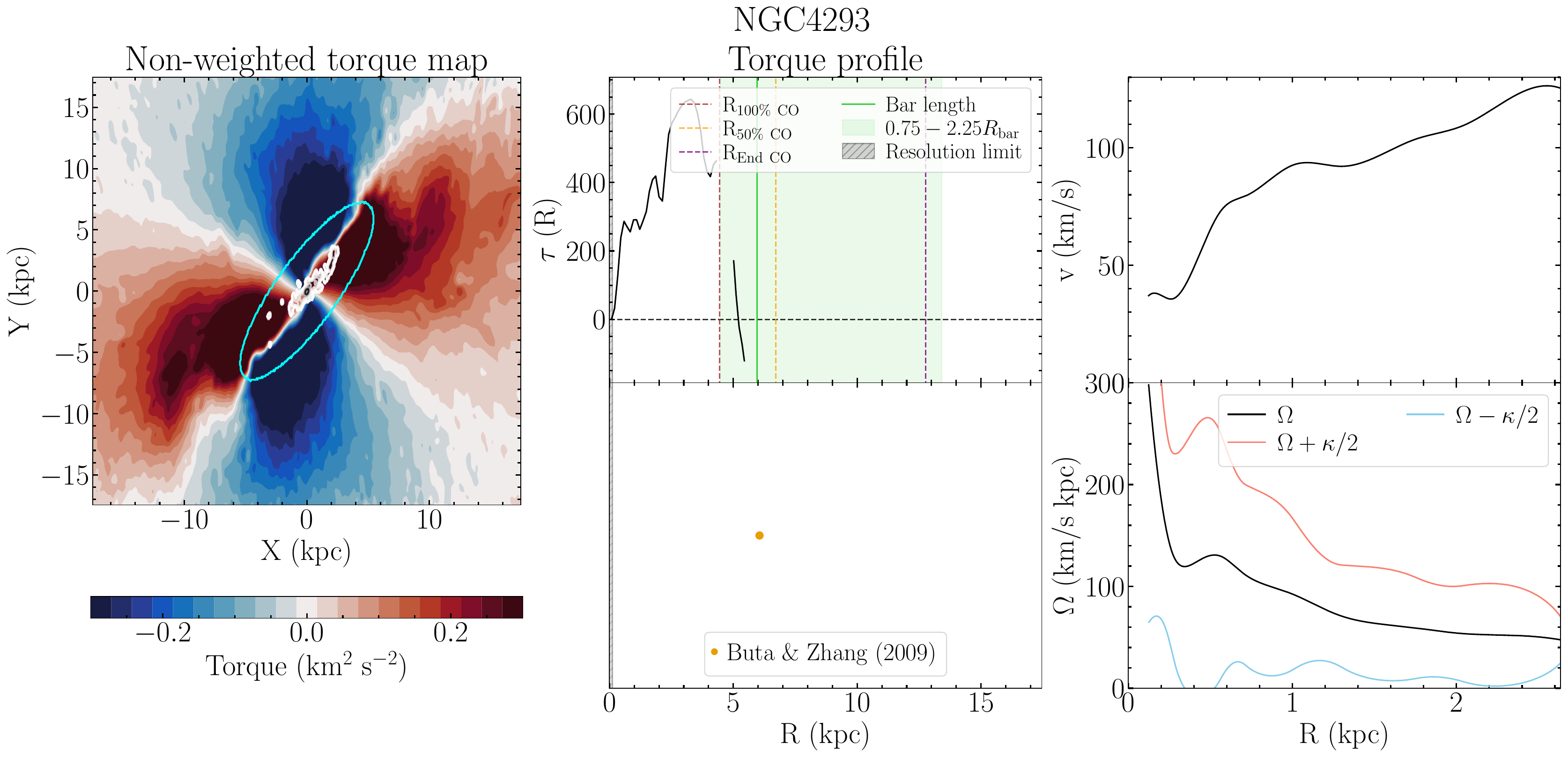}
    \end{center}
    \caption{\textbf{NGC\,4293} (SB, $QF=3$). Legend as in Figure \ref{fig:Appendix-All_galaxies-IC1954}.} %Large inclination \action{Working on it}}
    \label{fig:Appendix-All_galaxies-NGC4293}
\end{figure*}

\begin{figure*}[t]
\begin{center}
    
    \includegraphics[trim=0 0 0 0, clip,width=1\textwidth]{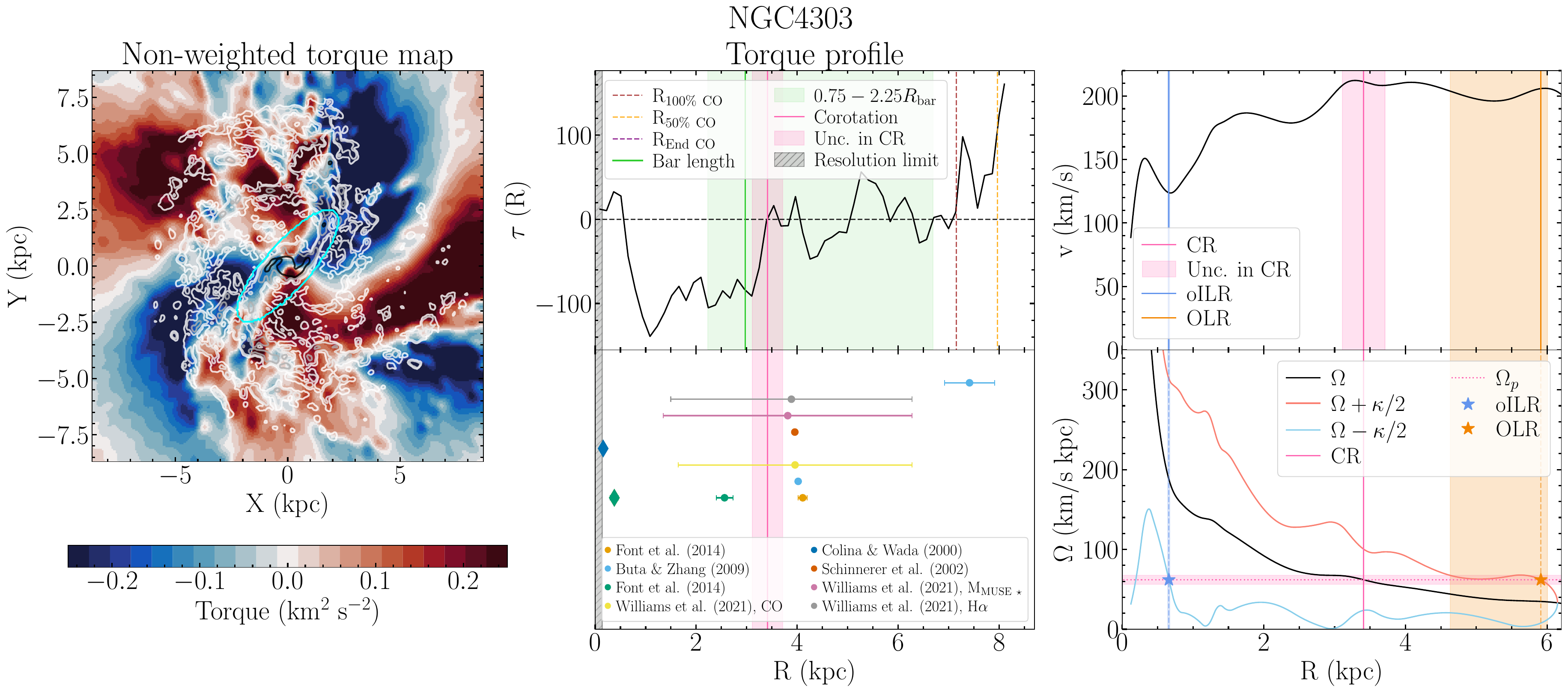}
    \end{center}
    \caption{\textbf{NGC\,4303} (SAB, $QF=1$). Legend as in Figure \ref{fig:Appendix-All_galaxies-IC1954}.}
    \label{fig:Appendix-All_galaxies-NGC4303}
\end{figure*}

\begin{figure*}[t]
\begin{center}
    \includegraphics[trim=0 0 0 0, clip,width=1\textwidth]{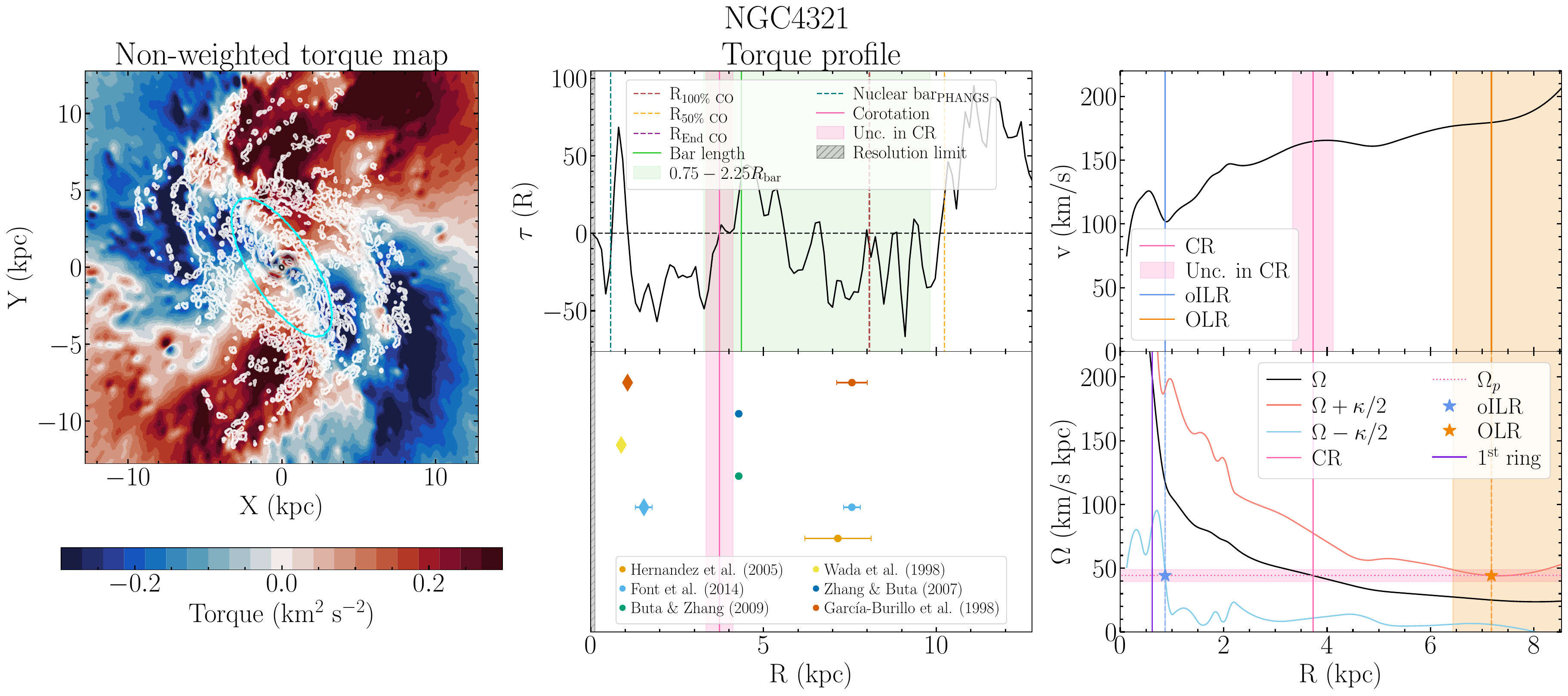}
    \end{center}
    \caption{\textbf{NGC\,4321} (SAB, $QF=2^{\ddagger}$). This galaxy has been marked as $QF = 2$ instead of $QF=3$ because we believe both the gas response and stellar potential behave as expected. Legend as in Figure \ref{fig:Appendix-All_galaxies-IC1954}.}
    \label{fig:Appendix-All_galaxies-NGC4321}
\end{figure*}

\begin{figure*}[t]
\begin{center}
    \includegraphics[trim=0 0 0 0, clip,width=1\textwidth]{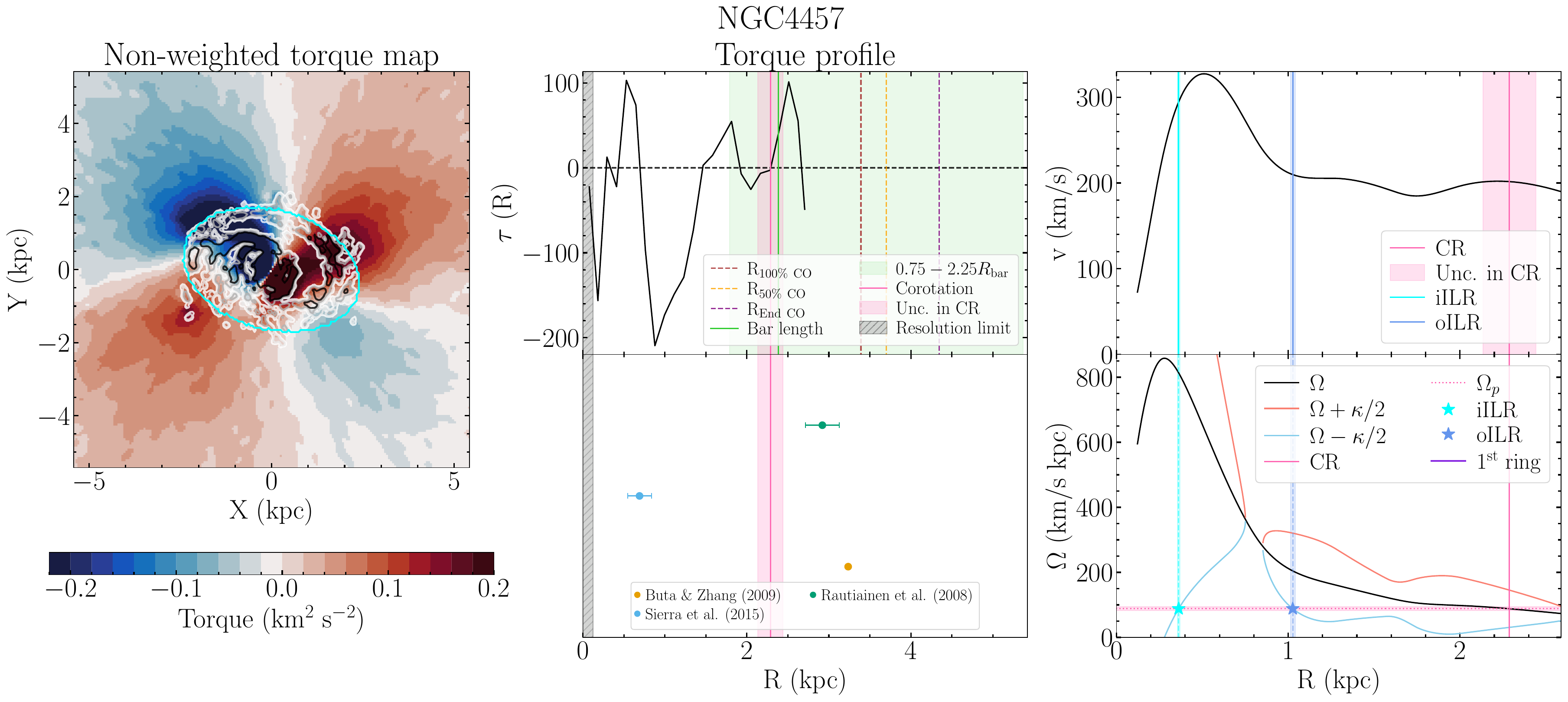}
    \end{center}
    \caption{\textbf{NGC\,4457} (SAB, $QF = 2^{\dagger}$). This galaxy has been marked as $QF=2$ instead of $QF = 1$ because we find a very bright point source in the center, which affects the stellar potential and gas response. Legend as in Figure \ref{fig:Appendix-All_galaxies-IC1954}.}
    \label{fig:Appendix-All_galaxies-NGC4457}
\end{figure*}

\begin{figure*}[t]
\begin{center}
    \includegraphics[trim=0 0 0 0, clip,width=1\textwidth]{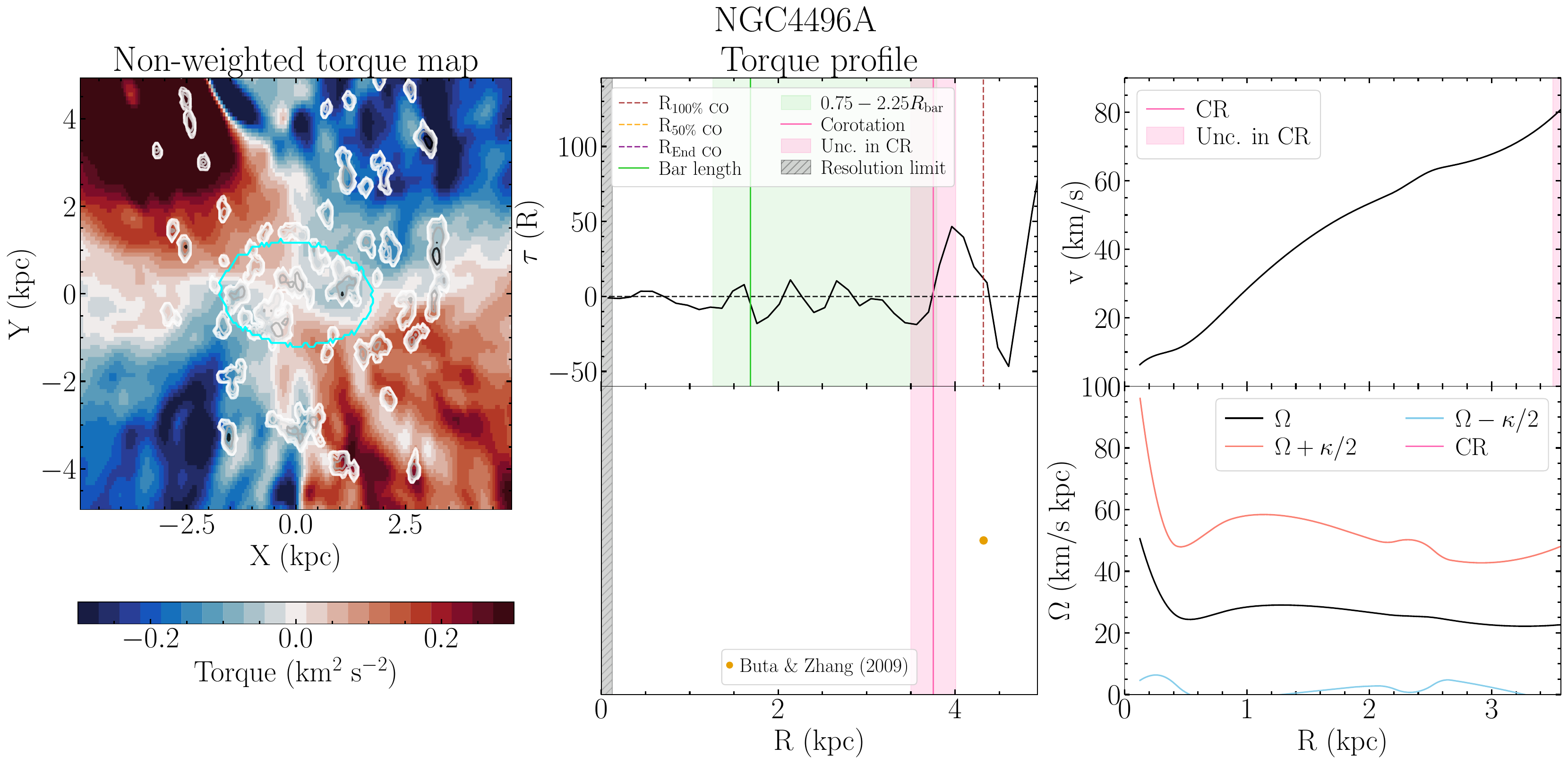}
    \end{center}
    \caption{\textbf{NGC\,4496A} (SB, $QF=3^{\dagger}$). This galaxy has been marked as $QF = 3$ instead of $QF = 2$ because there is few information of the gas response inside the bar, moreover the gas seems clumpy. Legend as in Figure \ref{fig:Appendix-All_galaxies-IC1954}.}
    \label{fig:Appendix-All_galaxies-NGC4496A}
\end{figure*}

\begin{figure*}[t]
\begin{center}
    \includegraphics[trim=0 0 0 0, clip,width=1\textwidth]{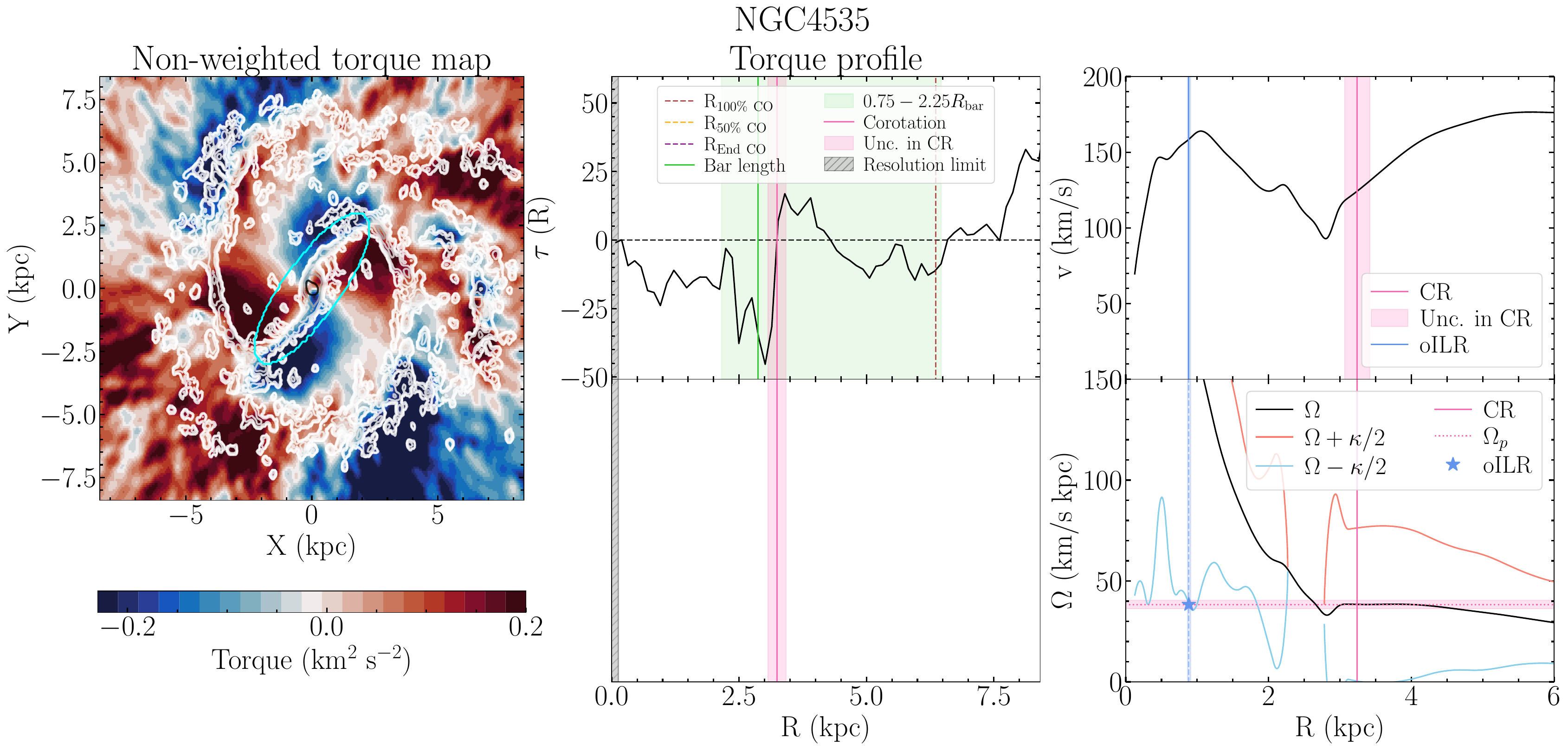}
    \end{center}
    \caption{\textbf{NGC\,4535} (SAB, $QF=1$). Legend as in Figure \ref{fig:Appendix-All_galaxies-IC1954}.}
    \label{fig:Appendix-All_galaxies-NGC4535}
\end{figure*}

\begin{figure*}[t]
    \begin{center}
    \includegraphics[trim=0 0 0 0, clip,width=1\textwidth]{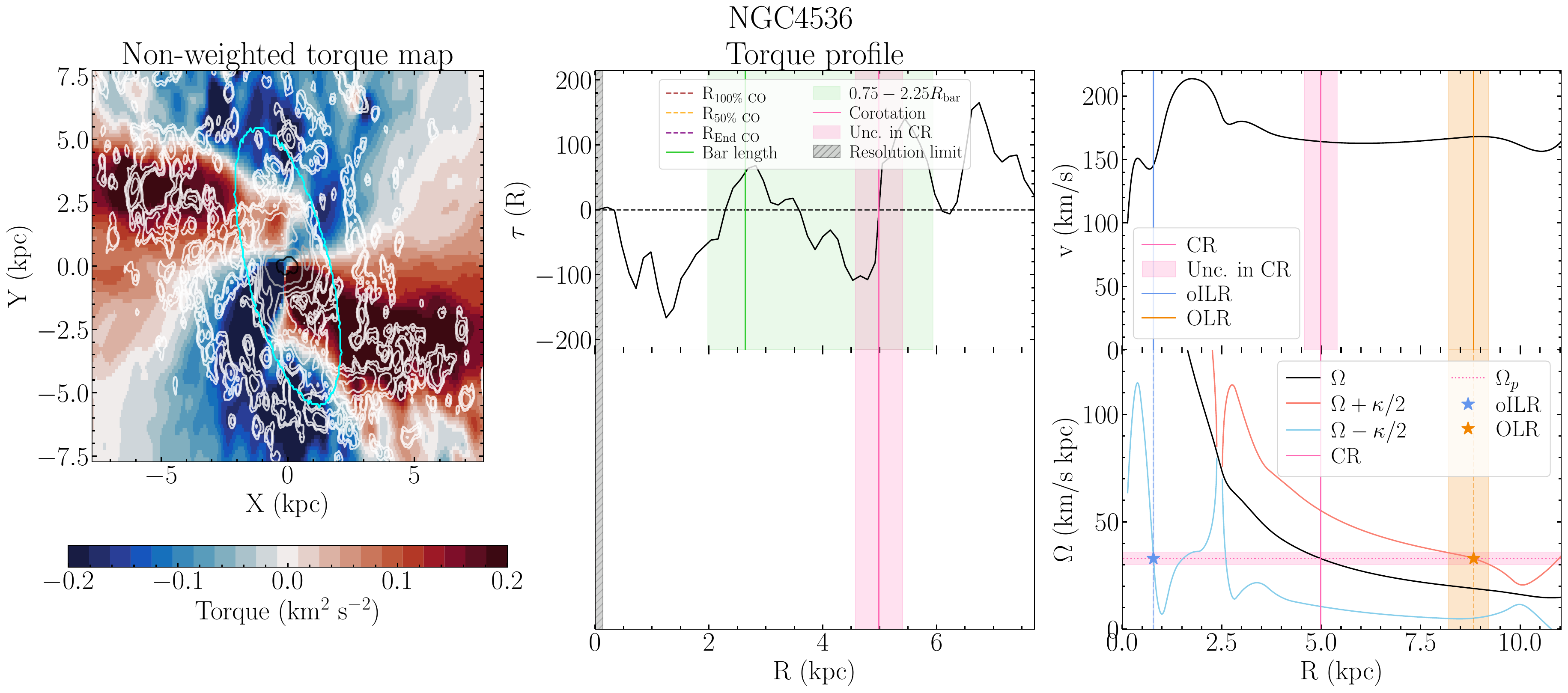} 
    \end{center}
    \caption{\textbf{NGC\,4536} (SAB, $QF=3$). Legend as in Figure \ref{fig:Appendix-All_galaxies-IC1954}.}
    \label{fig:Appendix-All_galaxies-NGC4536}
\end{figure*}

\begin{figure*}[t]
\begin{center}
    \includegraphics[trim=0 0 0 0, clip,width=1\textwidth]{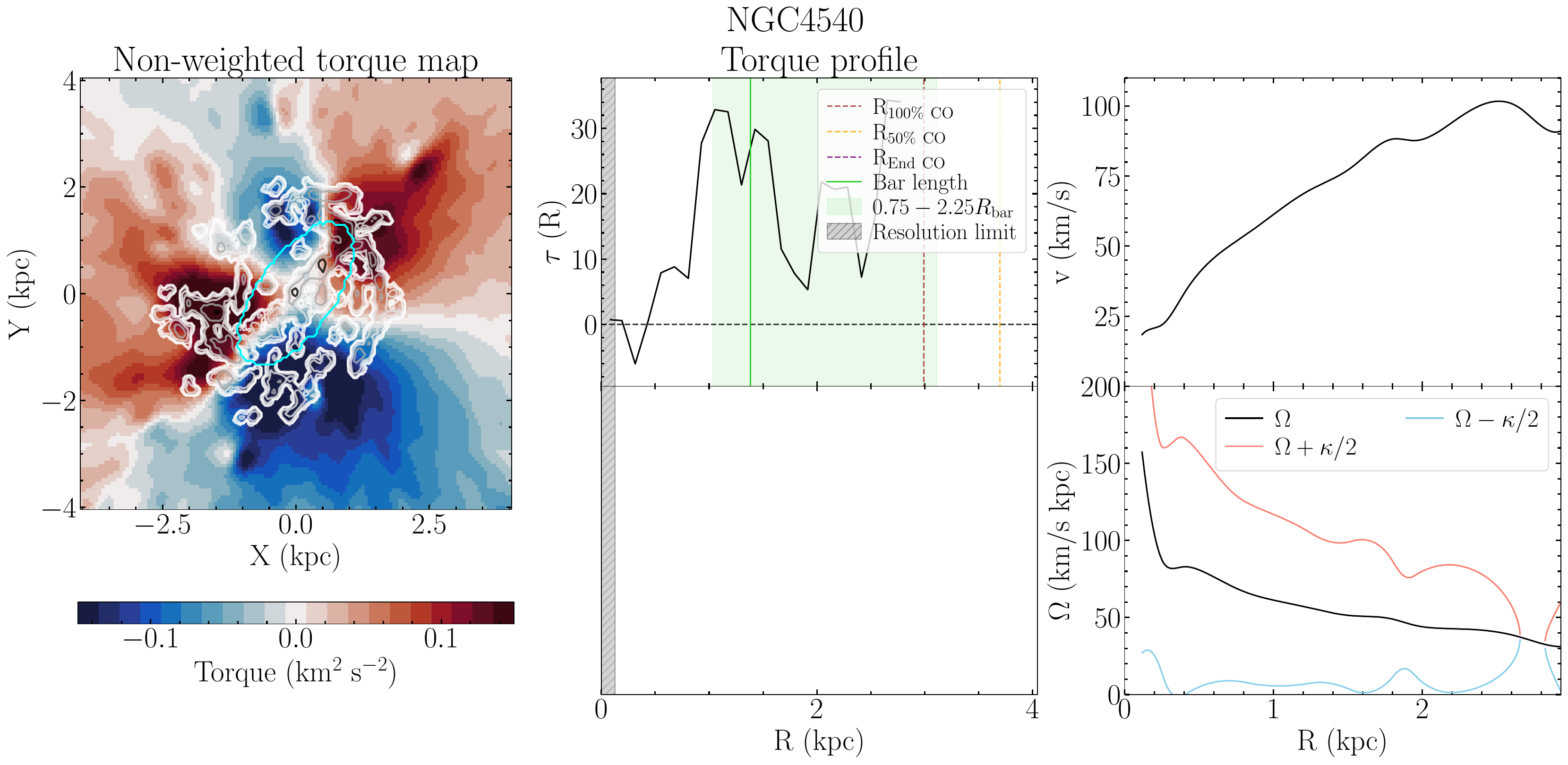}
    \end{center}
    \caption{\textbf{NGC\,4540} (SAB, $QF=3$). Legend as in Figure \ref{fig:Appendix-All_galaxies-IC1954}.}
    \label{fig:Appendix-All_galaxies-NGC4540}
\end{figure*}

\begin{figure*}[t]
\begin{center}
    \includegraphics[trim=0 0 0 0, clip,width=1\textwidth]{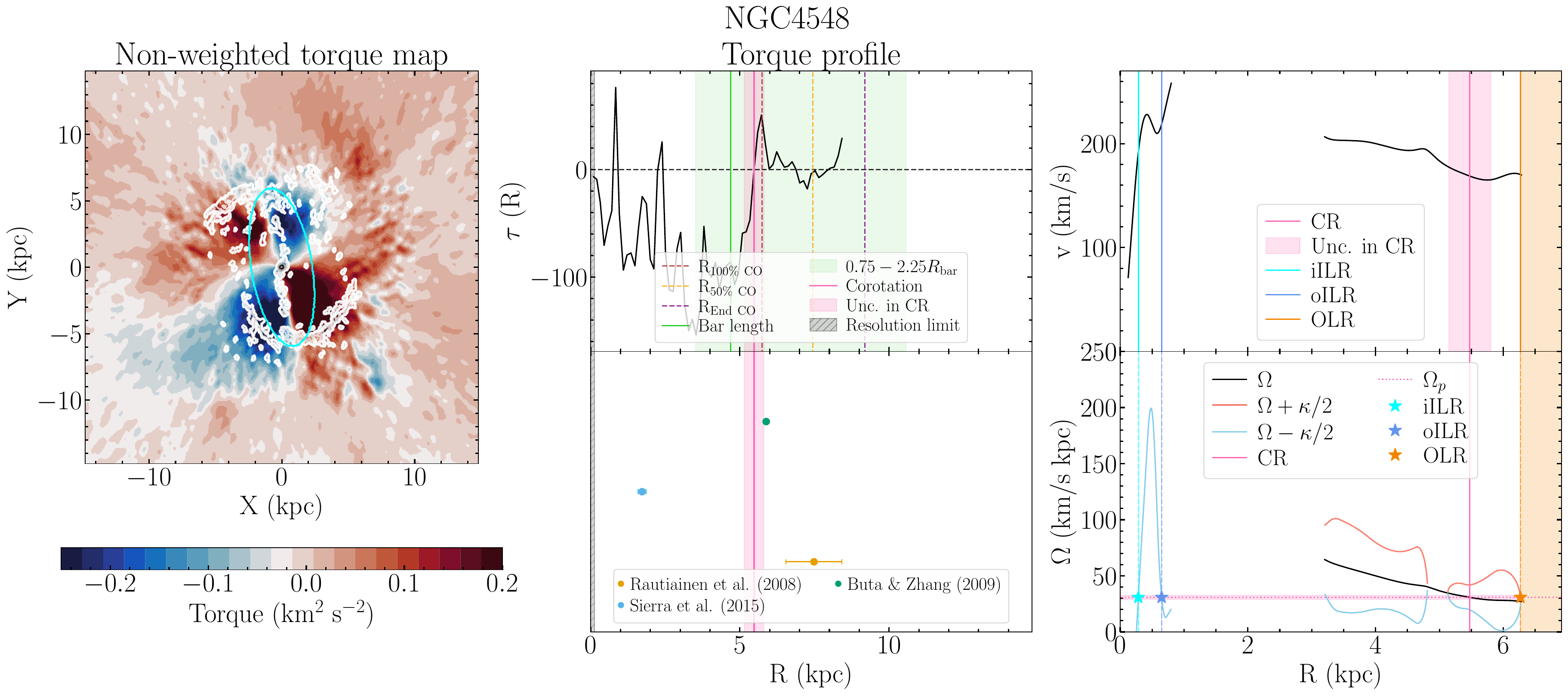}
    \end{center}
    \caption{ \textbf{NGC\,4548} (SB, $QF=1$). Legend as in Figure \ref{fig:Appendix-All_galaxies-IC1954}.}
    \label{fig:Appendix-All_galaxies-NGC4548}
\end{figure*}

\begin{figure*}[t]
\begin{center}
    \includegraphics[trim=0 0 0 0, clip,width=1\textwidth]{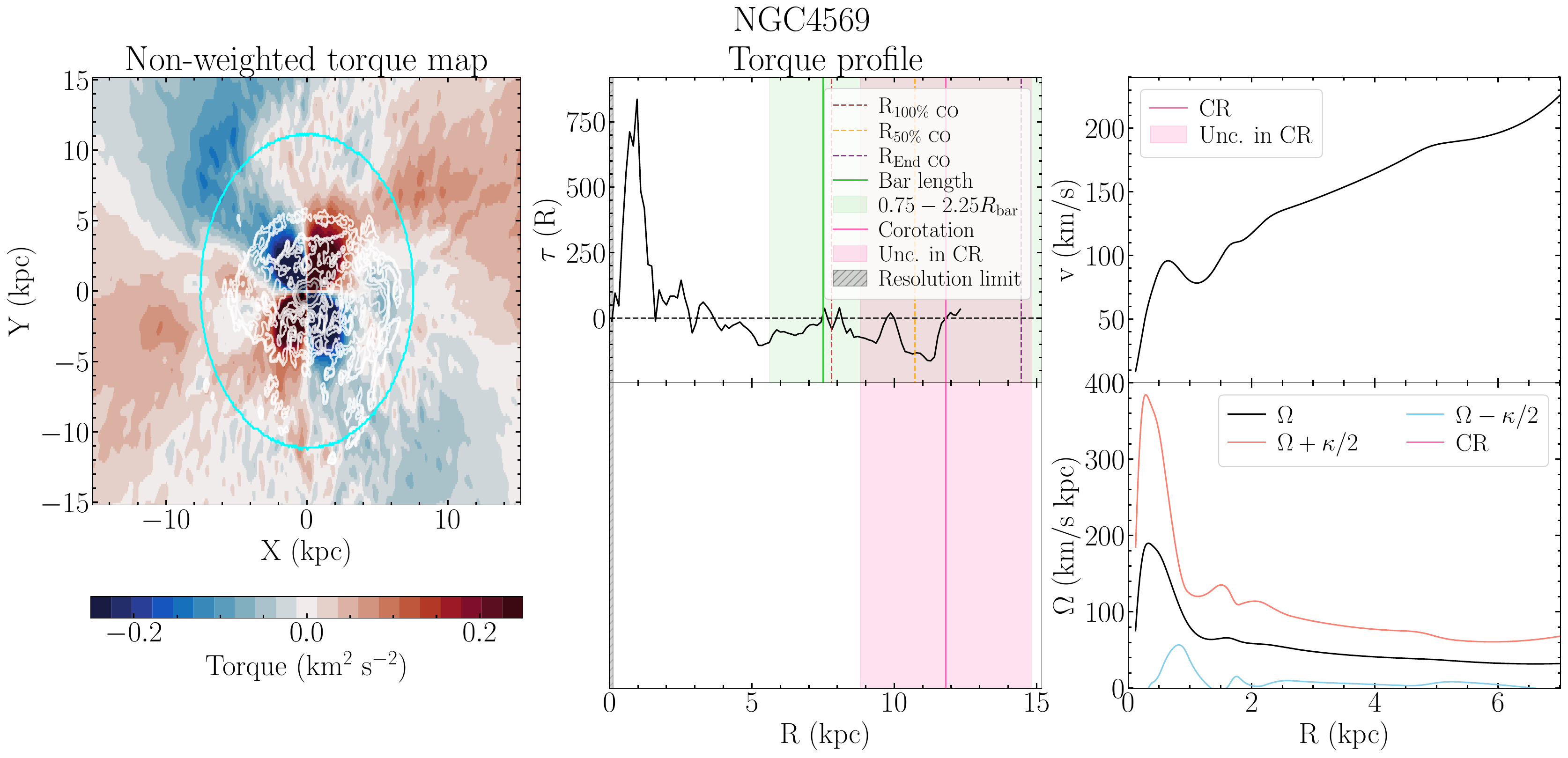}
    \end{center}
    \caption{\textbf{NGC\,4569} (SAB, $QF=3^{\dagger}$). This galaxy has been marked as $QF = 3$ instead of $QF = 2$ because there is few information of the gas response inside the bar. Legend as in Figure \ref{fig:Appendix-All_galaxies-IC1954}.}
    \label{fig:Appendix-All_galaxies-NGC4569}
\end{figure*}

\begin{figure*}[t]
\begin{center}
    \includegraphics[trim=0 0 0 0, clip,width=1\textwidth]{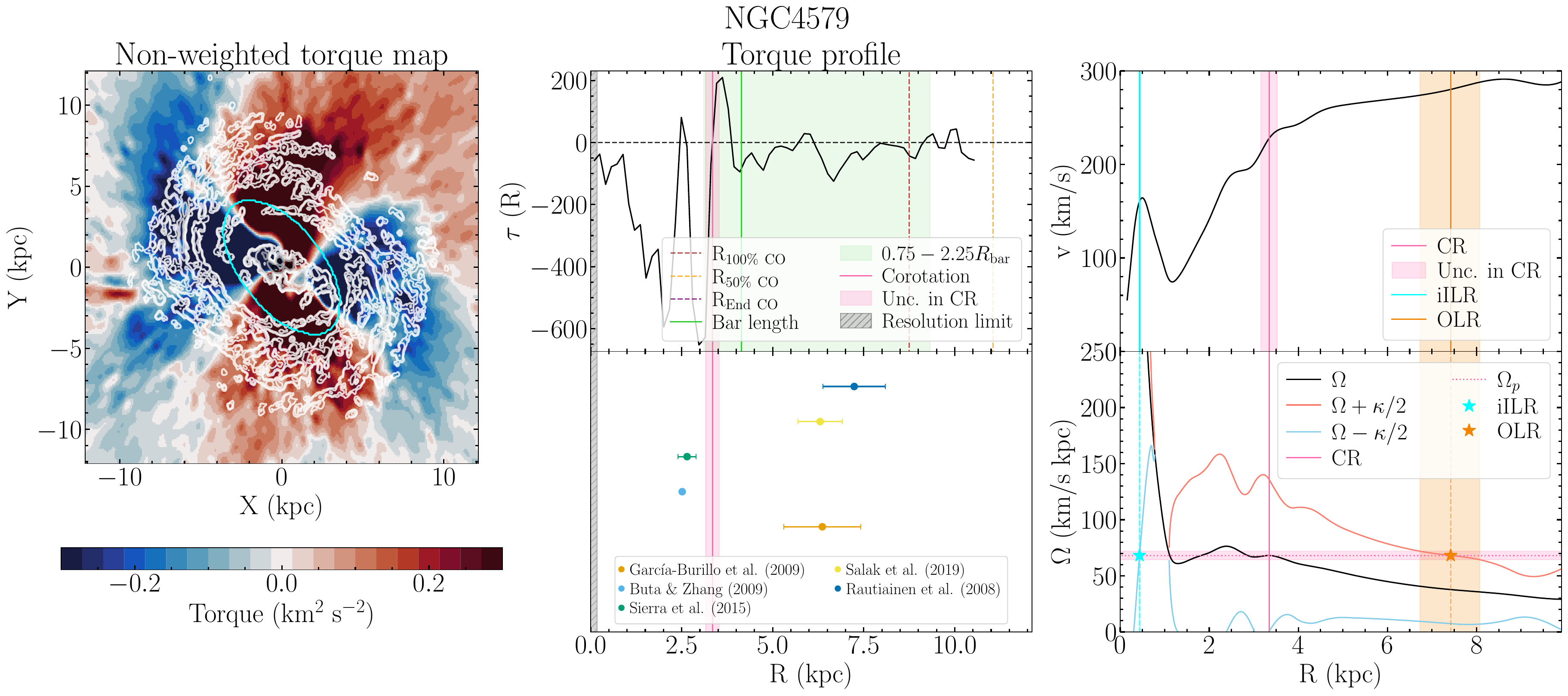}
    \end{center}
    \caption{\textbf{NGC\,4579} (SB, $QF=1$). Legend as in Figure \ref{fig:Appendix-All_galaxies-IC1954}.}
    \label{fig:Appendix-All_galaxies-NGC4579}
\end{figure*}

\begin{figure*}[t]
\begin{center}
    \includegraphics[trim=0 0 0 0, clip,width=1\textwidth]{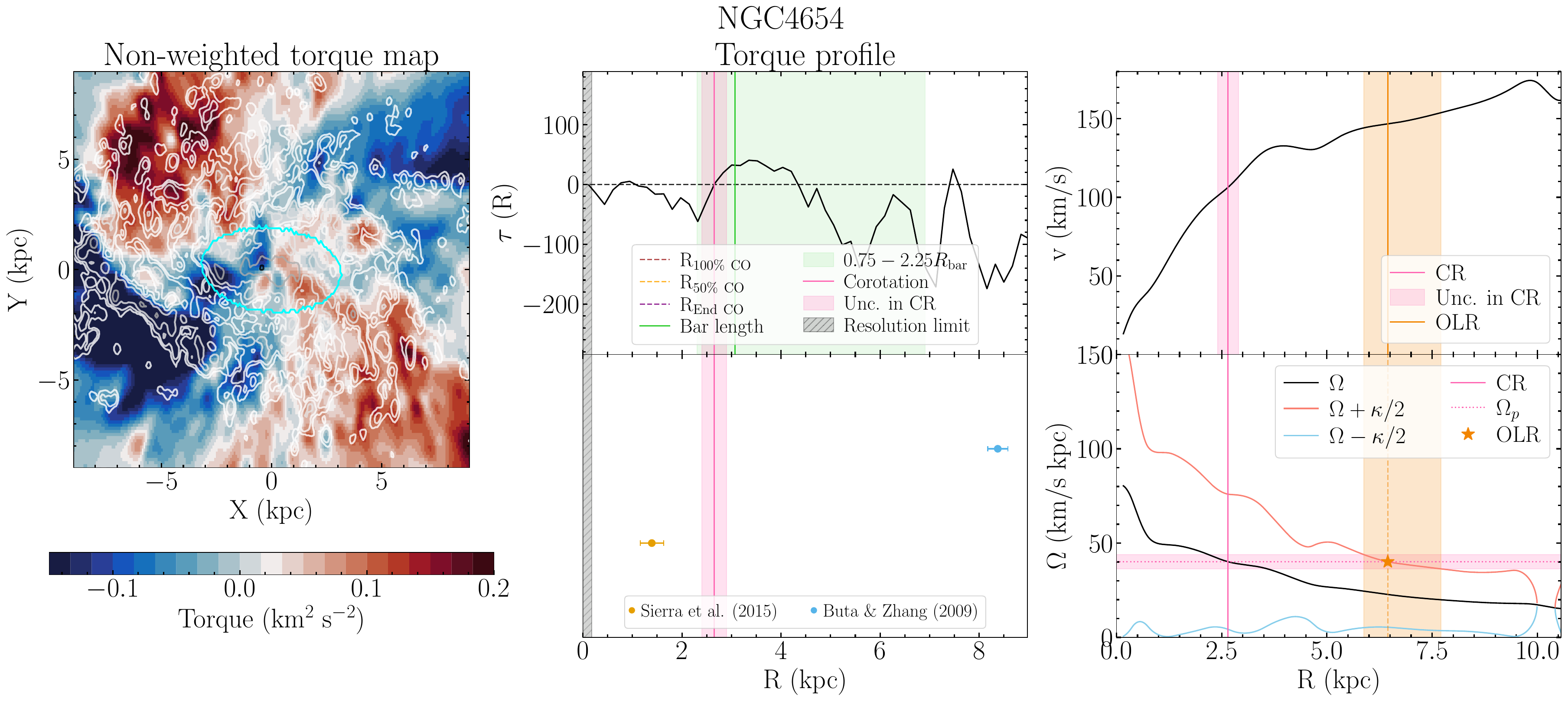}
    \end{center}
    \caption{\textbf{NGC\,4654} (SB, $QF=3$). Legend as in Figure \ref{fig:Appendix-All_galaxies-IC1954}.}
    \label{fig:Appendix-All_galaxies-NGC4654}
\end{figure*}

\begin{figure*}[t]
\begin{center}
    \includegraphics[trim=0 0 0 0, clip,width=1\textwidth]{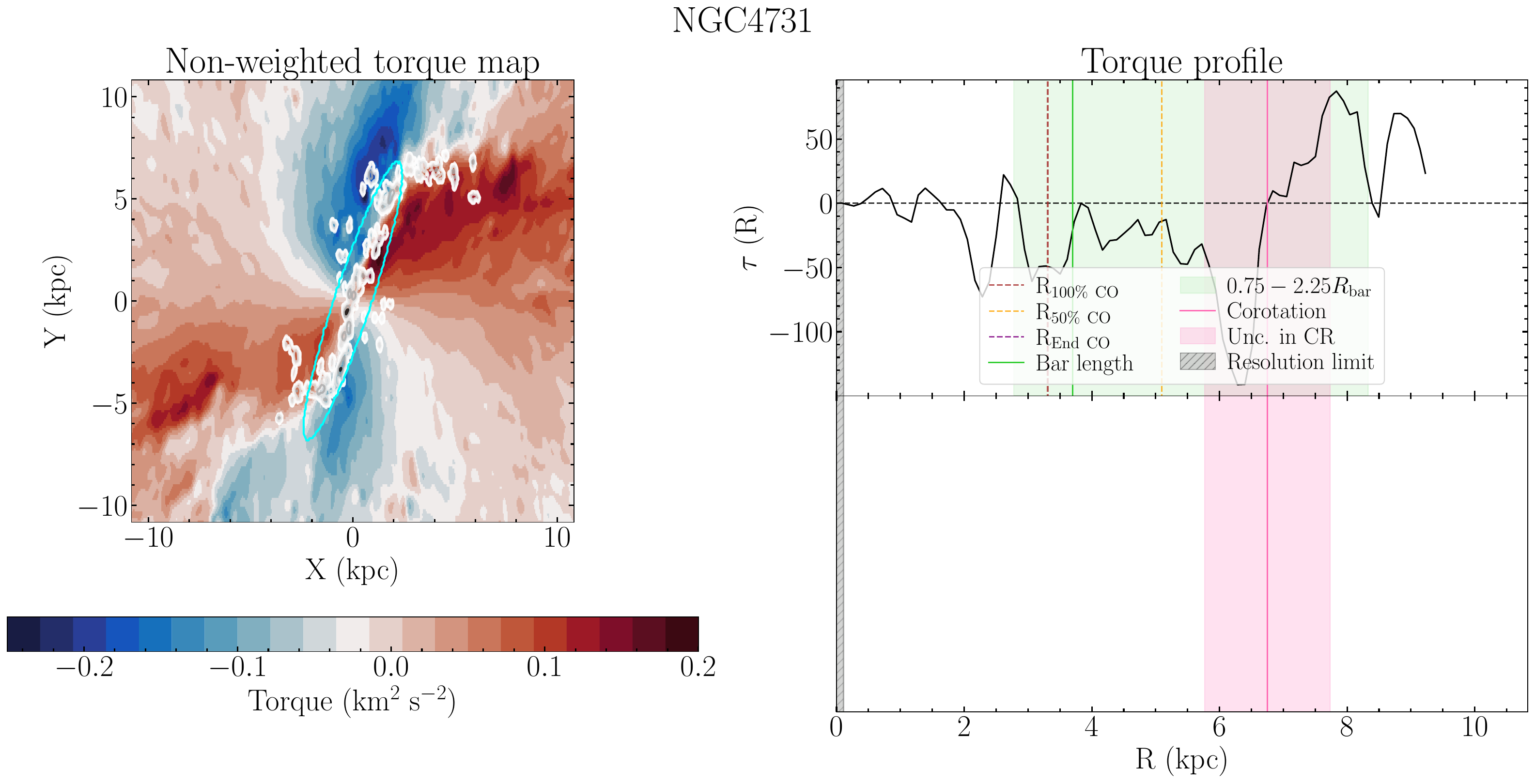}
    \end{center}
    \caption{\textbf{NGC\,4731} (SB, $QF=3^{\dagger}$). This galaxy has been marked as $QF=3$ instead of $QF=2$ because its inclination is large ($i = 64.0\pm6.0 \ \rm deg$). Legend as in Figure \ref{fig:Appendix-All_galaxies-IC1954}. } %And its bar thin
    \label{fig:Appendix-All_galaxies-NGC4731}
\end{figure*}

\begin{figure*}[t]
\begin{center}
    \includegraphics[trim=0 0 0 0, clip,width=1\textwidth]{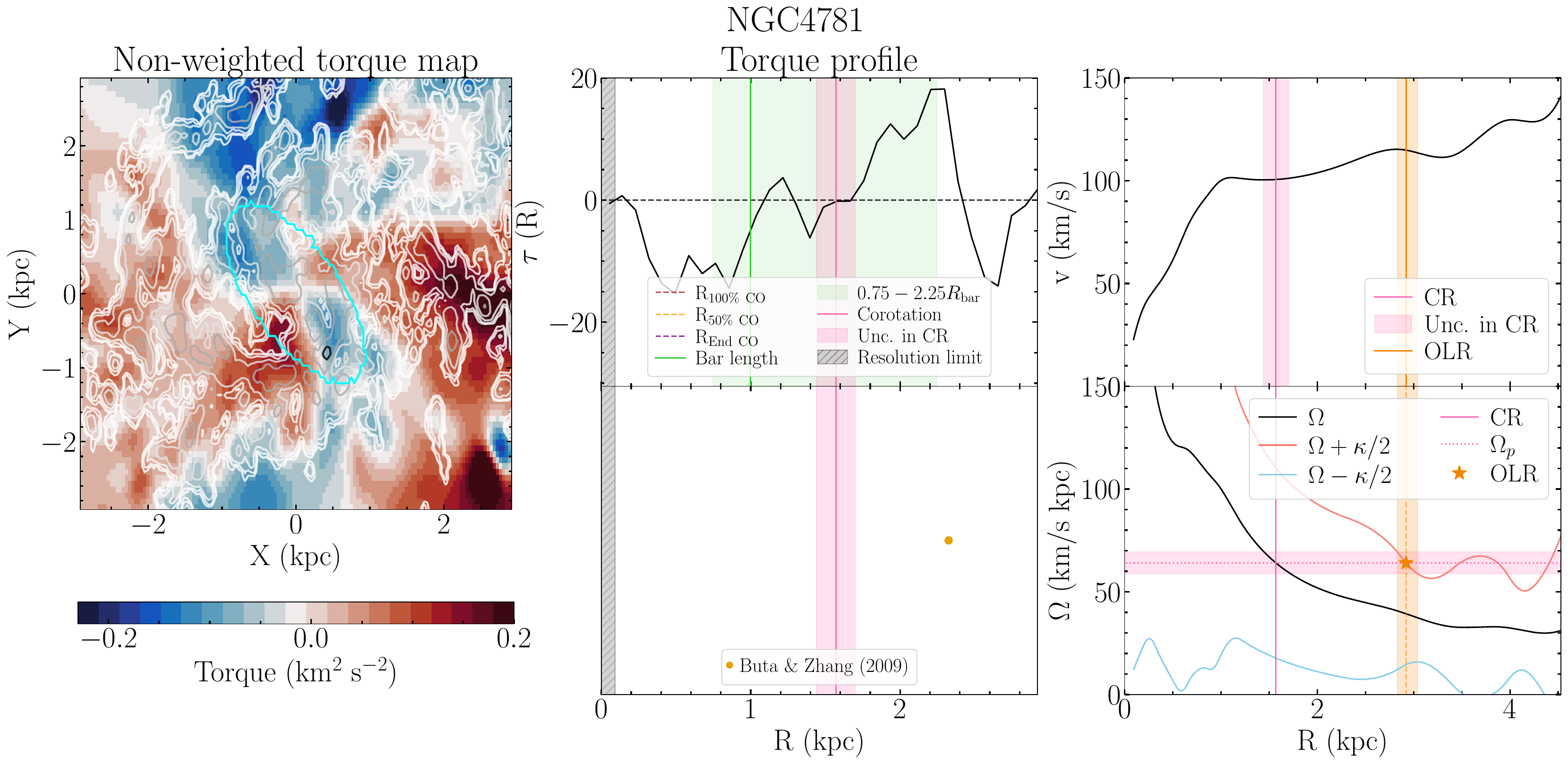}
    \end{center}
    \caption{\textbf{NGC\,4781} (SB, $QF=2$). Legend as in Figure \ref{fig:Appendix-All_galaxies-IC1954}. }
    \label{fig:Appendix-All_galaxies-NGC4781}
\end{figure*}

\begin{figure*}[t]
\begin{center}
    \includegraphics[trim=0 0 0 0, clip,width=1\textwidth]{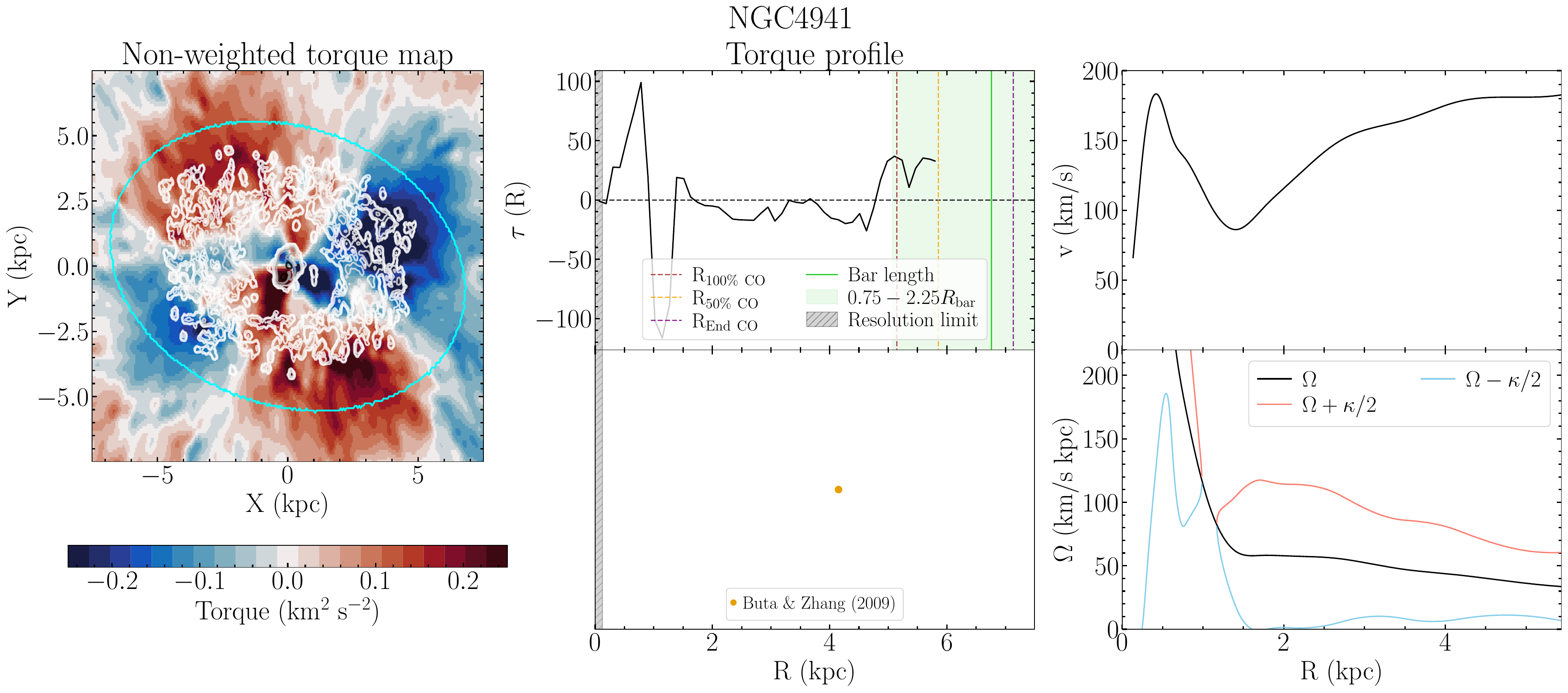}
    \end{center}
    \caption{\textbf{NGC\,4941} (SAB, $QF=3$). Legend as in Figure \ref{fig:Appendix-All_galaxies-IC1954}.}
    \label{fig:Appendix-All_galaxies-NGC4941}
\end{figure*}

\begin{figure*}[t]
\begin{center}
    \includegraphics[trim=0 0 0 0, clip,width=1\textwidth]{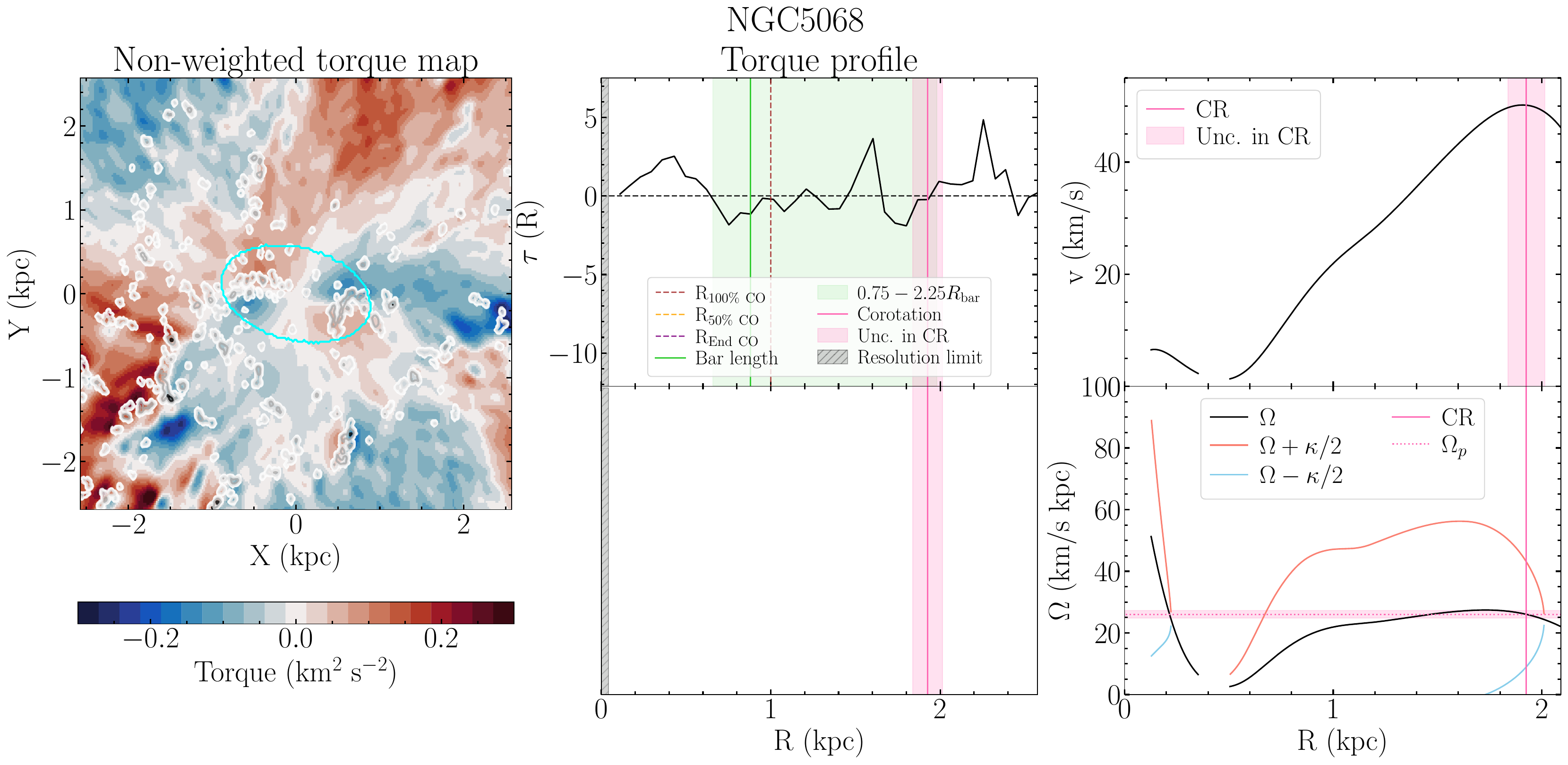}
    \end{center}
    \caption{\textbf{NGC\,5068} (SB, $QF = 3^{\dagger}$). This galaxy has been marked as $QF=3$ instead of $QF=2$ because there is insufficient CO coverage inside the bar region. Legend as in Figure \ref{fig:Appendix-All_galaxies-IC1954}.}
    \label{fig:Appendix-All_galaxies-NGC5068}
\end{figure*}

\begin{figure*}[t]
\begin{center}
    \includegraphics[trim=0 0 0 0, clip,width=1\textwidth]{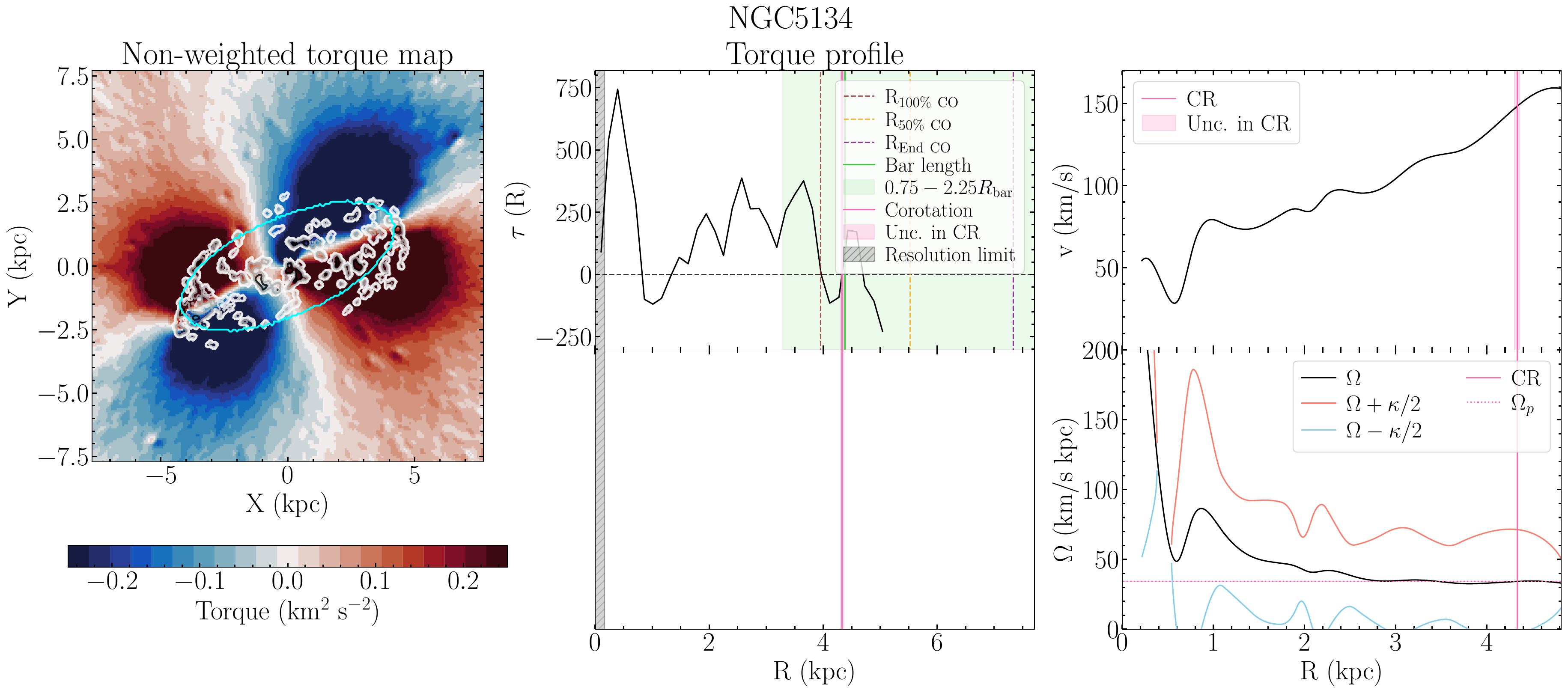}
    \end{center}
    \caption{\textbf{NGC\,5134} (SAB, $QF = 3^{\dagger}$). This galaxy has been marked as $QF = 3$ instead of $QF=1$ because there is insufficient gas and its response is not the expected one. Legend as in Figure \ref{fig:Appendix-All_galaxies-IC1954}.}
    \label{fig:Appendix-All_galaxies-NGC5134}
\end{figure*}

\begin{figure*}[t]
\begin{center}
    \includegraphics[trim=0 0 0 0, clip,width=1\textwidth]{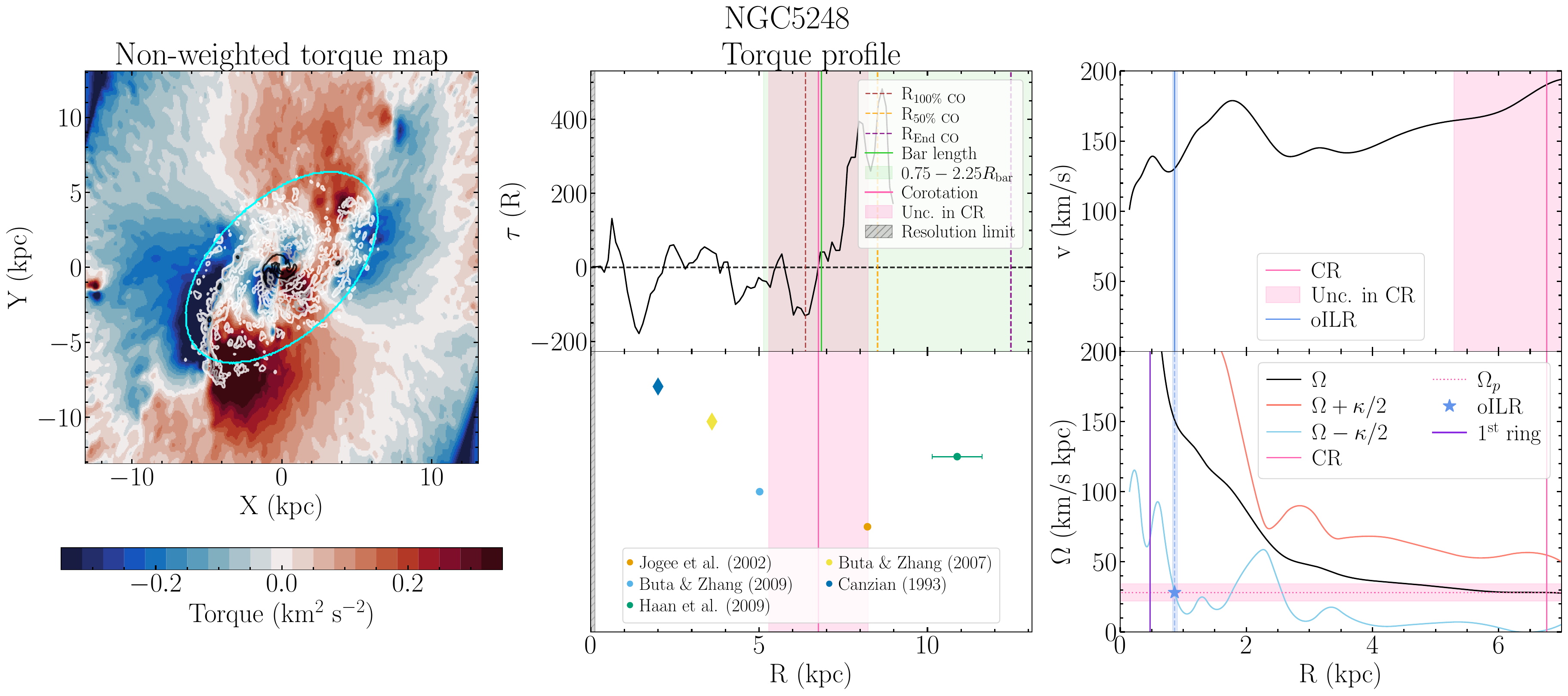}
    \end{center}
    \caption{\textbf{NGC\,5248} (SAB, $QF=2^{\dagger}$) This galaxy has been marked as $QF=2$ instead of $QF=1$ because the automatic CR has been found outside the full CO coverage, so there is no sufficient gas. Legend as in Figure \ref{fig:Appendix-All_galaxies-IC1954}.}
    \label{fig:Appendix-All_galaxies-NGC5248}
\end{figure*}

\begin{figure*}[t]
\begin{center}
    \includegraphics[trim=0 0 0 0, clip,width=1\textwidth]{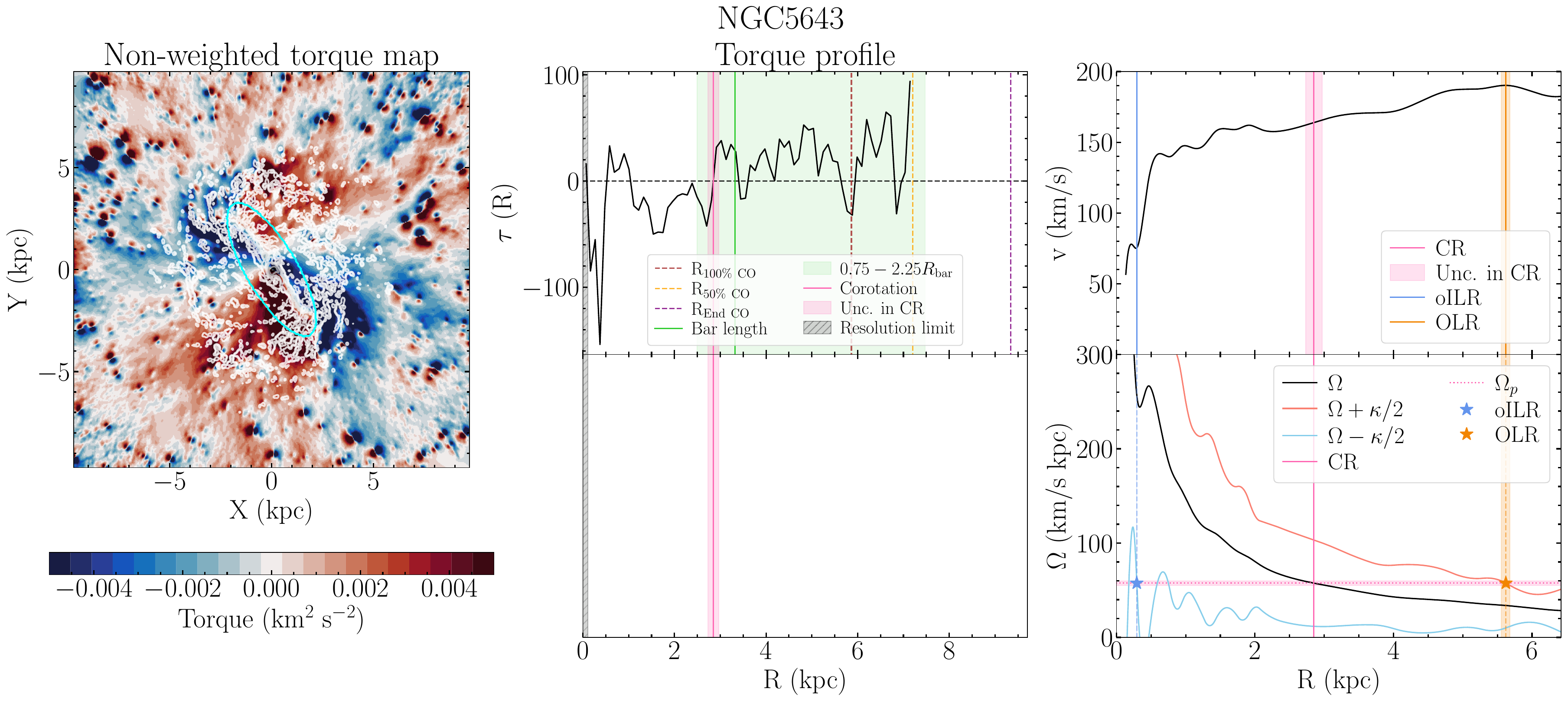}
    \end{center}
    \caption{\textbf{NGC\,5643} (SAB, $QF = 2^{\dagger}$). This galaxy has been marked as $QF = 2$ instead of $QF = 1$ because upon visual inspection, we do not believe the automatically selected CR. Therefore, we force the code to choose a different CR (the one shown here), which is more likely to be the CR of the bar. Legend as in Figure \ref{fig:Appendix-All_galaxies-IC1954}.}
    \label{fig:Appendix-All_galaxies-NGC5643}
\end{figure*}

\begin{figure*}[t]
\begin{center}
    \includegraphics[trim=0 0 0 0, clip,width=1\textwidth]{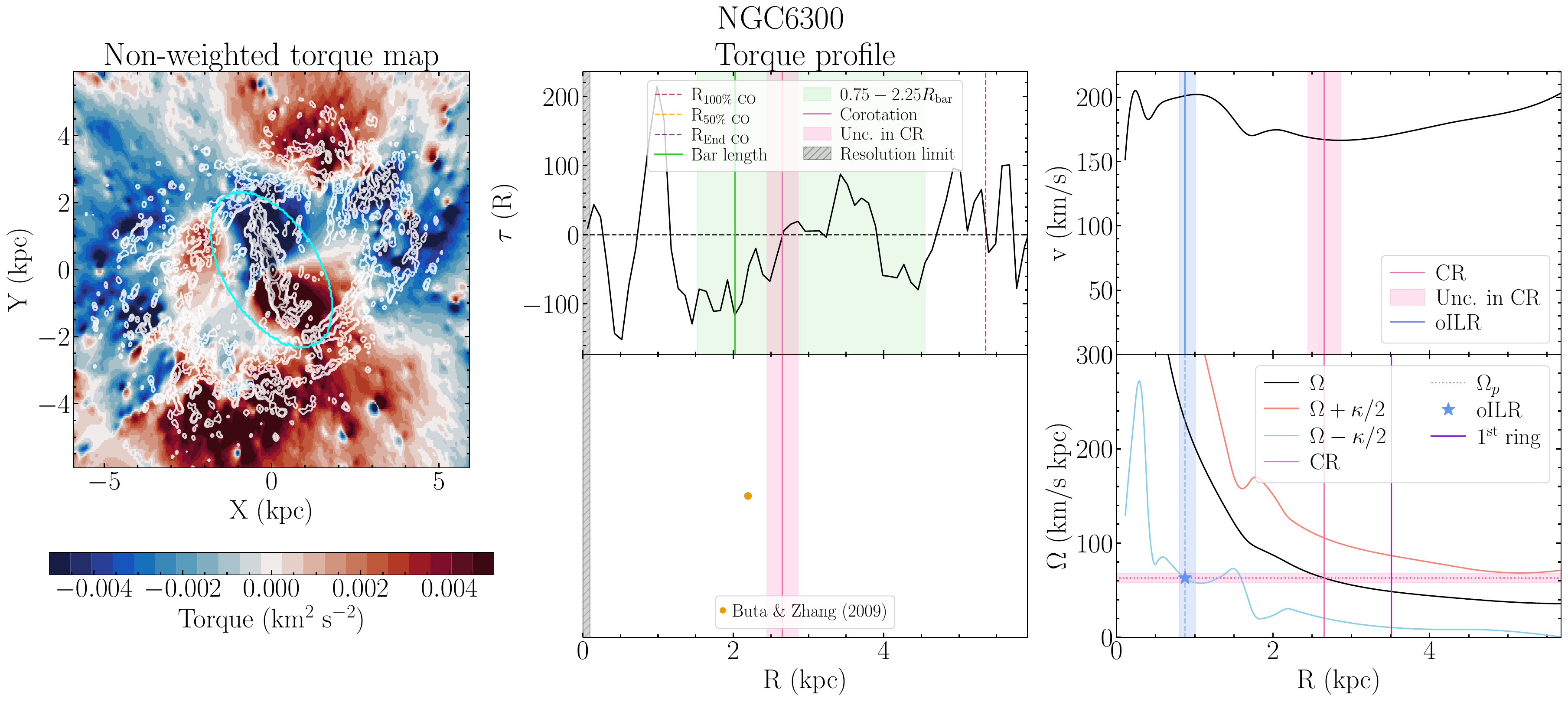}
    \end{center}
    \caption{\textbf{NGC\,6300} (SB, $QF = 3^{\dagger}$). This galaxy has been marked as $QF=3$ instead of $QF=1$ because the gas shows a behaviour different than the one expected. Legend as in Figure \ref{fig:Appendix-All_galaxies-IC1954}.}
    \label{fig:Appendix-All_galaxies-NGC6300}
\end{figure*}

\begin{figure*}[t]
\begin{center}
    \includegraphics[trim=0 0 0 0, clip,width=1\textwidth]{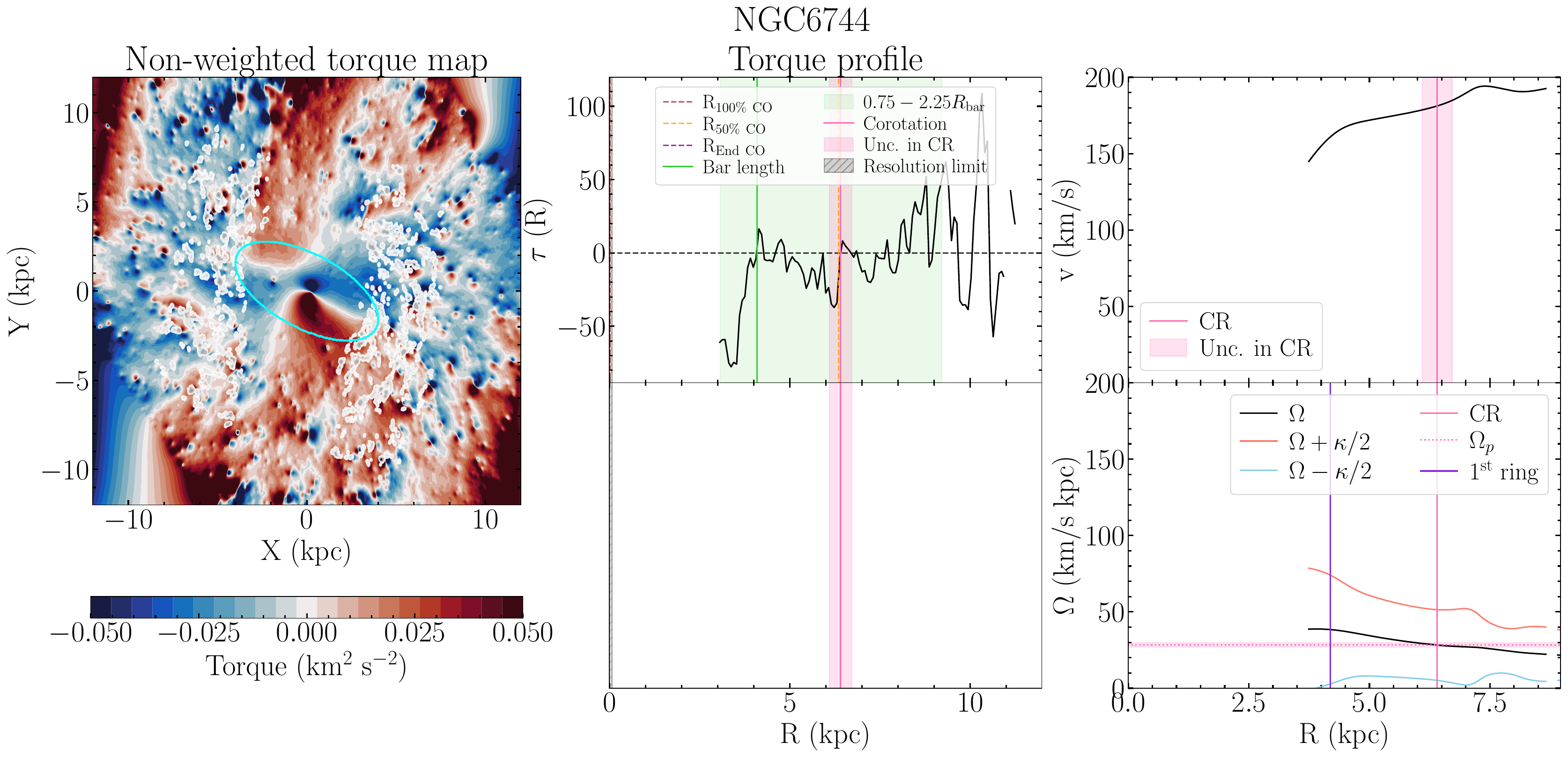}
    \end{center}
    \caption{\textbf{NGC\,6744} (SAB, $QF=3^{\dagger}$). This galaxy has been marked as $QF=3$ instead of $QF=1$ because there is insufficient CO coverage in the bar region. Legend as in Figure \ref{fig:Appendix-All_galaxies-IC1954}.}
    \label{fig:Appendix-All_galaxies-NGC6744}
\end{figure*}

\begin{figure*}[t]
\begin{center}
    \includegraphics[trim=0 0 0 0, clip,width=1\textwidth]{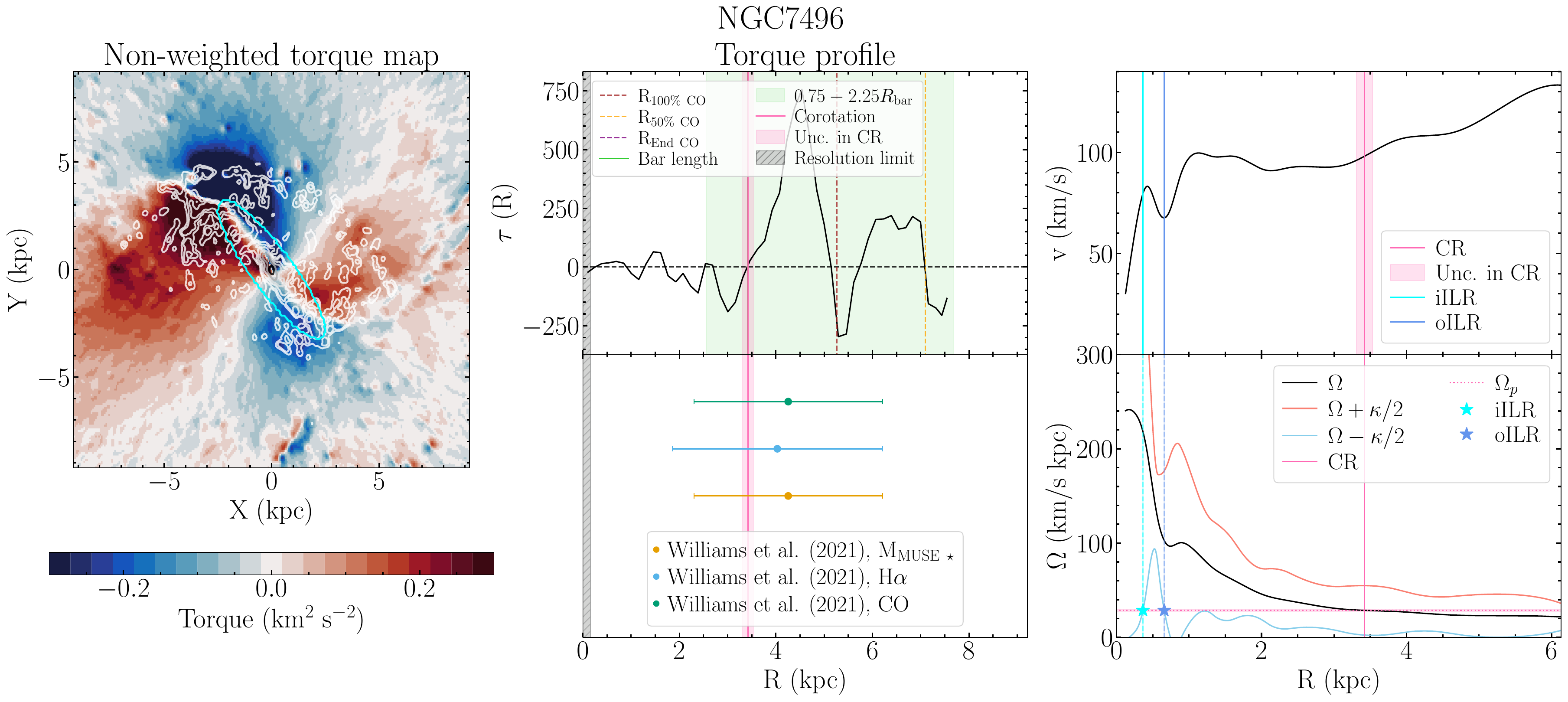}
    \end{center}
    \caption{\textbf{NGC\,7496} (SB, $QF=1$). Legend as in Figure \ref{fig:Appendix-All_galaxies-IC1954}.}
    \label{fig:Appendix-All_galaxies-NGC7496}
\end{figure*}

The complete sample of galaxies can be found below, in Figs.~\ref{fig:Appendix-All_galaxies-IC1954} to \ref{fig:Appendix-All_galaxies-NGC7496}, together with their non-weighted torque maps, torque profiles, literature comparison, rotation curves and dynamical resonances determinations. 

Galaxies marked with $\dagger$ have been downgraded, the $QF$ showing is the one established ``by hand'', while galaxies marked with $\ddagger$ have been upgraded, being the shown $QF$, the one established ``by hand''. The reason of this upgrading or downgrading can be found in the description of the galaxy. 

\end{appendix}

\end{document}